%% file: Main.tex
\documentclass[a4paper,12pt,twoside]{book}

\usepackage{natbib}
\usepackage[english]{babel}
\usepackage{graphicx}
\usepackage[small]{caption2}
\usepackage{epsfig}
\usepackage{fancyhdr} 
\usepackage{mathtools}
\usepackage{amssymb}
\usepackage{adjustbox}
\usepackage{rotating}
\usepackage{color}
\usepackage{hyperref}

\usepackage{longtable}
\usepackage[utf8]{inputenc}

\title{Exploiting X-ray spectroscopy to understand SNRs}
\author{Emanuele Greco}
\date{\today}

\begin{document} 
\thispagestyle{empty}
\begin{figure*}[!h]
\centering
\includegraphics[width=1.2\columnwidth]{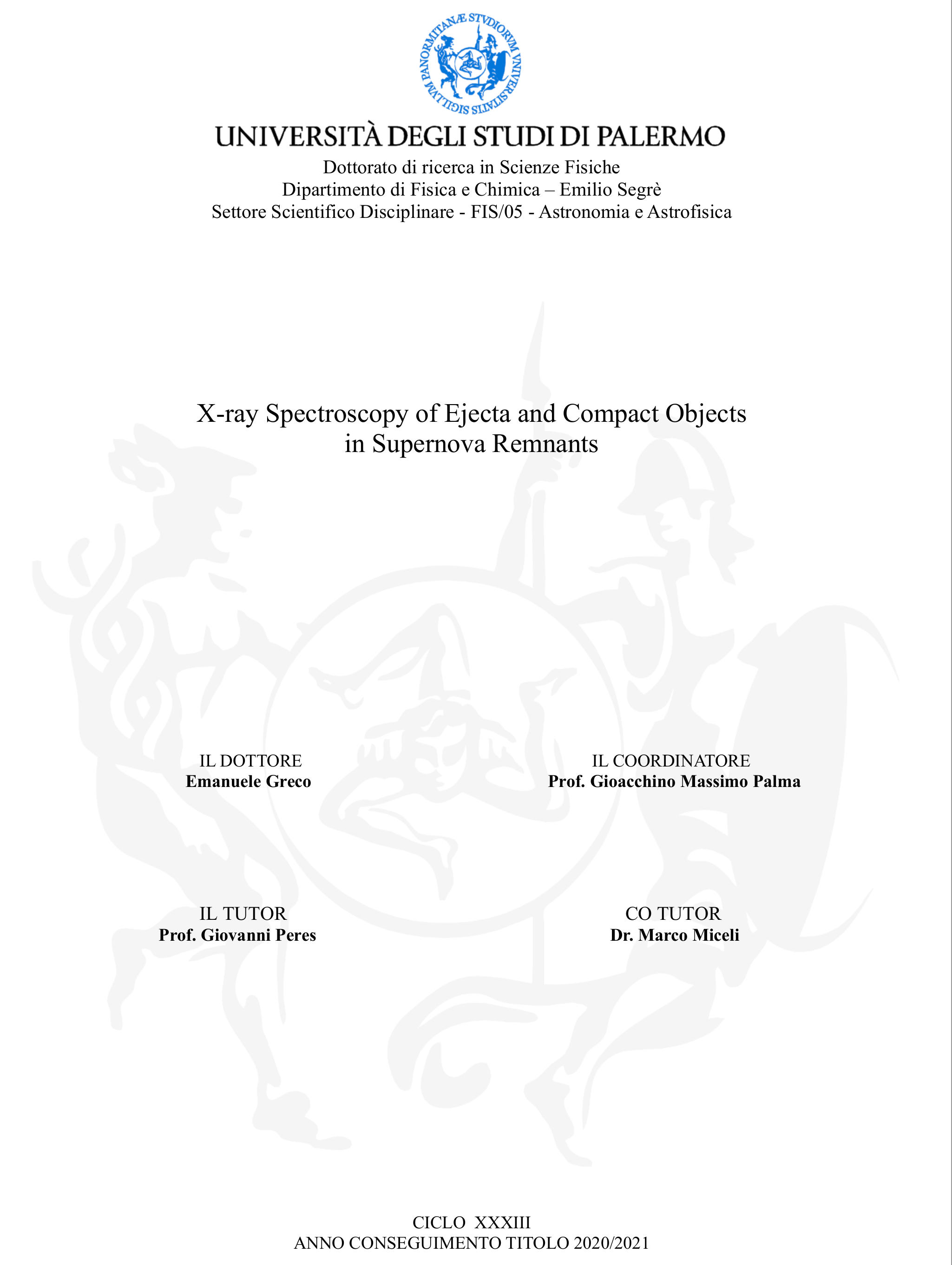}
\end{figure*}

\newpage
\clearpage
\thispagestyle{empty}

\newpage
\clearpage
\thispagestyle{empty}

\newpage
\clearpage
\thispagestyle{empty}

\tableofcontents

\newpage
\clearpage
\thispagestyle{empty} 

\include{Abstract}

\newpage
\clearpage
\thispagestyle{empty}

\include{Introduction}
\include{RRC}

\include{IC443}
\include{1987A}

\include{Conclusions}

\appendix
\include{Appendix1}

\include{Acronym}
\backmatter

\section*{Acknowledgements}
I would like to express my gratitude to my supervisor Prof. G. Peres for the significant help, wise indications and constant support through all my PhD studies and for improving this thesis with his suggestions and indications. I also want to especially thank Dr. S. Orlando for his insightful comments, discussion and contribution to the studies I have performed during my PhD. I sincerely acknowledge  Prof. J. Vink and members of his research group for the fruitful discussions, precious advices and kind hospitality they reserved to me. I extend my thanks to Dr. F. Bocchino, for his comments and ideas shared with me in these years. My sincere thanks also go to Dr. Barbara Olmi, Prof. S. Nagataki, Dr. M. Ono, and A. Dohi for their contribution to some of the work presented here. 
Finally, I want to express my gratitude to all the people at the INAF - Osservatorio Astronomico di Palermo for their assistance and for make me feel like I was home.

Oh, i was almost forgetting. \emph{Dulcis in fundo}, I want to particularly thank Dr. M. Miceli. In these tough and intense years he has been my landmark, when the things were going smoothly (rarely) and when everything seemed to mess up (almost always). When I was starting the PhD, I already knew Marco as a brilliant scientist and an extraordinary man. After these years, I can say without hesitation that he has been not only the best mentor I could ask for, but also one of my best friends. Grazie, Marco.

\bibliographystyle{aa}
\bibliography{references}

\end{document}

%% file: Abstract.tex
\chapter*{Abstract}
This thesis is devoted to the analysis of X-ray observations of supernova remnants (SNRs), exploiting the diagnostic potential provided by X-ray spectroscopy and complementing it with the synthesis of X-ray observables from multi-dimensional hydrodynamic (HD) and magneto-hydrodynamic (MHD) models. I tackle three open issues: the search for a spectral signature to correctly recover abundance and mass values of the stellar fragments ejected in the SN explosion through their X-ray spectra; the study of anisotropies and the origin of overionized plasma in the mixed-morphology SNR IC 443; the quest for the elusive compact object in SN 1987A.  

Investigating SNRs in the X-ray band is crucial to have a deep insight on the interaction between the shock wave generated in the supernova (SN) explosion and the interstellar medium (ISM). 
The shock waves (both forward and reverse shocks) that develop during the evolution of the system compress, accelerate and heat up to X-ray emitting temperatures the material encountered, namely the ISM and the stellar fragments expelled at supersonic speed, the ejecta. During the interaction with a SNR, the ISM is enriched with the elements synthesized during the life of the progenitor star and its explosive death. X-ray spectral analysis of shocked ejecta can be a powerful tool to recover the yields produced by SN explosions. However, current low resolution X-ray spectral data of SNRs are intrinsically affected by a degeneracy between the continuum and the line emission, causing a high uncertainty in the estimates of abundance and mass values.

I address this issue by performing a campaign of spectral simulations in the high-abundance regime, with the aim of identifying the spectral feature able to remove this degeneracy. I find that for chemical abundances $> 100$ the plasma enters in a pure-metal ejecta regime and bright radiative-recombination continua show up in the spectrum. I show that current charged-coupled device (CCD) cameras are not able to reveal these features because of their moderate spectral resolution, whereas future X-ray microcalorimeter spectrometers undoubtedly will. I also tested and verified the applicability of this diagnostic to a promising target for the future detection of pure-metal radiative recombination continua (RRC): the southeastern Fe-rich region of Cassiopeia A (Cas A). This SNR is characterized by a huge amount of dense clumps of ejecta, where we expect to find emission from pure-metal ejecta. Therefore, I synthesize X-ray spectra of Cas A from a state-of-the-art HD model and I show that future X-ray microcalorimeters spectrometers, like {\it XRISM}/Resolve, will be able to pinpoint the presence of pure-metal RRC and to provide correct estimates of both the mass and the chemical composition of the ejecta. 

RRCs can show up in the spectrum also because of a different physical condition of the plasma: overionization and I show how to discriminate the enhanced RRC associated with pure-metal ejecta from those originating from overionization. The physical scenario which causes the plasma to be overionized is still not completely clear. This peculiar ionization state of the plasma is only observed in mixed-morphology SNRs (MM-SNRs), characterized by shell-like radio emission and centrally peaked X-ray emission. MM-SNRs are typically observed to be very asymmetric and their shocked ISM distribution is quite inhomogenous (e.g. IC 443). X-ray spatially-resolved spectroscopy of these sources allows to scrutinize the physical and dynamical processes occurring in this complex class of SNRs.

I investigated the MM-SNR IC 443, a SNR placed in very inhomogeneous environment. It interacts with nearby atomic and molecular clouds and shows overionized (recombining) plasma, but the physical mechanisms responsible of overionization are still under debate. Either rarefaction (adiabatic expansion) or thermal conduction towards cold clouds are claimed as possible causes to account for overionization. The off-center position of the pulsar wind nebula (PWN) CXOUJ061705.3+222127 observed within IC 443 opens the question about the link between the two sources: the PWN may be the compact relic of the SN explosion but it also might be just a rambling object seen in projection on the remnant. 

I analyzed {\it XMM-Newton}/EPIC observations of IC 443 and I identified an elongated, jet-like structure: performing a spatially resolved spectral analysis, I found that it is made of Mg-rich overionized plasma. The jet is distorted by the molecular cloud in the north-west and its projection towards the remnant matches the position where the SN explosion occurred, estimated by taking into account the observed proper motion of the PWN and considering an age of the explosion $\sim 8000$ yr. This agreement points toward a relationship between the compact source and IC 443, indicating that the PWN CXOUJ061705.3+222127 is actually belonging to the remnant. I also propose a dynamical scenario for the plasma, based on the adiabatic expansion, which yields temperatures and degree of overionization compatible to values measured through X-ray spectral analysis.

As for SN 1987A, PWNe are typically observed within SNRs generated by a core-collapse SN. An exception is represented by SN 1987A, a hydrogen-rich core-collapse supernova discovered on 1987 February 23. Its dynamical evolution is deeply influenced by the highly inhomogeneous circumstellar medium  (CSM) made of a dense ring-like structure within a diffuse HII region. Despite the unique consideration granted to this SNR with deep and continuous observations, and despite the famous neutrinos detection indicating the formation of a neutron star (NS), the elusive compact object of SN 1987A is still undetected. The fruitless search for this object has led to different hypotheses on its nature. The most common one states that  dense and cold ejecta are absorbing its radiation, through photo-electric effect. The only hint of the existence of the compact object within SN 1987A is provided by a detection in radio {\it ALMA} data of a feature somehow compatible with the radiation of a proto-PWN, even though other scenarios are compatible with the observed emission. 

I tackle this 33-years old issue by analyzing archive observations performed with {\it Chandra} and {\it NuSTAR} between 2012 and 2014. I clearly detect non-thermal emission in the 10-20 keV energy band, due to synchrotron radiation. Two are the possible physical mechanisms powering this emission: diffusive shock acceleration (DSA) or emission from a heavily absorbed PWN. The two scenarios are very similar under a merely spectroscopic point of view, since, in both cases, the synchrotron radiation can be phenomenologically described with a power-law, apart from the presence of an additional absorbing component in the PWN case, the latter taking into account the absorption of the cold ejecta. To estimate the absorption power, I relate a state-of-the-art MHD simulation of SN 1987A to the actual data and I reconstruct the absorption pattern encountered by the radiation emitted by the possible PWN. I have found that the non-thermal excess, observed in the {\it NuSTAR} spectra between 10 and 20 keV, is most likely due to the absorbed emission arising from the PWN of SN 1987A, bringing crucial evidence of the compact object in SN 1987A.

%% file: Introduction.tex
\chapter{Introduction}
\label{ch:intro}

Supernova remnants (SNRs) are the vestige of the explosion of stars as supernovae (SN). They are usually composed of a diffuse part and a compact object, relic of the progenitor star. SNRs provide huge amounts of mass and energy to the interstellar medium (ISM) and their emission strongly depends on the interaction between the shock wave generated in the SN explosion and the ISM inhomogeneities. The material surrounding the place of the explosion and wiped out by the shock is heated, ionized and compressed but also mixed with the fragments of the progenitor star expelled at supersonic speed during the explosion, to form the so called ejecta. During this interaction, the ISM is enriched with the heavy elements produced during the life of the progenitor star and during the explosive nucleosynthesis, occurring during the very last instants of the stars.

Therefore, the study of shocked ISM and ejecta in SNRs is crucial in order to deal with different astrophysical topics, such as star formation, nucleosynthesis, ISM evolution and chemical enrichment and so on. One of the most powerful tool to dealt with all these topics is the X-ray spectroscopy. In fact, because of the heating due to the interaction with the shock, the material reaches temperatures of the order of 10$^6$-10$^7$ K, leading to significant X-ray emission (see Sect. \ref{sect:shocks+snr_evol} and \ref{sect:em_proc}) which can be detected through the modern in-orbit X-ray observatories such as {\it XMM-Newton}, {\it Chandra} and {\it NuSTAR}.

This thesis is aimed at studying physical and chemical properties of SNRs exploiting the diagnostic potential of X-ray spectroscopy through the analysis of X-ray observations and the synthesis of X-ray emission spectra from hydrodynamic (HD) and magneto-hydrodynamic (MHD) simulations.

This chapter is organized as follows: in Sect.\ref{sect:sn_exp} I shortly describe the SN explosion mechanisms; in Sect. \ref{sect:shocks+snr_evol} I discuss the physics of shocks and their effects on the evolution of SNRs; the classification of SNRs is described in Sect. \ref{sect:class}; in Sect. \ref{sect:em_proc} I review the emission processes which occur in SNR's X-ray emitting plasma, focusing on line emission (Sect. \ref{sect:line}), Bremsstrahlung (Sect. \ref{sect:bremss}), radiative recombination continua (Sect. \ref{sect:fb}), and on the effects of non-equilibrium of ionization (Sect. \ref{sect:nei}); in Sect. \ref{sect:contents} the contents and aims of the thesis are introduced; finally, in Sect. \ref{sect:outline} an outline of the thesis is presented.

\section{SN explosion}
\label{sect:sn_exp}

Right after the Big Bang, the universe was made only of H and He (aside from a small amount of Li, Be and B). Elements heavier than H up to Fe are synthesized via hydrostatic nuclear burning during the evolution of stars,  and even heavier elements are produced during the SN explosions. Studying these abrupt explosions is therefore critical to understand the chemical evolution of the whole universe.

SNe are classified on the basis of their optical spectra (\citealt{fil97}). Even if a very wide variety of physical properties are detected, the origin of the SN is related to two processes: white dwarfs in binary systems, undergoing a desctructive nuclear explosion leading to a Type Ia SN, and massive (8 $M_\odot  < M_{ZAMS}\footnote{Zero Age Main Sequence} \lesssim 130 M_\odot $) stars, undergoing a violent gravitational collapse and a subsequent explosion, responsible for all the other types of SNe\footnote{Stars more massive than 130 M$_{\odot}$ and with low metallicity are believed to explode as pair instability supernovae \citep{hir17}}) . 
The physical mechanism responsible for the explosion of massive stars is the collapse of the Fe core which can reach a density comparable to that of the atomic nuclei as a consequence of its downfall, thus releasing a gravitational energy of $\sim 10^{53}$ erg. The 99\% of this energy is released through the emission of neutrinos, while the remaining goes into kinetic energy. At the end of the process, a compact object (either a neutron star, NS, or a black hole, BH) is formed and the outer envelope is ejected in the surrounding medium.

Despite the complexity of the physics involved, which includes general relativity, relativistic fluid dynamics and nuclear physics, recently our physical understanding of the SNe explosion mechanism has improved, both on the Type Ia SNe \citep{it84,nty84,bks12} and on the particularly complex core-collapse SNe \citep{bmd03,wj05,jlm07}.


\section{SNR shocks and evolution}
\label{sect:shocks+snr_evol}

\subsection{Shocks in SNRs}
\label{sect:shocks}
The evolution of a SNR is ruled by the interaction of the shock wave generated by the SN with the ISM and/or with the circumstellar medium (CSM), namely the result of the wind activity and/or mass loss of the progenitor star \citep[e.g]{pls17}. The fundamental description of the shock physics is provided by the Rankine-Hugoniot equations of fluid dynamics described, for instance, by \cite{ll59}. By considering a shock as a planar discontinuity with infinitesimal width between two areas of the space, the hydrodynamic equations of consevation of mass, momentum and energy can be written in integral form in the reference frame of the shock as:
\begin{equation}
\rho_1 v_1=\rho_0 v_0
\label{eq:ranhug_1}
\end{equation}
\begin{equation}
p_1 + \rho_1 v_1^2=p_0 + \rho_0 v_0^2
\label{eq:ranhug_2}
\end{equation}
\begin{equation}
v_1(p_1+\rho_1 e_1 + \rho_1 v_1^2/2)=v_0(p_0+\rho_0 e_0 + \rho_0 v_0^2/2)
\label{eq:ranhug_3}
\end{equation}

where $v$ is the bulk velocity of the gas, $\rho$, $P$ and $e$ its density, pressure and internal energy per unit mass, and indexes 1 and 0 indicate the post- and pre-shock values, respectively.

If there is no jump in velocity between the two areas we refer to the discontinuity as a contact discontinuity; otherwise, we refer to the discontinuity as a shock. The shock strenght is indicated by its Mach Number $\mathcal{M}$ defined as the shock speed in units of the pre-shock sound speed. If $\mathcal{M} >> 1$, in the case of an adiabatic and strong shock, we have (\citealt{ll59}):
\begin{equation}
\frac{\rho_1}{\rho_0}=\frac{\gamma + 1}{\gamma - 1}
\label{eq:ratioro}
\end{equation}
\begin{equation}
\frac{P_1}{P_0} = \frac{2 \gamma \mathcal{M}^2}{(\gamma + 1) - \frac{\gamma - 1}{\gamma + 1} }
\label{eq:ratiop}
\end{equation}
\begin{equation}
\frac{T_1}{T_0} = \frac{(\gamma -1) P_1}{(\gamma + 1) P_0}
\label{eq:ratiot}
\end{equation}
 
 where $T$ is the temperature and $\gamma$ is the adiabatic index. Under these assumptions, the density ratio between the two sides of the shock does not depend on the shock speed and is equal to 4 if we are dealing with a monoatomic gas ($\gamma = 5/3$). This implies a velocity jump by a factor 1/4 in the shock reference frame (because of  Eq. \ref{eq:ranhug_1}). In the reference frame of the upstream material, the shock speed is $V_{\mathrm{s}} = 4/3 v_1$. Since the ejecta expand with an initial velocity\footnote{The shock speed decreases as the remnant evolves, as explained in Sect \ref{sect:snr_evol}} of the order of $10^4$ km/s, and the speed of sound is of the order of a few km/s (for T $\sim 10^3$ K), then $ \mathcal{M} \sim 1000$ and from Eqs. \ref{eq:ratiot} and \ref{eq:ratiop} we obtain a post-shock temperature $T_1$ of a few $10^8$ K, which causes X-ray emission. 
 
 \subsection{The evolution of SNRs}
 \label{sect:snr_evol}
 The evolution of a SNR can be described as undergoing four phases: the free expansion, the Sedov-Taylor expansion, the snow-plough (or pressure driven) phase and the merging phase.
 
During the first phase of the evolution of a SNR, the mass of interstellar medium swept up by the shock is much lower than the mass of the ejecta: the expansion is basically free and the initial conditions are the only critical parameters for its development. By studying a SNR in this early phase, it is possible to collect information about the physics of the explosion, the chemical composition and velocity of ejecta and so on. This early phase of evolution also offers the possibility of studying line emission originating from radiative isotopes synthesized in the SN explosion.
  
When the shocked mass becomes comparable to the mass of the ejecta, the SNR enters in the so-called Sedov-Taylor phase. In the hypothesis of uniform ambient medium, the evolution is described by a self-similar solution (\citealt{sed59,tay50}). Assuming adiabatic and radial expansion (i.e. spherically symmetrix explosion in a uniform ambient medium), the shock radius and the velocity can be derived on the basis of self-similar expansion and dimensional analysis:
\begin{equation}
R_{\mathrm{s}} \propto \left(\frac{Et^2 }{\rho_0}\right)^{1/5}
\label{eq:radius_sedov}
\end{equation}

\begin{equation}
V_{\mathrm{s}} = \frac{dR_{\mathrm{s}}}{dt} \propto \left(\frac{E}{\rho_0}\right)^{1/5}t^{-3/5}
\label{eq:velocity_sedov}
\end{equation}
where $E$ is the explosion energy, $t$ is the time and $\rho_0$ the ambient density.
Since the post-shock temperature $T_1 \propto v^2$ (Eq. \ref{eq:ratiot}), then $T_1 \propto R_{\mathrm{s}}^{-3}$. 

\begin{figure}[!ht]
\includegraphics[width=0.99\textwidth]{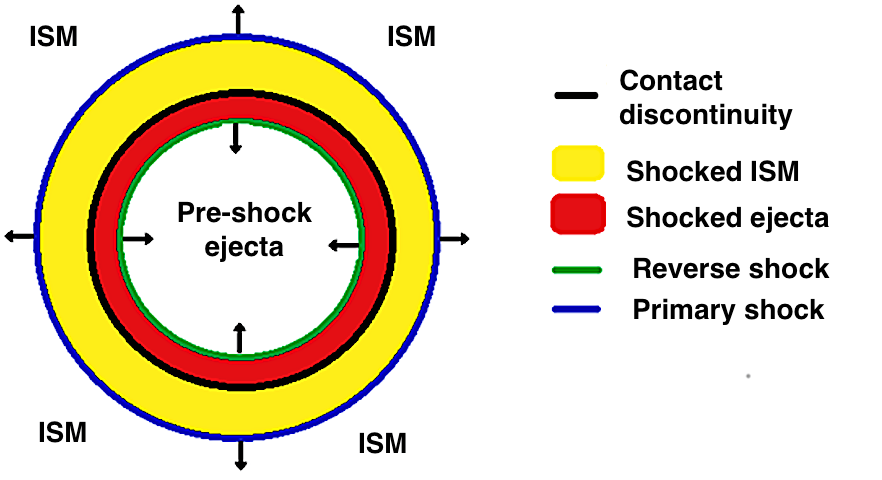}
\caption{Schematic representation of a SNR. The contact discontinuity divides the shocked ISM and the shocked ejected material which are heated by the primary and the reverse shocks, respectively.}
\label{fig:snr_scheme}
\end{figure}

While sweeping out material surrounding the progenitor star, the shock wave is followed by the expanding ejecta. 
The high pressure of the shocked ISM in front of the ejecta decelerates them and this leads to the formation of a shock which moves in the opposite direction of the primary shock wave, in the ejecta reference frame. For this reason, the shock wave which sweeps the ejecta is called reverse shock.
The detailed description of the formation of this wave is given by \cite{mck74}, here I just recall the main steps. While the shock front expands, the ejecta are decelerated by the positive gradient of the ISM pressure, because during the expansion a rarefaction wave moves back into the shell, thust lowering the internal pressure which is further lowered by the adiabatic expansion. 
The difference in pressure between the inner and the outer shells leads to the formation of a compression wave which rapidly transforms into a shock because of the low sound speed in the interior of the remnant. This reverse shock can heat the ejecta up to X-ray emitting temperatures. Fig. \ref{fig:snr_scheme} shows a schematic representation of the internal structure of a typical SNR.

Thus, by analysing X-ray observations it is possible to collect information about the shocked ISM/CSM and also about the heated ejecta. This is actually done through the analysis of X-ray spectra of SNRs. It is worth noticing that the assumptions we have made through this whole section are correct in general and for most cases but may not be valid in certain cases. For example, the CSM where the expansion takes place could be very inhomogeneous and the distribution of the ejecta may be very asymmetrical, presenting sort of collimated structures, such in the case of IC 443 which will be described in Chapter \ref{ch:ic443}.

When the radiative losses become important and the conservation of energy cannot be assumed, the SNR enters in the snow-plough (or pressure-driven) phase. The evolution of the shock wave is now governed by radial momentum conservation:

\begin{equation}
MV_{\mathrm{s}}=\frac{4\pi}{3} \rho_0 R^3_{\mathrm{s}} \frac{d R_{\mathrm{s}}}{dt} = constant 
\label{eq:mom_conservation}
\end{equation}

where the costant can be estimated by considering the momentum at the time $t_{\rm{rad}}$, namely the time when the SNR enters in the snow-plough phase. Generally, radiative losses are important when the temperature of shocked material is $\lesssim 5 \times 10^5$ K, corresponding to a shock velocity $V_{\mathrm{s}} = V_{\rm{rad}} = 200$ km/s \citep{wol72}. The age t$_{\rm{rad}}$ at which the SNR enters the radiative phase can be estimated by considering equations \ref{eq:radius_sedov} and \ref{eq:velocity_sedov}, since $V_{\rm{rad}}=R_{\rm{rad}}/t_{\rm{rad}}$. By imposing $V_{\mathrm{s}} = 200$ km/s:

\begin{equation}
t_{rad} \approx 44,600 \left(\frac{E_{51}}{n_{\rm{H}}}\right)^{1/3} \, \mathrm{yr}
\label{eq:t_rad}
\end{equation}

\begin{equation}
r_{rad} \approx 23 \left(\frac{E_{51}}{n_{\rm{H}}}\right)^{1/3} \, \mathrm{pc}
\label{eq:r_rad}
\end{equation}
where $E_{51}$ is the explosion energy in units of 10$^{51}$ erg and $n_{\rm{H}}$ the unshocked H density.

The last phase of the life of a SNR is the merging phase, i.e. when the shock speed becomes close to the sound speed. At this stage, the shock slowly disappears and expand at subsonically rate. The moment in which the shock disappears marks the end of the SNR but part of the material left behind may still emit as a hot plasma bubble. 

\section{SNR classification}
\label{sect:class}

The evolution of a SNR and its interaction with the surrounding CSM or ISM leads to many possible morphologies. However, four main classes of SNRs can be defined according to some generic features observed in the radio and the X-ray bands.

\begin{itemize}
\item \emph{Shell-like} SNRs show an almost spherical limb-brightened morphology. Emission coming from these objects can be either thermal, like in the Cygnus Loop, or nonthermal, like in SN 1006. 
\item \emph{Plerions} do not show any bright shell while being particularly luminous in the central area because of the presence of a rapidly rotating neutron star. Because of this fast rotation, the NS loses energy at a rate of $\dot{E} = I\dot{\Omega}\Omega$ where $I$ is the moment of inertia, $\Omega$ the angular frequency and $\dot{\Omega}$ its time derivative. The energy loss leads to a wind of relativistic electrons/positrons that terminates at a shock and accelerate particles to ultra-relativistic energies. The advection behind the shock causes the formation of a nebula emitting synchrotron radiation (see Sect \ref{sect:synchro}) and is called pulsar wind nebula (PWN). The best example of this kind of object is the Crab Nebula powered by the Crab Pulsar (also known as B0531+21). 
\item \emph{Composite} remnants are shell-like remnants which also contain a nonthermal plerion (e.g. Vela SNR)
\item \emph{Mixed-Morphology} (MM) SNRs present centrally peaked thermal X-ray emission and shell-like radio morphology (e.g. IC 443). This class of SNRs is one of the most interesting because MM-SNRs show a higher central density than that predicted by the Sedov model (\citealt{rp98}). The detection of X-ray emitting overionized plasma (see Sect \ref{sect:fb}) in MM-SNRs also challenges the typical evolutionary models of SNRs and provides an unexpected observational constrain to ascertain their puzzling physical origin.
\end{itemize}

\section{Emission processes in SNRs in the X-ray band}
\label{sect:em_proc}

The mechanisms that produce X-ray photons in SNRs can be divided in two main groups. The first one gathers the so-called thermal processes, i.e. the emitting electrons are described by Maxwellian energy distribution characterized by a certain temperature $T$. The second one gathers nonthermal ones, as synchrotron emission. 

The thermal X-ray emitting plasma in SNRs is optically thin and its emission depends on binary collisions between electrons and ions, putting SNRs in the group of sources for which the coronal model is valid.  The coronal model assumes that the ion fraction of a given ionization state depends on the temperature of the plasma and remains constant with time, leading to a configuration of Collisional Ionization Equilibrium (CIE), i.e. the ionizations are balanced by the recombination and viceversa. As we shall see (Sect \ref{sect:nei}) there are important exceptions to CIE.

 At a fixed temperature, T, and a given element abundance, the emissivity is determined by the Emission Measure (EM), defined as EM = $\int n_{\mathrm{e}} n_{\mathrm{H}} dV$ where $n_{\mathrm{e}}$ is the electron density, $n_{\mathrm{H}}$ is the ion density of hydrogen (H) and V is the volume of the X-ray emitting plasma (for solar abundances, $n_{\mathrm{e}} \simeq n_{\mathrm{H}}$). Thermal processes can be of various kinds: line emission (bound-bound process) or continuum emission, including bremsstrahlung (free-free process, hereafter FF), radiative recombination continua (free-bound process, hereafter RRC or FB) and two-photon emission (hereafter 2-$\gamma$).

\subsection{Line emission}
\label{sect:line}
In SNRs, line emission results from excitations due to the electron-ion collisions. The emissivity of line emission for a particular transition $j \rightarrow i$ of an element $z$ is (\citealt{mew99}):
\begin{equation}
P_{\mathrm{ji}} \propto A_{\mathrm{z}} n_{\mathrm{H}} n_{\mathrm{e}} F(T) \;\;\;\; \mathrm{photons \; cm^{-3}s^{-1}}
\label{eq:line_emiss}
\end{equation}
where $A_{\mathrm{z}}$ is the abundance of the element $z$ and $F(T)$ is a function, indicating the temperature dependence due to the combined effects of ionization and excitation, which sharply peaks at the characteristic line temperature.

\subsection{Bremsstrahlung}
\label{sect:bremss}

Bremsstrahlung emission is due to the acceleration of electrons interacting with ions. It is a free-free emission due to a transition between two unbound states and produces a continuum emission. The total bremsstrahlung emissivity is given by the sum of the emissivities of all ion species in the plasma(\citealt{mew99}):
\begin{equation}
\epsilon_{\mathrm{ff}} \propto  \sum_{\mathrm{i}} Z_{\mathrm{i}}^2 n_{\mathrm{e}} n_{\mathrm{i}} g_{\mathrm{ff,i}}(T)T^{-1/2} \rm{exp} \left( \frac{-h\nu}{kT} \right)  \;\; \mathrm{erg \; cm^{-3} s^{-1} Hz^{-1}}
\label{eq:brems_emiss}
\end{equation}
where $Z_{\mathrm{i}}$ and $n_{\mathrm{i}}$ are the effective charge and density of ion i and $g_{\mathrm{ff,i}} \approx 1$ is the gaunt factor. In a plasma with solar (or mildly enhanced) abundances, the main contribution to the bremsstrahlung emission originates from H ions and electrons stripped from H atoms, since this element is by far the most abundant one (values of the protosolar abundances in logarithmic units according to \citealt{lpg09} are shown in Table \ref{tab:ab_lodd}). 

\begin{table}[!ht]
\centering
\caption{Proto-solar abundances, expressed in logarithmic units}
\begin{tabular}{c|c}
\hline\hline
Element & Abundance (log units)\\
\hline
H& 12 \\
\hline
He& 10.987\\
\hline
C&8.443\\
\hline
O&8.782\\
\hline
Ne& 8.103\\
\hline
Mg& 7.599\\
\hline
Si& 7.586\\
\hline
S& 7.210\\
\hline 
Ar& 6.553\\
\hline 
Ca& 6.367\\
\hline 
Fe& 7.514\\
\hline
Ni& 6.276\\
\end{tabular}

H=12.0 by definition. Values taken from \citet{lpg09}.
\label{tab:ab_lodd}
\end{table}

\subsection{Radiative recombination continua}
\label{sect:fb}

Free-bound emission occurs when an electron is captured by an ion into an atomic shell and a photon with energy $h\nu= E_{\mathrm{e}} + \chi_{\mathrm{n}}$ is emitted (where $E_{\mathrm{e}}$ is the energy the free electron and $\chi_{\mathrm{n}}$ is the ionization potential of the shell $n$). The emissivity for this process is (\citealt{lie99}):

\begin{equation}
    \epsilon_{\mathrm{fb}} \propto n_{\mathrm{e}} n_{\mathrm{i+1}} \left(\frac{h\nu}{kT^{1/2}}\right)^3 \exp\left({\frac{-h\nu+\chi_{\mathrm{n}}}{kT}}\right)  \,\,\, \mathrm{erg \; \; cm^{-3} s^{-1} Hz^{-1}}
    \label{eq:fb}
\end{equation}

where $n_{\mathrm{i+1}}$ is the ion density of the recombining ion. The width of the RRC is $\Delta\nu\approx kT$ and if $ kT << h\nu $ the emission results in relatively narrow, line-like, emission peaks near the series limits of lines, called radiative recombination continua (RRC). The corresponding edge of recombination shows up in the spectra at the characteristic ionization energy of the given ion. Moreover, If $h\nu>>\chi_{\mathrm{n}}$ the RRC is wide and looks like a continuum distribution, similar in shape to the FF contribution.

\subsection{2-$\gamma$ emission}
\label{sect:2gamma}

Two-photon emission results from electrons in meta-stable levels, as the 2s level of an H-like ion. Since decay to the 1s level is forbidden (because $\Delta s = 0$), the meta-stable level can be de-excited by emitting two photons.Though the sum of energies of the individual photons is well determined, this emission process leads to a continuum spectrum. In fact, the energy of the individual photon can range between 0 and the energy gap in the interval between the metastable level and the ground level.  

\begin{figure}[!ht]
\includegraphics[scale=0.25]{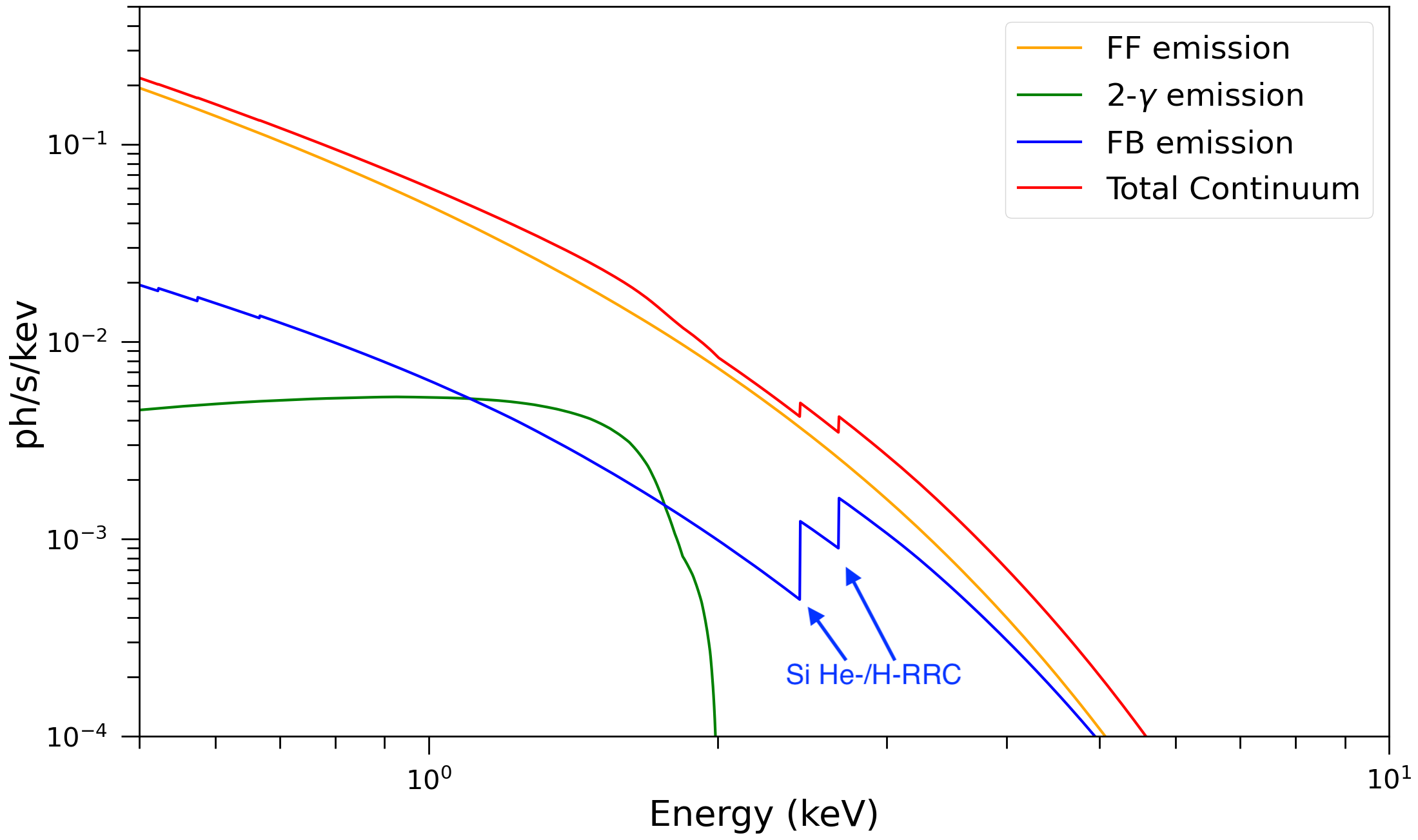}
\caption{Ideal spectra of the various contributions to the continuum emission for a Si-only plasma in CIE at a temperature of 1 keV.  Bremsstrahlung, labeled as FF, is in orange; 2-$\gamma$ emission is in green; radiative recombination continua, labeled as FB, is in blue. The total continuum contribution is shown in red.}
\label{fig:em_proc_spec}
\end{figure}

To give an idea of the contributions from each continuum emission process, I show in Fig. \ref{fig:em_proc_spec} the different types of continuum thermal emission in the X-ray spectra of a Si-only plasma in equilibrium of ionization at a temperature of 1 keV. 

\subsection{Non equilibrium of ionization}
\label{sect:nei}
SNR plasmas are often found in condition of non-equilibrium of ionization (NEI). SNR plasmas are in NEI because of their low densities: only a few ionizing collision per ion have occurred since the plasma was shocked and the ionization conditions are those of a plasma at lower electron temperature, i.e. the plasma is \emph{underionized}, or \emph{ionizing}. However, another and opposite configuration for the ionization state of the plasma is possible: the overionized, or recombining, plasma. If a plasma initially in CIE faces a very rapid cooling, then the degree of ionization of ions can be higher than that expected in CIE for the same electron temperature. This happens because not enough time has elapsed since the abrupt cooling in order to reach a new configuration of CIE. 

The parameter describing the state of ionization of the plasma is the so-called ionization age $\tau = \int n_{\mathrm{e}} dt$. The critical threshold on $\tau$ is of the order of 10$^{12} \, \mathrm{s/cm^3}$ \citep{sh10}: plasmas with lower $\tau$ are in NEI, otherwise the plasma reaches CIE conditions. The implication is that, considering a typical plasma density of 1 cm$^{-3}$, the X-ray emitting plasma in a SNR reaches CIE conditions in roughly 30 kyr. In this scenario, typical of SNRs, the plasma is called underionized or ionizing.
The consequences of underionization in plasma is that the highly ionized (e.g H-like) states at given temperature are less populated than in CIE scenario (\citealt{vin06}). The effects of NEI on the spectra can be appreciated by looking at the ratio between H-like and He-like lines of a given species. In fact, if the plasma is underionized the He-like to H-like line ratio is higher than that in the CIE at the same temperature. This effect is also responsible for the lowering of the RRC contribution in spectra of underionized plasma.

The same ionization parameter $\tau$, defined for the underionized plasma, can describe overionization, provided that the time is estimated since the onset of the cooling process. Up to now, overionized plasma has been observed in a large sample of mixed morphology SNRs during the last decade (e.g. \citealt{yok09}, \citealt{mtu17}, {\citealt{gmo18} for IC 443; \citealt{oky09}, \citealt{mbd10} for W49B, \citealt{zmb11}; \citealt{uky12,otu20} for W44; \citealt{sk12} for W28).  Two different scenarios can be invoked to explain the presence of overionized plasma and its rapid cooling, namely thermal conduction with closeby cold clouds and adiabatic expansion of the plasma. In Chapter \ref{ch:ic443} I will discuss the case of IC 443.

\subsection{Particle acceleration and nonthermal processes} 
\label{sect:synchro}

X-ray emission in SNRs can also be generated by electrons with a non-Maxwellian energy spectrum. These emission processes are called nonthermal and the most significant one for SNRs is synchrotron radiation. This radiation is observed in the outer shell of the SNR or in the PWNe embedded within the remnant. Considering an electron with energy E (E$_{100}$ in units of 100 TeV), its synchrotron emission is characterized by a sharp peak with a maximum at a frequency $\nu_{b}$ given by the relation \citep{gs65}:
\begin{equation}
h\nu_{b}= 13.9 \times B_{\bot,100} E^2_{100} \;\; \mathrm{keV}
\label{eq:nu_synch}
\end{equation}

where $B_{\bot,100} \approx \sqrt{2/3} B_{100}$ is the component of the magnetic field perpendicular to the motion of the electron expressed in units of 100 $\mu G$. 

An electron nonthermal distribution, typically a power-law, should be generated by some acceleration process. Such a process can be very relevant in the broader context of cosmic ray acceleration. A simple energy budget evaluation shows that SNR shocks are the most probable (if not the only possible) place where galactic cosmic rays (e.g. electrons below a few hundreds of TeV) are accelerated \citep{hil05}. For SNRs $B \sim 10-500 \mu$G \citep[and reference therein]{vin12}, and therefore synchrotron emission peaking at X-ray frequencies corresponds to electrons with energies of the order of $10-100$ TeV. 

Diffusive shock acceleration (DSA) \citep{bel78,bel278,bo78}, also known as first order Fermi-acceleration, an evolution of the second order Fermi acceleration \citep{fer49}, is the most common physical mechanism invoked to explain how cosmic rays can be accelerated by collisionless shocks. According to this model, the charged particles present in the shocked plasma can recross the shock front going back to the shock front because of the turbulence of the magnetic field. Once they are in the upstream plasma, they diffuse and can cross the shock again and the process can start again. DSA model predicts a power-law energy distribution of electrons, in agreement with what we measure on Earth from cosmic rays. In particular, it predicts a power-law index $q=2$  which is quite similar to that observed for Galactic cosmic rays . The corresponding radiation spectrum has a power-law distribution $\epsilon_{\mathrm{syn}} \propto \nu^{\alpha}$ where $\alpha=(q-1)/2$. In the X-ray band it is common to express the luminosity density in units of photons per seconds per unit energy with an associated photon spectral index of $\Gamma = \alpha + 1 = (q+1)/2$. 

It is important to notice that electron energy spectra present a cutoff at high energy. The physical reason for this cutoff is not clear \citep{rey08}. Different scenarios can be invoked to explain the cut-off in the spectrum: in the \emph{age-limited case}, the shock acceleration process cannot act for long enough time to accelerate electrons to higher energies leading to a sharp decrease at the cutoff energy $E_{\mathrm{cut}}$. This scenario can be described by including an high energy exponential cut-off to the power-law distribution \citep{rey98}:
\begin{equation}
N(E) \propto E^{-q} \mathrm{exp}(-E/E_{\mathrm{cut}})
\label{eq:age_limited}
\end{equation}

When the acceleration gains are comparable to the radiative losses, we enter in the \emph{loss-limited case}. The electrons lose energy through radiative losses at a rate equal to
\begin{equation}
\frac{dE}{dt} = -4.05 \times 10^{-7} E_{100}^2 B_{\bot,100}^2 \,\, \mathrm{erg/s}
\label{eq:loss_synch}
\end{equation}
and the time scale for the electrons to lose a substantial fraction of this energy is:
\begin{equation}
\tau_{\mathrm{synch}} = \left| \frac{E}{dE/dt} \right| = 12.5 \; E^{-1} B^{-2}  \;\; \mathrm{yr}
\label{eq:time:synch}
\end{equation}
It is common to include in this expression also the loss-rates due to inverse Compton, since the inverse Comption scattering loss rate \citep{lon11} is almost identical to the synchrotron one, apart from the term $B^2/8\pi$. In this case we replace $B^2$ in Eq. \ref{eq:time:synch} with $B^2_{\mathrm{eff}} \equiv B^2 + B_{\mathrm{rad}}^2$ where $B_{\mathrm{rad}} \equiv \sqrt{8\pi U_{\mathrm{rad}}}$. For the cosmic microwave background (CMB) $U_{\mathrm{CMB}} = 4.2 \times 10^{-13}$ erg cm$^{-3}$, therefore $B_{\mathrm{rad}}=3.2 \mu$G. In the loss-limited scenario, the cutoff is steeper than that in the time-limited case, being $\propto \mathrm{exp}(-E/E_{\mathrm{cut}})^2$ \citep{za07}.

Another possibility is the abrupt increase of the diffusion coefficient above the corresponding particle energy, resulting in particle escape upstream \citep{rey98}.

\section{Contents}
\label{sect:contents}
In this thesis I aim at investigating physical and chemical properties of X-ray emitting plasmas in SNRs by analyzing X-ray data collected by different telescopes and taking advantage of the high diagnostic potential provided by the X-ray spectroscopy. Different spectral features are related to different physical emission processes and a careful analysis of X-ray spectra permits to reconstruct a reliable scenario of the physics occurring in the SNRs investigated. Moreover, it is possible to study the chemical distribution of ejecta and of the surrounding medium. To get a deeper level of understanding of the complex phenomena involved and, therefore, a diagnostics, it is crucial to relate X-ray observations to HD/MHD simulations of the evolution of SNRs. The comparison between data collected by actual telescopes and synthetic observables (images, spectra) derived by the simulations further enriches our knowledge on the dynamical and chemical evolution of the hot material. A lot of open issues are strictly related to X-ray emission arising from SNRs. Here I focus on three main topics.

\begin{itemize}
\item In the CCD spectra of X-ray emitting SNRs it is common to face an high uncertainty in the estimate of the element abundances. This is due to the low spectral resolution of CCD detectors causing an entanglement between the continuum emission and the line emission. This high uncertainty on the abundances is reflected on the estimate of the ejecta mass and density, possibly invalidating any comparison with nucleosynthesis yields. A diagnostic tool able to solve this issue is needed to avoid any ambiguous result. I performed a campaign of spectral simulations by carefully investigating the characteristic emission processes of optically thin plasma in SNRs. I then related the results of the spectral simulations with the {\it Chandra} data of Cas A and with a state-of-the-art HD simulation of Cas A (\citealt{omp16}). I have developed a spectral diagnostic tool to recover the ejecta properties and show its applicability for the next generation of X-ray telescopes in Chapter \ref{ch:rrc} (see also \citealt{gvm20}). 

\item The class of MM-SNRs is one of the less understood classes of SNRs. In fact, there is no clear explanation which satisfyingly explains the observed X-ray morphology and anisotropies. Moreover, there is an ongoing debate about the origin of the overionization, detected only in SNRs belonging to this class, and the role that the nearby molecular/atomic clouds have in the formation of this peculiar status. I present the analysis of archive {\it XMM-Newton}/EPIC observations of IC 443. I aim at studying the distribution of chemical elements in the SNR, at clarifying the link between the PWN CXOU J061705.3+222127, hosting a NS, and the remnant itself and at studying the physical conditions of plasma in a jet-like structure that I discovered by analyzing the X-ray data. I investigated whether the presence of overionized plasma may be ascribable to thermal conduction or to rarefaction scenarios or a combination of the two effects. I also compared results of my analysis with a state-of-the-art HD simulation performed by \citet{uog20} obtaining a deeper insight on the peculiar morphology of the remnant. The results are presented in Chapter \ref{ch:ic443} (see also \citealt{gmo18}).

\item The characteristics of the explosion of SN 1987A are perfectly compatible with the formation of a NS. However, despite the unique consideration granted to this object with multi-wavelenghts observations, the detection of the elusive compact object is still missing. Many hypotheses are currently considered, and the most accepted one ascribes this non-detection to the presence of cold and dense ejecta, absorbing the radiation of the compact object. The possibility to directly study a young NS could shed the light on many topics related to the formation and early evolution of these extreme objects.  In Chapter \ref{ch:pwn_87A}, I show the results of the analysis of multi-epoch {\it Chandra} and {\it NuSTAR} observations of SN 1987A. In particular, I looked for X-ray signatures of emission, either thermal or nonthermal, arising from the still undetected compact object, which is expected to be present within the shell of SN 1987A. I took advantage of the state-of-the-art MHD simulation by \citet{oon20} to reconstruct the absorption pattern surrounding the expected position of the NS and to find strong indications for nonthermal emission likely originating in a PWN (see also \citealt{gmo21}).
    
\end{itemize}

\subsection{Identification of a spectral signature of pure-metal plasma: application to Cas A} 
\label{sect:Intro_RRC}

Spectral analysis of X-ray emission from ejecta in SNRs is hampered by the moderate spectral resolution ($R\sim 5-100$) of charged-coupled device (CCD) detectors, which typically causes a degeneracy between the best-fit values of chemical abundances and of  the plasma emission measure. Because of the low energy resolution, the blending between different emission lines can create a "false continuum", which makes it difficult to constrain the real continuum flux, especially below 4 keV.  Therefore, it is possible to describe a given X-ray spectrum either with high abundances and low emission measure, or vice versa. This degeneracy leads to big uncertainties in the mass estimates and may even hide the existence of pure-metal ejecta plasma in SNRs (\citealt{vkb96}). 

The combined contribution of shocked ambient medium and ejecta to the emerging X-ray emission further complicates the determination of the ejecta mass and chemical composition. This degeneracy leads to big uncertainties in mass estimates and can introduce a bias in the comparison between the ejecta chemical composition derived from the observations and the yields predicted by explosive nucleosynthesis models. Having information on these quantities is fundamental to knowing more about the progenitor star, the SN explosion, and the explosive nucleosynthesis processes.

The typical approach used to face this issue is to measure relative abundances between elements, typically by adopting Si or Fe as a reference (e.g., \citealt{wbv02}, \citealt{mdb06,mbr08}, \citealt{ksg12}, \citealt{lcs14}, \citealt{fbp15}, \citealt{zv18}, \citealt{zvs19}). However, this approach does not allow us to unambiguously derive absolute mass estimates for the yields in SNe. The comparison with theoretical nucleosynthesis yields (e.g., \citealt{ww95}, \citealt{tnh96}, \citealt{nht97}, \citealt{num99}, \citealt{sew16}) can only be performed through abundance ratios and may lead to a misunderstanding of the effective explosion mechanism and of the actual progenitor star properties. In order to have a fully reliable estimate of the abundances and of the mass of each element and to correctly compare these values with the theoretical predictions, a tool able to precisely estimate the absolute abundances of ejecta is badly needed. 

In Chapter \ref{ch:rrc} I present the study I performed to tackle this open issue, by identifying as a test case the SNR Cassiopeia A (Cas A). 

Cas A is one of the brightest and most studied SNR. It is a young (330 years old, \citealt{tfv01}) SN IIb-type remnant (\citealt{kbu08}) at a distance of 3.4 kpc (\citealt{rhf95}), which shows many asymmetries and an overall clumpiness (\citealt{rhf95}, \citealt{hl03}, \citealt{vl03}, \citealt{hlb04}, \citealt{byy05}, \citealt{drs10} \citealt{hl12}, \citealt{mf13}, \citealt{lph14}, \citealt{pf14}). Fig \ref{casa_intro} shows a composite map of Cas A in different spectral bands.
\begin{figure}[ht!]
\centering
\includegraphics[width=0.8\textwidth]{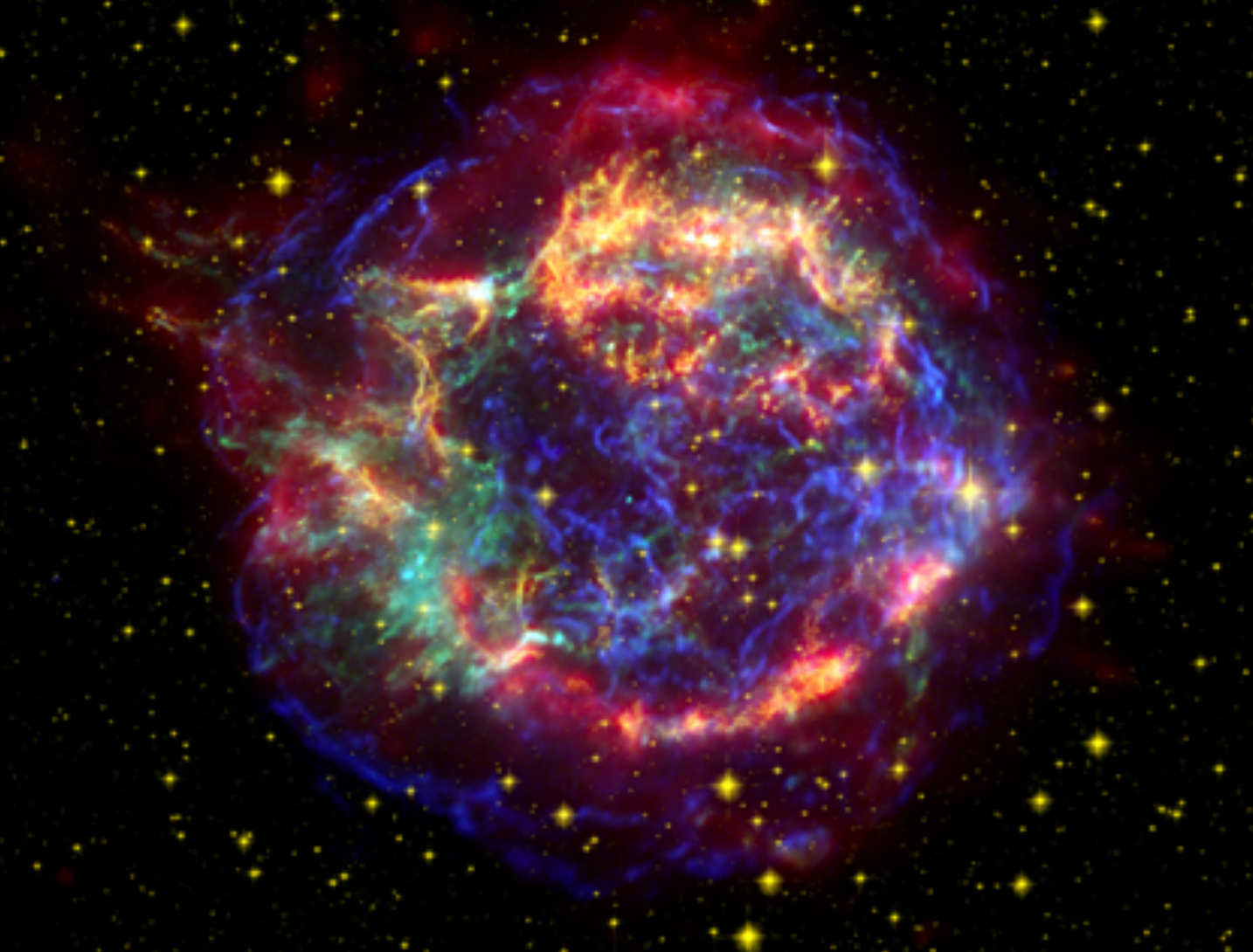}
\caption{Multi-wavelenght image of Cas A: [4.30-5.95] keV band in blue, [1.76-1.93] keV band in green, 24 $\mu$m image in red, optical (900 nm) in yellow. Credits to NASA/CXC for X-ray data, NASA/JPL-Caltech for infrared data, NASA/STScI for optical data.}
\label{casa_intro}
\end{figure}
\cite{hl12} performed a detailed survey of the ejecta distribution in Cas A, highlighting the presence of three large-scale Fe-rich clumps. They also confirmed the existence of an Fe-rich cloudlet, previously detected by \cite{hrb00} and \cite{hl03}, located within the southeastern clump. In this cloudlet the relative Fe$/$Si abundance is $\sim20$, while Fe$/$Si is $\sim5$ in the other Fe-rich regions of Cas A. The spectrum extracted from such cloudlet is dominated by the Fe L false continuum emission at energies around $\sim$ 1 keV.
\subsection{The supernova remnant IC 443}
\label{sect:Intro_IC443}
IC 443 (also called G189.1+3.0) belongs to the GEM OB1 association at a distance of 1.5 kpc \citep{ws03}. It is classified as a MM SNR, its radius is $\sim 20'$ while proposed values for its age span from $\sim 4000$ yr (\citealt{tbm08}) to $\sim 20000$ yr \citep{che99,bku08}. It is located in a very complex environment since it interacts with a molecular cloud in the northwestern (NW) and southeastern (SE) areas, and with an atomic cloud in the northeast (NE) (see Fig. \ref{fig:ic443_multilambda} for a multiwavelenght picture of the remnant). 
\begin{figure}[hb!]
\centering
\includegraphics[width=0.7\textwidth]{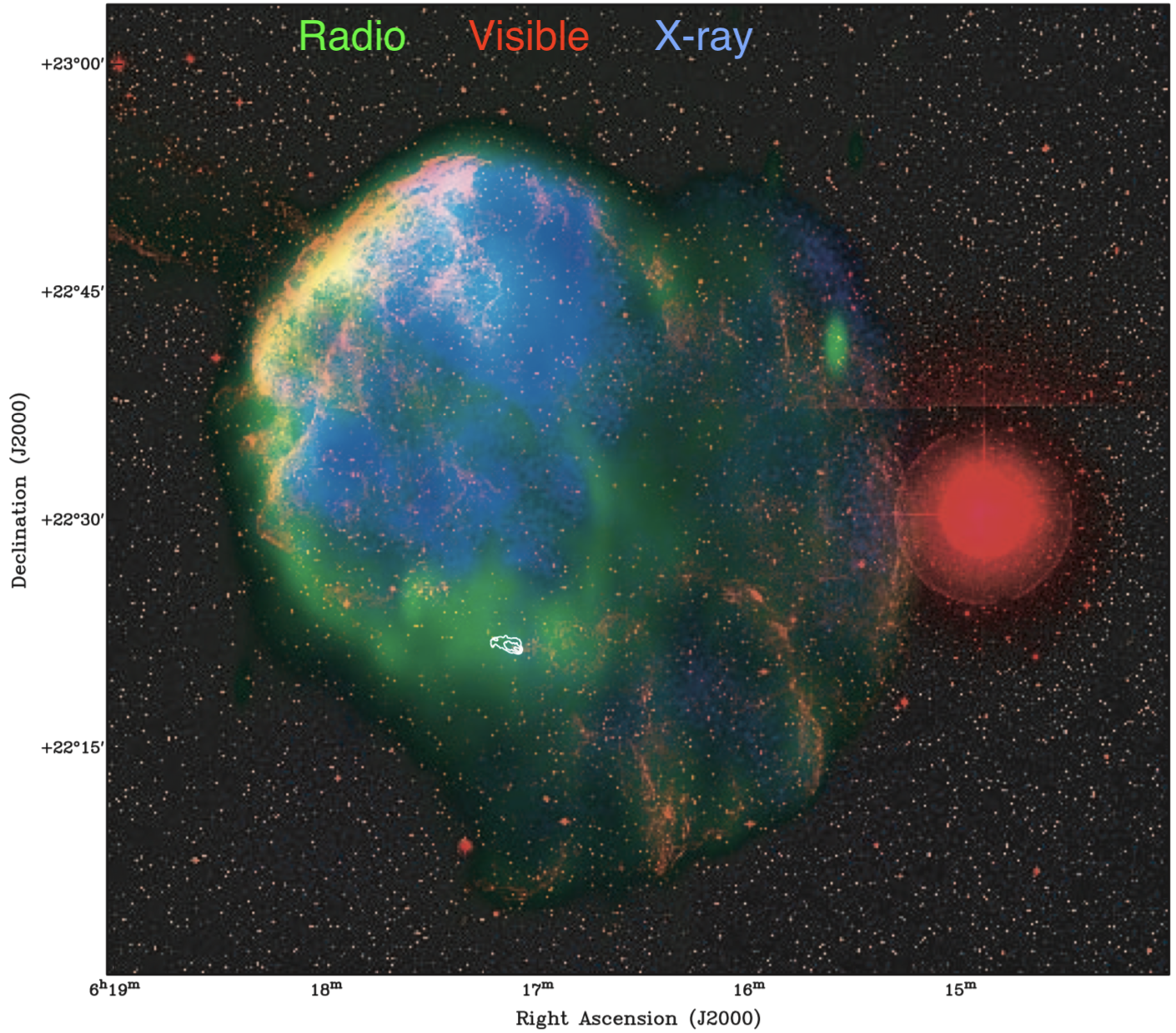}
\caption{Multiwavelength view of the SNR IC 443 and the PWN G189.22+2.90 produced by \citealt{gcs06}. Red shows 670 nm emission from the Second Palomar Observatory Sky Survey; green shows 1.4 GHz radio data taken by the DRAO Synthesis Telescope (\citealt{lea04b}), blue shows 0.1-2.4 keV X-ray data taken by the ROSAT PSPC (\citealt{aa94}). The white contours, traced starting from 8.5 GHz Very Large Array data, mark the position of PWN G189.22+2.90 (see \citealt{gcs06} for further details). The bright red star on the right is $\eta$ Gemini.}
\label{fig:ic443_multilambda}
\end{figure}

The dense molecular cloud was first identified by \cite{cck77}, and lies  in the foreground of IC 443 forming a semi-toroidal structure (\citealt{bgb88,tbr06,sfy14}) (see Fig. \ref{fig:mol_clouds_su}). In the NE the remnant is confined by the atomic HI cloud, discovered by \cite{den78}, which is well traced by optical, infrared and very soft X-ray emission (\citealt{tbr06}). 

\begin{figure}[!h]
\includegraphics[width=0.99\textwidth]{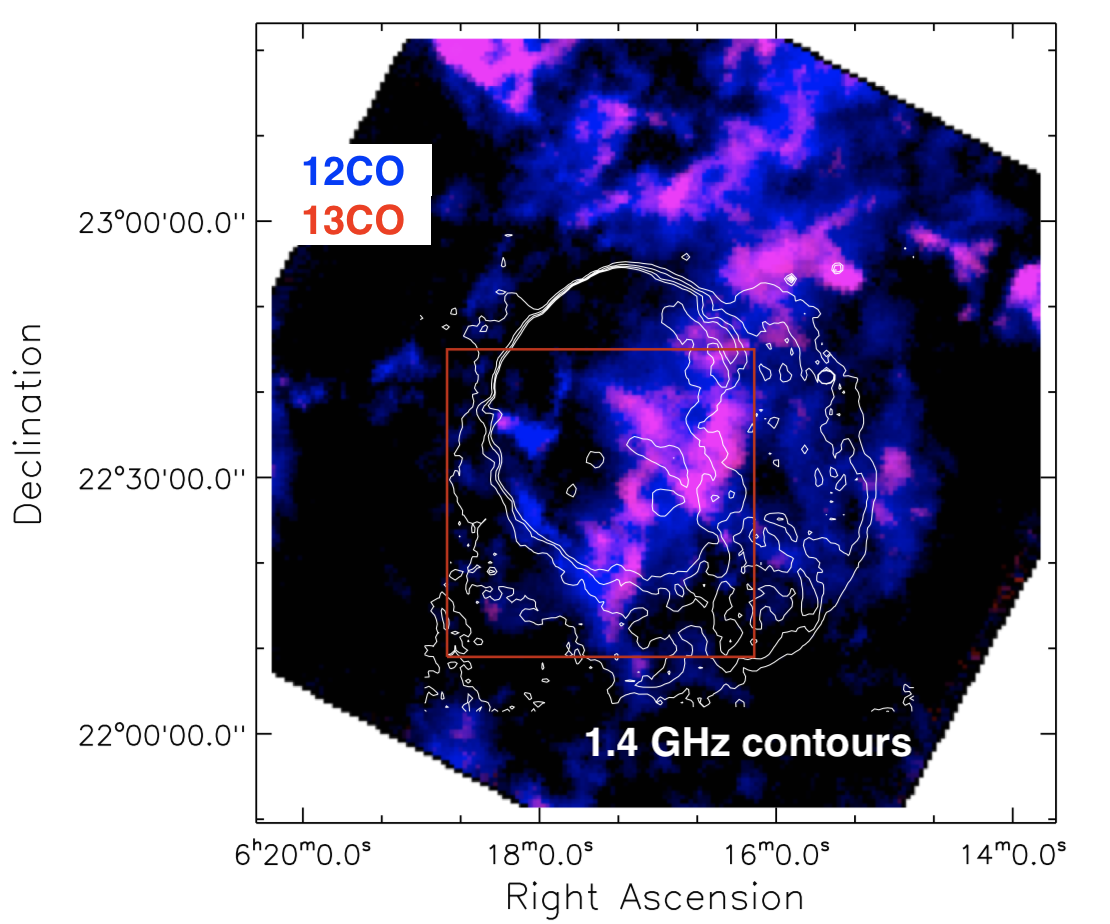}
\caption{$^{12}$CO (J = 1–0; blue) and $^{13}$CO (J = 1–0; red) intensity maps in the -10 km/s to 10 km/s interval with a square root scale toward SNR IC 443, overlaid with the 1.4 GHz radio continuum emission contours, obtained by \citet {sfy14}. The red box shows the area where the broadened molecular lines are detected (see the paper above for further details).}
\label{fig:mol_clouds_su}
\end{figure}

 For IC 443, marginal evidence of overionized plasma was detected in the inner region characterized by bright X-ray emission (\citealt{kon02}, \citealt{tbm08}), and later confirmed by Suzaku observations (\citealt{yok09}, \citealt{out14} and \citealt{mtu17}, hereafter M17). As mentioned in Sect \ref{sect:fb}, the physical origin of the overionized plasma is still unclear. 

The PWN CXOU J061705.3+222127 is observed within the remnant shell (see Fig. \ref{fig:ic443_multilambda}). However, since it is far away from the geometric center of the remnant (near the southern edge) and moves towards southwest (SW), it is not clear whether the PWN belongs to IC 443 (\citealt{gcs06}, \citealt{spc15}) or it is a rambling neutron star (NS) seen in projection on the remnant. In Sect. \ref{image} and \ref{discu} I discuss whether the off-centered position of the explosion has a role in the evolution of the remnant. 

\subsection{SN 1987A and its elusive compact object}
\label{sect:Intro_1987A}

SN 1987A in the Large Magellanic Cloud (LMC) was a hydrogen-rich core-collapse supernova (SN) discovered on 1987 February 23 \citep{wls87}. It occurred approximately 51.4 kpc from Earth \citep{pan99} and its dynamical evolution is strictly related to the very inhomogenous CSM, composed by a dense ring-like structure within a diffuse HII region \citep{sck05}, as shown in Fig. \ref{fig:img_1987A}. SN 1987A is the first naked-eye SN exploded since telescopes exist and has been closely monitored in various wavelengths since the SN event \citep{mcr93,mcf16}. In particular, the X-ray band is ideal to investigate the interaction of the shock front with the CSM and the emission of the expected central compact leftover of the supernova explosion.

\begin{figure}[h!]
\centering
\includegraphics[width=1\textwidth]{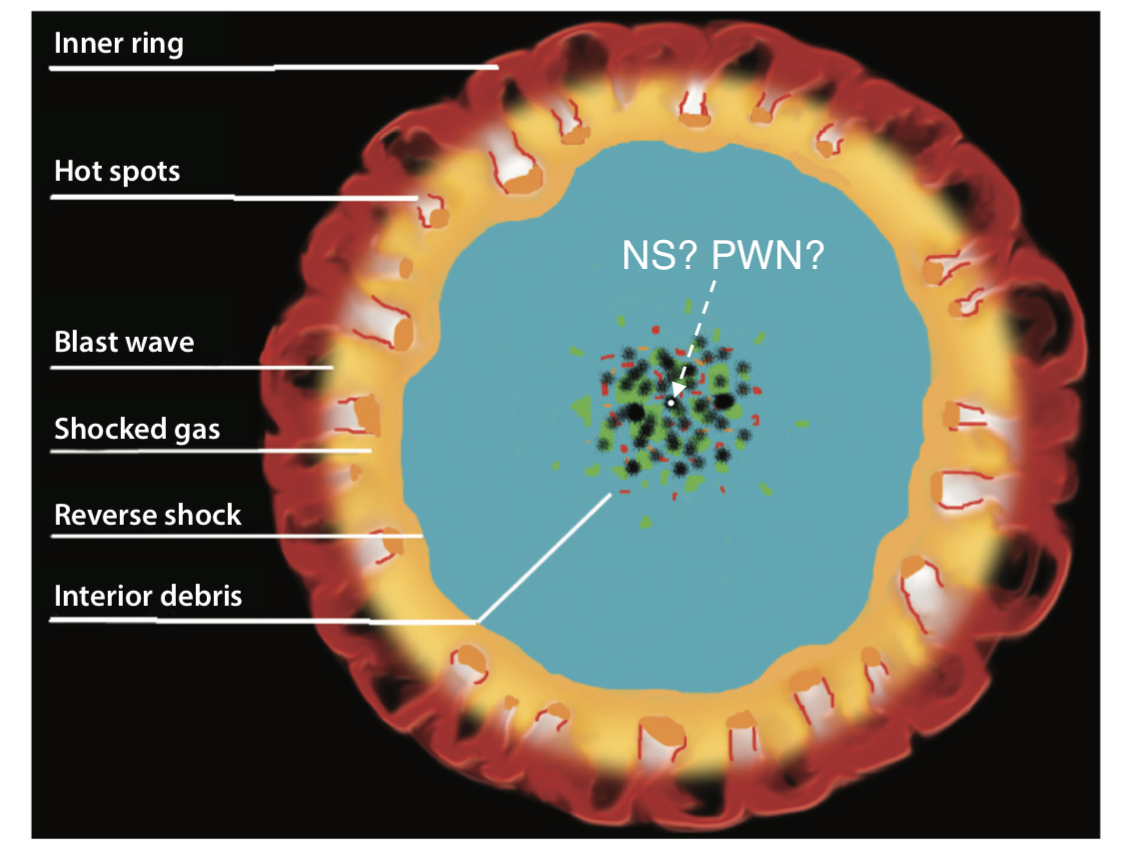}
\caption{Schematic picture by  \cite{mcf16} showing the interaction between the shocks, the CSM and the ejecta (interior debris) in SN 1987A, possibly hiding a putative NS/PWN.}
\label{fig:img_1987A}
\end{figure}

Despite of the unique consideration granted with deep and continuous observations and despite of the neutrinos detection \citep{bbb87}, strongly indicating the formation of a neutron star \citep{vis15}, the elusive compact object of SN 1987A is still undetected. The most likely explanation to this non-detection is ascribable to absorption due to the dense and cold material ejected by the supernova \citep{fcr87}, i.e., the ejecta: because of the young age of SN 1987A, the ejecta are still very dense and the reverse shock generated in the outer shells of the SNR has not heated the inner ejecta yet. Thus, photo-electric absorption due to these metal-rich material would hide the X-ray emission coming from an hypothetical compact object. During the last years, many works \citep{omp15,alf18a,erl18,pbg20} investigated the upper limit on the luminosity of the putative compact leftover in various wavelengths, considering the case of a NS emitting thermal (black-body) radiation and obscured by the cold ejecta and/or dust. However, the lack of strong constraints on the absorption pattern in the internal area of SN 1987A prevented either to further constrain the luminosity of the putative NS or to make predictions about its future detectability.

On the other side, the X-ray emission from a young NS may include a significant nonthermal component: synchrotron radiation arising from the PWN associated with the NS. Recently, {\it ALMA} images showed a {\it blob} structure whose emission is somehow compatible with the radio emission of a PWN \citep{cmg19}. However, the authors warned that this blob could be associated with other physical processes, e.g. heating due to Ti$^{44}$ decay.

\section{Thesis outline} \label{sect:outline}
During my PhD studies, I performed data analysis of X-ray observations and X-ray spectroscopy of various SNRs and complemented it with the analysis of synthetic observables derived from multi-D HD/MHD simulations of expanding SNRs. In particular, I focused on:  i) the identification ok	f an X-ray spectral signature to carefully estimate the ejecta contribution in SNRs and to obtain precise estimates of the nucleosynthesis yields in the ejecta (Ch. \ref{ch:rrc}); ii) the study of anisotropies produced by the explosion and induced by the inhomogeneous environment in the mixed morphology SNR IC 443 (Ch. \ref{ch:ic443}); iii) the quest for the compact object inside SN 1987A which allowed me to find strong indications for the presence of an X-ray emitting pulsar wind nebula (Ch. \ref{ch:pwn_87A}). At the end of my thesis, I draw my conclusions in Ch \ref{ch:final}.

%% file: RRC.tex
\chapter{Unveiling pure-metal ejecta emission through RRC}
\label{ch:rrc}

In this chapter, I report on the study of the main emission processes in SNRs in the high-abundance regime, by investigating in detail the behavior of bremmstrahlung, radiative recombination continua and line emission in a plasma highly enriched with heavy elements (like the ejecta). I performed a set of spectral simulations to identify a signature of pure-metal ejecta emission in the X-ray spectra of SNRs. I identified a novel diagnostic tool and successfully tested its applicability by adopting the Galactic SNR Cas A as a benchmark (see also \citealt{gvm20}).
Part of this work has been performed at the API-Anton Pannekoek Instituut for Astronomy of the University of Amsterdam, where I spent six months in the framework of a scientific collaboration with Prof. Jacco Vink and his research group. 

The chapter is divided as follows: in Sect. \ref{sect:spec_sim} I present the methods adopted to perform the spectral simulations and the corresponding results; in Sect. \ref{sect:allsynthesis} I describe the procedure developed to synthesize X-ray spectra and the resulting products; in Sect. \ref{sect:tool_synth} I describe the tool that self-consistently synthesizes spectra from any given HD/MHD simulation; in Sect. \ref{introCasA} I investigate in detail how my diagnostic tool can be adopted to detect pure-metal ejecta in Cas A; in Sect. \ref{final} I discuss the results; in Sect. \ref{sect:hd_anto} I shortly present another project in which the X-ray self-consistent synthesis tool was used to investigate spectral properties of a SNRs presenting various anisotropies; the conclusions of the chapter are drawn in Sect. \ref{sect:conc_rrc}.   
 
\section{X-ray spectral signatures of pure-metal ejecta}
\label{sect:spec_sim}
In  Section \ref{sect:em_proc}, I recall the physics involved in the thermal emission of an optically thin plasma. Here, I focus on the X-ray emission features of a pure-metal plasma.

I have performed spectral simulations using the X-ray spectral analysis code SPEX (version 3.04.00 with SPEXACT 2.07.00, \citealt{spex}) with the aim of investigating the behavior of the X-ray emission processes at high abundances.
Considering a plasma in CIE, I studied the dependence of the flux on the abundance of two elements: silicon (Si) and iron (Fe). For these simulations, I consider the energy ranges reported in Table \ref{tab:bands}\footnote{The S abundance, as well the abundance of all other elements, is fixed to 1, i.e. solar abundance. Thus, even if the S XVI line is present at 2.623 keV, it does not affect the results.}.
\begin{table}[!ht]
\centering
\caption{Energy bands adopted for the X-ray fluxes in Fig. \ref{Flux_comparison}.}
\begin{tabular}{c|c}
\hline\hline
Band name & Energy range (keV)\\
\hline
SiLine & 1.79-2.01 \\
SiCont & 2.47-2.67 \\ 
\hline
FeLine & 1.2-1.4 \\
FeCont & 2.05-2.15 \\
\hline
\end{tabular}
\label{tab:bands}
\end{table}
The SiLine band includes the Si XIII (1.865 keV) and Si XIV (2.006 keV) emission lines, while the SiCont band has been chosen to cover the emission at energies slightly above that of the Si XIII recombination edge (2.438 keV) and excluding the Si XIV recombination edge (2.673 keV). The FeLine band includes the forest of Fe L-lines of ions Fe XVII-XXIII, while the FeCont band includes the emission at energies above the Fe XXIV recombination edge (2.023 keV). The values of the ionization energy are taken from \citet{lid03} and the line energies are taken from the AtomDB WebGuide\footnote{http://www.atomdb.org/Webguide/webguide.php.}.

In the following, I present the results obtained for Si and Fe in two different subsections. In any case examined, the results obtained for Fe are analogous to those obtained for Si, with the only difference that Fe emission shows a very complex line pattern up to energies close to those of its RRC.

\subsection{Spectral simulations for Si-rich ejecta}
\label{sect:ss_Si}
\begin{figure}[!ht]
    \centering
    \includegraphics[width=0.85\columnwidth]{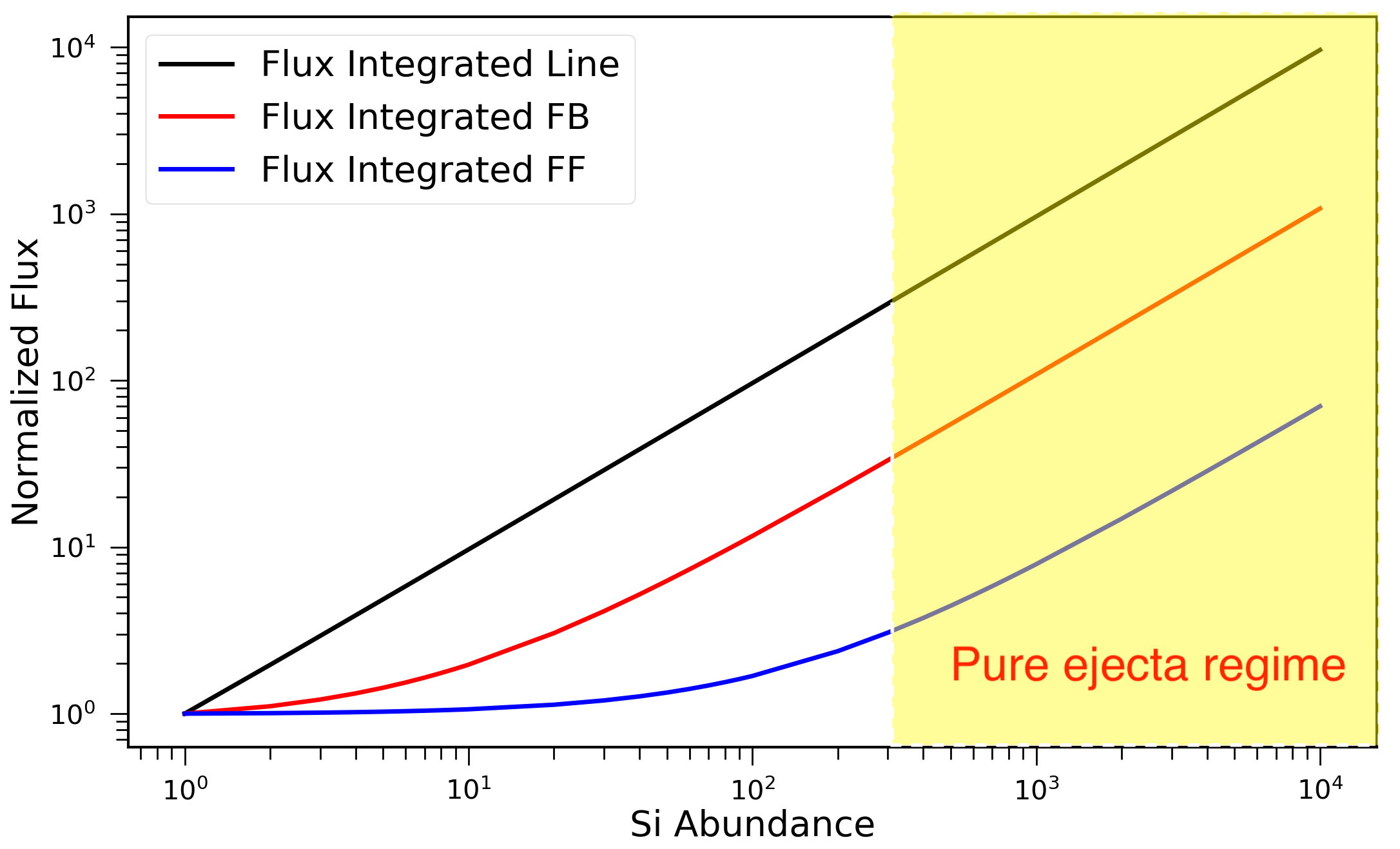}
    \caption{X-ray fluxes of different emission processes in the corresponding energy bands (see Table \ref{tab:bands}) as a function of Si abundance for a plasma temperature of 1 keV. All values are normalized to those obtained for solar abundances. CIE conditions are assumed \citep{gvm20}.}
\label{Flux_comparison}
\end{figure}

Figure \ref{Flux_comparison} shows the flux for each emission process (normalized to that obtained for solar abundances) as a function of Si abundance. To produce this plot, I assumed a plasma temperature of 1 keV and calculated the flux of each process in the corresponding energy band, namely SiLine band for line emission and SiCont band for FF and FB processes. The flux of line emission increases linearly with the abundance (as predicted by Eq. \ref{eq:line_emiss}). The total FB emission shows a weak increase for abundance values between 1 and 10, because the FB emission associated with Si is only a fraction of the total FB emission. For $ A_{\mathrm{Si}} \gtrsim $ 10, the total FB emission is due mainly to the Si FB and  the FB flux depends linearly on the Si abundance (as predicted by Eq. \ref{eq:fb}). 

The FF emission is substantially insensitive to the increasing abundance until values of a few hundreds are reached; then, it increases linearly with the abundance like the FB and the line processes. This is because for abundance values $A_{\mathrm{Si}}\lesssim 10^2$, the FF emission produced by electrons originally belonging to H (hereafter H-electrons) overcomes that of electrons stripped from Si (Si-electrons). For a solar abundance, the number of H-electrons is about $10^4$ times the number of Si-electrons (Table \ref{tab:ab_lodd} and Eq. \ref{eq:brems_emiss}). Therefore, if the Si abundance increases only slightly, the global contribution to the FF emission is still mainly associated with H-electrons (and H ions) and bremsstrahlung emission is not affected by the Si abundance.

\begin{figure}[!ht]
    \centering
    \includegraphics[width=0.99\columnwidth]{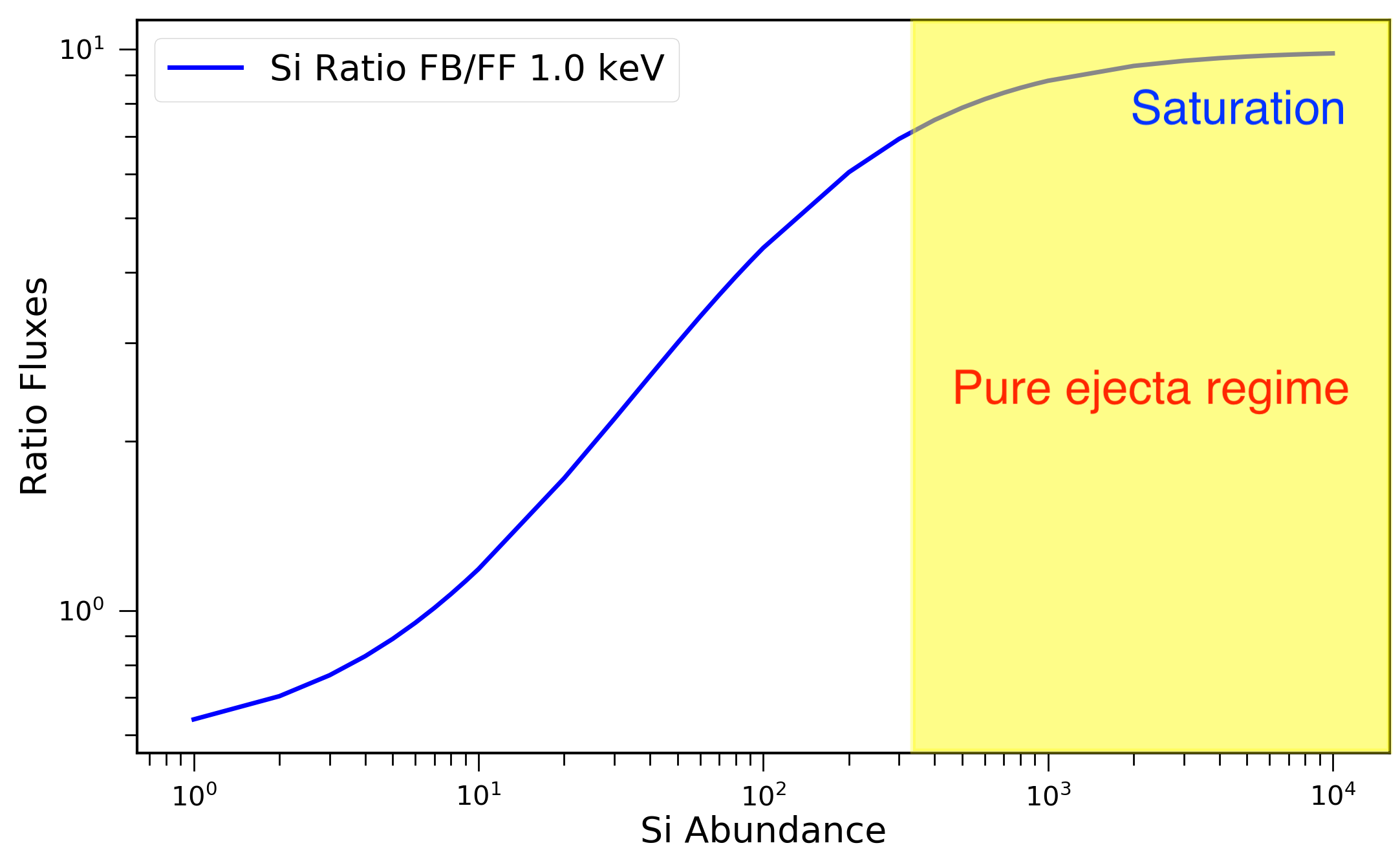}
    \caption{Ratio between the FF and the FB flux values shown in Fig. \ref{Flux_comparison} as a function of the Si abundance for a plasma temperature of 1 keV in CIE.}
    \label{RatioFBFF}
\end{figure}

If abundances of the order of a few hundreds are reached, the number of Si-electrons is not negligible with respect to the number of H-electrons. Moreover, in this regime, which I call the \emph{pure-metal ejecta regime}, the contribution of Si ions to the electron scattering becomes important. For this and higher Si abundance, therefore, the term including the Si contribution becomes the dominant one in the summation of  Eq. \ref{eq:brems_emiss} (given the dependence on $Z_{\mathrm{i}}^2$, with Z=12 for He-like Si) and we thus observe the expected linear increase with the abundance. Figure \ref{RatioFBFF} shows the FB over the FF flux ratio: the observed flattening of the flux ratio reflects the pure-metal ejecta regime discussed so far.

\begin{figure}[!ht]
    \centering
    \includegraphics[width=0.99\columnwidth]{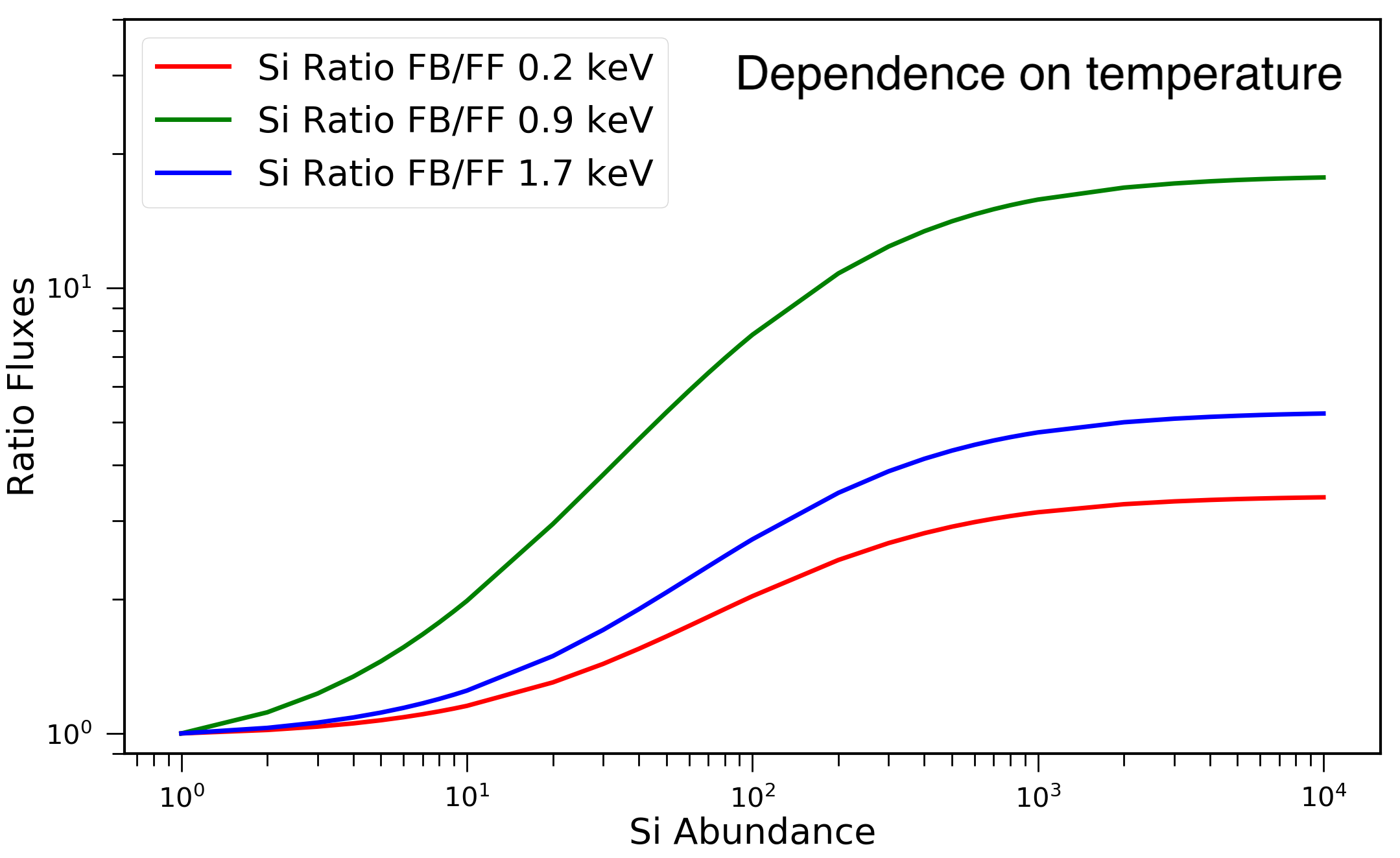}
    \caption{Same as Fig. \ref{RatioFBFF} for plasma temperatures of 0.2 keV (red line), 0.9 keV (green), and 1.7 keV (blue).}
    \label{RatioFBFF2}
\end{figure}

Since the FB-to-FF flux ratio depends also on the plasma temperature, I repeated the simulations described above by exploring energies in the range $kT=0.2-3$ keV (separated with a step of 0.1 keV). Figure \ref{RatioFBFF2} shows the FB to the FF flux ratios for three different temperatures, namely $kT=0.2$ keV, $kT=0.9$ keV, and $kT=1.7$ keV. By increasing the temperature in the range 0.2 to 0.8 keV, the slope $\sigma$ of the FB to FF ratio as a function of Si abundance increases; that is, for a high plasma temperature, the FB contribution becomes higher than that of the FF. However, at $kT\sim$ 0.9 keV, $\sigma$ reaches its maximum and a further increase in temperature leads to lower values of $\sigma$. The observed trend at energies below 0.9 keV is due to the increasing degree of Si ionization and the subsequent higher number of free electrons combined with the increasing width of the RRC, which is spread out at higher temperatures thus reducing the effect in the 2.47-2.67 keV band considered here. On the other hand, at energies above this threshold, the electrons are so energetic that they can escape recombination while still increasing the FF emission. 

The value of the temperature threshold depends on the element considered, because of the different degree of ionization at the given temperature and of the corresponding number of vacancies in the ion itself. Heavier elements have higher thresholds (see Fig. \ref{fluxfe} for the Fe case) also because of the stronger electrostatic field. I found that, in the Si case, the FB to FF ratio has its maximum when the electron temperature is in the range $kT=0.6-1.2$ keV. In this wide temperature range, a high metallicity can act as a boost for FB emission, by contributing more than the thermal bremsstrahlung. 

\newpage
\subsection{Spectral simulations for Fe-rich ejecta}
\label{sect:ss_Fe}
\begin{figure}[!ht]
    \centering
    \includegraphics[width=0.75\columnwidth]{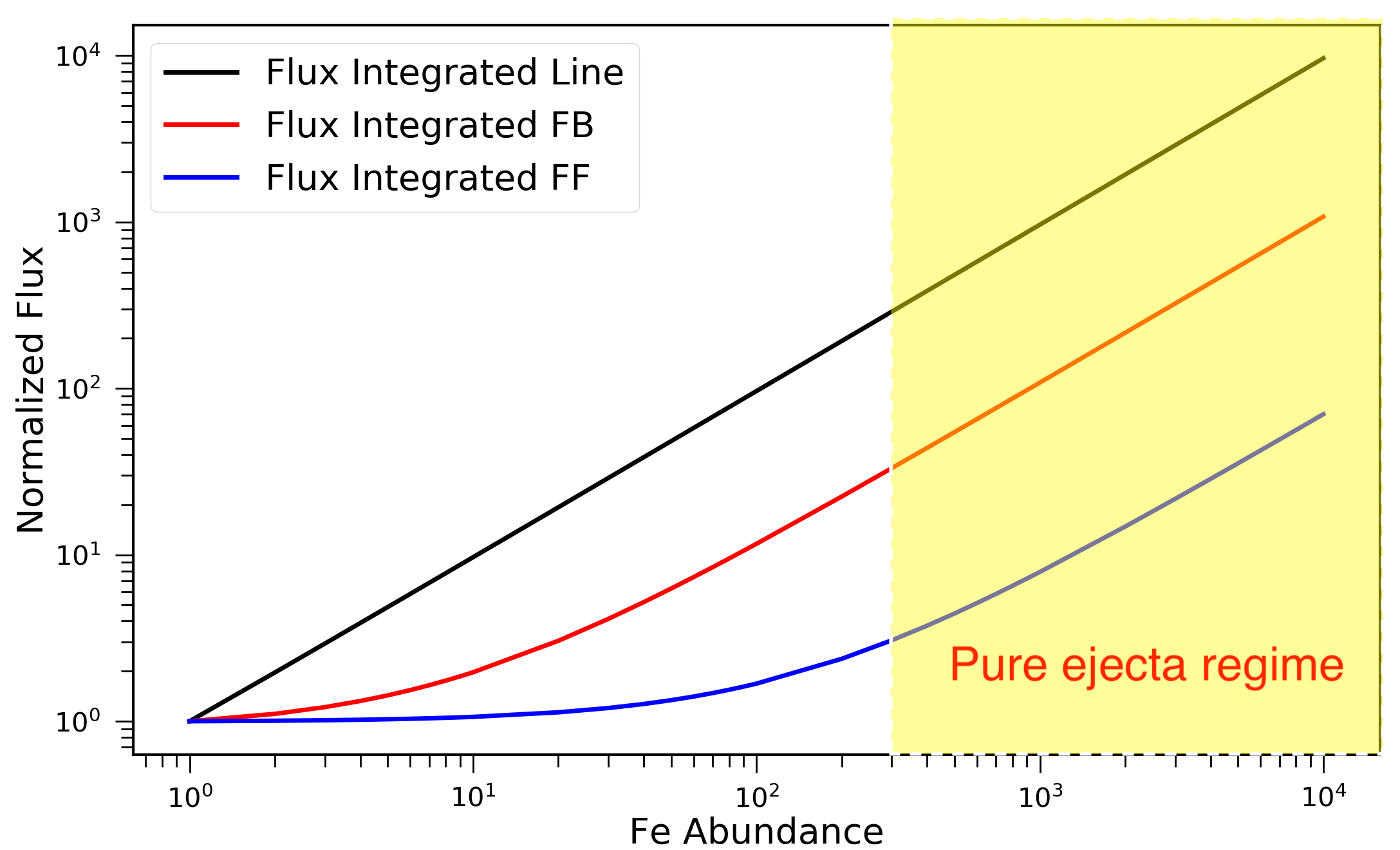}
    \includegraphics[width=0.75\columnwidth]{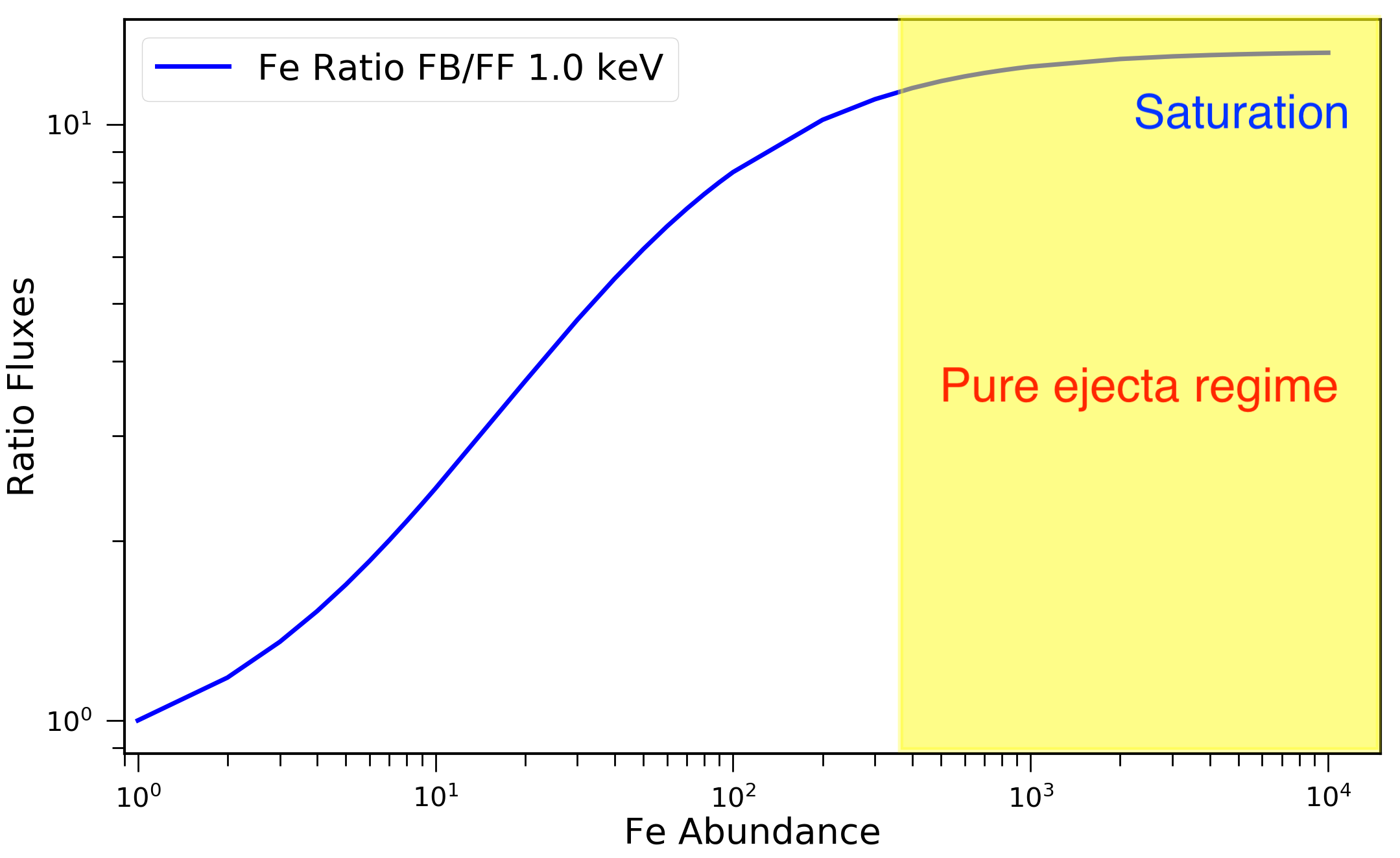}
    \caption{X-ray fluxes of different emission processes in the corresponding energy bands (see Table \ref{tab:bands}) as a function of Fe abundance. \emph{Upper panel}: Same as Fig. \ref{Flux_comparison} but for Fe. \emph{Lower panel}: Ratio between the FF and the FB flux values shown in the upper panel as a function of Fe abundance for a plasma temperature of 1 keV.}
    \label{fluxfe}
\end{figure}

In this section, I show the results obtained performing the analysis described in the previous section also for Fe-rich ejecta. Figure \ref{fluxfe} shows the flux for each emission process (normalized to that obtained for solar abundances) as a function of Fe abundance. To produce this plot, I assumed a plasma temperature of 1 keV and calculated the flux of each process in the corresponding energy band, namely FeLine band for line emission and FeCont band for FF and FB processes. 

\begin{figure}[!ht]
    \centering
    \includegraphics[width=0.99\columnwidth]{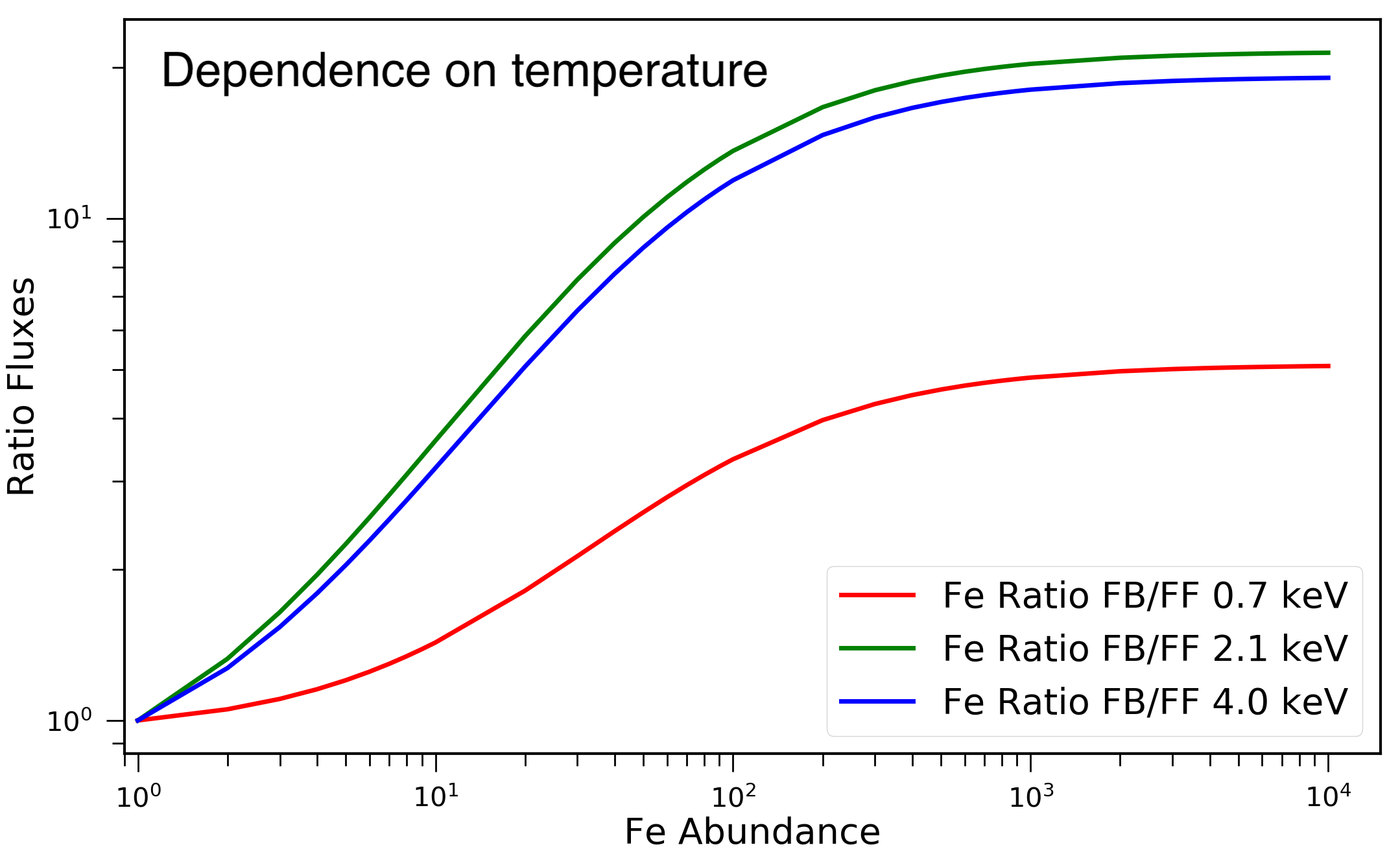}
    \caption{Same as lower panel of Fig. \ref{fluxfe} for plasma temperatures of 0.7 keV (red line), 2.1 keV (green), and 4.0 keV (blue).}
    \label{TdependFe}
\end{figure}

The trend is the same as that discussed for the Si case with saturation occurring when the Fe abundance is close to 100. Also for this element, I investigate how the FB to FF flux ratio depends on the temperature (Fig. \ref{TdependFe}). I repeated the simulations of the previous section by exploring temperatures in the range $kT = 0.5-5$ keV (with a step of 0.1 keV). The slope $\sigma$ of the ratio increases for $kT < 1.5$ keV and decreases for $kT > 3.5$ keV. 

Here, the threshold values are higher than those found for Si because of the stronger electrostatic field, requiring higher temperature to further ionize the Fe-ion and higher energy for the electrons to escape from recombination.
 I also notice that there are different recombination edges, corresponding to the different Fe ionization states, and that the brightest RRC depends on the temperature and on the degree of ionization of the plasma. In any case, the conclusions obtained for a specific Fe RRC are also valid for the other ones.
As for the case of Si, also high elemental abundance raises the contribution of FB emission more than that of FF radiation.
In conclusion, I expect that a spectral signature of pure-metal ejecta might be related to the enhanced FB emission. Therefore, I carried a thorough study of radiative recombination continua and edges to develop a diagnostic tool aimed at revealing pure metal ejecta in SNRs, as described in Sect. \ref{sect:allsynthesis}.

\section{Synthetic X-ray spectra}
\label{sect:allsynthesis}

To investigate the observability of pure-metal ejecta emission in SNRs, I produced synthetic spectra by folding the spectral models with the response matrix of actual detectors.
To mimic actual conditions, I also included the contribution of the ISM X-ray emission. The main idea is that pure-metal ejecta emission should reveal itself through an enhanced RRC emission. Pure-metal ejecta can be distributed on large, expanding, shells as well as concentrated in dense clumps embedded in an environment of shocked ISM, where the mixing between ISM and ejecta is less remarkable (as \citealt{hl03} found for the Fe cloudlets in Cas A). In any case, the ejecta emission is always superimposed onto the emission stemming from the shocked ISM.

To synthesize the X-ray spectra, I used the response matrices of {\it Chandra}/ACIS-S, a CCD detector on board the {\it Chandra} X-ray telescope\footnote{https://heasarc.gsfc.nasa.gov/docs/chandra/chandra.html.} and Resolve, a microcalorimeter that will be on board {\it XRISM}\footnote{http://xrism.isas.jaxa.jp.}, the JAXA/NASA X-ray telescope to be launched in Japanese fiscal year 2022. I point out that using the {\it XMM-Newton}/MOS response matrix instead of the {\it CHANDRA}/ACIS-S leads to analogous results. 
{\it XRISM} will have lower effective area and spatial resolution than {\it Chandra} but, thanks to the microcalorimeter-based focal plane detector, its spectral resolution will be better than that of {\it Chandra} by a factor of fifty (for a more detailed description of the two telescopes see Appendix \ref{app:telescopes}). In this section, I assume a generic value for the ISM column density equal to $n_{\mathrm{H}}=5\times10^{21}$ cm$^{-2}$.


I will discuss the case of Si-rich ejecta in Sect. \ref{sect:synth_Si} and then the Fe-rich ejecta scenario in Sect. \ref{sect:synth_Fe}.

\subsection{Synthesis of pure-Si spectra}
\label{sect:synth_Si}
\subsubsection{Synthesis of {\it Chandra}/ACIS-S spectra}
\label{chandrasynth}
\begin{figure}[!ht]
    \centering
    \includegraphics[width=0.9\columnwidth]{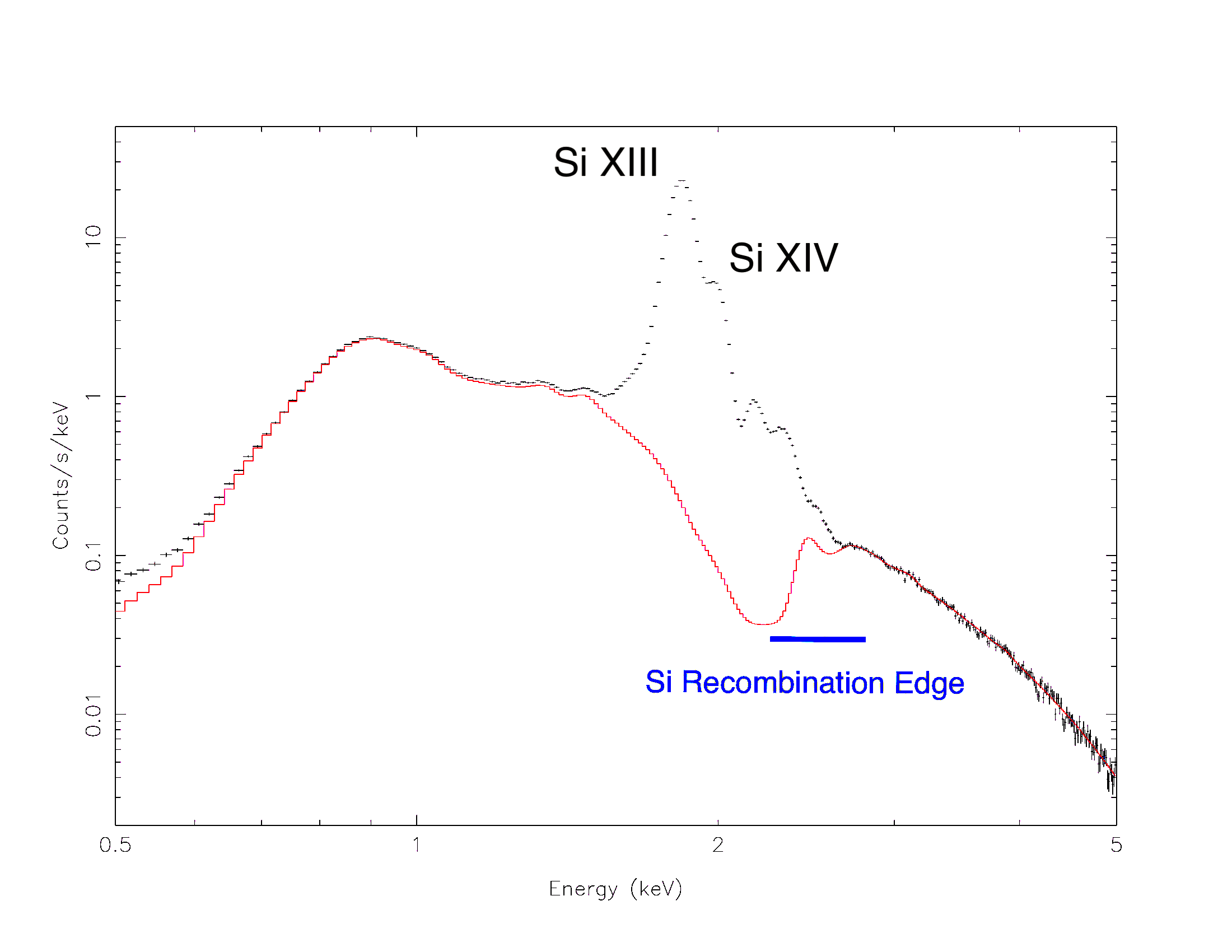}
    \caption{Black: Synthetic {\it Chandra}/ACIS-S spectrum of a CIE plasma with abundance of all elements set to 3, with the exception of the Si abundance set to 300; $kT=0.8$ keV and EM=1.5$\times10^{55}$ cm$^{-3}$ (pure-metal model in Table \ref{bestsim}). Red solid: Same spectrum without the Si line emission (Si contributions to the FF and FB continuum are still included.)}
    \label{lineonly}
\end{figure}

I simulated a {\it Chandra}/ACIS-S synthetic spectrum (assuming 1 Ms of exposure time) of a plasma in CIE with element abundance set to 3, with respect to the solar one, for all the elements except for Si, which was set to 300, and an electron temperature of 0.8 keV (Fig. \ref{lineonly}, pure-metal model in Table \ref{bestsim}). The Si abundance was so high that the corresponding emission lines dominate the whole spectrum. Even if this scenario is not realistic, because I have not included the contribution of the ISM X-ray emission yet, it allows us to notice that the corresponding spectrum does not show any spectral signatures related to the RRC emission, which is instead expected to be present on the basis of the study of the fluxes discussed in the previous section.
This is because the poor resolution of the CCD spectrometer makes the He-Si and H-Si lines heavily broadened. The spill-over of the line emission due to the coarse spectral resolution blurs the recombination edges, thus hiding the spectral signature of pure-metal ejecta emission. In fact, if we remove the Si line emission from the synthetic spectrum (Si contributions to the FF and FB continuum are still included), the resulting spectrum (in red in Fig. \ref{lineonly}) shows, as expected, a prominent edge of recombination at the characteristic energy of Si-RRC. 

\begin{figure}[!ht]
    \centering
    \includegraphics[angle=270,width=0.7\columnwidth]{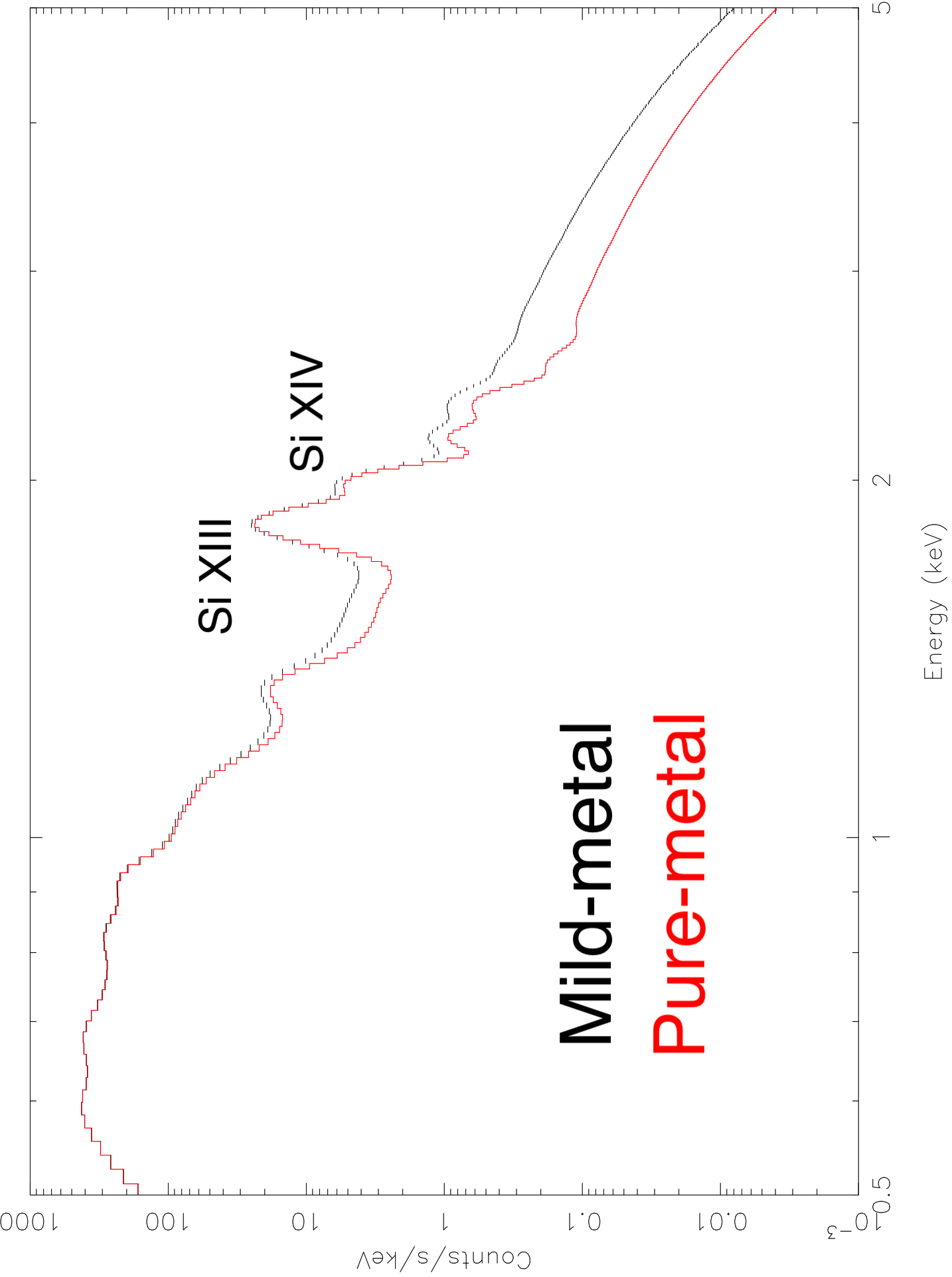}
    \caption{Synthetic {\it Chandra}/ACIS-S spectra of ISM+pure-Si (pure-metal model, red) and ISM+Si-rich (mild-metal model, black) plasmas.}
    \label{SimMOS}
\end{figure}

The measurement of the enhanced FB emission is further hampered by the contamination from shocked ISM in the spectra, as shown below. As an example, here I show a synthetic SNR spectrum by including ejecta and ISM emission (a more realistic simulation, performed for a specific case, is presented in Sect \ref{introCasA}). I considered a spherical clump of Si-rich ejecta (Si abundance set to 300, as before) with radius $R_{\mathrm{clump}}=0.5$ pc and temperature $kT_{\mathrm{clump}}=0.8$ keV, surrounded by a colder ISM with temperature $kT_{\mathrm{ISM}}=0.15$ keV. 
I assume a particle density $n_{\mathrm{e}} = 3$ cm$^{-3}$ and pressure equilibrium between the clump and the ISM and extract the spectrum from a box corresponding to a region of $8~\rm{pc} \times 8 ~\rm{pc}$ in the plane of the sky and extending $8~\rm{pc}$ along the line of sight. Under these assumptions, the ISM emission measure is four orders of magnitude larger than that of the clump (see Table \ref{bestsim} for details). 
\begin{table}[!ht]
    \centering
      \caption{Parameters of the ISM (in CIE), highly Si-rich ejecta (pure-metal) and of the mildly Si-rich ejecta (mild-metal) models (in CIE) adopted for the spectral synthesis, with the corresponding Si and ejecta masses.}
    \begin{tabular}{c|c|c|c}
    \hline\hline
    Parameter & ISM & Pure-metal& Mild-metal\\
    \hline
    EM (cm$^{-3}$) & $1.6 \cdot 10^{59}$ & $1.5 \cdot 10^{55}$ & $1.5 \cdot 10^{57}$\\
    kT (keV)& 0.15 & 0.8& 0.8 \\
    Si Abundance& 1& 300& 3 \\ 
    \hline
    Si mass (M$_{\odot}$)&  /& 0.015 & 0.0016 \\
    Ejecta mass (M$_{\odot}$)& / & 0.06 & 0.6 \\
    \hline
    \end{tabular}
    \label{bestsim}
\end{table}
I synthesized the {\it Chandra}/ACIS-S X-ray spectrum using the ISM+pure-metal ejecta model described in Table \ref{bestsim}. I assumed a distance of 1 kpc and an unrealistically high exposure time of 10$^8$ s in order to highlight the features. In Sect. \ref{introCasA} I will discuss cases with a more realistic exposure time. Figure \ref{SimMOS} shows the resulting synthetic {\it Chandra}/ACIS-S spectrum.

As a comparison, I also produced a spectrum starting from the mild-metal model described in Table \ref{bestsim} (third column), namely a model with the same parameters as those of the pure-metal model except for the Si abundance, reduced by a factor $f=100$ and set to 3 (instead of 300), and for the ejecta EM, enhanced by a factor $f$ and set to $1.5 \times 10^{57}$ cm$^{-3}$.
This new spectrum is shown in Fig. \ref{SimMOS}. A comparison between the two spectra does not reveal any clear difference that could be related to the pure-metal ejecta emission (even with an unrealistically high exposure time of 10$^8$ s). By fitting the spectrum, synthesized from the pure-metal model, with the Si abundance free to vary, it is not possible to recover univocally the Si abundance input value. A more detailed analysis is presented in Sect. \ref{synthdata}. 

Spectra in Fig. \ref{SimMOS} confirm that the degeneracy between abundance and emission measure is a serious issue, which is intrinsically due to the instrumental characteristics of the CCD cameras and does not depend on the statistics of the observation.

The ability to identify the presence of FB contributions offers a unique diagnostic tool to assess whether the spectrum is stemming from a highly enriched, but still hydrogen-dominated plasma, or from a pure-metal ejecta plasma.

\subsubsection{Synthesis of {\it XRISM}/Resolve spectra}
Here, I further show that the degeneracy between abundance and emission measure is intrinsically due to the instrumental characteristics of the CCD cameras and does not depend on the statistics of the observation. I repeated the spectral simulations discussed above by folding the pure-metal model and the mild-metal model with the {\it XRISM}/Resolve response matrix\footnote{{\it XRISM}/Resolve response and ancillary files used are xarm\_res\_h5ev\_20170818.rmf and xarm\_res\_flt\_fa\_20170818.arf, available on https://heasarc.gsfc.nasa.gov/docs/xrism/proposals/.}. 

\begin{figure}[!ht]
    \centering 
    \includegraphics[width=0.9\columnwidth]{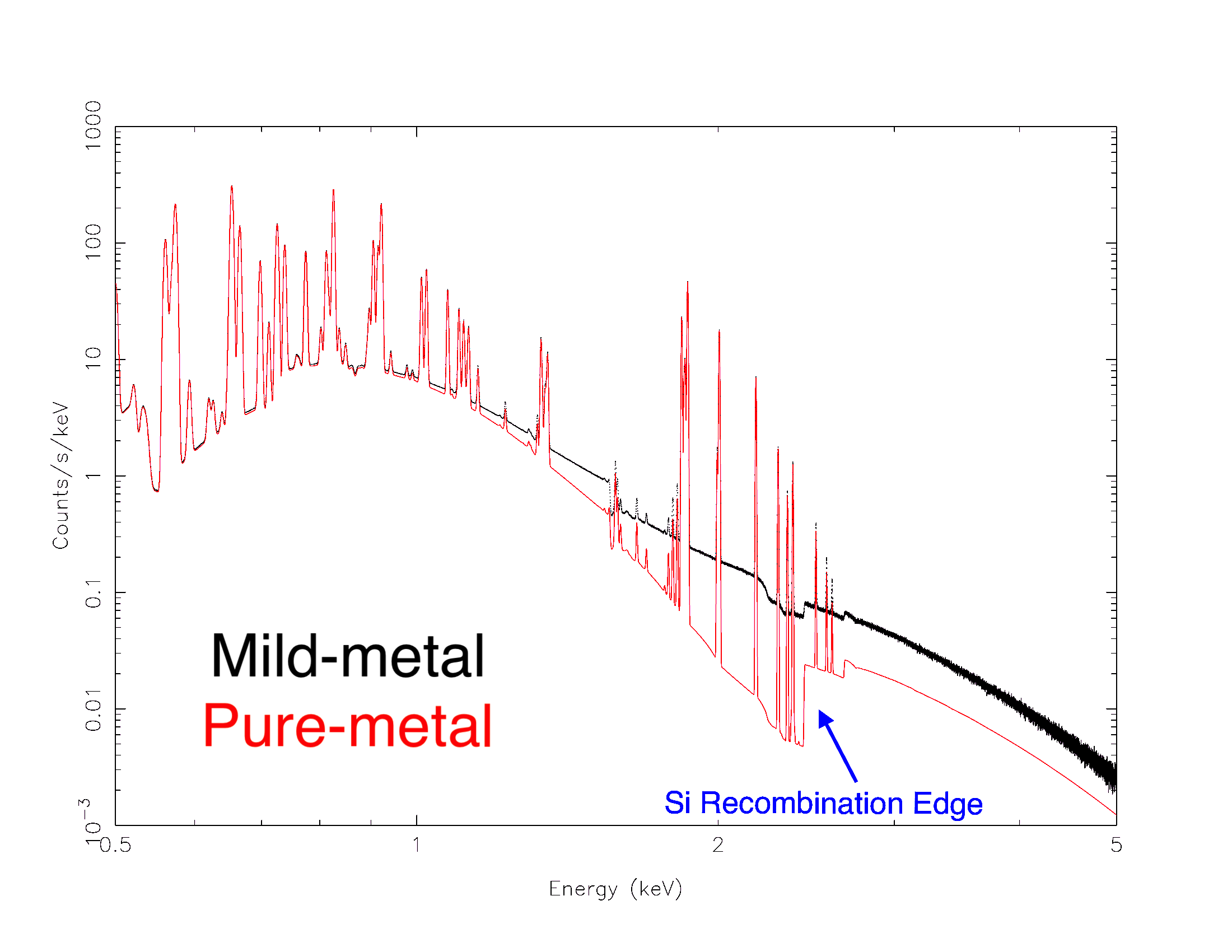}
    \caption{Same as Fig. \ref{SimMOS} but the model is folded through the {\it XRISM}/Resolve response matrix.}
    \label{SimMicro}
\end{figure}

Figure \ref{SimMicro} shows the pure-metal case and the mild-metal case, again assuming an exposure time of 10$^8$ s. 
Thanks to the high spectral resolution of the microcalorimeters, it is now possible to observe a clear spectral difference between the two scenarios. As expected, on the basis of the study presented in Sect. 2, a bright edge of recombination shows up at $\sim2.5$ keV (i.e., the He-like Si RRC typical energy) when the abundance of Si is 300. Figure \ref{SimMicro} also clearly shows that the recombination edge and the RRC are much dimmer in the mild-metal case.
I stress that, in the pure-metal ejecta regime, even if the bremsstrahlung emission from the shocked ISM enhances the continuum emission and strongly reduces the equivalent width of emission lines, the RRC still emerges above the FF emission at energies $>2.5$ keV. Therefore, according to these simulations, the enhanced FB emission is a better tracer of pure-metal ejecta than the line equivalent width.

High resolution spectrometers like {\it XRISM}/Resolve (and, in the future, {X-IFU on board the Advanced Telescope for High-Energy Astrophysics, ATHENA) are therefore capable of pinpointing the enhancement of the FB emission associated with a plasma with extremely high metallicity.

\subsection{Synthesis of pure-Fe spectra}
\label{sect:synth_Fe}

 I synthesized a pure-Fe spectrum with abundances of all elements, except Fe, set to 3, Fe abundance set to 300, kT=1.5 keV, and a number density of 3 particles per cm$^3$. I considered a distance of 1 kpc and an absorbing column density of $n_{\mathrm{H}}= 5 \times 10^{21} \rm{cm}^{-2}$. The resulting spectrum (black crosses in Fig. \ref{onlyfe}), synthesized assuming that the emission originates from a clump with radius $R_{\mathrm{clump}}=$1.6 pc, shows the same issues faced with Si synthesis, due to the limited spectral resolution of the CCD detectors. 
 
 \begin{figure}[!ht]
    \centering
    \includegraphics[width=0.9\columnwidth]{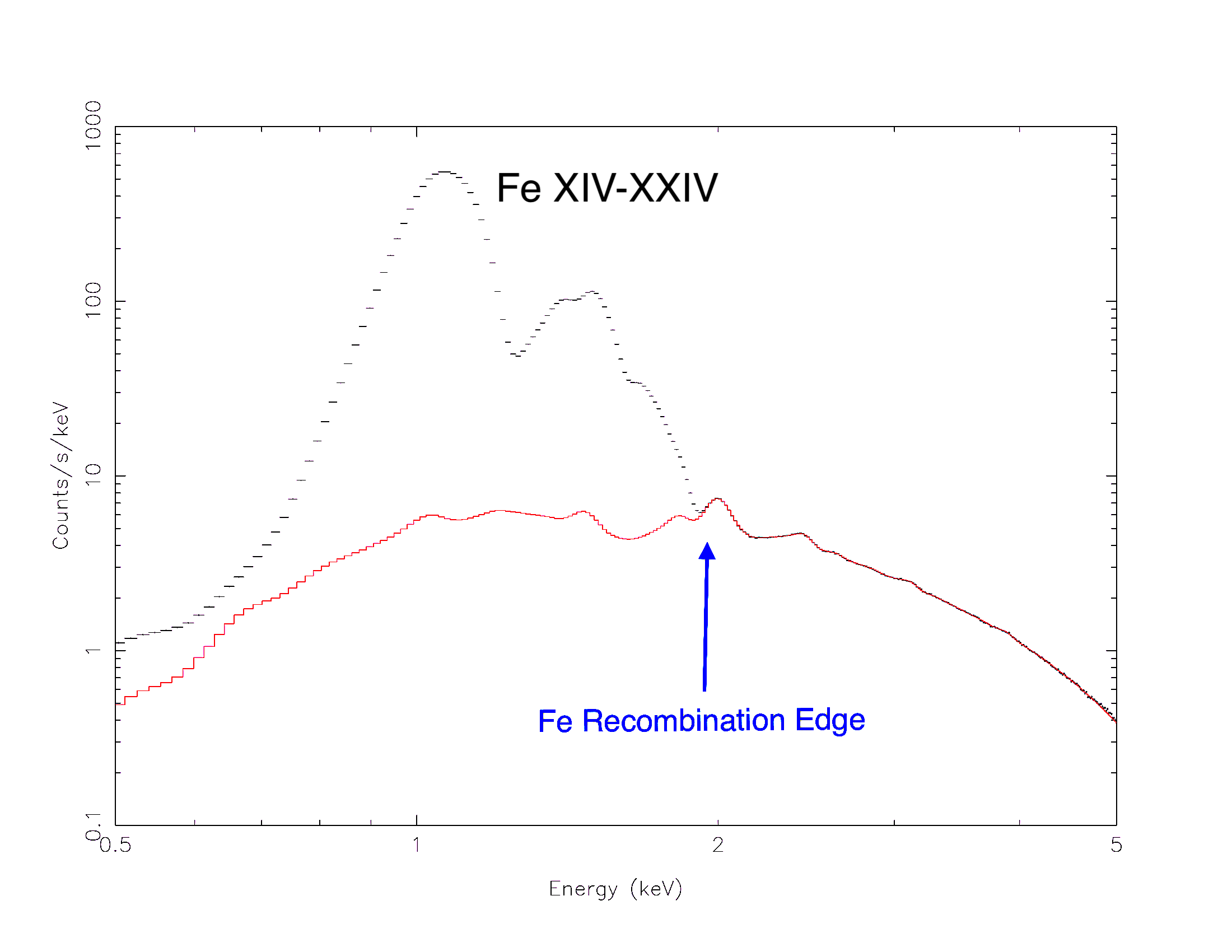}
    \caption{Synthetic {\it Chandra}/ACIS-S spectra with high Fe abundance. Black: Synthetic {\it Chandra}/ACIS-S spectrum of a CIE plasma with abundance of all elements but Fe set to 3, Fe abundance set to 300, $kT=1.5$ keV, and EM=$1.5\times10^{56}$ cm$^{-3}$. Red solid: Same spectrum but considering only continuum emission.}
    \label{onlyfe}
\end{figure}
 
Here, I chose a clump larger than that of the Si case because the Fe RRC lies in a part of the spectrum where the ISM emission is more significant (i.e., the ejecta FB emission is less visible). A more realistic case, aimed at studying the Fe-rich ejecta in Cas A, is presented in Sect. \ref{introCasA}. In addition, the scenario is even more complex because of the large amount of Fe lines (from Fe XIV to Fe XXIV) at energies around 1 keV. Fig \ref{onlyfe} shows that the Fe RRC is definitely undetectable in the CCD spectra. As for the case of Si-rich ejecta, the Fe RRC sticks out only by removing the line emission from the synthetic spectra. In particular, the red solid line in Fig. \ref{onlyfe} shows the continuum emission only, and reveals a recombination edge at the characteristic energy of Fe XXIV (2.023 keV). By ignoring the line emission of all the elements, I notice that the effective continuum contribution (red solid line of Fig. \ref{onlyfe}) to the emission shows a recombination edge at the characteristic energy of the Fe XXIV RRC. 

I then produced synthetic spectra by adding another CIE component related to the ISM emission, considering the same configuration as that described in Sect. \ref{chandrasynth} for the Si case. I chose $T_{\mathrm{ISM}}$= 0.23 keV and $T_{\mathrm{clump}}$=1.5 keV. Even though the high temperature chosen for the clump maximizes the FB emission (as discussed above), I here show that CCD spectrometers cannot reveal the recombination edge. The parameters used for this pure-metal model and the corresponding Fe and total ejecta mass are summarized in Table \ref{bestsimfe}. The produced spectra, folded with {\it Chandra}/ACIS-S are shown in the upper panel of Fig. \ref{Fesim}.

\begin{table}[!ht]
    \centering
     \caption{Parameters of the ISM (in CIE), highly Fe-rich ejecta (pure-metal) and of the mildly Fe-rich ejecta (mild-metal) models (in CIE) adopted for the spectral synthesis with the corresponding Fe and ejecta masses.}
    \begin{tabular}{c|c|c|c}
    \hline\hline
    Parameter & ISM & Pure-metal & Mild-metal\\
    \hline
    Emission measure (cm$^{-3}$) & $1.6 \cdot 10^{59}$ & $1.5 \cdot 10^{56}$ & $1.5 \cdot 10^{58}$  \\
    Temperature (keV)& 0.23 & 1.5 & 1.5 \\
    Fe abundance& 1& 300& 3 \\ 
    \hline
    Fe mass (M$_{\odot}$) & / & 0.3& 0.04\\
    Ejecta mass (M$_{\odot}$)& / & 0.6& 6\\
    \end{tabular}
    \label{bestsimfe}
\end{table}

I also produced a spectrum considering the mild-metal scenario (Table \ref{bestsimfe}, third column) in which the Fe abundance is set to 3 (instead of 300) and the ejecta EM is set to $1.5 \times 10^{58} \, \mathrm{cm}^{-3}$. This spectrum is shown in Fig. \ref{Fesim}. The results of the simulations performed on Fe are analogous to those obtained for Si and confirm that the instrumental line broadening completely hides the enhanced RRC related to the pure-metal ejecta emission.

I then synthesized the spectra by considering the {\it XRISM}/Resolve response. Lower panel of Fig \ref{Fesim} shows the pure-metal and mild-metal cases, assuming an exposure time of 10$^8$ s. Thanks to the high spectral resolution of {\it XRISM}/Resolve, the Fe RRC shows up a 2.02 keV in the pure-metal case. As for the case of Si-rich ejecta, here I find that Fe-rich ejecta can be revealed with microcalorimeters thanks to their enhanced RRC emission.

\begin{figure}[htb!]
    \centering
    \includegraphics[angle=270,width=0.8\columnwidth]{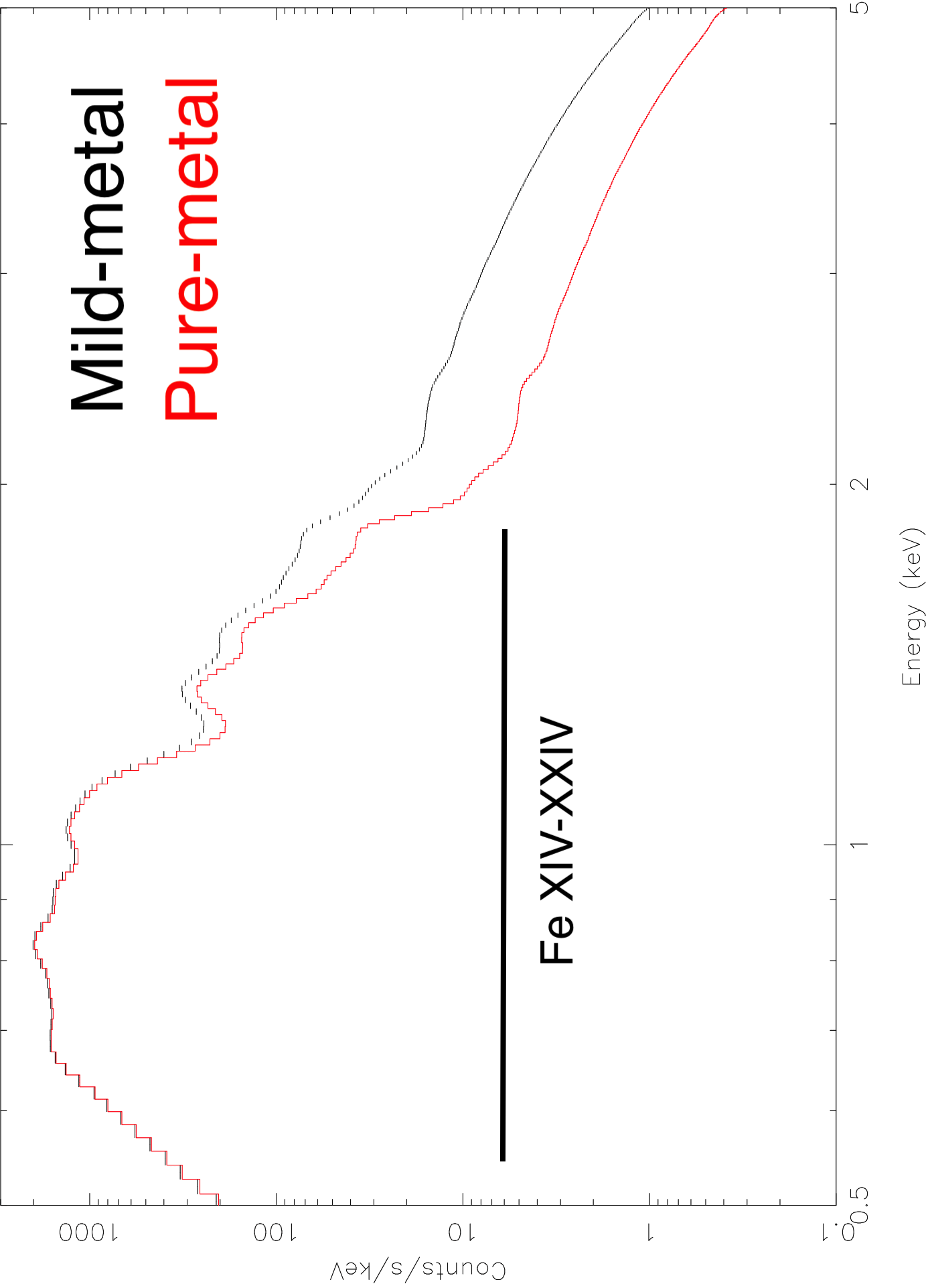}
    \includegraphics[width=0.9\columnwidth]{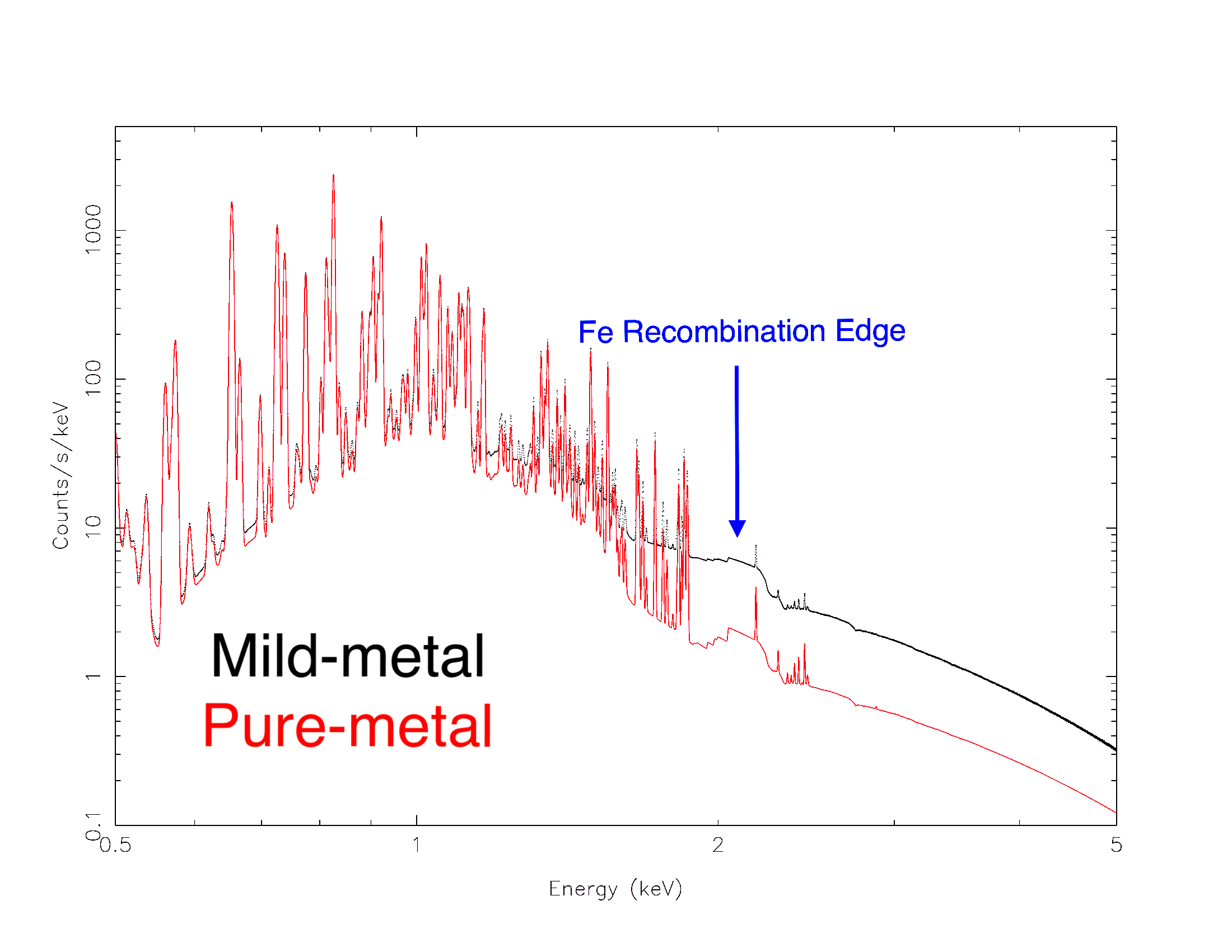}
    \caption{\emph{Upper panel}: Synthetic {\it Chandra}/ACIS-S spectra of ISM+pure-Fe (pure-metal model, red) and ISM+Fe-rich (mild-metal model, black) plasmas. \emph{Lower panel}: Same as upper panel but the model is folded with the {\it XRISM}/Resolve response matrix.}
    \label{Fesim}
\end{figure}

\newpage

\section{Self-consistent X-ray synthesis tool}
\label{sect:tool_synth}
I developed a tool to self-consistently synthesize thermal X-ray emission from HD/MHD simulations of SNRs. In each cell of the computational domain, I extract the local value of temperature, electron density, ionization age $\tau$, total mass, and mass tracer of each element. The mass tracer describes the mass of a given species as a fraction of the total mass in the cell. I use these values as input parameters for the non-equilibrium of ionization, optically thin plasma model \emph{neij} (\citealt{kj93}) based on the atomic database SPEXACT 2.07.00, within SPEX (\citealt{spex}). In particular, for each atomic species $i$, I estimate the local value of ion and electron density and synthesize the corresponding pure species spectrum in each computational cell\footnote{The latter step is done by setting the abundance of all elements, except $i$ equal to 0, while $i$ abundance is set to 1}. I then sum the resulting spectra over all the species by weighting each term for the corresponding emission measure. Finally, I sum the spectra of each cell within a given region of the domain to extract its global spectrum. The tool also allows to set column density and distance appropriate to the given source. Each spectrum can be folded through any desired X-ray instrument response matrix. The resulting spectra are binned using the optimal bin tool present in SPEX (\citealt{kb16}). I stress that this tool does not require any assumption on the abundances set (as in \citealt{mob19}, for instance), since these are self-consistently recovered from the HD/MHD model.

Beside of the application to the SNR Cas A, described in Sect. \ref{introCasA}, I also used this tool to synthesyze X-ray spectra of an evolving SNR characterized by numerous anisotropies (Sect \ref{sect:hd_anto}), and to compare synthetic and actual X-ray spectra of the SNR IC 443 (Sect. \ref{sect:ic443_hd}).
\newpage
\clearpage
\section{Pure-metal ejecta in Cas A}
\label{introCasA}

 The spectral analysis described in Sect \ref{sect:allsynthesis} shows that the enhancement in the RRC emission can be a strong signature of pure-metal ejecta and that such a spectral signature can be detected with high resolution spectrometers, while being almost impossible to observe with CCD detectors. I here apply this diagnostic tool to a real case, by focusing on Cas A (see Sect. \ref{sect:Intro_RRC}). In particular, I aim at understanding whether it will be possible to pinpoint pure-metal ejecta emission with the {\it XRISM}/Resolve spectrometer.
 

Here, I take advantage of the 3D HD simulation of Cas A performed by \cite{omp16} (hereafter O16). This state-of-the-art simulation models the evolution of Cas A from the immediate aftermath of the supernova to the complex interaction of the remnant with the ambient environment. In particular, I adopted the model configuration that best describes the observed ejecta distribution (run CAS-15MS-1ETA in O16, see O16 for the list of isotopes included in the simulation). This model reproduces the observed average expansion rate of the remnant and the shock velocities, and constrains the post-explosion anisotropies responsible for the observed structure and chemical distribution of ejecta. The model can reproduce very well the  shocked Fe distribution (both on large and relatively small spatial scales) while the Si (and S) mass seem to be slightly underestimated. I therefore focused on the Fe emission and adopted the HD simulation as a reference template to synthesize the expected X-ray emission. I point out that the remnant evolution modeled by O16 clearly shows that large regions of Cas A are expected to be filled with pure Fe-rich ejecta. 

I self-consistently produced synthetic {\it Chandra}/ACIS and {\it XRISM}/Resolve spectra of the southeastern Fe-rich clump in Cas A from the 3D HD simulation by using the X-ray spectral synthesis tool described in Sect. \ref{sect:tool_synth}. I assumed $n_{\mathrm{H}}=1.5 \times 10^{22}$ cm$^{-2}$ \citep{hl12} and a distance of 3.4 kpc (\citealt{rhf95}). I compare the synthetic spectra with those observed by {\it Chandra} and make predictions for future {\it XRISM}/Resolve observations. 

\subsection{Data analysis}
\label{data}
To compare synthetic products with actual data, I analyzed the {\it Chandra}/ACIS observation of Cas A with ID 114 (PI Holt), performed on 30/01/2000, by adopting the spectral tool XSPEC \citep{arn96}.
I used the tool \emph{fluximage} to produce a count-rate image of Cas A with a bin size of $1''$. Figure \ref{casadata} shows the count-rate image in the $0.3-7$ keV energy band, together with the regions selected for the spectral extraction. 
Within the southeastern Fe-rich clump, non-thermal emission from the Cas A reverse shock has been detected (\citealt{gkr01}) and mapped accurately (\citealt{vh08}). Since here I focus on thermal emission, I carefully selected the large box in the southeastern part of the shell to extract the Chandra X-ray spectrum, by excluding the reverse shock (the region is indicated by a black box in Fig. \ref{casadata}). I also considered a small extraction region centered on a bright Fe-rich knot (hereafter cloudlet), identified by \citet{hl03} and indicated by a red box in Fig. \ref{casadata}.
\begin{figure}[!ht]
    \centering
    \includegraphics[width=.9\columnwidth]{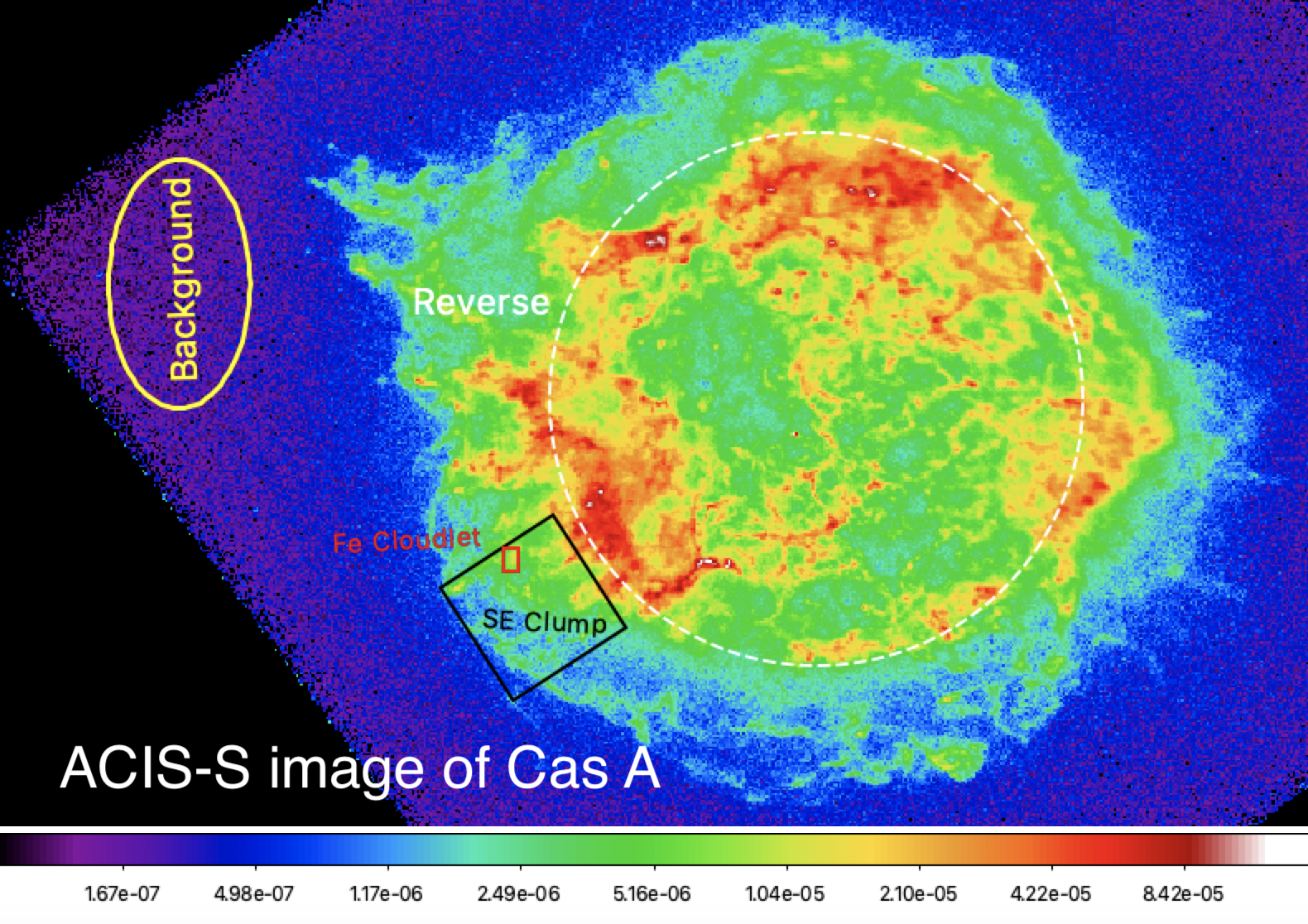}
    \caption{{\it Chandra} count-rate image of Cas A in the $0.3-7$ keV energy band with a logarithmic color scale. shown at the bottom of the image. The small red box marks the Fe-rich cloudlet, the black box indicates the region selected in the southeastern Fe-rich clump, the yellow ellipse shows the region chosen for the background, and the dashed white circle marks the nominal position of the reverse shock.}
    \label{casadata}
\end{figure}

I described the cloudlet spectrum by adopting the same model as \citet{hl03} and found best-fit parameters in good agreement with theirs, including an Fe/Si abundance ratio equal to $\sim 20$. In addition, I found that the absolute Fe abundance is not well constrained. In fact, two statistically equivalent fits with $\chi^2_{\mathrm{red}}=1.55$ (136 d.o.f.) can be obtained with $A_{\mathrm{Fe}} = 29$ (and EM $= 1.4 \times 10^{55}$ cm$^{-3}$ for the ejecta component) and $A_{\mathrm{Fe}} = 290$ (EM = $1.6 \times 10^{54}$ cm$^{-3}$).
 The cloudlet spectrum is shown in the upper panel of Fig. \ref{sinthspec}: the wide and bright spectral structure at energy $\sim$ 1 keV, reveals the presence of a remarkable complex of Fe L lines, but the Fe RRC is not visible. As explained in Sect. \ref{sect:allsynthesis}, the absence of this feature can be related to the instrumental characteristics of {\it Chandra}/ACIS (and of CCD detectors in general). 
 Upper panel of Fig. \ref{sinthspec} shows the spectrum obtained from the black box of Fig. \ref{casadata}, which presents a very strong Fe emission line complex, but with the addition of bright Si and S emission lines (the Fe/Si abundance ratio is lower than that in the Fe-rich cloudlet, being only $\sim 5$). This suggests the presence of both Fe-rich and Si-rich ejecta though I cannot exclude that the observed silicon emission may be somehow enhanced by dust scattering in this region.  

\subsection{Synthesis of Cas A spectra}
\label{synthdata}

From the O16 simulation, I selected a region in the southeastern Fe-rich clump with the same size as the black box chosen from the actual {\it Chandra} data. The resulting synthetic {\it Chandra}/ACIS-S spectrum, obtained assuming an exposure time of 1 Ms, is shown in the lower panel of Fig. \ref{sinthspec}. The spectrum clearly shows signatures of bright Fe line emission, similar to those actually observed in the corresponding region. At odds with the observations, the synthetic spectrum does not show very bright emission lines from Si and S ions, thus indicating that intermediate mass elements are somehow under-represented in the simulation in this particular region. 

\begin{figure}[!htb]
    \centering
    \includegraphics[width=0.58\columnwidth, angle=270]{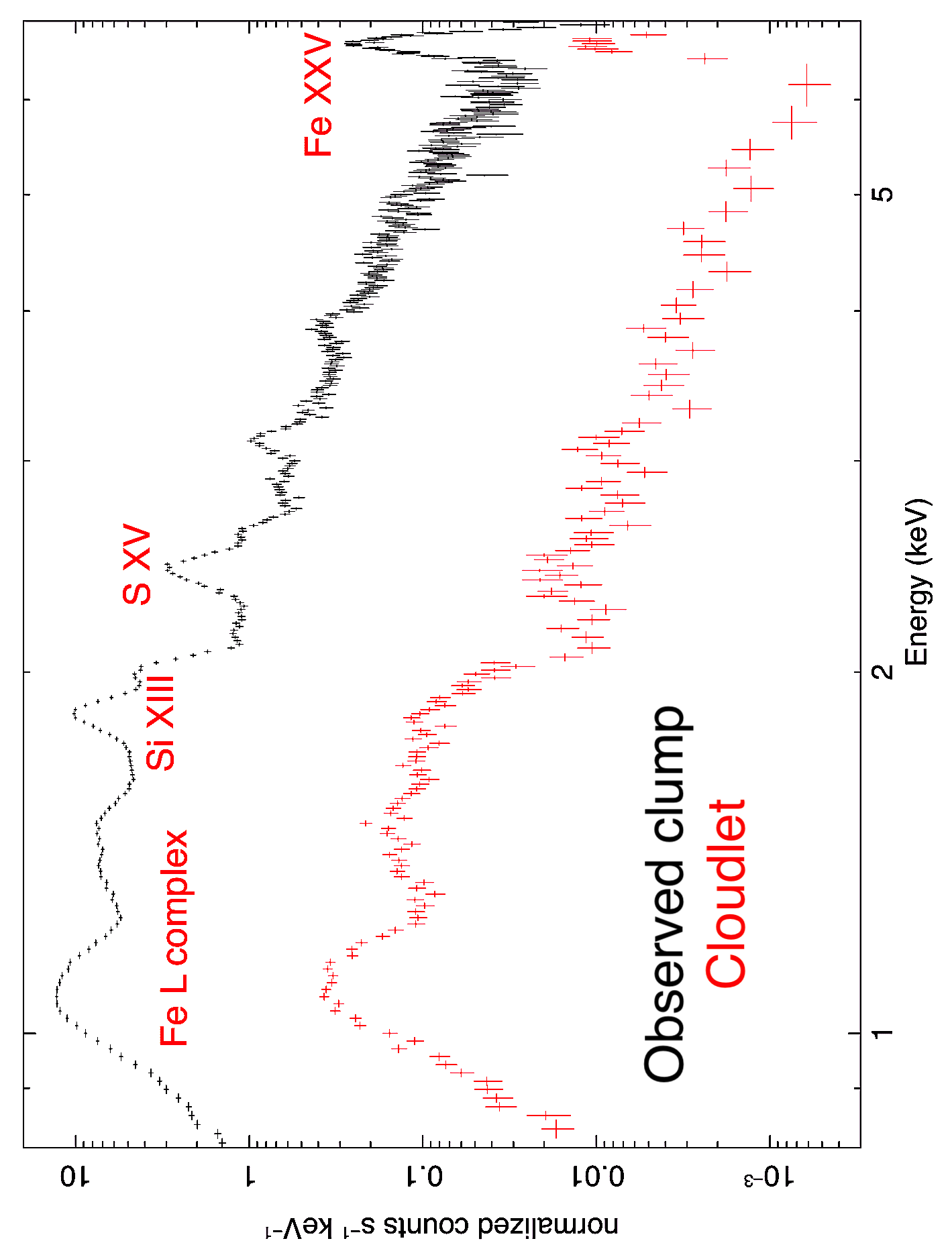}
    \includegraphics[width=0.58\columnwidth, angle=270]{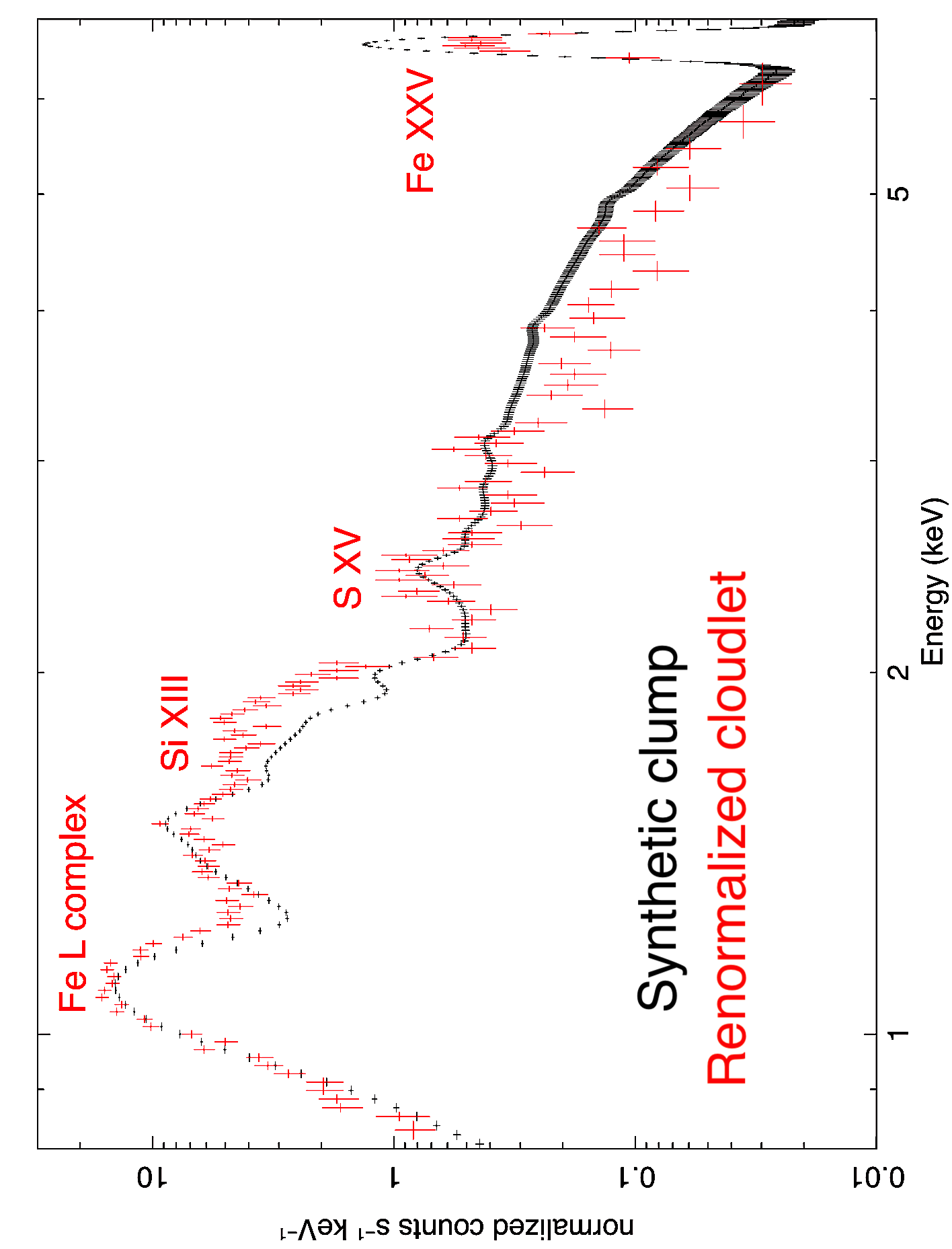}
    \caption{\emph{Upper panel}: {\it Chandra}/ACIS-S spectra extracted from the southeastern Fe-rich clump (black box in Fig. \ref{casadata}) and from the Fe-rich cloudlet (red box in Fig. \ref{casadata}) shown by black and red crosses, respectively. \emph{Lower panel}: {\it Chandra}/ACIS-S synthetic spectrum  derived from the HD simulation of the southeastern Fe-rich clump (black crosses). Red crosses show the renormalized observed spectrum of the Fe-rich cloudlet shown in the upper panel for easy comparison with the synthetic spectrum.}
    \label{sinthspec}
\end{figure}

I point out that I am not aiming to find a perfect agreement between synthetic and observed spectra, but I am interested in providing reliable and robust predictions of the X-ray emission from Fe-rich ejecta I show below that the lack of bright Si and S emission lines in the synthetic spectra does not affect the conclusions. I here notice that the synthetic spectrum is a good proxy of the real Fe-rich ejecta emission, given that it is extremely similar to the actual spectrum of the Fe-rich cloudlet (red box in Fig. \ref{casadata}). This is shown in Fig. \ref{sinthspec}, where the observed and renormalized spectra of the cloudlet can be compared with that derived from the HD simulation.

I only note a few discrepancies in the brightness of the Fe XXV emission line, which is slightly overestimated in the synthetic spectrum. This occurs because, in this region, the simulated temperature is slightly ($\sim30\%$) higher than that observed. However, this excess does not affect the conclusions since both actual and modeled temperatures lie in the range in which the Fe FB to FF ratio has its maximum (see Sect \ref{sect:ss_Fe}). I only expect a slightly narrower edge of Fe RRC in the actual spectra with respect to that derived from the simulation. In my analysis I did not include effects due to the ejecta expansion. If, in the region under analysis, there are two different knots of ejecta moving in opposite direction along the line of sight, the emission lines could undergo to a doppler broadening of the order of $\sim 50$ keV, leading to some blending of emission lines. However, this broadening is less significant in the outer part of the remnant, where the projected velocities are lower, such as in the box considered in this thesis, and the doppler broadening can be neglected.

As in the actual data, I found that it is not possible to constrain the Fe abundance in the {\it Chandra}/ACIS-S spectrum of this region. In fact, two statistically equivalent fits that have $\chi^2_{\mathrm{red}}=1.06$ (138 d.o.f.) can be obtained with $A_{\mathrm{Fe}} = 45$ (and EM $=1.3 \times 10^{57}$ cm$^{-3}$ for the ejecta component) and $A_{\mathrm{Fe}} = 1200$ (EM $= 6 \times \,10^{55}$ cm$^{-3}$). The corresponding Fe mass is therefore highly uncertain, spanning from $M_{\mathrm{Fe}}=2$ $\times 10^{-4}$ M$_{\odot}$ to $M_{\mathrm{Fe}}=4 \times 10^{-3}$ M$_{\odot}$.

I then synthesized the {\it XRISM}/Resolve spectrum from the same large southeastern region, by assuming an exposure time of 1 Ms (the spatial resolution of {\it XRISM} is not good enough to resolve the Fe-rich cloudlet marked in red in Fig. \ref{casadata}). I selected a box with a dimension of $\sim 1$ arcmin, comparable with the {\it XRISM} PSF, to limit contamination from surrounding emission. A precise estimate of the contamination cannot be performed before the effective launch of {\it XRISM} and this effect must be analyzed case by case. The resulting spectrum is shown in Fig. \ref{synthxrism}. The microcalorimeter spectrometer provides a superior spectral resolution, allowing us to clearly identify all the emission lines and spectral features. Here I show that, by analyzing the {\it XRISM}/Resolve synthetic spectrum, it is possible to unambiguously identify the enhanced RRC emission from the recombination of Fe ions, thus revealing the presence of pure-metal ejecta (as predicted in Sect. \ref{sect:allsynthesis}).

 I fitted the spectrum with two isothermal components of an optically thin plasma in non-equilibrium of ionization, associated with the shocked ISM and ejecta, respectively (hereafter NEI$+$NEI model). I considered two different scenarios: mild-metal and pure-metal ejecta. In the first case, I kept the Fe abundance fixed to ten in the ejecta component (the spectrum with the best-fit model and residuals is shown in the upper panel of Fig. \ref{synthxrism}); in the second case I left the Fe abundance of the ejecta component free to vary (mid panel of Fig. \ref{synthxrism}). Best-fit parameters are shown in Table \ref{xrismspecfit}. 

\begin{figure}[!ht]
    \centering
    \includegraphics[width=0.6\columnwidth,angle=270]{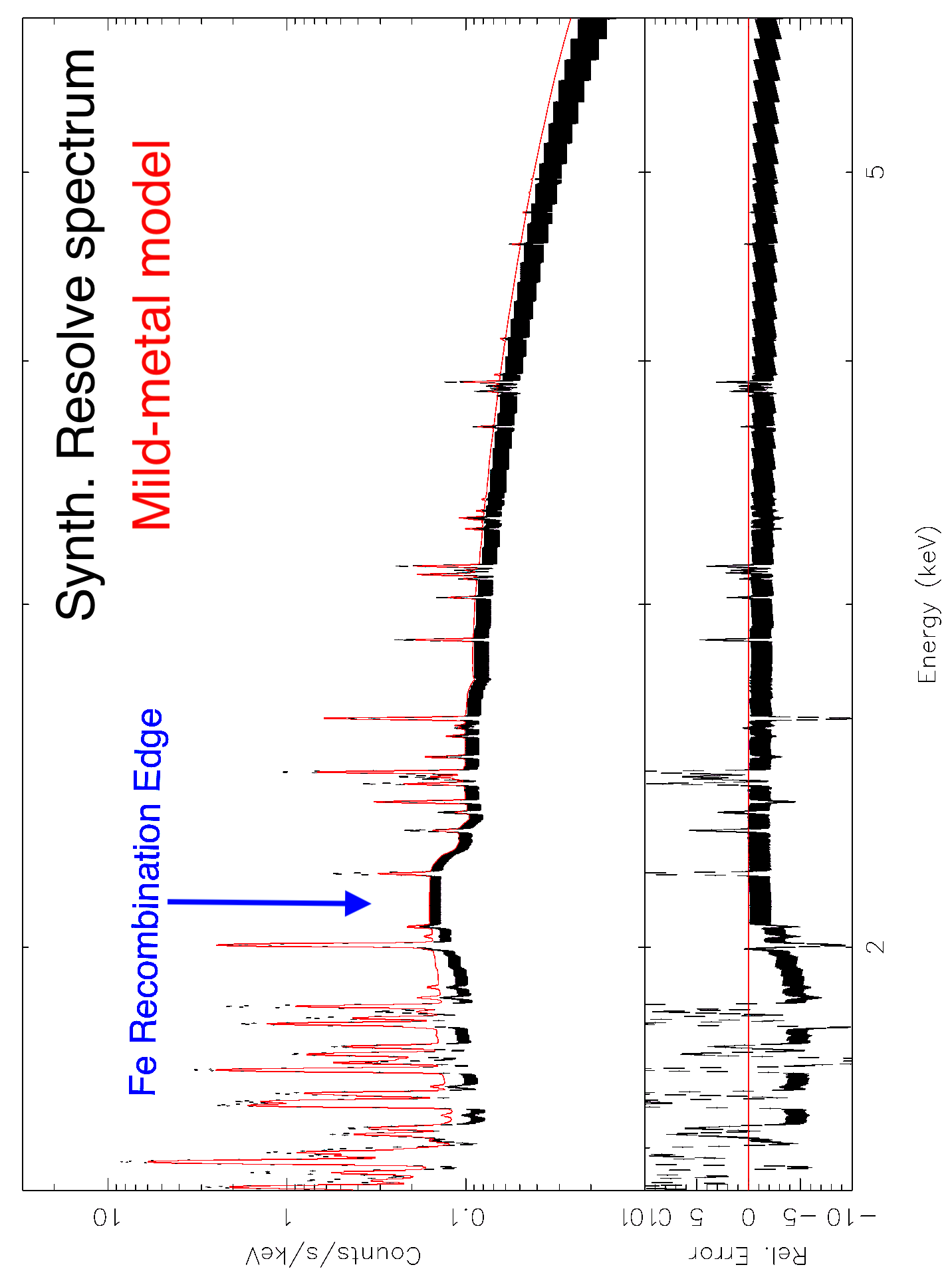}
    \includegraphics[width=0.6\columnwidth,angle=270]{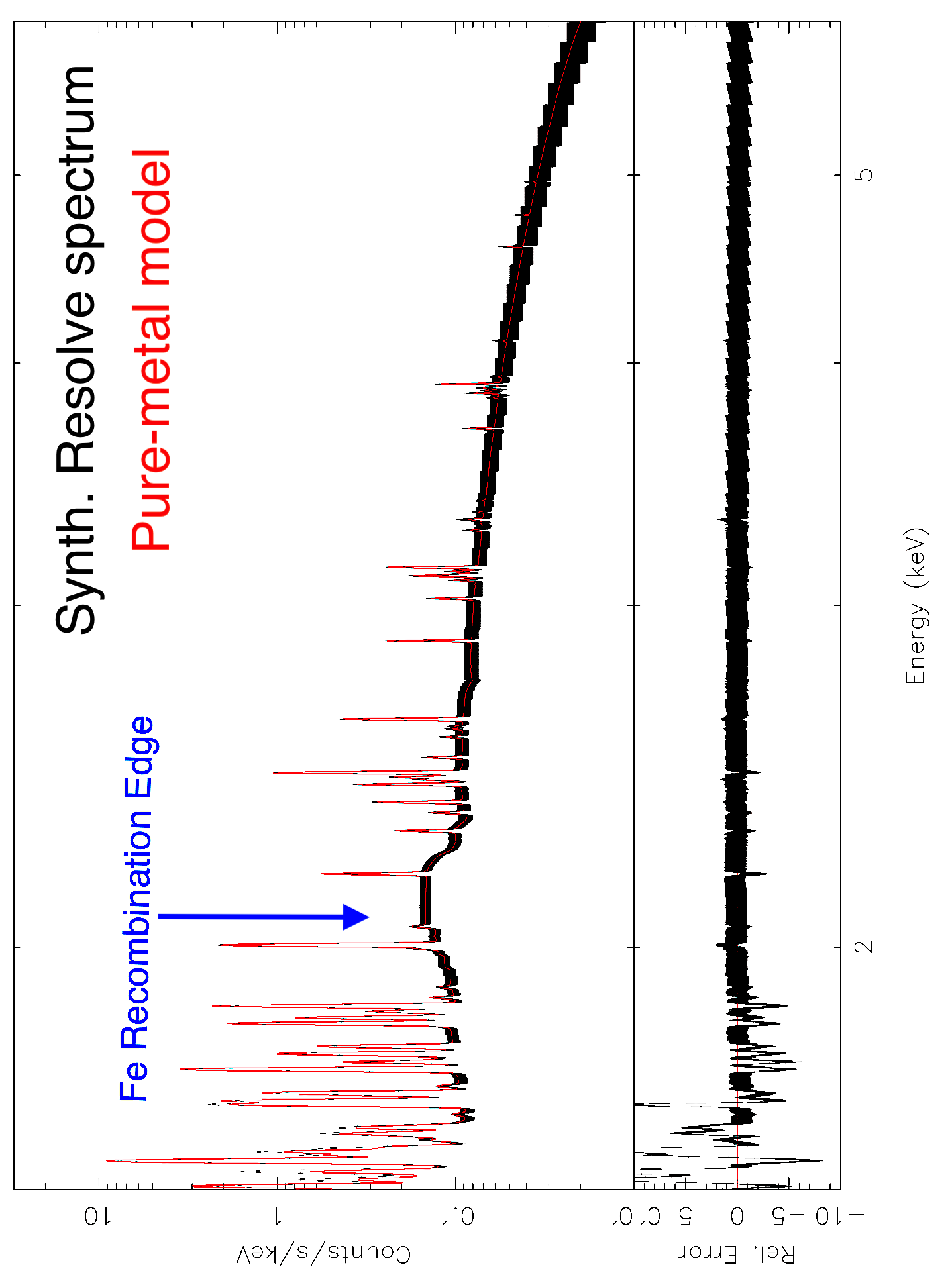}
    \caption{Synthetic {\it XRISM}/Resolve spectra of the Fe-rich southeastern clump with the corresponding best-fit models and residuals obtained for the mild-metal case (upper panel) and the pure-metal ejecta case (lower panel).}
    \label{synthxrism}
\end{figure}

\begin{figure}[!ht]
\centering
\includegraphics[angle=270,width=.9\columnwidth]{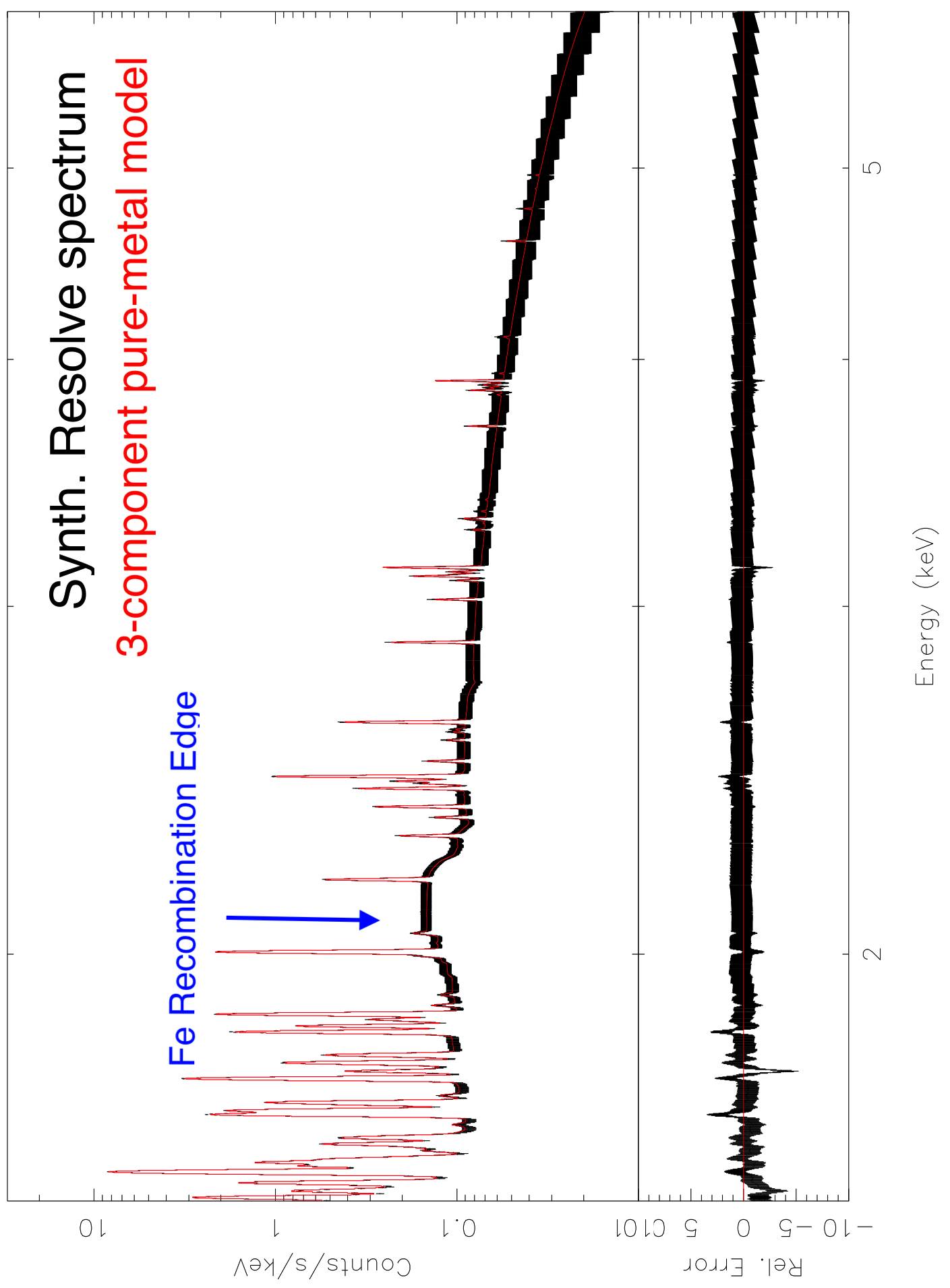}
\caption{Same as lower panel of Fig. \ref{synthxrism} but the best-fit model is composed by three NEI components.}
\label{fig:multikt}
\end{figure}

\begin{sidewaystable}[!ht]
    \centering
    \caption{Best-fit values for the fitting of the synthetic {\it XRISM}/Resolve spectrum of the Cas A southeastern region by adopting two thermal components, in the mild-metal and pure-metal ejecta scenarios.}
    \begin{tabular}{c|c|c|c}
    \hline\hline
Parameter & NEI+NEI (mild-metal)& NEI+NEI (pure-metal ejecta)& NEI+NEI (pure-metal ejecta) 250ks \\
 \hline
 n$_{\mathrm{H}}$ (10$^{22}$cm$^{-2}$) & \multicolumn{3}{c}{1.5(frozen)} \\
 \hline
 kT (keV)& 0.5$_{-0.001}^{+0.3}$ & 1.96$\pm$ 0.05& 1.62$_{-0.06}^{+0.11}$ 
\\
 $\tau$ (10$^{11} $s/cm$^{3})$ &1.10$_{-0.05}^{+0.08}$ & 1.13$\pm$ 0.04 & 1.8$_{-0.2}^{+0.3}$\\ 

  EM (10$^{58}$ cm$^{-3}$)& 0.19$_{-0.17}^{+0.002}$ & 0.150$ \pm$ 0.004 & 0.20$_{-0.01}^{+0.02}$ \\
            \hline
  kT (keV)& 3.9$_{-0.15}^{+0.3}$  & 2.90$\pm$ 0.01& 2.89$\pm 0.03$\\ 
    $\tau$ ($10^{11}$ s/cm$^3)$ & 1.53$_{-0.08}^{+0.05}$& 2.51$\pm $ 0.02 & 2.51$_{-0.03}^{+0.05}$\\ 
    EM (10$^{58}$ cm$^{-3}$)& 0.34$_{-0.001}^{+0.04}$& 0.056$\pm$ 0.005 & 0.002$^{+0.0003}_{-0.001}$\\
   Si  &0.415$_{-0.011}^{+0.003}$ & 0.77$\pm$ 0.08& 1 $\pm 1$\\
    S & 0.377$_{-0.013}^{+0.005}$  & 0.93$_{-0.07}^{+0.08}$& 2.0$_{-0.2}^{+2}$\\
    Ar &0.25 $\pm$ 0.02  & 0.4$\pm 0.2 $& 1$_{-1}^{+15}$\\
    Ca & 0.15$_{-0.03}^{+0.02}$& 0.49 $\pm$ 0.2& 1$_{-1}^{+7}$\\  
    Fe & 10 (frozen) &  119$_{-9}^{+11}$ &300$^{+400}_{-100}$\\
    \hline
    $\chi^2_{\mathrm{red}}$ (d.o.f.) &11.85 (5005) & 2.11 (5004) &  0.65 (4709) \\
    \hline
    Counts & \multicolumn{2}{c}{1.4$\cdot 10^6$} & \\
    \hline
    \end{tabular}
   
    The abundance values of the first component (associated with the ISM) are all frozen to 1. The rightmost column shows the best-fit values obtained from the 250 ks spectrum.
    \label{xrismspecfit}
\end{sidewaystable}

The spectral residuals and $\chi^2$ show that a Fe abundance $>100$  is necessary to properly fit the Fe RRC. The mild-metal scenario leads to strong residuals at the recombination edge energy and to a global misrepresentation of the spectrum. The high spectral resolution provided by the microcalorimeters allows us to reveal the enhanced RRC emission of pure-metal ejecta and this clearly removes the degeneracy between abundance and emission measure in the fitting procedure, leading to a correct estimate of the absolute Fe mass in this area. 

As explained above, the HD simulation provides an accurate description of the actual distribution of Fe-rich ejecta in Cas A and is able to reproduce the spectral features observed with {\it Chandra}. However, the moderate spectral resolution of {\it Chandra}/ACIS (and CCD detectors in general) does not allow us to confirm that pure-metal ejecta are present in Cas A, as predicted by the O16 model. If pure-Fe ejecta are actually present in the southeastern limb of Cas A, it will be possible to pinpoint their presence with {\it XRISM} and to correctly derive their mass (see Sect. \ref{final}). 

As already mentioned, I expect that the actual {\it XRISM}/Resolve spectrum will show more Si and S line emission than that predicted by the simulation (see Fig. \ref{sinthspec}). Therefore, we should observe brighter Si and S lines and a bright RRC from Si (at energies $E>2.5$ keV, see Fig. \ref{SimMicro}) and S ($E>3$ keV). These features will not mask out the Fe RRC, thus not affecting our conclusions. In fact, the Si and S lines will be well resolved by the {\it XRISM} spectrometer without contaminating the Fe RRC and the Si (and S) RRC; also, edges show up at energies higher than that of the Fe XXIV recombination (2.023 keV). 

I also wish to note that the residuals at low energy in the lower panel of Fig. \ref{synthxrism} are due to the simplified spectral model that includes only one isothermal component for the ejecta. Indeed, the HD simulation I used shows a relatively broad distribution of plasma temperatures and ionization parameters in the region selected, but I am fitting the spectrum with only two isothermal components. A multi-component model provides a much better fit to the emission line complexes (Fig. \ref{fig:multikt}), but this is beyond the scope of this project. In any case, even with this more complex spectral model, I verified that a Fe abundance $>100$ is always required to fit the spectrum.

\begin{figure}[!ht]
    \centering
    \includegraphics[width=0.6\columnwidth,angle=270]{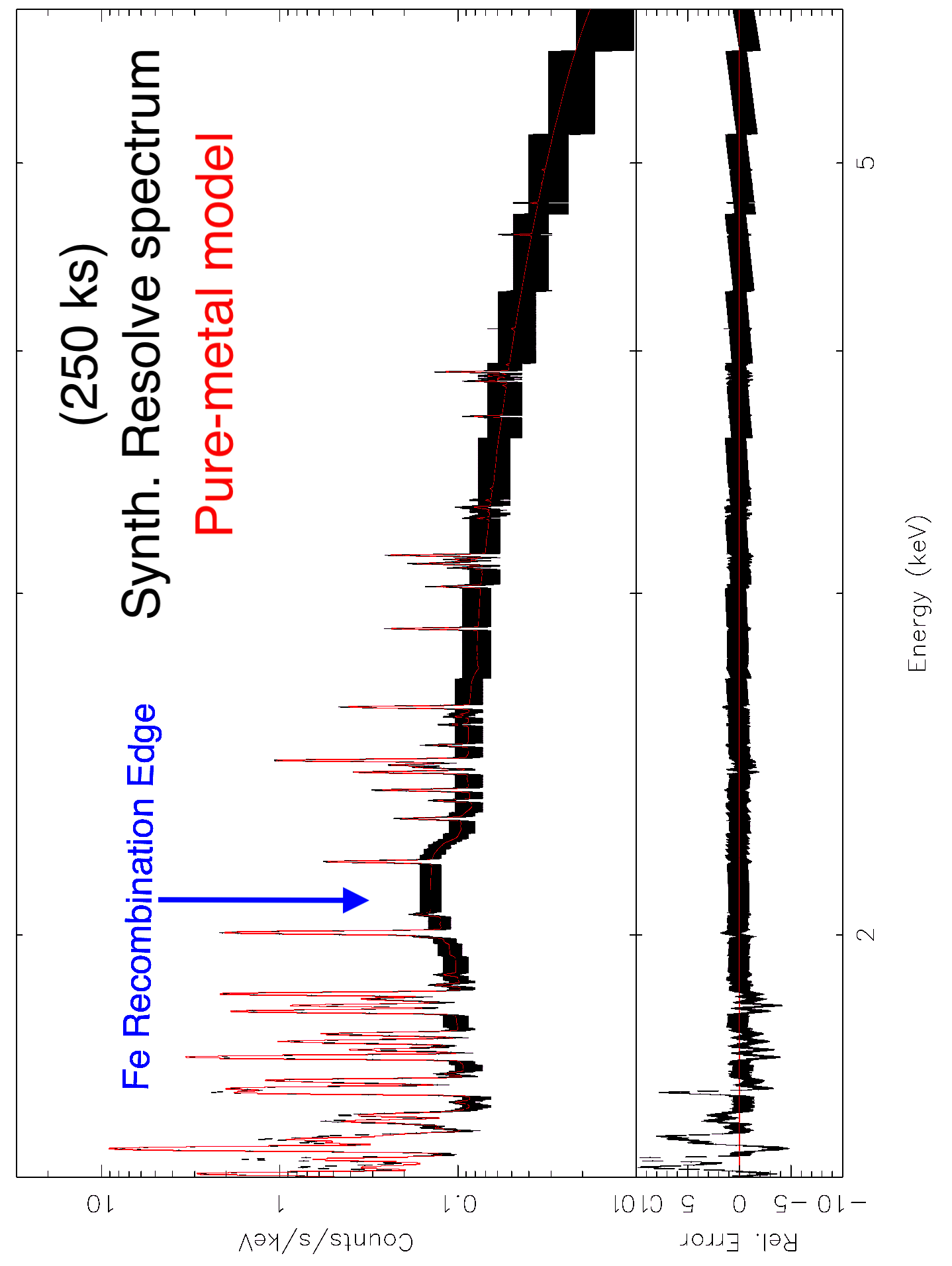}
    \caption{Same as lower panel of Fig. \ref{synthxrism} but with an exposure time of 250 ks.}
    \label{synthxrism_250}
\end{figure}

So far, I have shown the synthetic spectra produced assuming an exposure time of 1 Ms. I am aware that it may not be possible to observe this region of Cas A for such a large amount of time. Therefore I synthesized the {\it XRISM}/Resolve spectrum of the southeastern Fe-rich knot by assuming an exposure time of 250 ks (Fig. \ref{synthxrism_250}). The enhanced Fe RRC remains visible in the spectrum and the pure-metal ejecta model describes the spectrum significantly better than the mild-metal model. An exposure time shorter than 250 ks may not be sufficient to unambiguously detect the pure-metal ejecta. 

\newpage
\clearpage

\section{Discussion}
\label{final}

\noindent \subsection{Implications for SNRs with overionized ejecta components}

The RRC features discussed here have a different physical origin with respect to those associated with overionized plasmas that have been detected so far in spectra of mixed-morphology SNR, namely those remnants that show bright radio emission in the outer shells and peaked X-ray emission in the internal ones (see Sect. \ref{sect:fb}).
 The physical mechanism that causes the rapid cooling of the plasma is still debated and both thermal conduction with nearby molecular clouds (e.g., \citealt{otu20}) and adiabatic expansion (e.g., \citealt{zmb11}, \citealt{mbd10}, \citealt{yok09}) have been proposed as viable cooling processes (even for the same object, see \citealt{mtu17} and \citealt{gmo18} for IC 443). I verified that the enhanced recombination features related to pure-metal ejecta can be distinguished from those associated with overionization. The upper panel of figure \ref{ovpure} shows the continuum emission of an overionized plasma with $\tau_{\mathrm{rec}}=10^{11}$ s/cm$^3$, cooled down to $kT_{\rm fin}=$ 0.5 keV from an initial temperature $kT_0=5$ keV  (typical values for overionized plasmas in SNRs, \citealt{mtu17,gmo18}), and the continuum emission of ejecta in the pure-Si regime (Si abundance set to 300). The RRC produced by the overionized plasma is much stronger than that of pure-metal ejecta. 
 This is because in the pure-metal ejecta regime the ratio FB/FF does not increase arbitrarily with the abundance, but reaches a saturation value (see Fig. \ref{RatioFBFF}), which is smaller than that obtained for the case of overionization. Moreover, in the case of overionization, the Si-edge is actually composed of two different edges: the one at lower energy is related to the He-like Si, while the edge at higher energy, which is also the brightest, is associated with H-like Si. This is a characteristic feature of overionized plasmas, since the high degree of ionization leads to an increase in the H-like ions at the expense of the He-like ions (see Sect. \ref{sect:nei}). Therefore, the shape and the intensity of both the RRC and edges make it possible to discriminate between pure-metal ejecta and overionized plasmas.
\begin{figure}[!ht]
   \centering
    \includegraphics[width=0.6\columnwidth,angle=270]{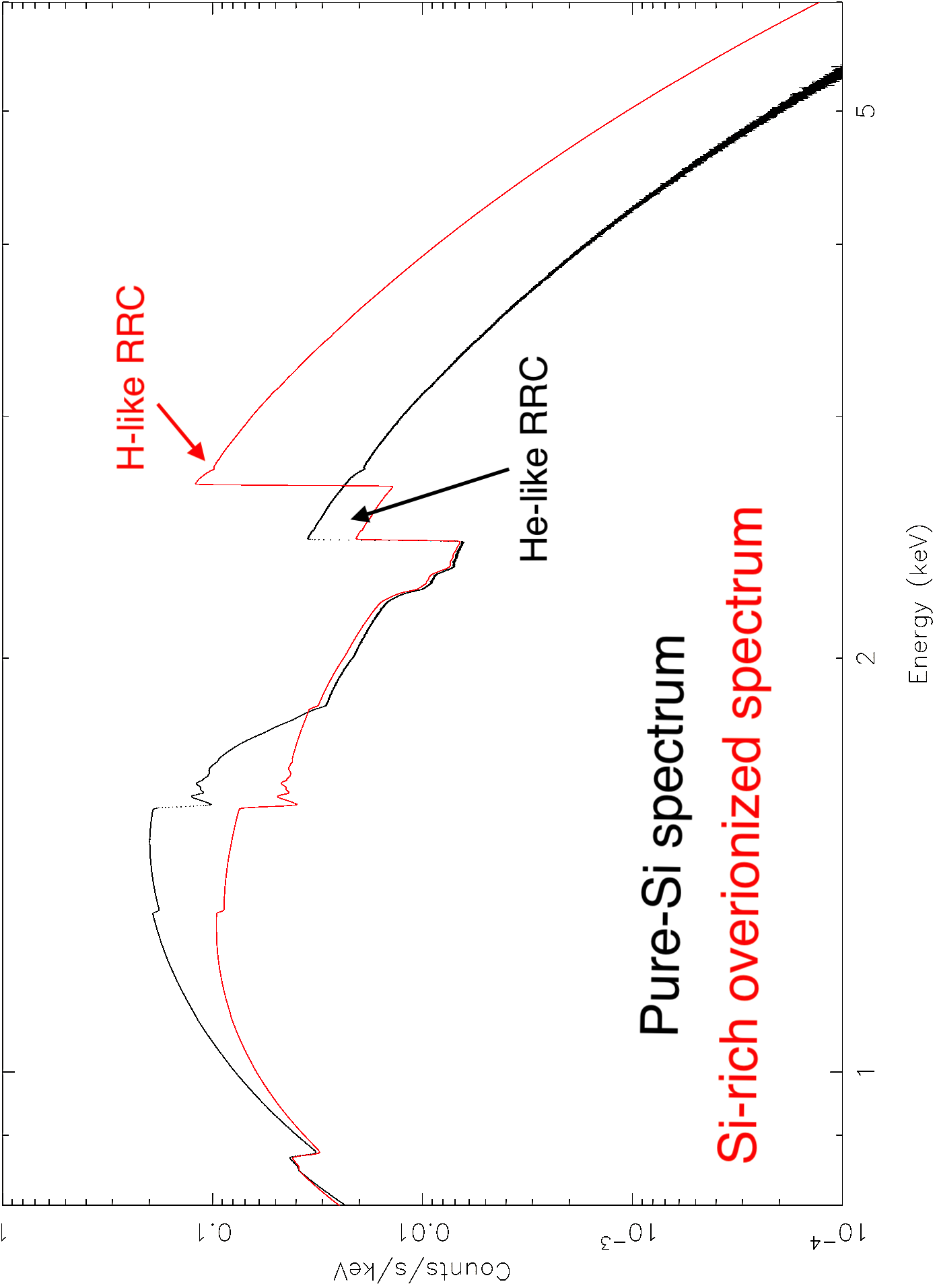}
    \includegraphics[width=0.8\columnwidth]{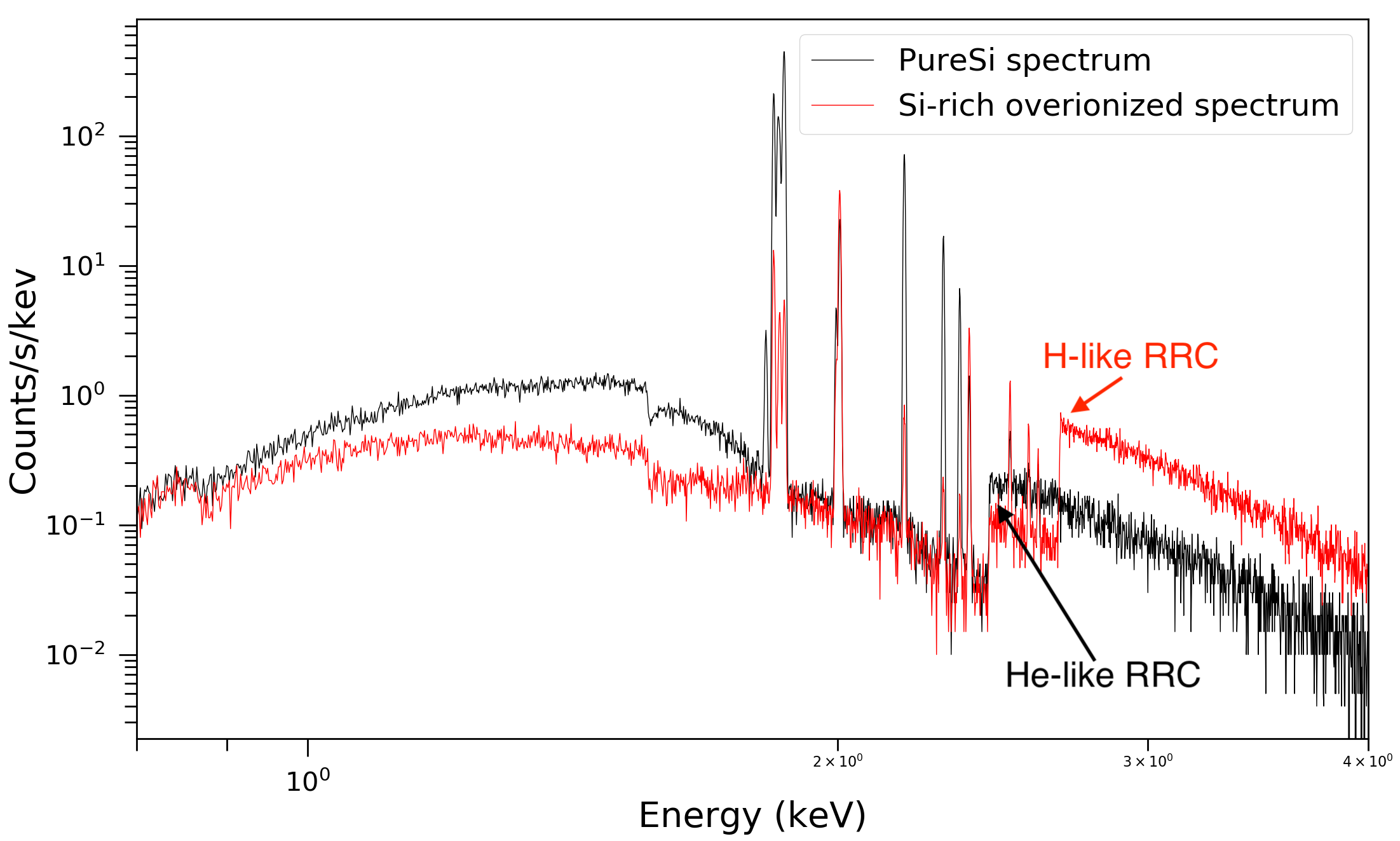}
    \caption{Comparison between spectra produced by a pure-metal plasma and an overionized plasma. \emph{Upper panel}: In red, continuum-only spectrum of a typical overionized SNR (Si abundance set to 3, $\tau_{\mathrm{rec}}=10^{11}$ s cm$^{-3}$, $kT_{\rm fin}$=0.5 keV and kT$_{0}$=5 keV); in black, continuum-only spectrum of a typical pure-metal ejecta plasma in CIE  (Si abundance set to 300 and kT=0.5 keV). Both spectra are synthesized assuming an exposure time of 10$^8$ ks. \emph{Lower panel}: Same as upper panel but assuming an exposure time of 100ks and including the line emission.}
    \label{ovpure}
\end{figure}

Nevertheless, the case of a pure-metal ejecta overionized plasma also needs to be considered. This scenario may be realistic for the Galactic SNR W49B (\citealt{oky09}, \citealt{mbd10}, \citealt{zmb11}, \citealt{lpr13}, \citealt{zv18}). The combination of the two effects (pure-metal plasma and overionization) further increases the contribution of the RRC emission and needs to be taken into account to correctly estimate the ionization parameter. The line emission has a crucial role in this scenario. In the lower panel of Fig. \ref{ovpure}, I show the same synthetic {\it XRISM}/Resolve spectra as that shown in the upper panel of Fig. \ref{ovpure}, assuming an exposure time of 100 ks and including the line emission.
Even with this realistic exposure time, I can easily distinguish the different shapes of the RRCs and discriminate between pure-metal ejecta and overionized plasma. Moreover, the line emission is completely different in the two spectra. In the recombining overionization scenario the Si H-like lines dominate the spectrum and the emission lines of He-like S are significantly fainter. At the same time, the overall brightness of Si lines is higher in the pure-metal scenario, as expected because of the high abundance. In conclusion, a careful analysis of the line emission is needed to correctly distinguish between the pure-metal ejecta and the recombining overionization scenarios.
\subsection{Recovering the absolute mass with {\it XRISM}/Resolve spectra}
The capability of detecting pure-metal ejecta emission is crucial to recover the correct mass of the atomic species in the ejecta. I here show that from the synthetic {\it XRISM}/Resolve spectrum of Cas A (discussed in Sect. \ref{synthdata}) it is possible to properly derive the Fe mass. I compared the Fe mass calculated from the best-fit models of the synthetic {\it XRISM}/Resolve and {\it Chandra}/ACIS spectra with the correct value, known from the HD simulation. In order to estimate the filling factor of ejecta and ISM from the analysis of the synthetic spectra, I followed the recipe by \cite{bms99} under the assumption of pressure equilibrium. As explained in Sect. \ref{synthdata}, by fitting the {\it Chandra} synthetic spectrum, I found that the errors on Fe abundance are extremely large and can vary from a few tens to more than $10^3$. By fitting the spectrum synthesized with {\it XRISM}, I instead obtained $M_{\mathrm{Fe}}=4.8_{-0.4}^{+0.5} \times 10^{-3}$ M$_{\odot}$. From the simulation, I know that the correct value of the Fe mass in the region is 4.6 $\times 10^{-3}$ M$_{\odot}$. Despite the lower statistics, by fitting the 250 ks {\it XRISM} spectrum I obtained $M_{\mathrm{Fe}}=6_{-2}^{+5} \times 10^{-3}$, which is consistent with the correct value. I stress that an exposure time shorter than 250 ks could lead to an ambiguous detection of the pure-metal ejecta feature. By analyzing the {\it XRISM}/Resolve spectra, remarkably I recover the real mass of the Fe-rich ejecta without the degeneracy present in the {\it Chandra}/ACIS-S analysis. The analysis of CCD spectra can be misleading in estimating the Fe mass by up to a factor of 20. This comparison further confirms that in order to correctly evaluate the ejecta mass it is crucial to accurately study the RRC of the elements, and high-resolution spectrometers are needed. This is critical since the element mass is used for comparisons with theoretical nucleosynthesis yields.

\section{X-ray synthetic spectra of a large-scale anisotropy in an evolving SNR}
\label{sect:hd_anto}
I applied the self-consistent synthesis tool described in Sect \ref{sect:tool_synth} to a set of HD simulations performed by \citet{tom20}. In this project, the authors show that the presence of anisotropy in the supernova explosions can lead to the formation of large-scale jet-like structure. The matter-mixing strongly depends on the physical and chemical properties of the pristine anisotropies. We adopted a forward modeling approach to ascertain how X-ray spectra synthesized from the model convey the information on the evolution of the system. I selected three different boxes from the HD domain of the run Fe-R5-D750-V7 (Fig. \ref{fig:reg_anto}), corresponding to three significantly different environments: the ejecta bullet protrudring beyond the remnant outline (the clump), the low-density stripped region behind the clump (the wake) and an area of the shell not affected by the anisotropy. I assumed a distance of 1 kpc and a column density $n_{\mathrm{H}} = 2 \times 10^{21} \mathrm{cm}^{-2}$. The resulting synthetic spectra are shown in Fig. \ref{fig:spec_anto} in red, black and green, respectively. 
\begin{figure}[!ht]
\centering
\includegraphics[width=0.6\columnwidth]{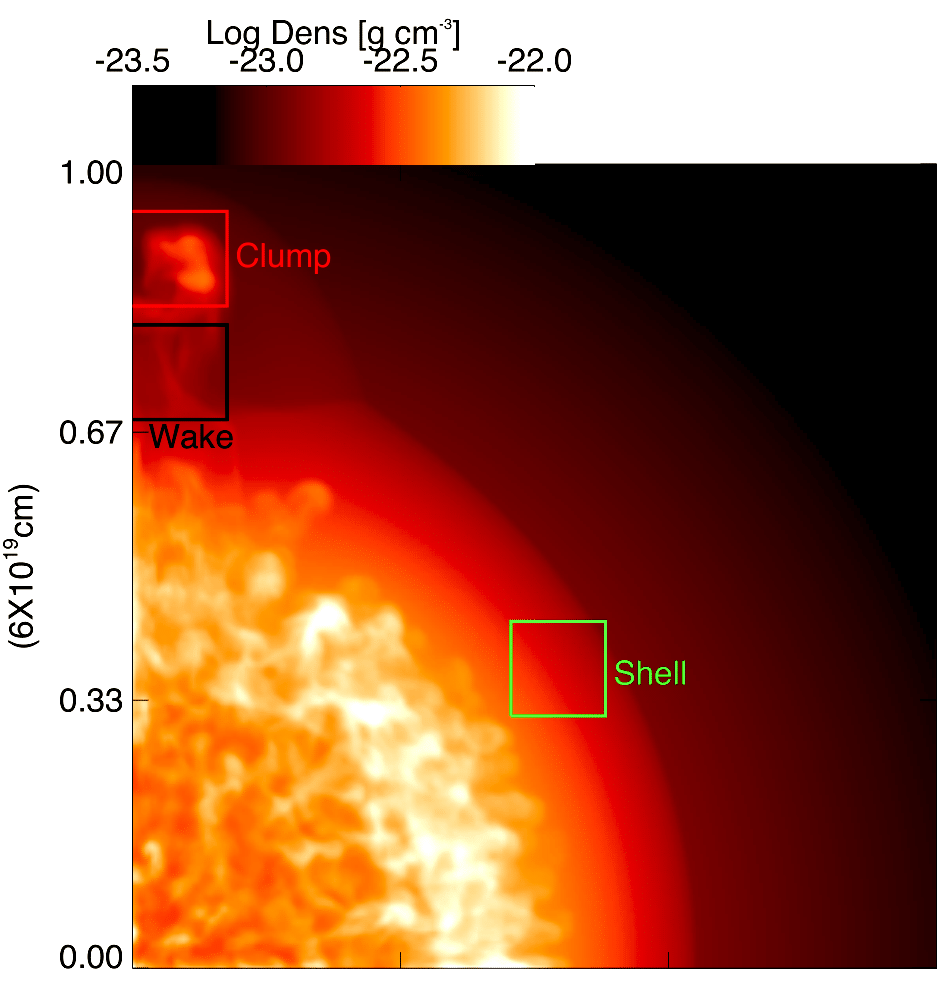}
\caption{Density map integrated along the line of sight which shows the extraction regions adopted to produce synthetic X-ray spectra: the clump region is marked in red, the wake region is indicated in black and the area of the shell not involved by the motion of the clump is identified by the green box (\citealt{tom20}).}
\label{fig:reg_anto}
\end{figure}

\begin{figure}[h!t]
\centering
\includegraphics[width=0.7\columnwidth]{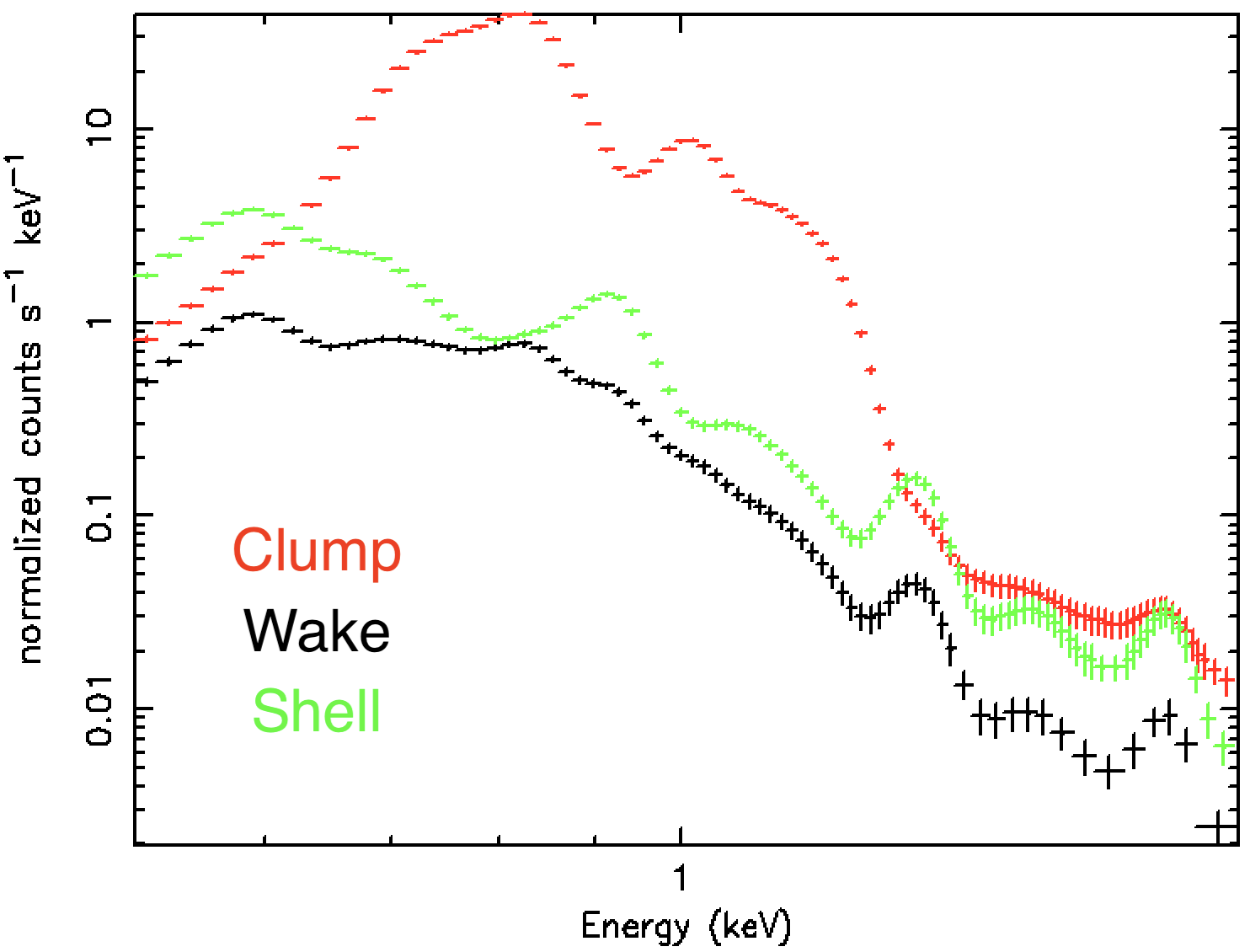}
\caption{X-ray {\it Chandra}/ACIS-S synthetic spectra in the 0.5-2 keV band for the clump (red), the wake (black) and the shell (green). The extraction regions are shown in Fig. \ref{fig:reg_anto}. (\citealt{tom20}).}
\label{fig:spec_anto}
\end{figure}

The Fe false continuum emission due to the emission of Fe-rich ejecta is prominent in the spectrum extracted from the Fe-rich clump whereas the lines of other elements are fainter, because of the protrusion of the clump. On the other hand, the spectrum extracted from the wake shows bright Mg and Si lines with an overall lower emission measure. This is due to the low density of this region, as evident from Fig. \ref{fig:reg_anto}. The spectra show that the matter mixing, induced by the large-scale anisotropy, can be reconstructed thanks to the X-ray analysis of the synthetic spectra. For more details, see the full paper by \citet{tom20}.

\section{Conclusions}
\label{sect:conc_rrc}

In this chapter I have shown that a spectral analysis carried out with current CCD X-ray detectors suffers from an intrinsic degeneracy in measuring the abundances using line-dominated spectra from SNRs. I studied the characteristics of the X-ray emission processes at high abundances, identifying the enhanced RRC as the spectral signature of pure-metal ejecta. I showed that the low energy resolution of current CCD detectors limits our understanding of the spectra, in particular hiding the characteristic  RRC. 
I found that future microcalorimeter spectrometers, such as Resolve, onboard the {\it XRISM} telescope, can address this issue. I also presented a very promising case for the detection of pure-Fe emission in the southeastern clump of Cas A, by using both {\it Chandra} data and a 3D HD simulation to derive quantitative predictions of the expected {\it XRISM}/Resolve spectrum. The resulting synthetic spectra show bright Fe RRC around 2 keV and the fits performed on such spectra can definitely pinpoint the presence, if any, of pure-metal ejecta, recovering the correct mass of the elements. Finally, I showed that the pure-metal ejecta RRC have a very different shape from those observed for overionized plasma.

%% file: IC443.tex
\chapter{A jet-like overionized structure in the SNR IC 443}
\label{ch:ic443}

In this chapter, I present my analysis of deep {\it XMM-Newton} observations of IC 443 (see Sect. \ref{sect:Intro_IC443}), which revealed for the first time a jet-like structure of overionized plasma in the NW part of the remnant (see also \citealt{gmo18}). The identification of this structure is interesting because X-ray emitting collimated structures of ejecta have been discovered and analyzed only in other two core-collapse SNRs, namely Cas A (\citealt{hlb04}, \citealt{fhm06}) and, recently, Vela SNR (\citealt{gsm17}). This kind of structures is not well understood and some authors suggested that they may be deeply related to the physical mechanism of the SN explosion (\citealt{wil85}, \citealt{jms16}, \citealt{gs17}). 

The chapter is organized as follows: observations and data reduction are described in Sect. \ref{obs}; image and spectral analysis are illustrated in Sect. \ref{res}; I discuss the results in Sect. \ref{discu} and summarisze them in Sect. \ref{sect:conc_ic443}.

\section{The data }
\label{obs}

IC 443 has been observed several times with the European Photon Imaging Camera (EPIC) on board of {\it XMM-Newton} (see Appendix \ref{app:telescopes} for details about the instruments adopted). I have analyzed nine observations in order to build complete images of IC 443, but only two of them (ID 0114100201 and 0114100501, PI Fred Jansen) were used for the spectral analysis since they are the only ones which have the jet within their field of view (FOV). Table \ref{osservazioni} summarizes the main information about the data analyzed. 

\begin{table}[!h]
\centering
\caption{Main information about the {\it XMM-Newton observations} analyzed in this chapter.}
\begin{tabular}{c|c|c|c}
\hline\hline
OBS ID& Camera& $t_{\mathrm{exp}}$ U (ks)& $t_{\mathrm{exp}}$ F (ks)\\
\hline
0114100101& MOS1& 22.9 & 9.5 \\ 
0114100101& MOS2& 22.9 & 9.5 \\
0114100101& pn& 26.7 & 3.7 \\
\hline
0114100201& MOS1& 5.4& 5\\ 
0114100201& MOS2 &5.4& 5.2\\
0114100201& pn& 3& 2.6\\
\hline
0114100301& MOS1& 25.9 & 21.3 \\
0114100301& MOS2& 25.6& 21.5 \\
0114100301& pn& 23.2 & 17.7 \\
\hline
0114100401& MOS1& 29.9 & 23.5\\
0114100401& MOS2& 29.9 & 23.8\\
0114100401& pn& 27.8 & 18.7\\
\hline
0114100501& MOS1& 24.9& 18.1\\
0114100501& MOS2& 24.9& 18.9\\
0114100501& pn& 22.5& 12.8\\
\hline
0114100601& MOS1& 6.3 & 4.8\\
0114100601& MOS2& 6.3 & 5.3\\
0114100601& pn& 3.9 & 3.2\\
\hline
0301960101& MOS1& 81.5& 56.3\\
0301960101& MOS2& 81.5& 58.9\\
0301960101& pn& 77.4& 51.2\\
\hline
0600110101& MOS1& 90.2 & 32.3\\
0600110101& MOS2& 90.3 & 37.5\\
0600110101& pn& 84.4 & 22.2\\
\end{tabular}

U stands for unfiltered, F for filtered
\label{osservazioni}
\end{table}

I used the Science Analysis System (SAS), version 16.1.0, to perform the data reduction and analysis and I used the task \emph{ESPFILT} to eliminate contamination due to soft protons, reducing the effective exposure time (fourth column of Table \ref{osservazioni}). 
For the whole analysis I considered only data with FLAG=0 and PATTERN$<$13,$<$5 for {\it XMM-Newton}/MOS,pn cameras respectively, as suggested by the {\it XMM-Newton} handbook\footnote{https://xmm-tools.cosmos.esa.int/external/xmm-user-support/documentation/uhb/}. 

I generated background-subtracted images of IC 443 in two different bands: Soft (0.5-1.4 keV) and Hard (1.4-5.0 keV), same as those adopted by \citet{tbm08} for their analysis of IC 443. I used a double subtraction procedure in order to subtract background contribution and correct for the vignetting, similar to that adopted by \cite{mbo17}. I subtracted the non-photonic background from the data and from the {\it XMM-Newton}/EPIC high signal-to-noise background event files\footnote{Such files are available at https://www.cosmos.esa.int/web/xmm-newton/blank-sky} by scaling the {\it XMM-Newton}/EPIC Filter Wheel Closed files\footnote{Such files are available at https://www.cosmos.esa.int/web/xmm-newton/filter-closed} data. 
Before subtraction, I used the ratio of count-rates in the corners of the CCDs, namely the part of the cameras outside of FOV, as a scaling factor. Then, I subtracted the photonic background from the pure photonic data. Moreover, I combined the observations using the task \emph{emosaic}. Finally, I adaptively smoothed the mosaicked image (with the task \emph{asmooth}) to a signal to noise ratio of 25. 

I used the SAS tool \emph{evigweight} to correct the vignetting effect in the spectra. I applied tasks \emph{rmfgen} and \emph{arfgen} obtaining the  response and ancillary matrices and I binned the spectra to obtain at least 25 counts per bin. The spectral analysis has been performed with XSPEC (version 12.9, \citealt{arn96}) in the energy range 0.6-5 keV for all cameras. {\it XMM-Newton}/MOS1,2 and pn spectra of the two observations were fitted simultaneously.
I selected two different regions outside the SNR shell to extract background spectra and I verified that the best fit results do not depend on the choice of the background region. Errors are at 2$\sigma$ confidence levels and $\chi^2$ statistics has been used.

I also generated a {\it Chandra}/ACIS-S image of the PWN CXOU J061705.3+222127 located near the southern rim of the remnant in the 0.5-7 keV energy band. I analyzed observations 13736 and 14385 (PI Weisskopf), for a total exposure time of 152 ks, through the software CIAO version 4.7. After reprocessing the data with the tool \emph{chandra\_repro}, I used the CIAO task \emph{fluximage} to create exposure-corrected images for the observations (merged with the task \emph{merge\_obs}).

Scales in pc shown in the images are calculated assuming a distance of 1.5 kpc, according to \cite{ws03}.

\section{Results of the X-ray analysis} 
\label{res}

\subsection{Image analysis}
\label{image}

\begin{figure}[!htb]
\centering
\begin{minipage}{0.49\columnwidth}
\includegraphics[width=.95\columnwidth]{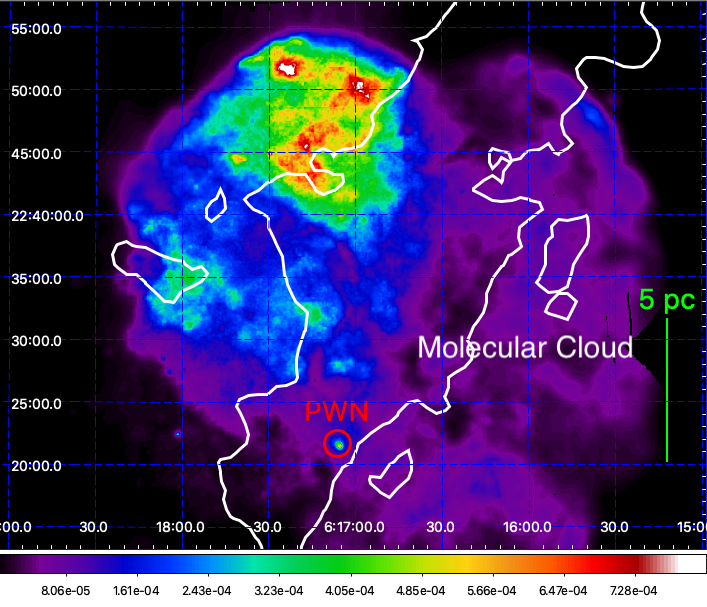}
\end{minipage}
\hfill
\begin{minipage}{0.49\columnwidth}
\includegraphics[width=.95\columnwidth]{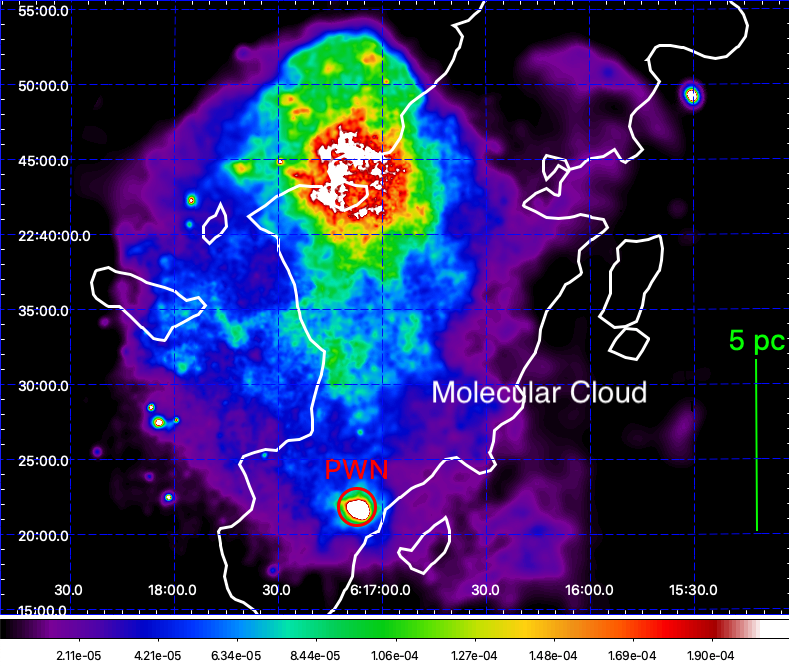}
\end{minipage}
\begin{minipage}{0.49\columnwidth}
\includegraphics[width=.95\columnwidth]{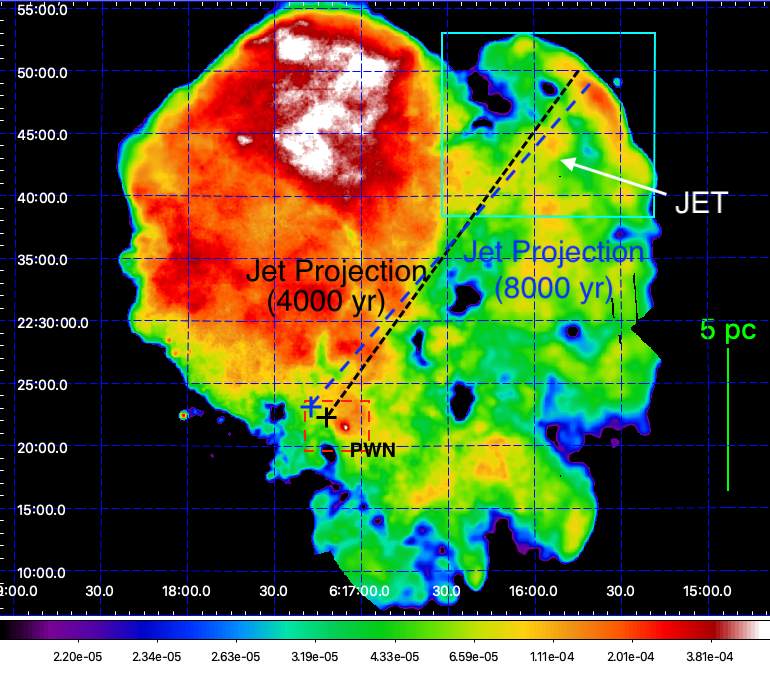}
\end{minipage}
\hfill
\begin{minipage}{0.49\columnwidth}
\includegraphics[width=.95\columnwidth]{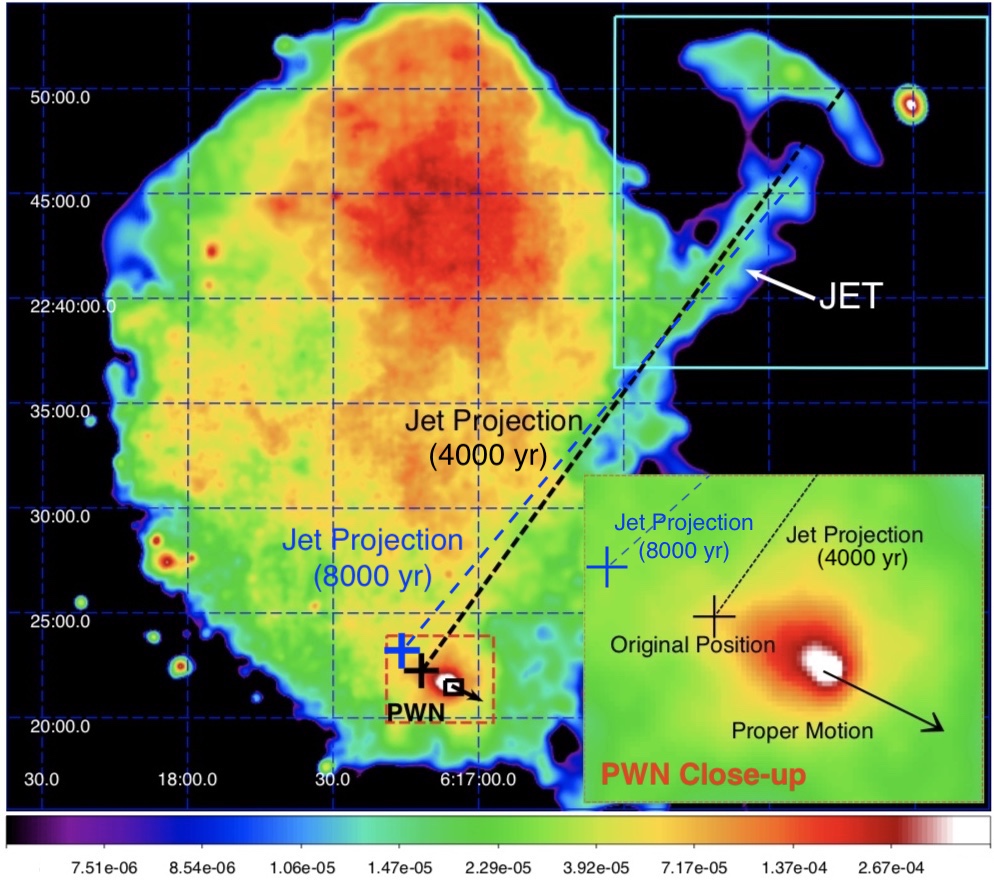}
\end{minipage}
\caption{Background-subtracted and vignetting-corrected count-rate images of IC 443 in the Soft (0.5-1.4) and the Hard (1.4-5 keV) bands obtained combining all the {\it XMM-Newton}/EPIC cameras. The images are adaptively smoothed to a signal-to-noise ratio 25 and the bin size is $4''$. \emph{First row}: linear color-scale images of IC 443 in the Soft band (left panel) and Hard band (right panel). The red circle marks the PWN and the white contours trace the molecular cloud as reported by \cite{bgb88}. \emph{Last row}: logarithmic color-scale images of IC 443 in the Soft band (left panel) and Hard band (right panel). 
The cyan box marks the jet (and the field of view of Fig. \ref{dettagliojet}), the dashed black and blue lines are the projections of the jet towards the PWN assuming an age of 4000 yr \citep{tbm08} or 8000 yr \citep{uog20}, respectively; the red dashed box indicates the field of view of Fig. \ref{PWN}, the black and blues crosses indicate the original position of the PWN assuming an age of 4000 yr and 8000 yr, respectively; the black arrow marks the proper motion of the PWN. The close-up view of the Hard image shows a detail of the PWN area with the original position assuming an age of 4000 yr or of 8000 yr (black and blue crosses, respectively) and the proper motion marked by the black arrow, respectively.}
\label{ic443andjet}
\end{figure}

Fig. \ref{ic443andjet} shows the count-rate image of IC 443 in the Soft (0.5-1.4 keV) and the Hard energy bands (1.4-5 keV). The white contour in the upper panels traces the semi-toroidal molecular cloud mentioned in Sect. \ref{sect:Intro_IC443} (\citealt{bgb88}). The hard X-ray map clearly reveals an extended and elongated jet-like feature, approximately aligned along the SE-NW direction.

\begin{figure}[!htb] 
\centering
\includegraphics[width=0.65\columnwidth]{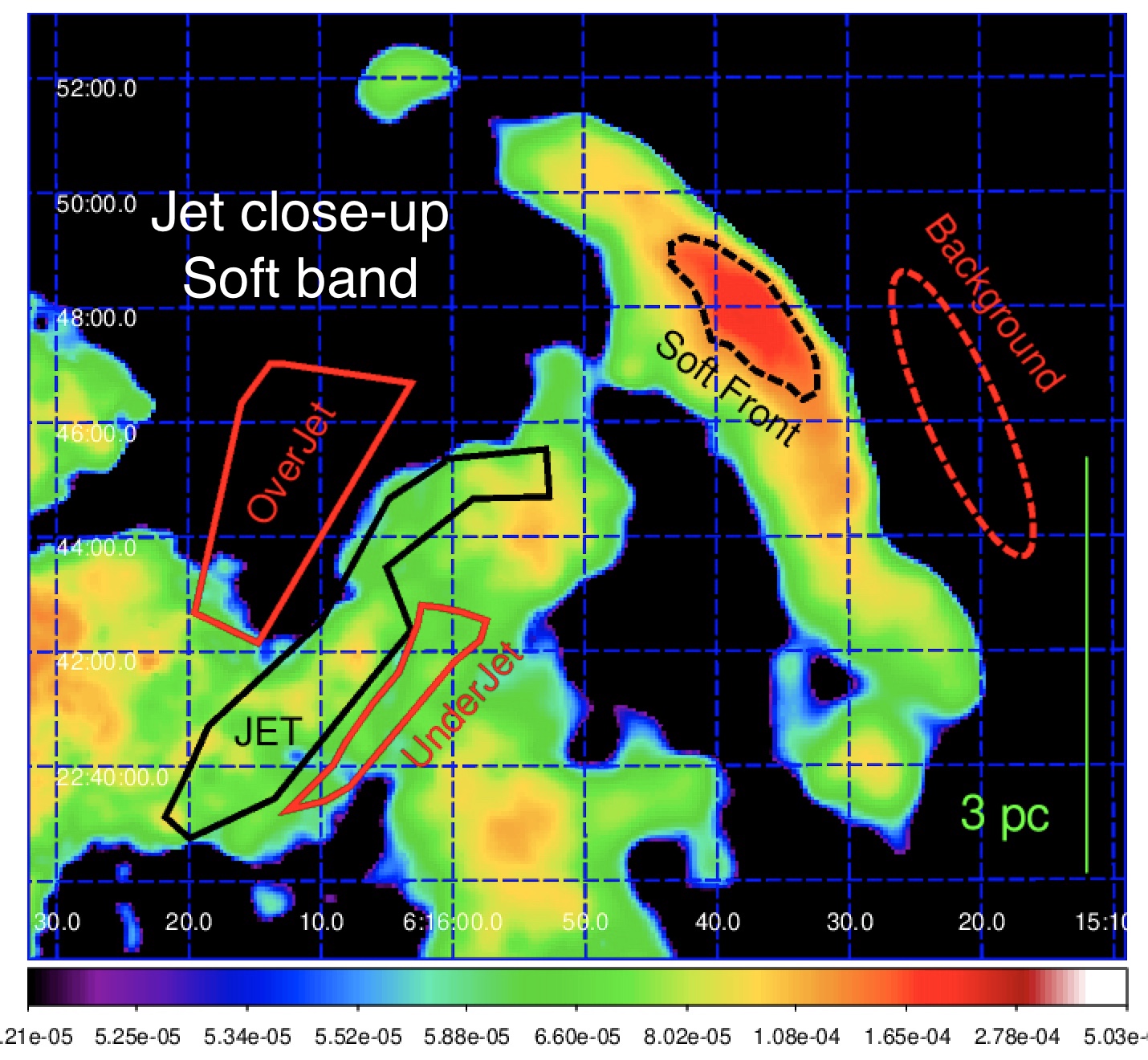}
\includegraphics[width=0.65\columnwidth]{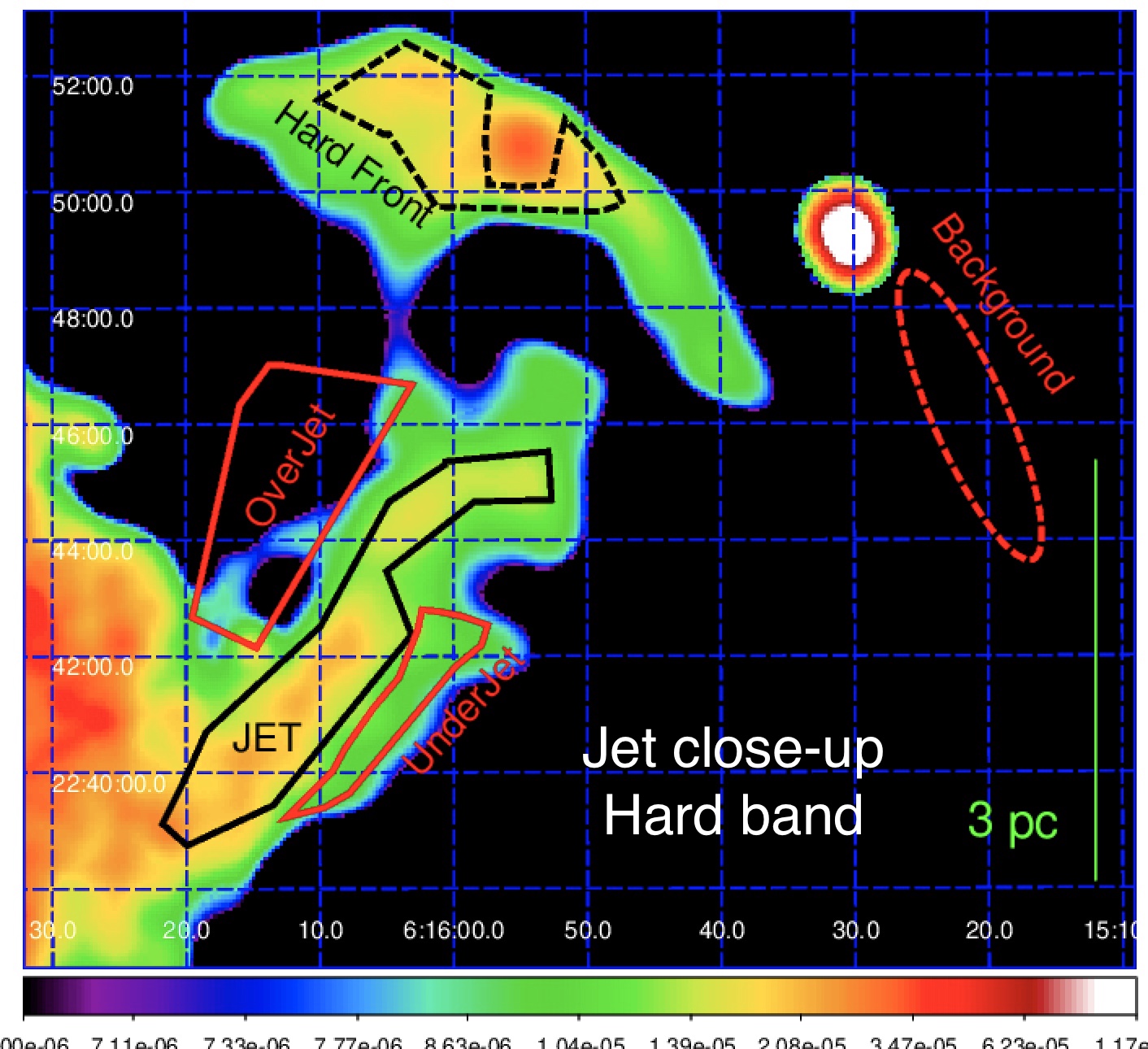}
\caption{Close-up view of the jet area marked by the cyan box in Fig. \ref{ic443andjet} in the Soft band (0.5-1.4 keV, top) and in the Hard band (1.4-5 keV, bottom). The image is adaptively smoothed to a signal-to-noise ratio 25, the bin size is $6''$ and the color scale is logarithmic. Regions selected for the spectral analysis are shown: Jet region is marked in black, Soft Front and Hard Front are marked in dashed black, background region in dashed red and OverJet and UnderJet regions in red.}
\label{dettagliojet}
\end{figure}

Until now, only the internal area of IC 443 has been studied in detail with {\it XMM-Newton} (\citealt{tbr06}), while the elongated jet-like structure, clearly evident within the cyan box (bottom panels), in the NW has not been investigated yet. By taking a closer look at this area (Fig. \ref{dettagliojet}), we can see that the jet, $\sim$ 3 pc long and indicated by the central black region, is the brightest feature in the Hard energy band. The jet is characterized by a relatively narrow, well-collimated, morphology which is also visible in the Soft band, though with a lower contrast.

By taking into account the NS proper motion ($\sim$ 130 km/s towards SW, \citealt{gcs06}) and multiplying it by the age of remnant ($\sim$ 8000 yr, see Sect. \ref{sect:ic443_hd}), I can derive the position of the neutron star immediately after the explosion, as shown by the black cross in the right panel of Fig. \ref{ic443andjet}.
We can see that extending the jet towards the SE directions, it crosses the position which the NS had at the time of the explosion (Fig \ref{dettagliojet}, the figure also shows the position of the explosion site by assuming an age of 4000 yr, as \citet{tbm08}). 

\begin{figure}[!htb]
\centering
\includegraphics[width=0.75\columnwidth]{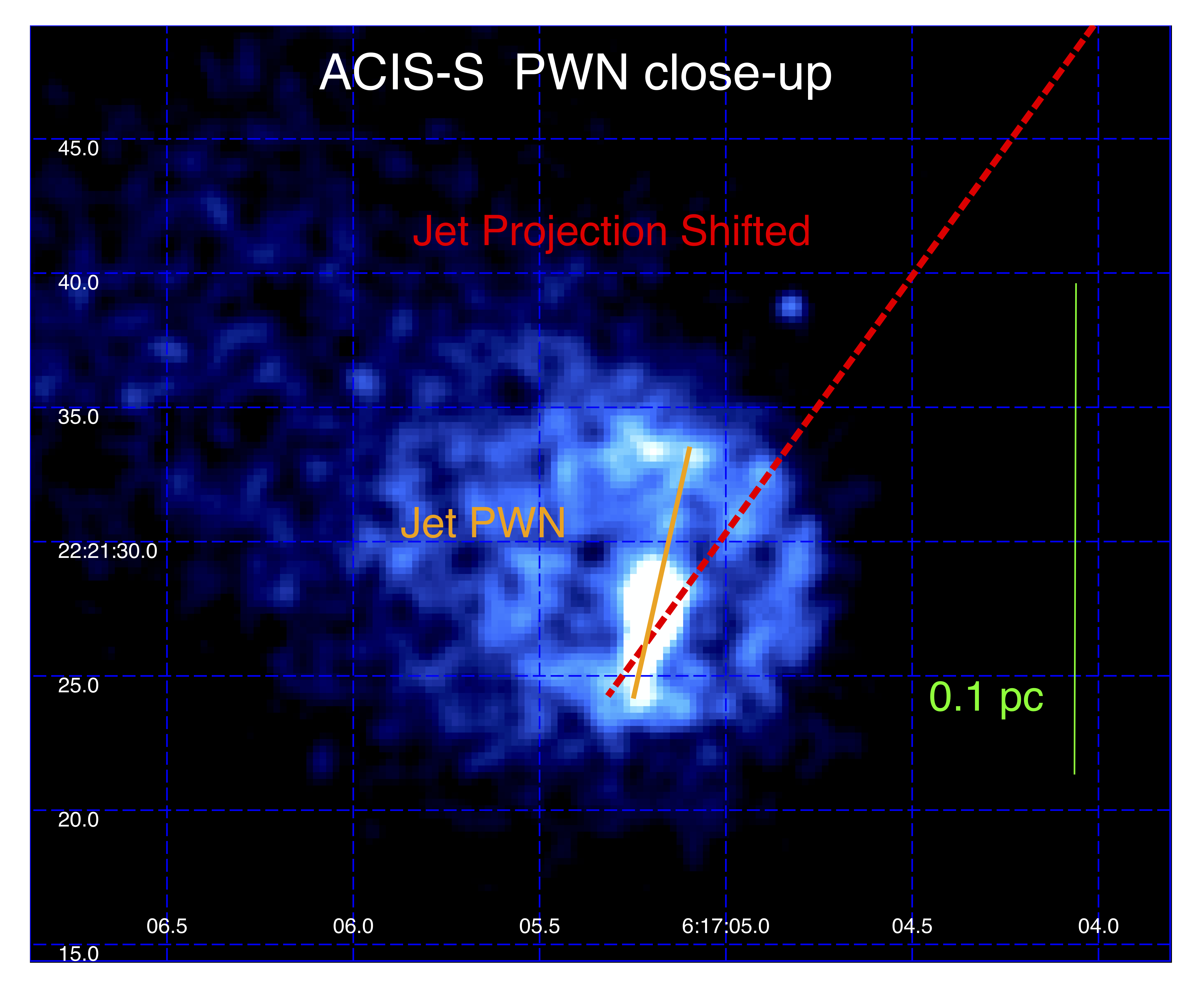}
\caption{{\it Chandra}/ACIS-S image of the area near the PWN which provides a comparison between the SNR jet and the PWN jet in the 0.5-7 keV energy band.  The image is smoothed with a Gaussian with $\sigma=2$ pixel, the pixel size is $0.246''$ and the color scale is linear. The projection of the jet has been shifted and superimposed on the PWN in order to compare the direction of the two jets.}
\label{PWN}
\end{figure}

This result is a clear indication of the link between the PWN CXOU J061705.3+222127 and IC 443, given that the jet is made of ejecta expelled by the IC 443 progenitor, as I will show in Sect. \ref{spectra}. I also created an image of the area near the PWN using data from the {\it Chandra} telescope (Fig. \ref{PWN}). 
This image was produced by rebinning the events files to the half pixel size ($0.246''$), as in \cite{spc15}. In Fig \ref{PWN} I superimpose the direction of the ejecta jet to the PWN map, by shifting it to the current position of the NS, to ease the comparison with the PWN jet detected by \cite{spc15}. The directions of the two jets are somehow similar and this may suggest some physical link between their origins.
I measured an angle of $\sim$ 20$^\circ$ between the two jets. The collimated structure detected by \cite{gsm17} in Vela SNR is not as clear as that observed in IC 443 but it is still possible to measure the angle between the PWN jet and the SNR jet. In the Vela SNR the angle is of $\sim$ 90$^\circ$, as it can be derived by comparing the jet detected by \citet{gsm17} with the PWN jet showed by \citet{ptk03}.

\subsection{Spatially resolved spectral analysis}
\label{spectra}
 
\begin{figure}[!h]
\centering
\includegraphics[width=0.85\columnwidth]{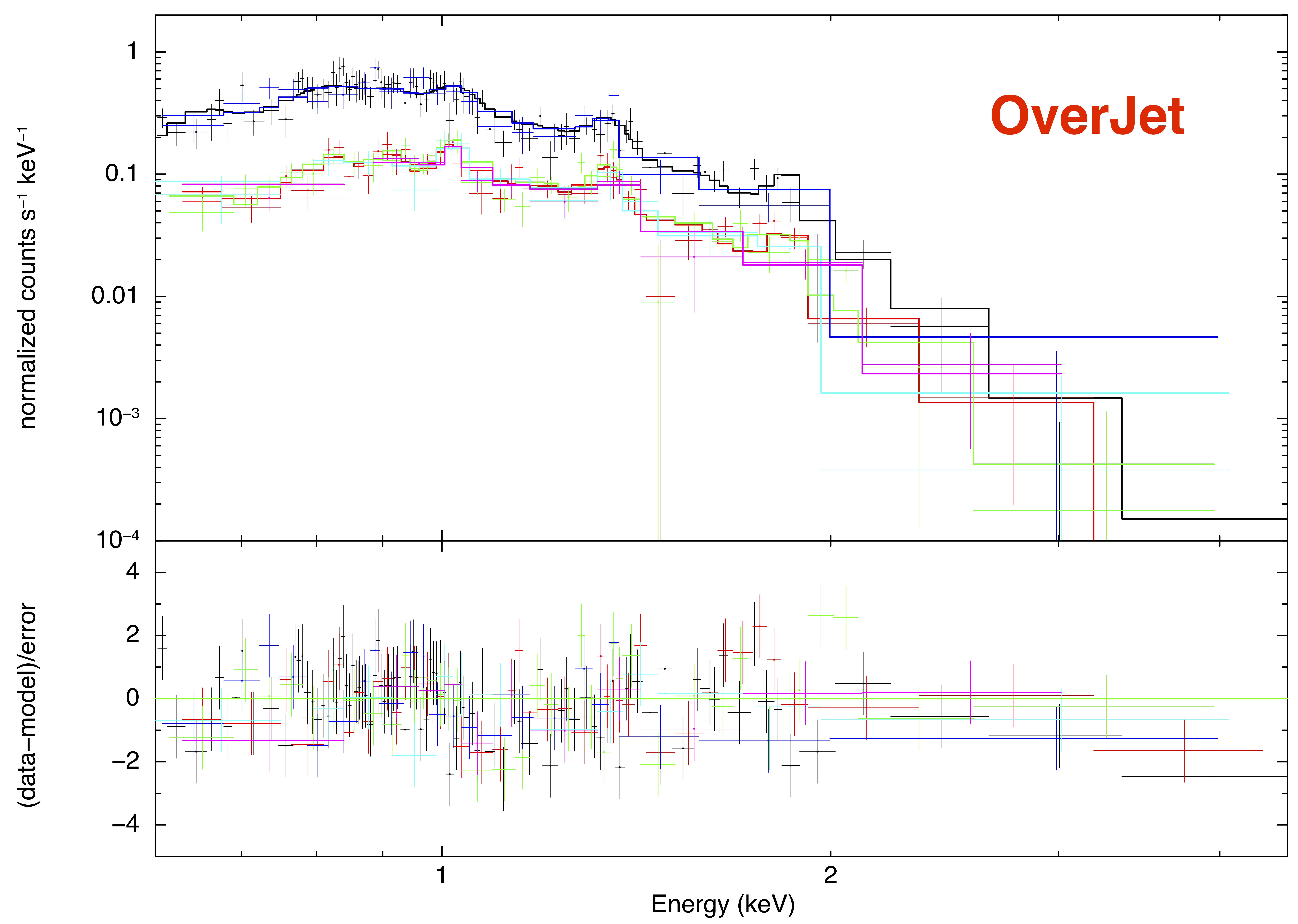}
\includegraphics[width=0.85\columnwidth]{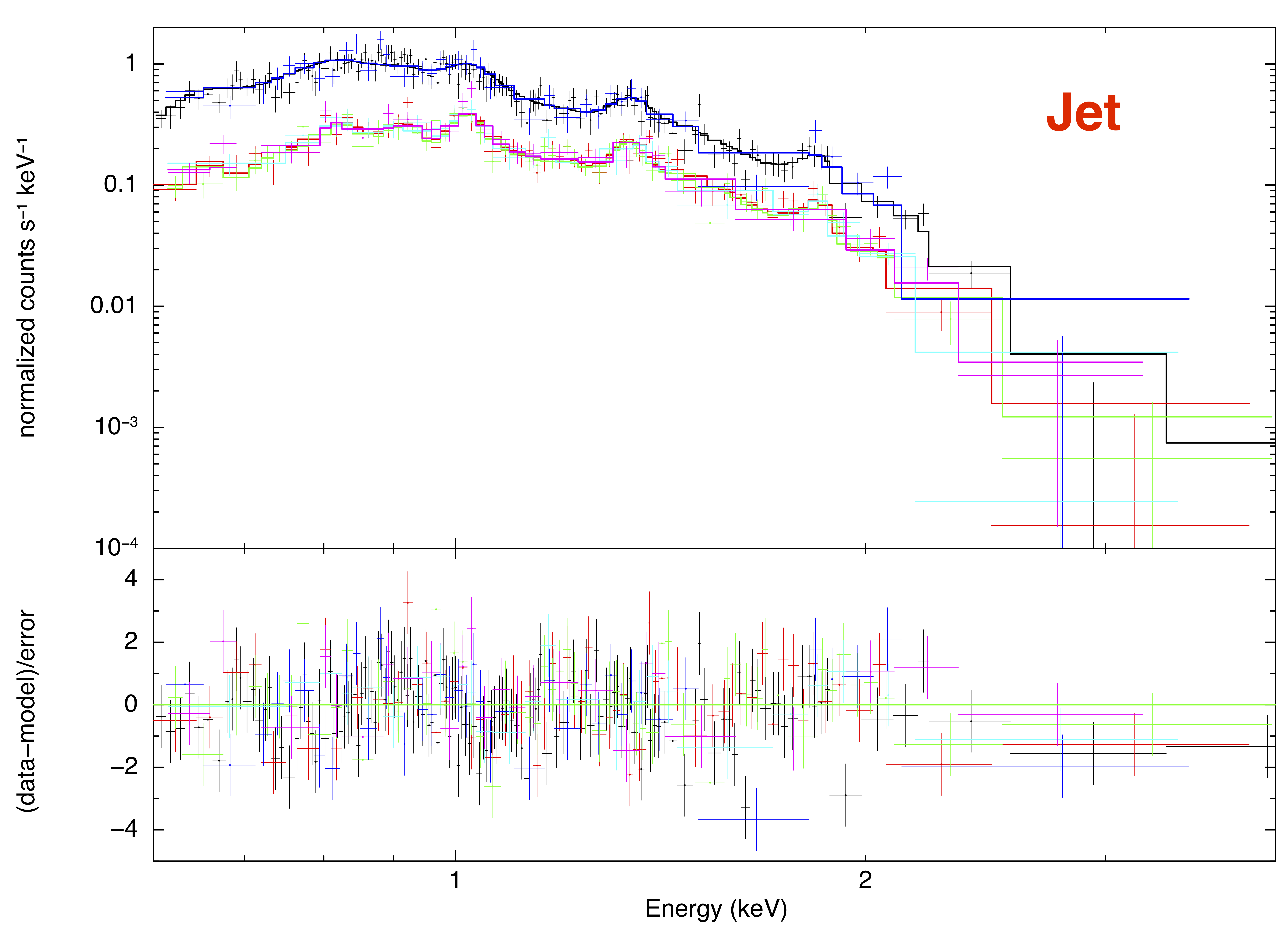}
\caption{\emph{Upper panel}: {\it XMM-Newton}/EPIC spectra of the OverJet region with the corresponding best-fit models and residuals obtained by simultaneously fitting spectra from six cameras with a VAPEC model (best fit parameters are in Table \ref{spettrioverundersoft}). \emph{Lower panel}: same as is the upper panel but for the Jet region described with a VAPEC+VRNEI model (best fit parameters are in Table \ref{risultatifit}).}
\label{spettritot}
\end{figure}

Fig. \ref{dettagliojet} clearly shows that the jet is much brighter than the adjacent regions, labelled OverJet and UnderJet (shown in red in the figure), particularly in the Hard band. I investigated the nature of the related plasma through spectral fits and found that the spectra extracted from regions OverJet and UnderJet are both satisfyingly described by a model of plasma in collisional ionization equilibrium (VAPEC model). 
The spectra of the OverJet region is shown in Fig. \ref{spettritot}. The best-fit parameters for the two regions are all consistent within 2 sigmas, except for the $n_{\mathrm{H}}$. The latter result is not surprising since the molecular cloud, indicated by the white contours in Fig. \ref{ic443andjet}, lies on the foreground and completely covers the OverJet region (thus providing additional absorption for the X-rays). On the other hand, the UnderJet region is only partially covered by the cloud, being close to its border, in projection. Since both regions are well described with a VAPEC model with either solar or under-solar abundances, it is natural to affirm that X-ray emission is associated with shocked ISM in OverJet and UnderJet. 

\begin{table}[!h]
\centering
\caption{Best fit results for OverJet, UnderJet and Soft Front.}
\begin{tabular}{c|c|c|c}
\hline\hline
Parameter & \multicolumn{3}{c}{Region}\\
\hline
& OverJet& UnderJet& Soft Front \\
\hline
$n_{\rm{H}}(10^{22}$ cm$^{-2})$ & 0.65$^{+0.04}_{-0.07}$& 0.50$^{+0.06}_{-0.07}$ & 0.57 $ \pm 0.04$ \\
\hline
kT (keV)& 0.39$^{+0.05}_{-0.01}$ & 0.33$^{+0.02}_{-0.01}$& 0.26 $\pm 0.01$ \\
O& 2.0$^{+0.4}_{-0.3}$& 1.4$ \pm 0.3$ & 0.78$^{+0.11}_{-0.10}$ \\ 
Mg& 0.7 $\pm$ 0.1& 0.7$\pm 0.1 $& 0.6$^{+0.1}_{-0.2}$  \\
Si& 0.8 $\pm$ 0.2& 0.9$ \pm 0.3 $& 1.1$ \pm 0.5 $ \\
Fe& 0.29$^{+0.07}_{-0.05}$&0.42$^{+0.09}_{-0.07}$ & 0.41$^{+0.06}_{-0.07}$  \\
$ n^2 l^{*}$ ($10^{18}$ cm$^{-2}$) &2.4$^{+0.3}_{-0.4}$ &2.4$^{+0.6}_{-0.3}$ & 6$^{+2}_{-1}$ \\
\hline
Counts& 12000 & 13000 & 14000\\
$\chi^2$ (d.o.f)& 400.24 (357) &412.79 (367)&  366.82 (345)\\
\hline
\end{tabular}

$^*$Emission measure per unit area. S abundance is fixed equal to the Si one.
\label{spettrioverundersoft}
\end{table}

A second thermal component is required to model the spectra extracted from the jet. In particular, a single VAPEC component gives an unacceptable $\chi^2 = 643.88$ with 506 d.o.f. The best fit parameters of the soft thermal component are consistent within 2$\sigma$ with those of the UnderJet and OverJet regions. I then fixed the abundances of this soft component to those of the OverJet region and I forced the temperature of this component to vary in the range 0.38-0.44 keV, as in the OverJet region, letting only the emission measure free to vary. By adding a hard component in CIE (VAPEC model in XSPEC) the quality of the fit has drastically improved ($\chi^2$ = 613.77 with 506 d.o.f.). Since this new component shows over-solar abundances (Table \ref{risultatifit}), I associate it with shocked ejecta while the quite homogeneous soft X-ray emission in all the three regions is due to shocked ISM.   

A problem that has emerged is the impossibility to estimate absolute abundances. Analogously to what I discussed in Chapter \ref{ch:rrc}, since spectra are dominated by line emission and the jet area is quite faint, it is impossible to estimate precisely the continuum and the intensity of each emission line with CCD spectrometers, like those of the {\it XMM-Newton}/EPIC cameras: I can fit the data by enhancing all the abundances by the same factor $f$ and reducing the emission measure by $f$ without significant variation of $\chi^2$, as showed in Sect. \ref{sect:allsynthesis}. In order to solve partially this degeneracy between abundances and emission measure, I considered relative abundances calculated with respect to Mg, the element which shows the most prominent line in the spectrum (see Fig. \ref{spettritot}). With this approach, I satisfyingly constrained abundances (provided that Mg abundance is fixed during the error calculation). Error on Mg abundance is calculated fixing the Fe abundance.   

\begin{figure}[!ht]
\centering
\includegraphics[width=0.9\columnwidth]{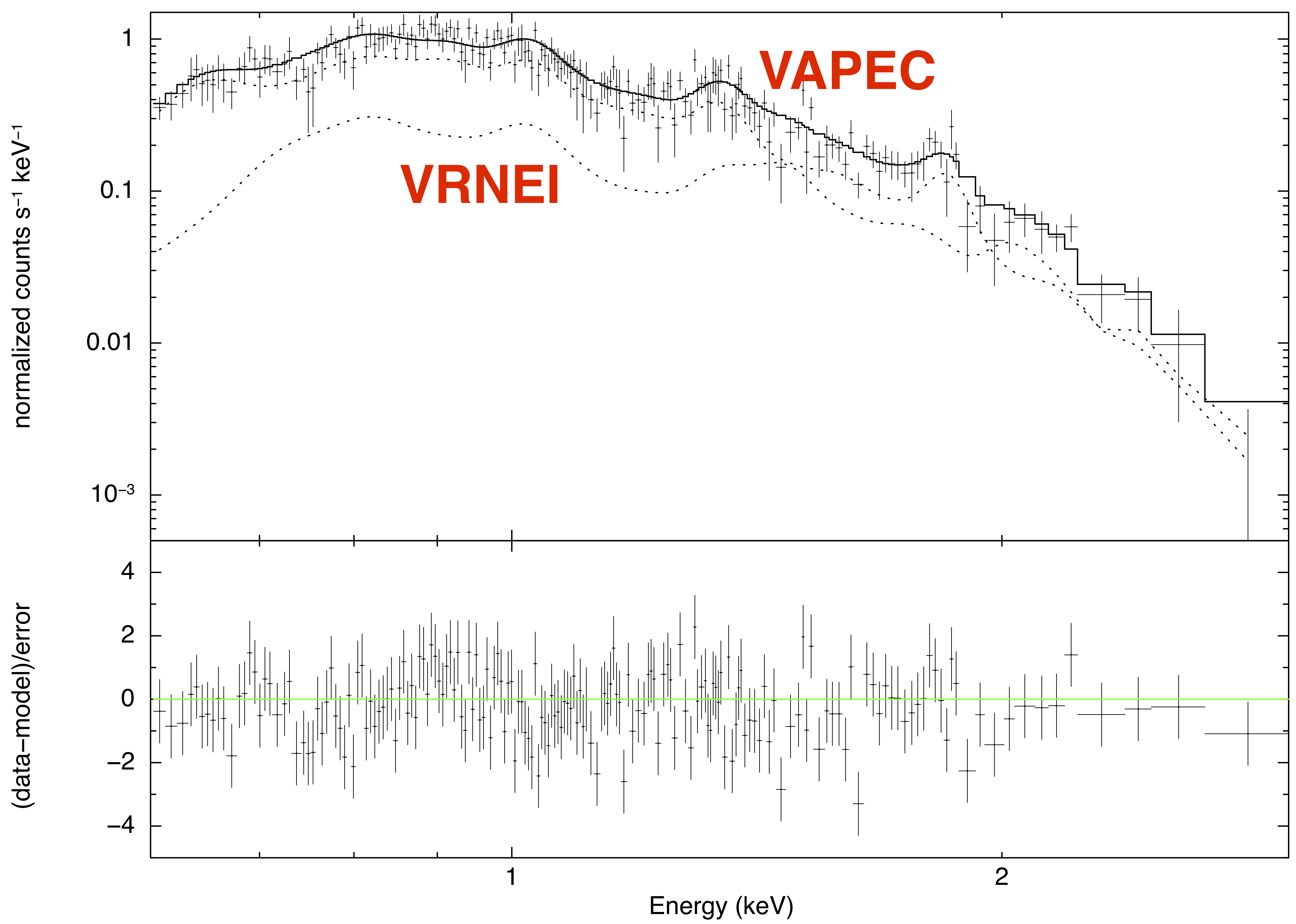}
\caption{{\it XMM-Newton}/EPIC pn spectrum of the jet extracted from observation 0114100501 showing two different components of the VAPEC+VRNEI model.  The lower panel shows the residuals.}
\label{spettrojetsolopn}
\end{figure}

Since previous works (\citealt{out14}, \citealt{yok09}, \citealt{mtu17}, hereafter Ma17) pointed out the presence of overionized plasma in IC 443, I investigated whether the jet could show some indication of overionization associated with a rapid cooling of the plasma (see Sect. \ref{sect:nei}). I replaced the hotter component of the two-components VAPEC model with a \emph{VRNEI} component, used to describe plasma in recombination phase. 
To this end, I assumed an initial temperature of 5 keV (in agreement with Ma17). The $\chi^2$ significantly decreased from 613.77 (506 d.o.f ) to 575.05 (506 d.o.f.). This result suggests that ejecta in the jet are overionized. Table \ref{risultatifit} also shows the best fit results for this VAPEC+VRNEI model.
Fig. \ref{spettrojetsolopn} shows the {\it XMM-Newton}/EPIC pn spectrum extracted from observation 0114100501 with the contribution of the two spectral components. Below $\sim$ 1.4 keV, the ISM component is dominant while the VRNEI component dominates at higher energies. Fig. \ref{spettritot} shows the spectra obtained with all cameras of both observations from regions OverJet and Jet. It is clear that the spectrum of the region OverJet is significantly softer than that of the jet region.

As it appears from Fig. \ref{dettagliojet}, the morphology of the upper part of the jet (labeled as Front in Fig. \ref{dettagliojet}), is different according to the energy band we are looking at, i.e. it is shifted to NE in the Hard band (region "Hard Front" in the figure)  with respect to the Soft band (region "Soft Front").
I analyzed spectra extracted from these regions, finding that the Soft Front has chemical properties consistent within 2 $\sigma$ with those of the OverJet and UnderJet regions, being well described with a single VAPEC component with solar or under-solar abundances (Table \ref{spettrioverundersoft}). This indicates that the emission in the Soft Front does not originate from the ejecta. Therefore, I interpret the enhanced brightness of this region as due to the interaction between the jet and the molecular cloud, as I will discuss in detail in Sect. \ref{discu}.
 On the other hand, the spectrum of the Hard Front needs to be fitted with a two component VAPEC+VRNEI model providing a $\chi^2=346.33$ with 342 dof and the recombining component shows abundances consistent with those of the jet within 2$\sigma$ (the Hard Front can also be described with a VAPEC+VAPEC model providing a $\chi^2$ = 342.57 with 342 dof). Hard Front plasma showed higher values of the ionization parameter ($\tau = 9 \pm 1 \times 10^{11}$ vs $\tau = 3^{+2}_{-1} \times 10^{11}$ s/cm$^3$ found for the jet) and of the Si to Mg abundance ratio (Si/Mg $\sim 0.75$), which is  $\sim$ 10 times greater than in the jet (Si/Mg $\sim$ 0.08). This may be indicative of spatial inhomogeneities in the physical and chemical properties of the jet. Nevertheless, the low surface brightness and the poor statistics do not allow us to address accurately this issue.

\begin{table}[!h]
\centering
\caption{Best fit results for the jet and the Hard Front.}
\begin{tabular}{c|c|c}
\hline\hline
\multicolumn{3}{c}{Jet}\\
\hline
Parameter & \multicolumn{2}{c}{Model}\\
\hline
 & VAPEC+VAPEC& VAPEC+VRNEI\\
\hline
$n_{\mathrm{H}} (10^{22} \mathrm{cm}^{-2})$ & 0.57 $\pm$ 0.03   & 0.60$^{+0.03}_{-0.01}$ \\
\hline
kT (keV)& 0.44& 0.38\\
$ n^2 l^{*}$ ($10^{18}$ cm$^{-2}$) &  5 $\pm 1$ & 4$\pm 3$ \\
\hline
 kT$_{\mathrm{init}}$ (keV) & - & 5 (frozen) \\
 kT (keV) & 0.76$^{+0.08}_{-0.05}$& 0.21$^{+0.04}_{-0.03}$ \\
 O& 11$^{+7}_{-5}$ & 1 (frozen) \\ 
 Ne&  4$^{+2}_{-3}$ & 11$^{+8}_{-5}$ \\
 Mg& 4$^{+1}_{-2} $& 13$^{+11}_{-7}$ \\
 Si& 0.8$^{+0.4}_{-0.6} $& 1$^{+2}_{-1}$ \\
 Fe& 0.8$\pm{0.3}$ & 10$^{+10}_{-6}$ \\
 $\tau$ (s/cm$^3$) & - & 3$^{+2}_{-1} \times 10^{11}$ \\
 $ n^2 l^{*}$ ($10^{18}$ cm$^{-2}$) & 0.6$\pm 0.2$& 4.2$^{+0.9}_{-1.2}$ \\
\hline
$\chi^2$ (d.o.f)& 613.77 (506)&  575.05 (506)\\
 Counts & \multicolumn{2}{c}{20000}  \\
\hline\hline
\multicolumn{3}{c}{Hard Front}\\
\hline
$n_{\mathrm{H}} (10^{22} \mathrm{cm}^{-2})$ & 0.73$\pm$0.05 & 0.99$^{+0.08}_{-0.10}$ \\
\hline
kT (keV) & 0.44& 0.38\\
$ n^2 l^{*}$ ($10^{18}$ cm$^{-2}$) & 4$^{+1}_{-2}$ & 4$^{+4}_{-3.9}$ \\ 
\hline
 kT$_{\mathrm{init}}$ (keV) & - & 5 (frozen) \\
 kT (keV) &0.73$\pm$ 0.6 & 0.25 $\pm$ 0.03 \\
 O& 7$^{+8}_{-4}$& 1.6 $\pm$ 0.9 \\ 
 Ne& 2$^{+3}_{-2}$ & 3.6 $^{+1.3}_{-0.8}$ \\
 Mg& 3.1$^{+1.2}_{-1.4}$& 2.5 $\pm 1.2$ \\
 Si& 0.9$^{+0.7}_{-0.8}$ & 2.0$^{+0.8}_{-0.1}$ \\
 Fe& 1.8$^{+1.4}_{-0.6}$ & 2$^{+2}_{-1}$ \\
 $\tau$ (s/cm$^3$) & - & 9$^{+2}_{-1} \times 10^{11}$ \\
 $ n^2 l^{*}$ ($10^{18}$ cm$^{-2}$) & 0.8$^{+0.5}_{-0.4}$& 1.3 $\pm$ 0.5 \\
\hline
$\chi^2$ (d.o.f)& 342.57 (342)&  346.33 (342)\\
 Counts & \multicolumn{2}{c}{11000} \\
\hline
\end{tabular}

S abundance is fixed equal to the Si one. Abundance of the first component are fixed to the best fit values shown in Table \ref{spettrioverundersoft} and $kT$ of the ISM component can vary only in the range 0.38-0.44 keV. $^*${Emission measure per unit area.}
\label{risultatifit}
\end{table}
\newpage
\clearpage
\section{Discussion}
\label{discu}

\subsection{Comparison with HD simulations}
\label{sect:ic443_hd}
\begin{figure}
\centering
\includegraphics[width=.99\columnwidth]{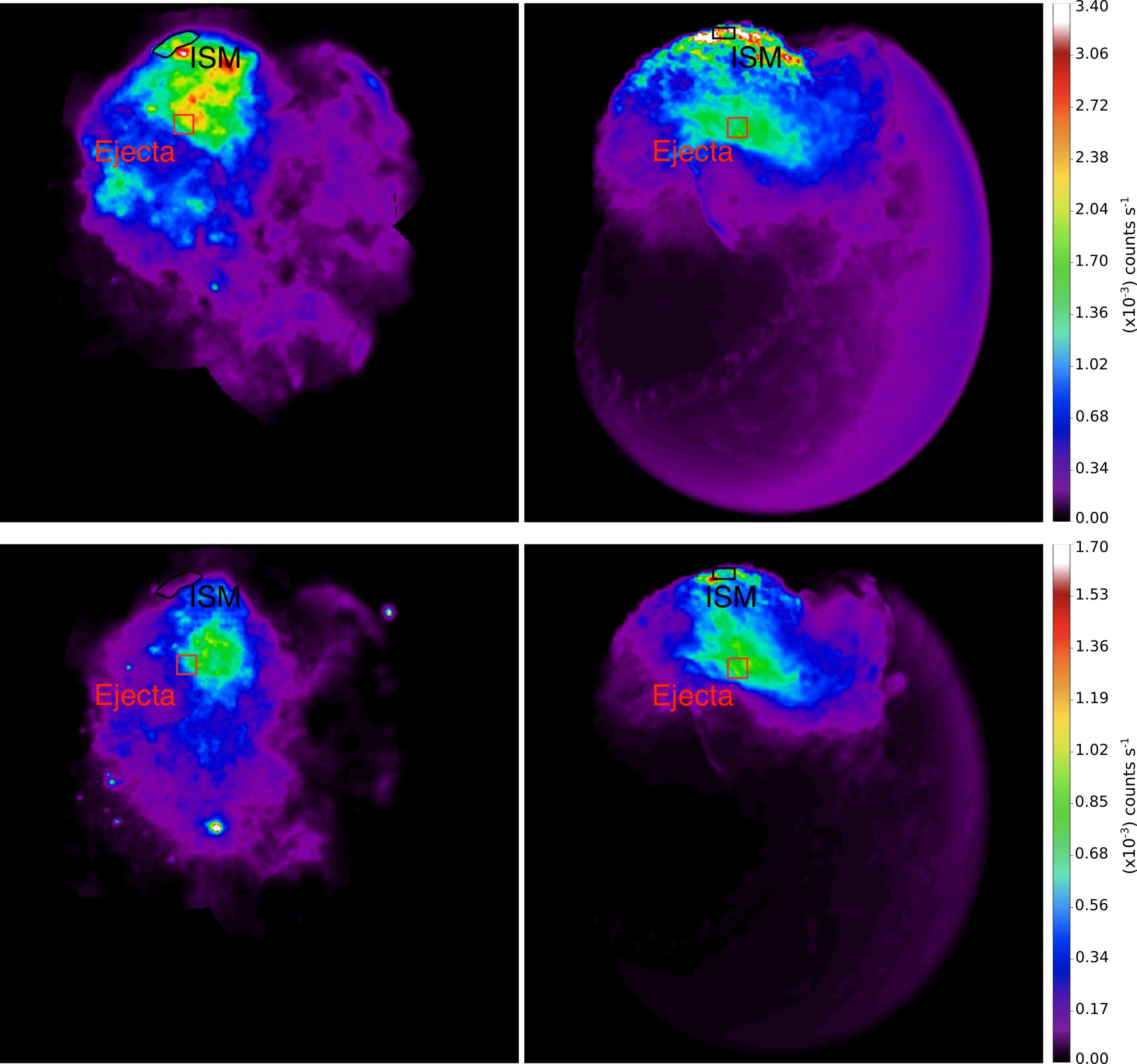}
\caption{Comparison between images generated from actual data and synthetic images derived from the HD model presented by \citet{uog20}.  All images are adaptively smoothed to a signal-to-noise ratio of 25, the bin size is $11''$ and the color scale is linear. The red boxes mark the regions associated with the shocked ejecta emission. The black regions identify the shocked ISM emission. \emph{Left panels}: Background subtracted and vignetting-corrected count-rate image of IC 443 in the Soft (top panel) and Hard (bottom panel) energy bands. \emph{Right panels}: synthetic images derived from the HD model in the Soft (top panel) and Hard (bottom panel) (see \citealt{uog20}).}
\label{fig:reg_sabina}
\end{figure}

\begin{figure}
\centering
\includegraphics[width=.85\columnwidth]{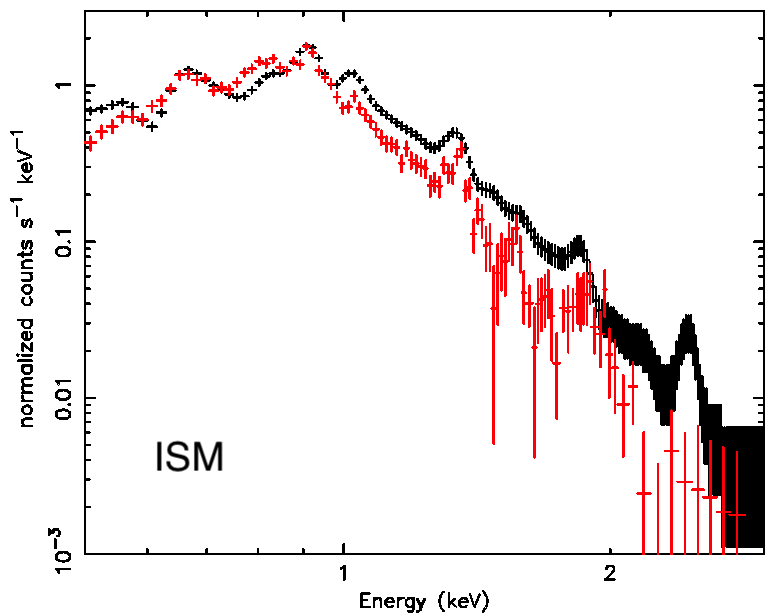}
\includegraphics[width=.85\columnwidth]{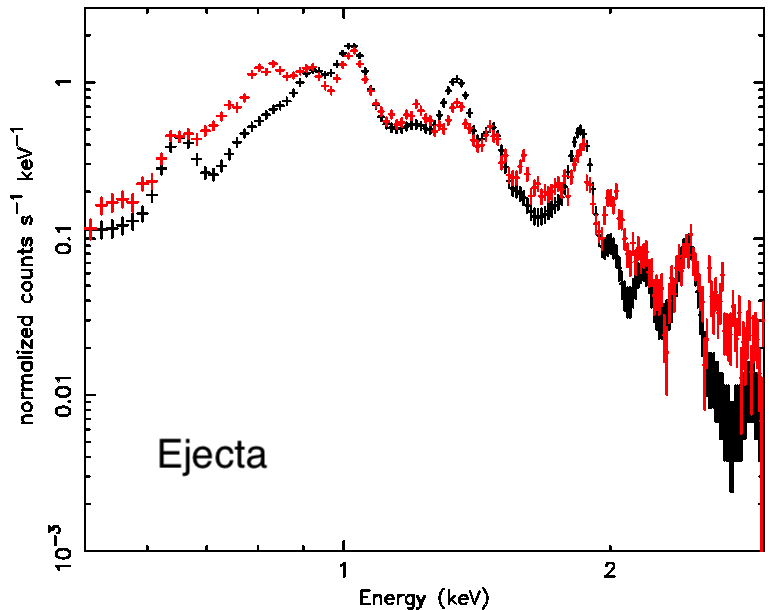}
\caption{{\it XMM-Newton}/MOS spectra extracted from the limb (upper panel) and ejecta regions (lower panel) shown in Fig. \ref{fig:reg_sabina}. In black, the synthetic spectra derived from the HD model; in red, spectra extracted from the actual {\it XMM-Newton}/EPIC observations.}
\label{fig:spec_sabina}
\end{figure}

The link between the PWN and IC 443 is further supported by a set of HD simulations performed by \citet{uog20}, where the authors explain the origin of the complex X-ray morphology of IC 443 and reproduce various observables taking into account the position of the explosion. In particular, the model which best reproduces the observations predicts an age $\sim$ 8000 yr and an explosion siste consistent with the position of the PWN at that time. Scaling the position of the PWN by 8000 yr, I notice that the projected jet is compatible with the position where the progenitor star exploded. 
I actively contributed to this project by sinthesizing X-ray spectra from the HD simulation, analogously to what I described for the HD simulation of Cas A in Chapter \ref{ch:rrc}, and by comparing the synthetic spectra with actual data. In particular, I selected two regions which showed similar observed and synthetic count rates, one corresponding to an area of IC 443 where the emission is dominated by ejecta (black boxes in Fig. \ref{fig:reg_sabina}), and the other one near the northern limb where the ISM emission dominates the X-ray spectrum (white regions in Fig. \ref{fig:spec_sabina}). This comparison shows that the HD model is remarkably able to reproduce the main features of the spectra (see Fig. \ref{fig:spec_sabina}) such as the continuum and the line emission. For more details on the HD model see \citet{uog20}.  

\subsection{Origin of the overionization}

The results of the spatially resolved spectral analysis help to infer a reliable scenario which can explain the observed morphology of the jet region (see Fig \ref{dettagliojet}.) The molecular cloud has an important role in the dynamical evolution of IC 443: after the SN explosion, ejecta expanded and encountered the NW molecular cloud which has a density more than 100 times higher (\citealt{cck77}), thus representing a wall for the ejecta. In particular, the jet cannot pass through the cloud and is therefore distorted, the distorted part being the region Hard Front (which is actually the upper part of the jet-like structure). The Soft Front emission is likely due to heating, caused by the impact of the jet onto the cloud.

The jet-like structure of IC 443 is the first known structure of this type which clearly presents emission of overionized plasma. Until now other two jets have been inferred by \cite{wbv03}, \cite{hlb04} and \cite{fhm06} in Cas A and \cite{gsm17} in Vela SNR, but in both cases the plasma is underionized. I investigated possible causes of this overionization and I devised the following scenario. 

Ma17 studied the distribution of overionized plasma in IC 443 by analyzing a set of Suzaku observations. They found overionized plasma in the region surrounding the PWN, where the ionization parameter is extremely low ( $ \tau \approx 4 \times 10^{11}$ s/cm$^3$), indicating a plasma very far from conditions of collisional ionization equilibrium (CIE). Instead, they found CIE conditions in the NW jet region (which is very distant from the SE cloud) and concluded that overionization in the jet is due to thermal conduction with the cold SE cloud. I here proved that, though a large area of the NW shell is in CIE (i.e. the part where the ISM emission dominates), the ejecta at NW are overionized. The ejecta in this region are concentrated in the collimated jet that is relatively narrow (the thickest part is 1.5$'$ large) and therefore unresolved by the large PSF of the Suzaku mirrors. I also found that the degree of overionization in the NW jet ($\tau= 3^{+2}_{-1} \times 10^{11}$ s/cm$^3$) is perfectly consistent with that measured by Ma17 at SE ($\tau = 4.2^{+0.2}_{-0.1} \times {10^{11}}$ s/cm$^3$, see Table 3 of Ma17). This result seems to exclude that the overionization is due to thermal conduction with the SE cloud and may instead suggest an adiabatic cooling associated with the expansion of the ejecta. Before the expansion, the ejecta may have been heated by the reflected shock generated by the impact of the forward shock front with the SE cloud (which is very close to the explosion site, marked by the original position of the PWN in Fig. \ref{ic443andjet}).

\begin{figure}[!htp]
\centering
\includegraphics[width=0.7\columnwidth]{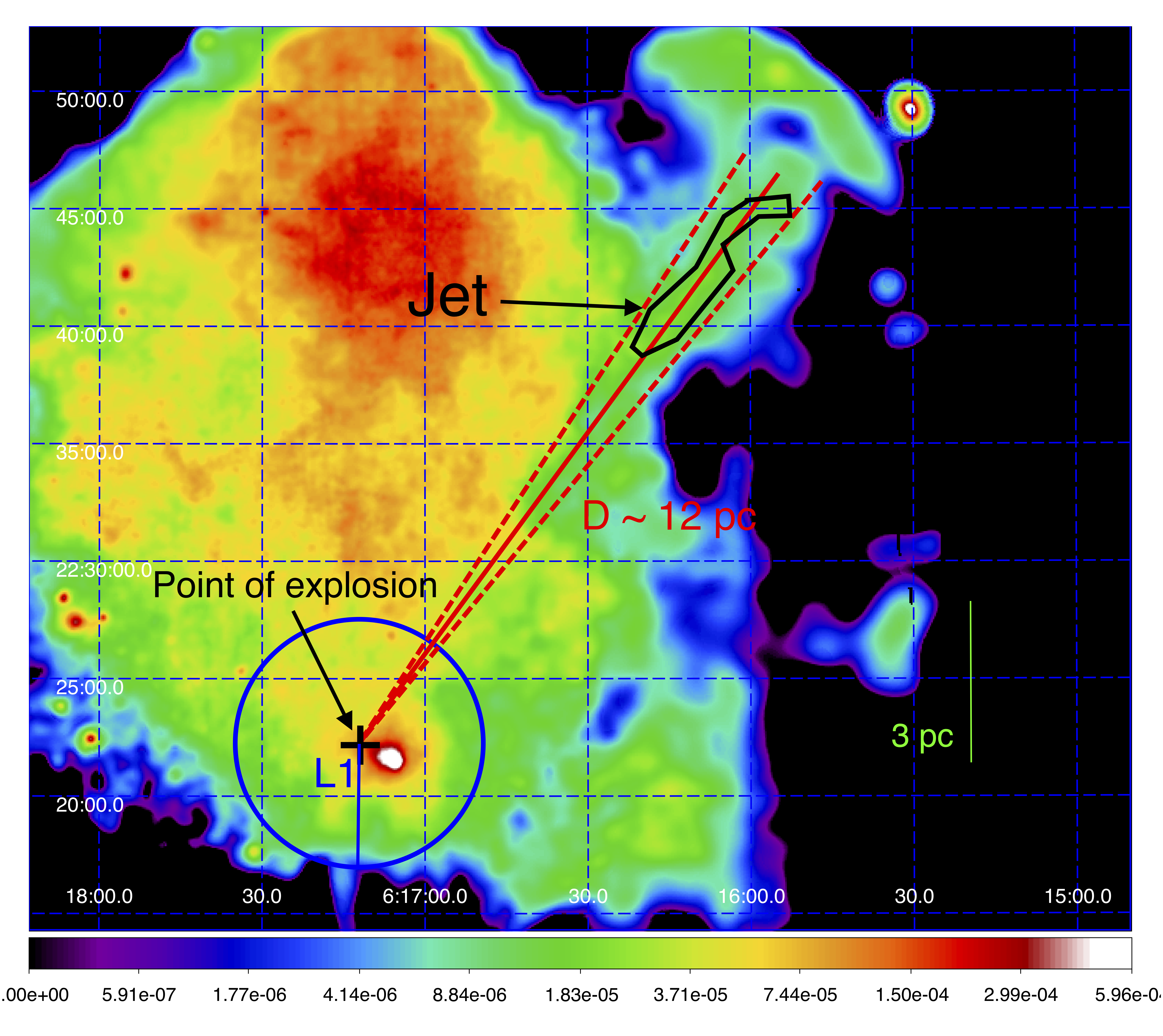}
\caption{Count-rate image of IC 443 in the Hard energy band (1.4-5 keV). Overview of the scheme used to estimate the volume V1 of the jet at time $t_0$, namely when the forward shock interacts with the molecular cloud.}
\label{stimatauT}
\end{figure}

 To verify this scenario, I roughly estimated temperature and ionization parameter of the plasma by assuming adiabatic expansion. I considered an initial temperature of 5 keV, which is the fixed value of kT$_{\mathrm{init}}$ in the best fit model of the jet, with the VAPEC+VRNEI model. In order to get overionized plasma, it is necessary that a hot plasma in CIE undergoes a rapid cooling. 
 For a gas in adiabatic expansion $T_1V_1^{\gamma-1}=T_2V_2^{\gamma-1}$ where $T_2$  and $V_2 \sim 10^{56}$ cm$^3$ are the current temperature and volume of the jet, and $T_1 = 5$ keV, $V_1$ are the jet temperature and volume at the time of the interaction of the expanding forward shock with the molecular cloud in the south (see Fig. \ref{stimatauT}). 
 
$V_1$ can be estimated in the following way. The ejecta are likely heated by the reflected shock generated by the interaction of the primary shock with the molecular cloud in the south. If $L_1$ $\approx$ 2 pc is the distance between the area of the explosion and the SE cloud \citep{uog20} and $v_{\mathrm{jet}}$ is the expansion velocity (assumed constant) of the jet, then $t_0=L1/v_{\mathrm{jet}}$ is the time when the hot ejecta are in CIE, immediately before their expansion. 
Then, I need to evaluate the volume of the jet at $t_0$ assuming that it has a conic shape, as shown in Fig. \ref{stimatauT}. Considering the projection (marked with the letter D in Fig. \ref{stimatauT} and $\sim$ 12 pc long) of the current jet (which has volume $V_2$) towards the PWN, it is possible to calculate $V_1 \sim 10^{54}$ cm$^3$ and then $T_2 \sim 0.5$ keV. 

 The ionization parameter $\tau$ is defined as the time integral of the electron density $n_{\mathrm{e}}$,  $ \tau= \int n_{\mathrm{e}} dt$. $n_{\mathrm{e}}$ is defined as $c/V$ where $c=m_{\mathrm{jet}}/\mu$ is the ratio between the jet mass and the average atomic mass $\mu=2.1\times10^{-24}$ g and $V$ is the volume occupied by the plasma. Thanks to the assumption of a conic shape for the current jet it is possible to write $V=V_1+\alpha t^3$ where $\alpha=v_{\mathrm{jet}}^3 \pi \tan^2(\theta)/3$ which gives $\tau \approx 2 \times 10^{11}$ s/cm$^3$ and $\theta$ is the opening angle of the cone. 
 The values of kT and $\tau$ are similar to those measured through the spatially resolved spectral analysis, further confirming the goodness of the adiabatic scenario (similar to that proposed for W49B by \citealt{zmb11}) which needs to be confirmed through dedicated hydrodynamic models. 

Finally, I estimated the mass and the kinetic energy associated with the jet and the Hard Front and I obtained M$\sim 0.03$ M$_{\odot}$ and K $\sim 4 \times 10^{48}$ erg, respectively; these are values intermediate between those found by \cite{wbv03} and \cite{omp16} for Cas A (M $\sim 0.4$ M$_{\odot}$ and K $\sim 10^{49}$ erg) and those found by \cite{gsm17} for the Vela SNR (M $\sim 0.008$ M$_{\odot}$ and K $\sim 10^{47}$ erg). The jet of IC 443 is then a sort of intermediate jet structure in terms of kinetic energy and mass.

\section{Conclusions}
\label{sect:conc_ic443}

In this Chapter, I described the study of a Mg-rich jet-like structure detected in the NW area of IC 443 whose emission is mainly due to overionized plasma. Adiabatic expansion is the most likely scenario that caused the overionization in the jet. The jet projection towards the PWN crosses the position of the NS at the time of the explosion of the progenitor star, strongly indicating that the PWN belongs to IC 443 and that the collimated jet has been produced by the exploding star. A comparison between a set of 3D HD simulations of IC 443 and the results from the analysis of the observations, further confirms that the explosion occurred in the place where the progenitor star was $\sim 8000$ years ago.

%% file: 1987A.tex
\chapter{Indications of a pulsar wind nebula within SN 1987A}
\label{ch:pwn_87A}

In this chapter I report on my analysis of multi-epoch {\it Chandra} and {\it NuSTAR} observations of SN 1987A, covering a wide X-ray energy band (0.5-20) keV. I detected synchrotron emission in the (10-20) keV band. The possible scenario powering such emission is twofold: DSA or emission arising from an absorbed PWN. By relating a state-of-the-art magneto-hydrodynamic simulation of SN 1987A to the actual data, I reconstruct the absorption pattern of the PWN embedded in the remnant and surrounded by cold ejecta. I find that, even though the DSA scenario cannot be firmly excluded, the most likely scenario that well explains the data is the PWN emission (see also \citep{gmo21}).

The chapter is organized as follows: in Sect. \ref{sect:data_87a} I present the data analysis procedure; in Sect. \ref{sect:ray_tracing} I explain how I reconstructed the absorption pattern starting from the MHD simulation of SN 1987A; in Sect. \ref{sect:x-ray_data} I show the results of spectral analysis; in Sect. \ref{sect:pwnvsdsa} I discuss the physical implications for the possible processes powering the synchrotron radiation and in Sect. \ref{sect:conc_1987A} I draw the conclusions.
\section{The data}
\label{sect:data_87a}

SN 1987A has been continuously monitored through annual dedicated observations in a very wide range of wavelengths. In particular, here I use data collected in 2012, 2013 and 2014 with {\it Chandra}/ACIS-S and {\it NuSTAR}/FPMA,B detectors (I analyzed all the publicly available {\it NuSTAR} observations of SN 1987A. The main characteristics of the two telescopes are presented in Appendix \ref{app:telescopes}. Details of the observations can be found in Table \ref{tab:obs}\footnote{I limited my analysis to these three epochs because of the availability of {\it Chandra} and {\it NuSTAR} public observations}.
\begin{table*}[!ht]
    \centering
     \caption{Summary of the main characteristics of the analyzed observations.}
    \begin{tabular}{c|c|c|c|c}

     Telescope &OBS ID& PI& Date (yr/month/day) & Exposure time (ks)  \\
     \hline\hline
     & 13735& Burrows& 2012/03/28 &48\\
     & 14417& Burrows& 2012/04/01&27\\
     {\it Chandra}&14697& Burrows& 2013/03/21 &68\\
     &14698& Burrows& 2013/09/28 &68 \\
     & 15809& Burrows & 2014/03/19 &70\\ 
     & 15810& Burrows& 2014/09/20 &48\\
     \hline
     & 40001014003& Harrison& 2012/09/08&136\\
     & 40001014004& Harrison& 2012/09/11&200\\
     & 40001014007& Harrison& 2012/10/21&200\\
     {\it NuSTAR}& 40001014013& Harrison& 2013/06/29 &473\\
     & 40001014018& Harrison& 2014/06/15&200\\
     & 40001014020& Harrison& 2014/06/19&275\\
     & 40001014023& Harrison& 2014/08/01&428\\
    \end{tabular}
    \label{tab:obs}
\end{table*}

{\it Chandra} data were reprocessed with the CIAO v4.12.2 software, using CALDB 4.9.2. I reduced the data through the task {\it chandra\_repro} and extracted the ACIS-S spectra by using the tool {\it specextract} which also provided the corresponding ancillary and response files. 

{\it NuSTAR} data were reprocessed with the standard pipelines provided by the {\it NuSTAR} data analysis software NuSTARDAS\footnote{https://heasarc.gsfc.nasa.gov/docs/nustar/analysis/nustar\_swguide.pdf} by using {\it nupipeline} and {\it nuproducts}.

\begin{figure}
\centering
\begin{minipage}{0.49\textwidth}
  \includegraphics[width=\textwidth]{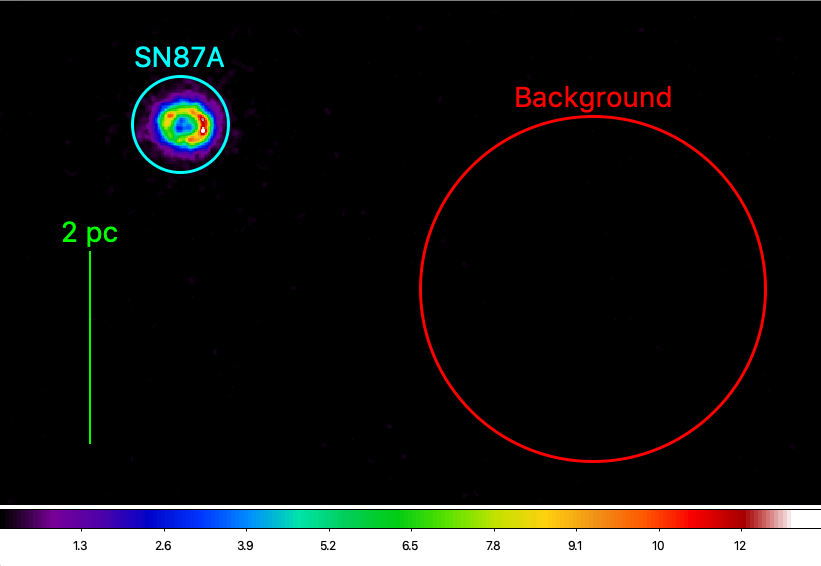}
 \end{minipage}
  \begin{minipage}{0.47\textwidth}
  \includegraphics[width=\textwidth]{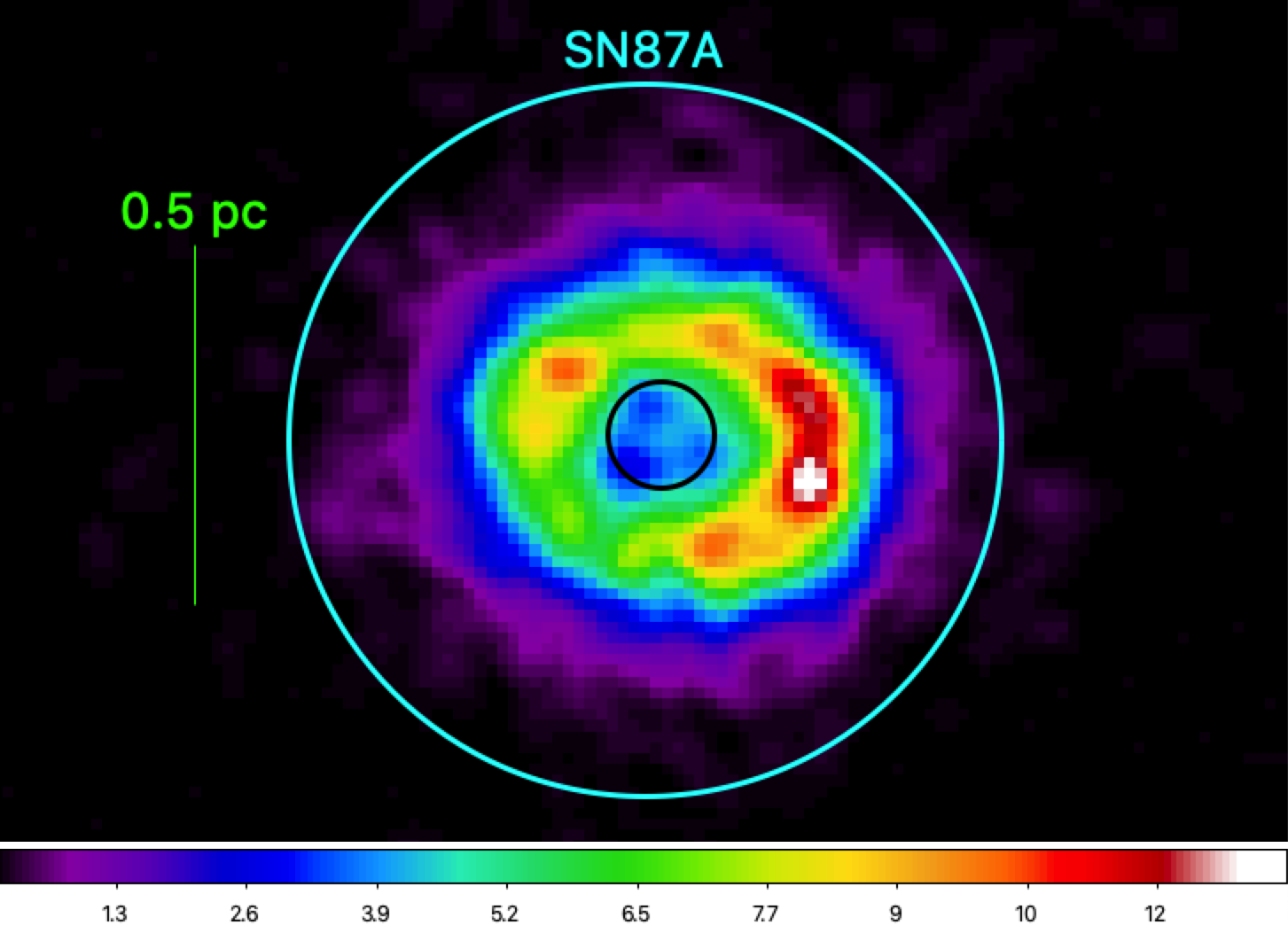}   \end{minipage}
  \includegraphics[width=.97\textwidth]{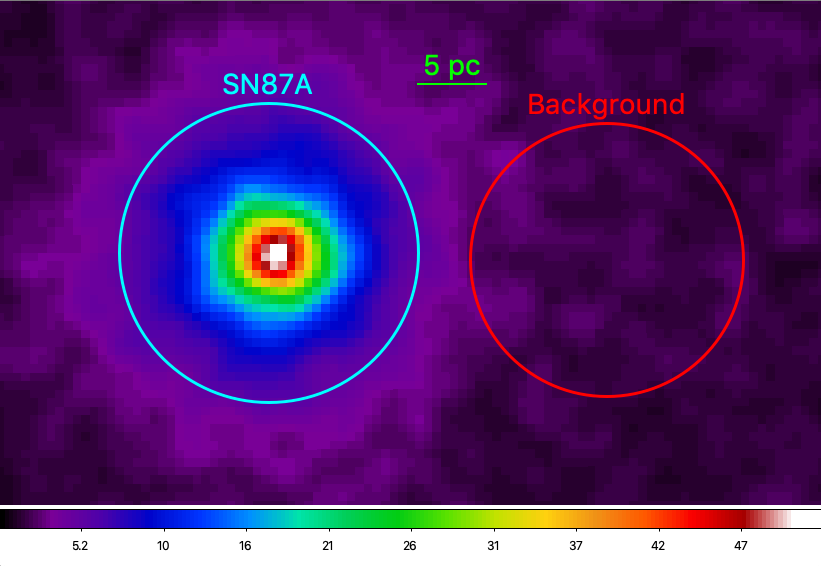}
    \caption{ {\it Chandra}/ACIS-S (top) and {\it NuSTAR}/FPMA (bottom) count images of SN 1987A in 2014. All the images are smoothed with a Gaussian having a sigma of 1.5 pixel. \emph{Top left panel}. {\it Chandra}/ACIS-S image in the $0.5-8$ keV energy band with a bin-size of $0.06''$.  Cyan and red circles mark the regions selected to extract the source and the background spectra, respectively. \emph{Top right panel}. Closeup view of the top left panel. The black circle marks the faint central region of SN 1987A. \emph{Bottom panel.} {\it NuSTAR}/FPMA image in the 3-30 keV band with a binsize of $2.5''$. The cyan and red circles identify the $43''$ region selected to extract the source spectra and the corresponding background for the {\it NuSTAR} data, respectively.} 
    \label{fig:image_obs}
\end{figure}

Fig.~\ref{fig:image_obs} shows the image regions that I selected to extract the spectrum of the whole remnant. In particular, I adopted a circular region centered at $\alpha = 5^h 35^m 28^s$ and  $\delta = -69^{\circ}16' 10''$ with a radius of $2''$ and $43''$ for {\it Chandra} and {\it NuSTAR} data, respectively. The different radii of the two circular regions reflect the different PSFs of the {\it Chandra} and {\it NuSTAR} telescopes.

Spectral analysis was performed with XSPEC (v12.11.1, \citealt{arn96}) in the $0.5-8$ keV and $3-20$ keV bands for \emph{Chandra} and \emph{NuSTAR}, respectively. All spectra were rebinned adopting the optimal binning procedure described in \citet{kb16}, and the background spectrum to be subtracted was extracted from a nearby region immediately outside of the source (see Fig. \ref{fig:image_obs}). I also verified that my results are not affected by the choice of the background regions.

\newpage
\section{X-ray absorption from cold ejecta} \label{sect:ray_tracing}

As mentioned in Sect. \ref{sect:Intro_1987A}, the internal shells of SN 1987A are mainly composed by cold and dense ejecta, which heavily absorb any possible radiation emitted by a putative compact object. Therefore, I adopted the 3D MHD simulation of SN 1987A by \citet{oon20} (hereafter Or20) to estimate the absorption pattern of the cold ejecta. The simulation reproduces most of the features observed in the remnant of SN 1987A in various spectral bands; it also links the SNR with the properties of the asymmetric parent SN explosion \citep{onf20} and with the nature of its progenitor star \citep{uru18}. The model provides all the relevant physical quantities in each cell of the 3D spatial domain (MHD variables, plasma composition of the CSM and of the ejecta vs time). 
The most notable quantities for my purposes are: the electron temperature, the ion density for several chemical species ($^{1}$H, $^{3}$He, $^{4}$He, $^{12}$C, $^{14}$N, $^{16}$O, $^{20}$Ne, $^{24}$Mg, $^{28}$Si, $^{32}$S, $^{36}$Ar, $^{40}$Ca, $^{44}$Ti, $^{48}$Cr, $^{52}$Fe, $^{54}$Fe, $^{56}$Ni, and the decay products, including $^{56}$Fe), and the ionization age (see \ref{sect:nei}). The model also predicts that the putative NS, relic of the supernova explosion, has received a kick towards the observer and towards the north with a lower limit to the kick velocity of $\approx 300$~km/s, as a result of the highly asymmetric explosion \citep[Or20]{onf20}. 

For the present study, I assumed a kick velocity of 500 km/s and checked that the results do not change significantly for values ranging between 300 and 700 km/s. Given the kick velocity and the direction of motion of the NS as predicted by the model \citep[Or20]{onf20}, and orienting the modeled remnant as it is observed in the plane of the sky (Or20), I established the position of the NS in the 3D spatial domain of the simulation for each year analyzed in this work. 
Considering that the extension of the putative radio PWN is of the order of $<1000$ AU \citep[see Sect \ref{sect:Intro_1987A}]{cmg19}, that the X-ray PWN is expected to be smaller than its radio counterpart, and that the spatial resolution of the MHD simulations is of $\approx 180$ AU, I can consider the central source as point-like in my procedure.

I reconstructed the absorption pattern encountered by the synchrotron X-ray emission of the putative PWN through the subsequent absorbing layers of metal-rich, cold ejecta along the line of sight. The physical effect responsible for the absorption is the photo-electric effect, since the material surrounding the putative central source is cold ($T < 100$ K). I extracted values of temperature, column density and abundances associated with each absorbing layer of the 3D domain of the model and I included these parameters in the spectral analysis through the XSPEC photo-electric absorption model {\it VPHABS}. 

For the years considered in this work, the absorption due to cold ejecta is comparable with an equivalent H column density$> 10^{23}$ cm$^{-2}$. This indicates that a potential signature of a PWN emission must be searched in the high energy part ($\gtrsim$ 10 keV) of the X-ray spectra, less affected by absorption.

\section{Data analysis} \label{sect:x-ray_data} 

{\it Chandra} spectra of SN 1987A are usually fitted with two isothermal components of optically thin plasma in NEI (e.g. \citealt{zmd09}). Therefore, I fitted the 2012, 2013 and 2014 {\it Chandra}/ACIS-S spectra with a model composed by two \emph{VNEI} (i.e., isothermal optically thin plasma out of equilibrium of ionization) components and a interstellar (Milky Way + Large Magellanic Cloud) absorption component (\emph{TBabs} model in XSPEC). The column density $n_{\mathrm{H}}$ in the \emph{TBabs} model is kept fixed to $2.35 \times 10^{21}$ cm$^{-2}$ \citep{pzb06}. In the fitting procedure, I also included a constant factor between the two data sets of any given year to take into account cross-calibrations, finding always factors $< 2\%$. 

\begin{figure}[!ht]
 \begin{minipage}{0.5\textwidth}
  \includegraphics[width=.7\textwidth,angle=270]{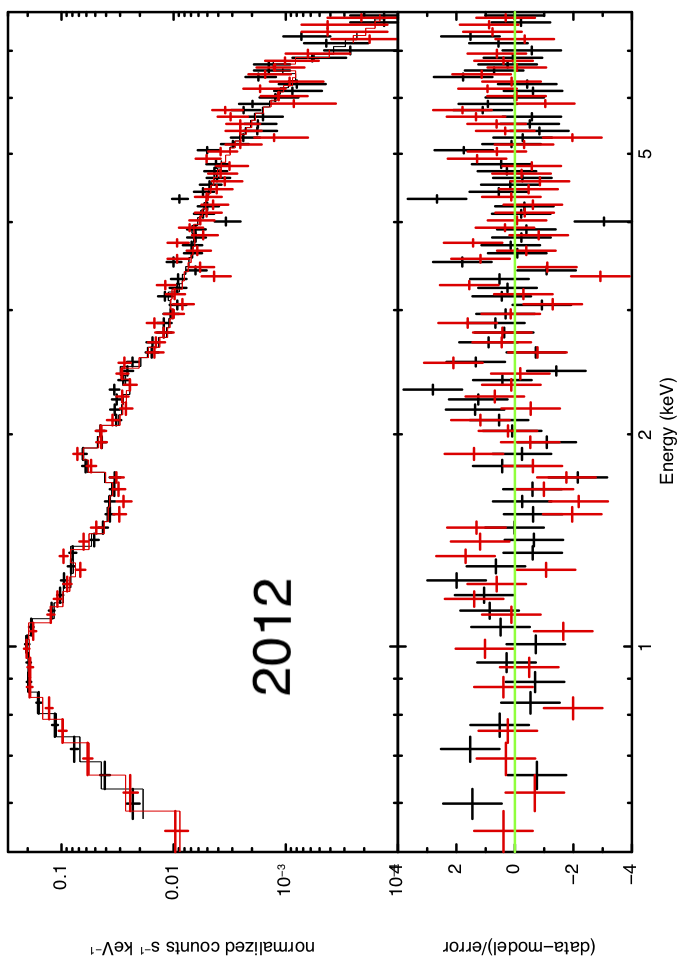}
  \end{minipage}
  \hfill
  \begin{minipage}{0.5\textwidth}
  \includegraphics[width=.7\textwidth,angle=270]{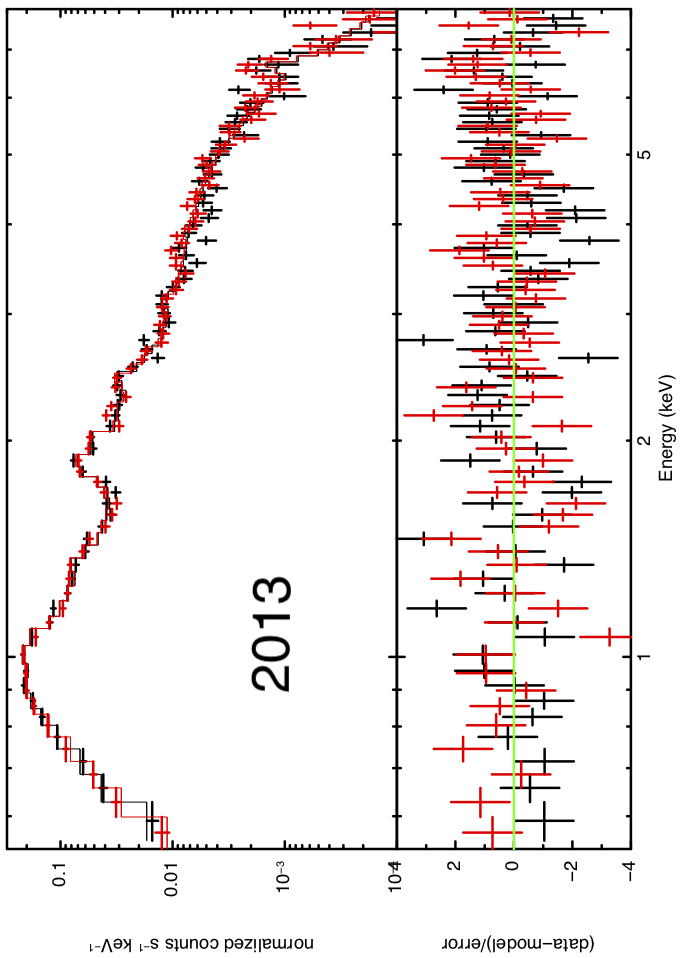}
  \end{minipage}
  \begin{minipage}{0.5\textwidth}
  \centering
  \includegraphics[width=.7\textwidth,angle=270]{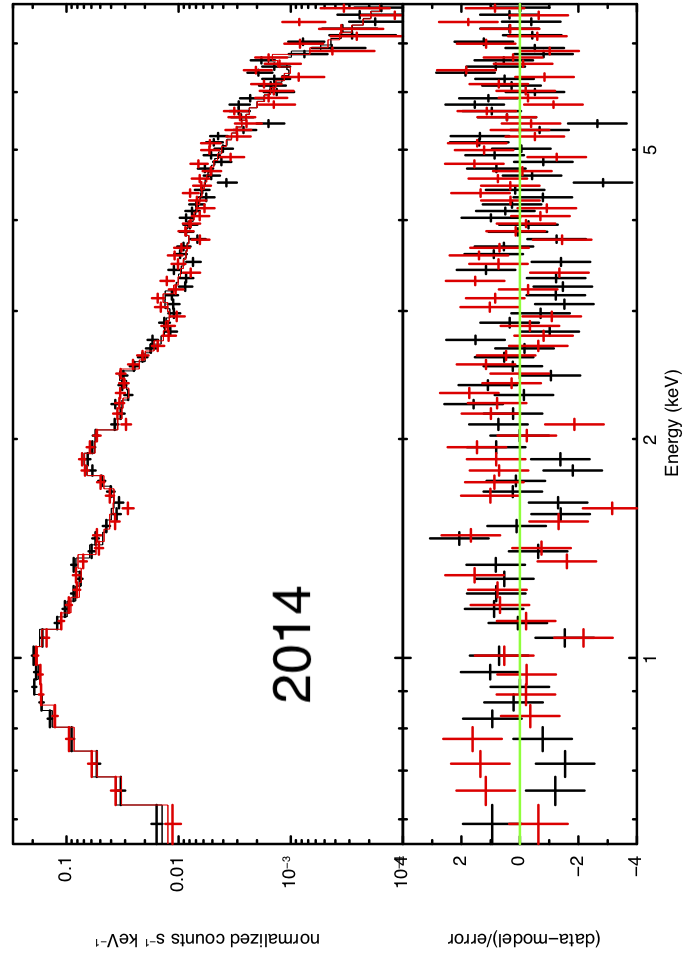}
  \end{minipage}
  \begin{minipage}{0.5\textwidth}
  \centering
  \includegraphics[width=.7\textwidth,angle=270]{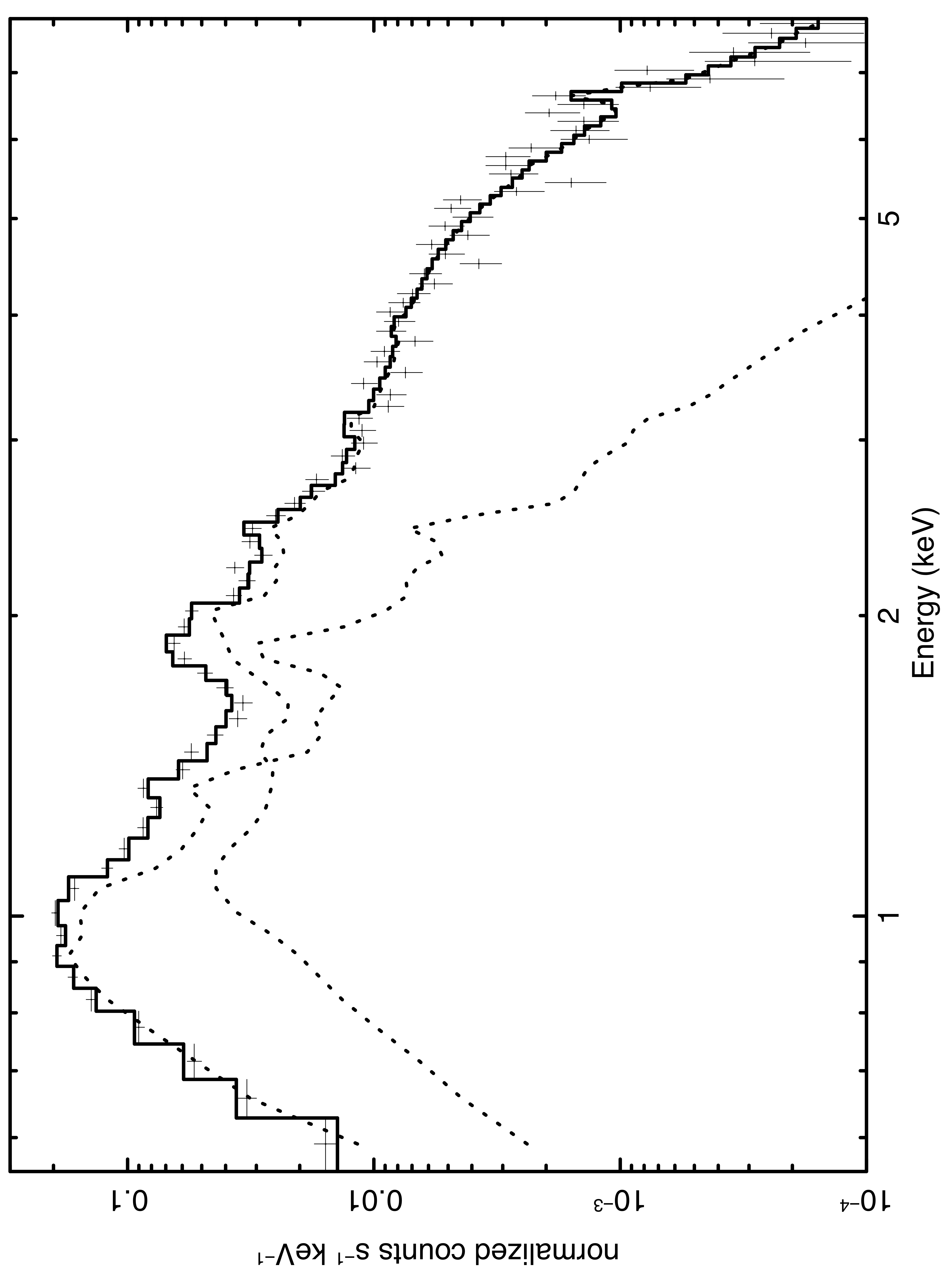}
  \end{minipage}
\caption{Chandra/ACIS-S spectra  of SN 1987A in 2012 (top left), 2013 (top right) and 2014 (bottom left) with the corresponding soft thermal model (two VNEI components) and residuals. Lower right panel shows the spectrum extracted from obsID 15809 (see Table \ref{tab:obs}) and the contribution of the two different VNEI components.}
\label{fig:chandra_spectra}
\end{figure}
For each thermal component, the free parameters of the fitting process are temperature, ionization parameter, emission measure and O, Ne, Mg, Si and S abundances. I kept fixed the other abundance values to those found by \citet{zmd09} since leaving free to vary these additional parameters does not significantly improve the quality of the fit. 

In agreement with previous works \citep[Or20]{omp15, mob19}, I found that the soft ($0.5-8$ keV) X-ray emission consists of thermal X-rays originating in the shocked CSM, with a variation in time of the abundance values consistent with zero. Overall, my best-fit values of temperatures, $kT$, and ionization parameters, $\tau$, are compatible with previous studies \citep{zmd09}. The spectra with the corresponding best-fit model, hereafter {\it soft thermal model}, and residuals are shown in Fig. \ref{fig:chandra_spectra}. Best-fit parameters are shown in Table \ref{tab:fit_soft}.
\begin{table}[!h]
    \centering
    \caption{Best-fit parameters of the soft thermal model.}
    \begin{tabular}{c|c|c|c|c}
    \hline\hline
    Component& Parameter & 2012 & 2013 & 2014 \\
    \hline
    TBabs & n$_{\rm{H}}$ (10$^{22}$ cm$^{-2}$) & \multicolumn{3}{c}{0.235 (fixed)} \\
    \hline
    & kT (keV) & 2.4$\pm 0.4$ & 2.6$_{-0.2}^{+0.5}$ & 2.4$\pm 0.2$ \\
    & O& 0.3$\pm 0.2$ & 0.23$^{+0.17}_{-0.1}$& 0.21$_{-0.09}^{+0.11} $ \\
    & Ne& 0.62$_{-0.11}^{+0.13}$& 0.68$^{+0.11}_{-0.13}$& 0.69$_{-0.11}^{+0.12}$\\
    VNEI & Mg &0.42$_{-0.10}^{+0.11}$& 0.42$^{+0.09}_{-0.10}$& 0.48$^{+0.10}_{-0.09}$\\
    & Si& 0.43$_{-0.10}^{+0.14}$& 0.50$_{-0.09}^{+0.08}$& 0.47$_{-0.07}^{+0.09}$\\
    & S& 0.8$\pm 0.2$ & 0.70$_{-0.12}^{+0.07}$& 0.70$\pm 0.13$\\
    & $\tau$ (10$^{11}$ s/cm$^3$)& 3$_{-2}^{+9}$& 1.9$_{-0.5}^{+0.6}$& 2.5$_{-0.5}^{+0.9}$\\
    & EM ($10^{58}\, \mathrm{cm}^{-3}$) & 8.0$\pm 0.2$ & 8.4$_{-1.0}^{+0.9}$ & 10.7$\pm 1.1$\\
    \hline
    & kT (keV)& 0.56$^{+0.07}_{-0.08}$ & 0.56${^{+0.11}_{-0.07}}$& 0.52$^{+0.06}_{-0.07}$\\
    VNEI & $\tau$ (10$^{11}$ s/cm$^3$) & 2.2$_{-0.6}^{+1.5}$ & 1.9$^{+0.9}_{-0.5}$& 2.1$^{+1.6}_{-0.5}$ \\
    & EM ($10^{58}\, \mathrm{cm}^{-3}$) & 33$\pm 4$ & 29$_{-3}^{+4}$& 30$_{-3}^{+7}$\\
    \hline
    \multicolumn{2}{c|}{$\chi^2$ (d.o.f.)} & 156 (137)& 215 (153) & 174 (150) \\
    \hline
    \end{tabular}
    
    Abundance values of other elements are kept fixed to those found by \citet{zmd09}. Uncertainties are estimated at 90\% confidence level.
    \label{tab:fit_soft}
\end{table}

I then included the {\it NuSTAR}/FPMA,B data in the analysis, thus extending the energy range under investigation up to $20$ keV. I simultaneously analyzed {\it Chandra}/ACIS-S and {\it NuSTAR}/FPMA,B spectra for each year considered by adopting the soft thermal model used to describe the {\it Chandra}/ACIS-S spectra. By simultaneously fitting data collected from both detectors in the (3-8) keV band, I found cross-calibration factors $<8\%$ between {\it Chandra}/ACIS-S and {\it NuSTAR}/FPMA,B, compatible with the characteristic corrections between {\it NuSTAR} cameras and other telescopes. Similar to what I measured for the {\it Chandra}/ACIS-S analysis, I report no significant variations in the chemical abundances in the time lapse considered.

\begin{figure}[!ht]
\centering
 \begin{minipage}{0.32\textwidth}
  \includegraphics[width=.75\textwidth,angle=270]{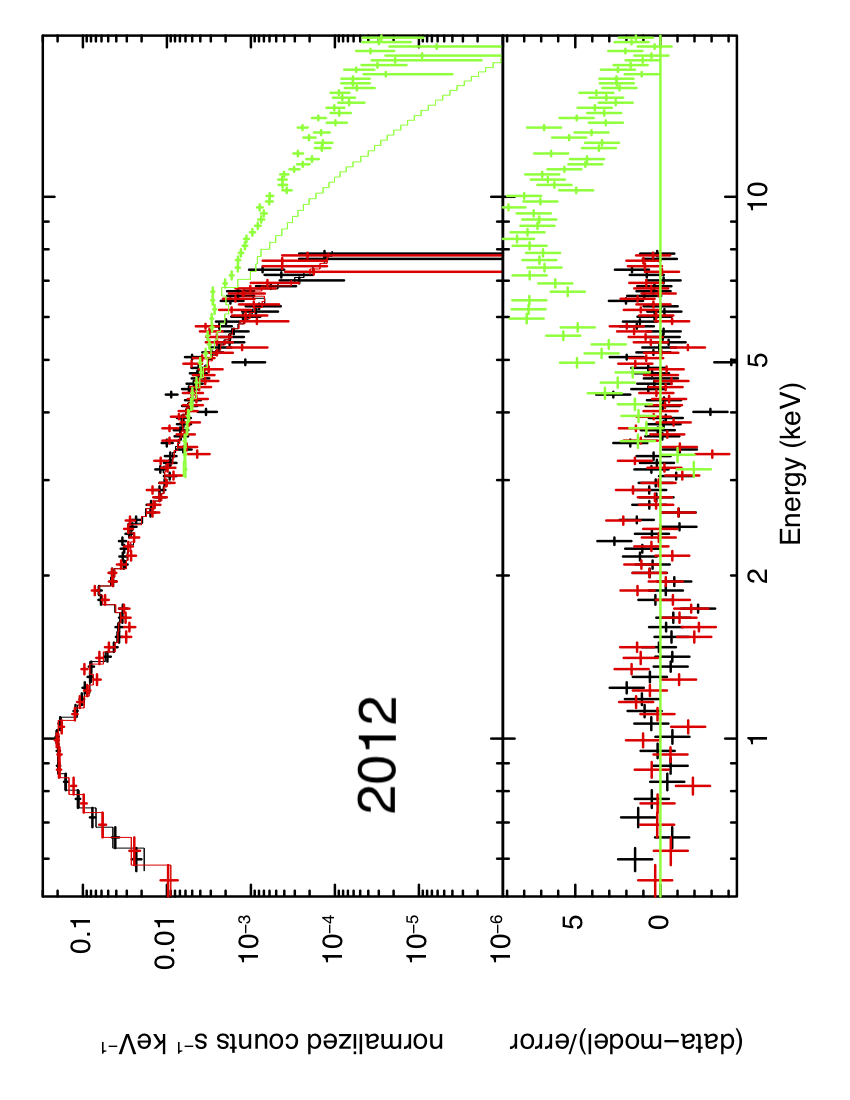}
 \end{minipage}
 \hfill
 \begin{minipage}{0.32\textwidth}
  \includegraphics[width=.75\textwidth,angle=270]{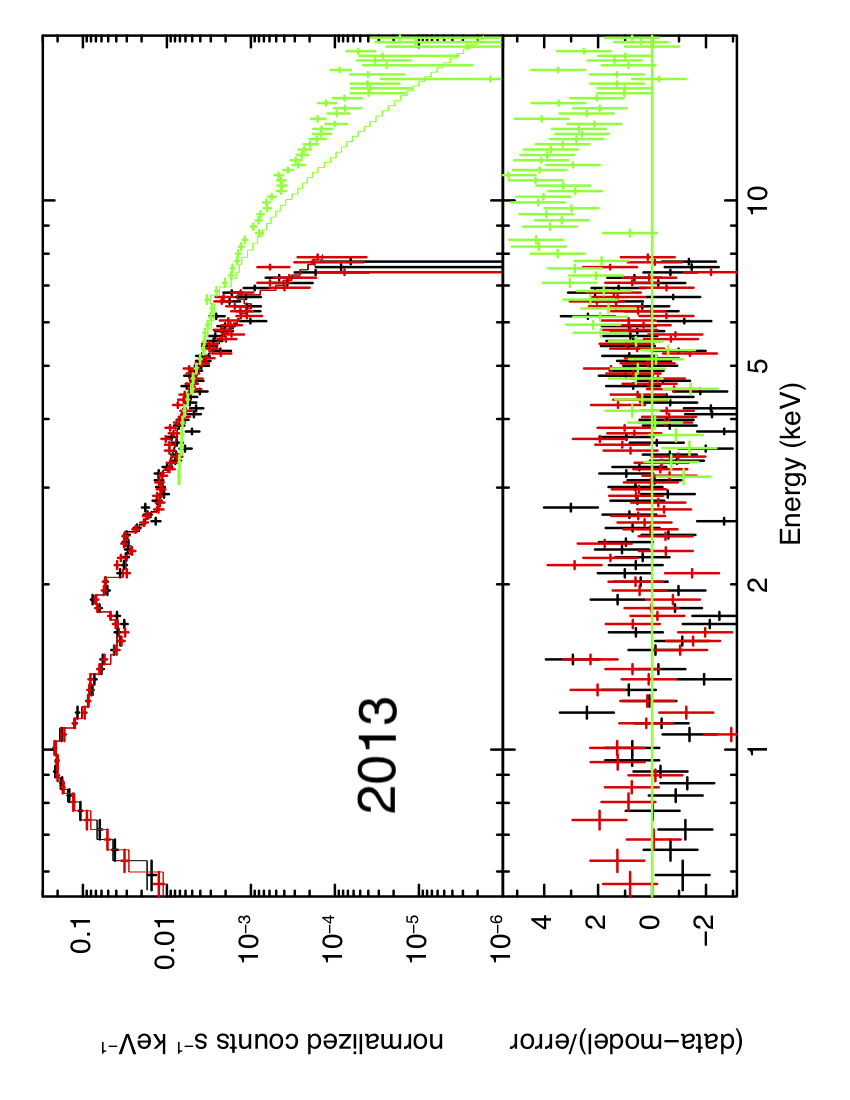}
 \end{minipage}
 \hfill
 \begin{minipage}{0.32\textwidth}
  \includegraphics[width=.75\textwidth,angle=270]{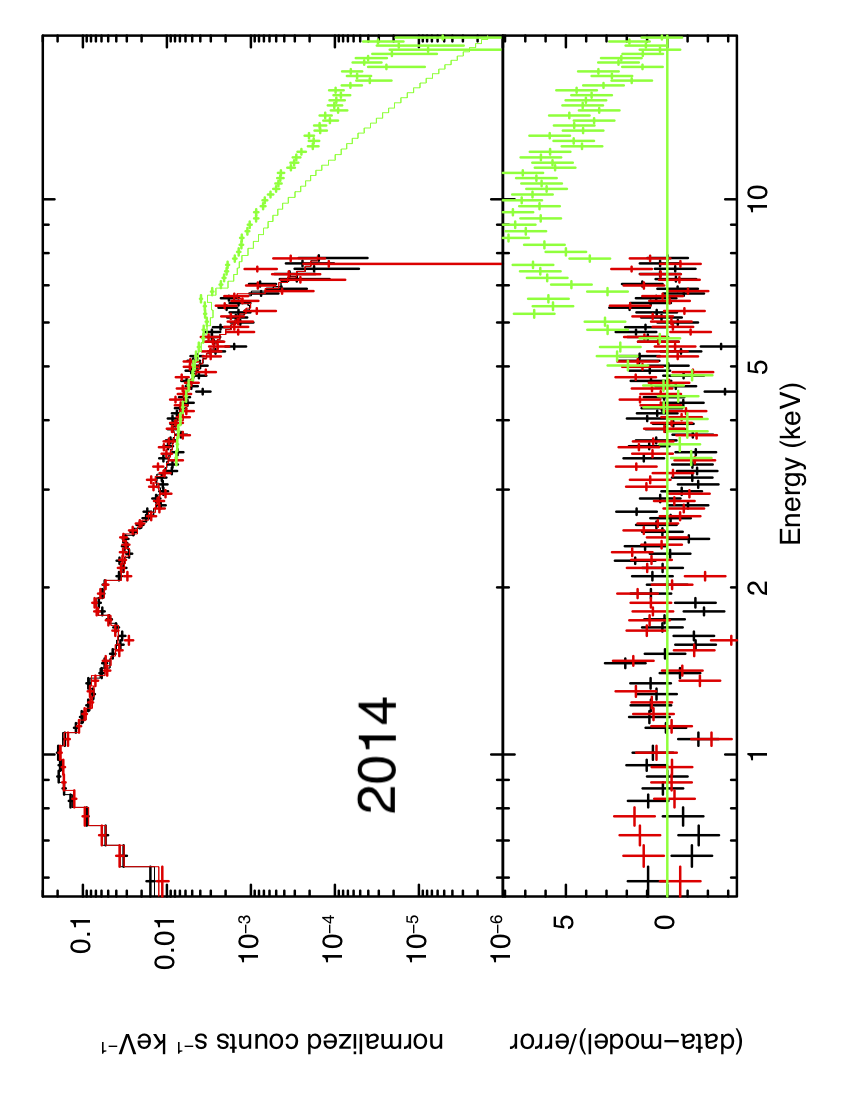}
 \end{minipage}
  \caption{{\it Chandra}/ACIS-S and {\it NuSTAR}/FPMA,B spectra of SN 1987A as observed in 2012 (left), 2013 (center) and 2014 (right), with the corresponding best-fit thermal model (two isothermal components with interstellar absorption) and residuals. A different color is associated with each data set: black and red for {\it Chandra}/ACIS-S and green for {\it NuSTAR}/FPMA,B. {\it NuSTAR}/FPMA,B spectra of the same year are summed for presentation purposes.}
  \label{fig:chandra+nu_spectra_res}
  \end{figure}

I found strong residuals in all {\it NuSTAR}/FPMA,B spectra at energies $> 10$ keV, clearly showing that an additional component must be added to the model to properly describe the hard X-ray emission (see the best-fit model and the residuals in Fig.~\ref{fig:chandra+nu_spectra_res}). The bottom right panel of Fig. \ref{fig:chandra+nu_spectra_res} shows the contribution of the various components of the spectral model adopted, for the deep observation 15809 ({\it Chandra}/ACIS).  This additional component cannot be associated with thermal emission from the shocked plasma, since otherwise unrealistically high temperatures would be necessary ($kT\sim20$ keV) to fit the hard X-ray excess.

Nonthermal emission may arise from a compact object, most likely a NS and its PWN (hereafter PWN87A), which emits synchrotron radiation (see Sect. \ref{sect:synchro}). I model the synchrotron emission with a power-law (as explained in Sect. \ref{sect:synchro}) but the main issue in tackling this scenario is to isolate the radiation coming from this object. In fact, the putative PWN is embedded within the dense and cold ejecta which heavily absorb the radiation.

 However, I can estimate the absorbing power of the cold ejecta in the year considered thanks to the MHD model by Or20. Therefore, I couple the power-law component with an absorbing component, modeled with {\it VPHABS} within XSPEC, and generated consistently to the MHD model, as described in Sect. \ref{sect:ray_tracing}. While the photon index $\Gamma$ and the normalization of the power-law are free to vary, the parameters of the \emph{VPHABS} component are kept fixed during the fitting procedure since they are derived from the MHD model.

\begin{figure}[!ht]
\centering
 \begin{minipage}{0.49\textwidth}
  \includegraphics[width=.79\textwidth,angle=270]{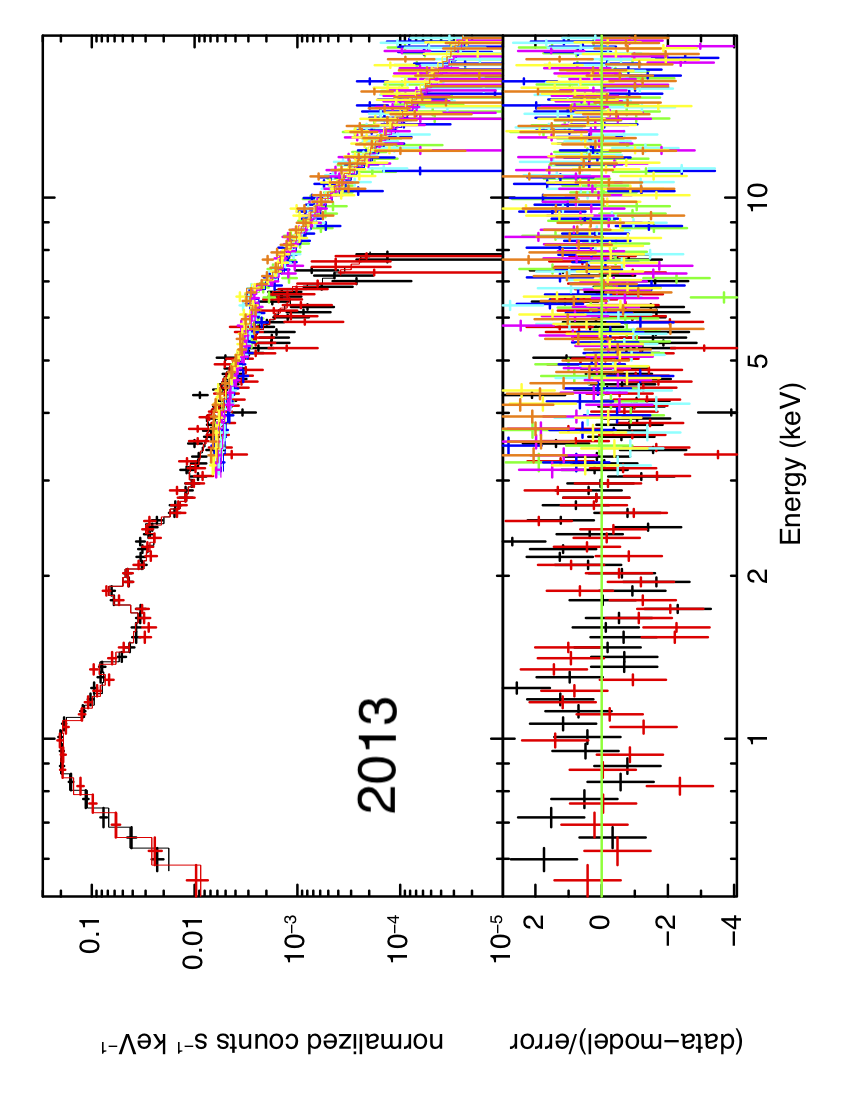}
 \end{minipage}
 \hfill
 \begin{minipage}{0.49\textwidth}
  \includegraphics[width=.79\textwidth,angle=270]{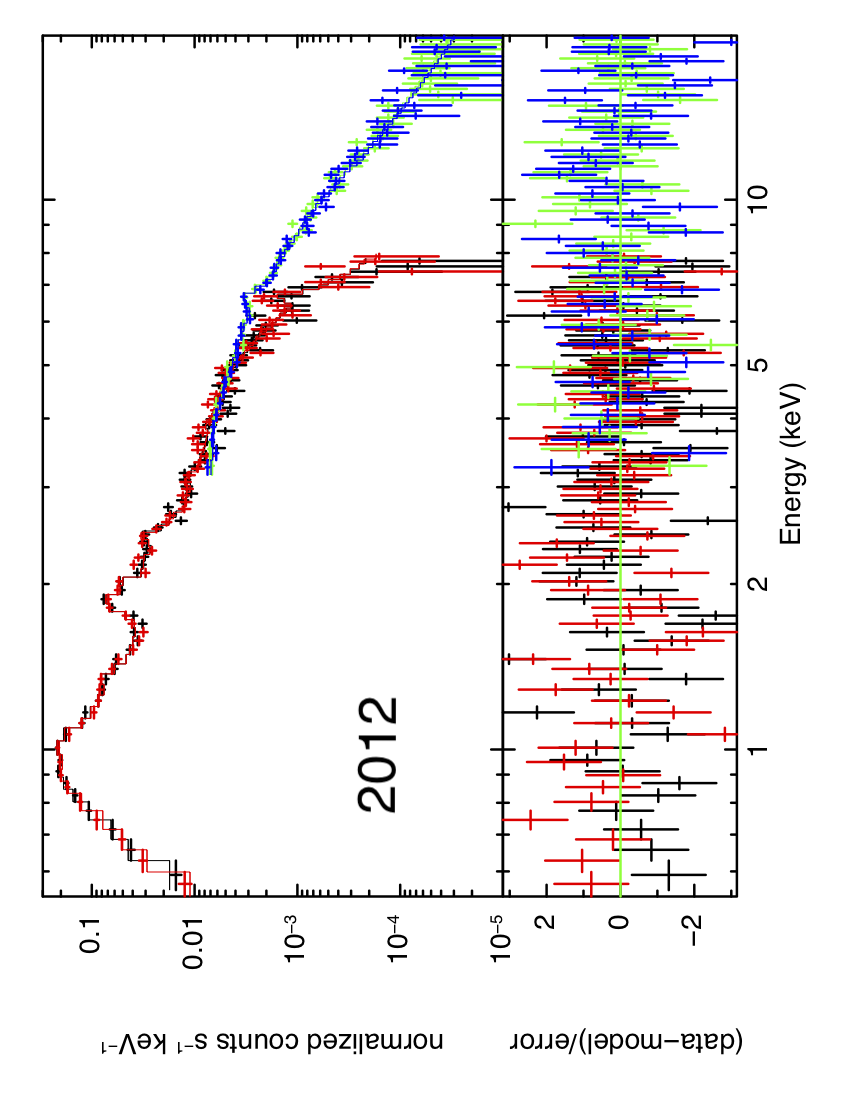}
 \end{minipage}
 \begin{minipage}{0.49\textwidth}
  \includegraphics[width=.79\textwidth,angle=270]{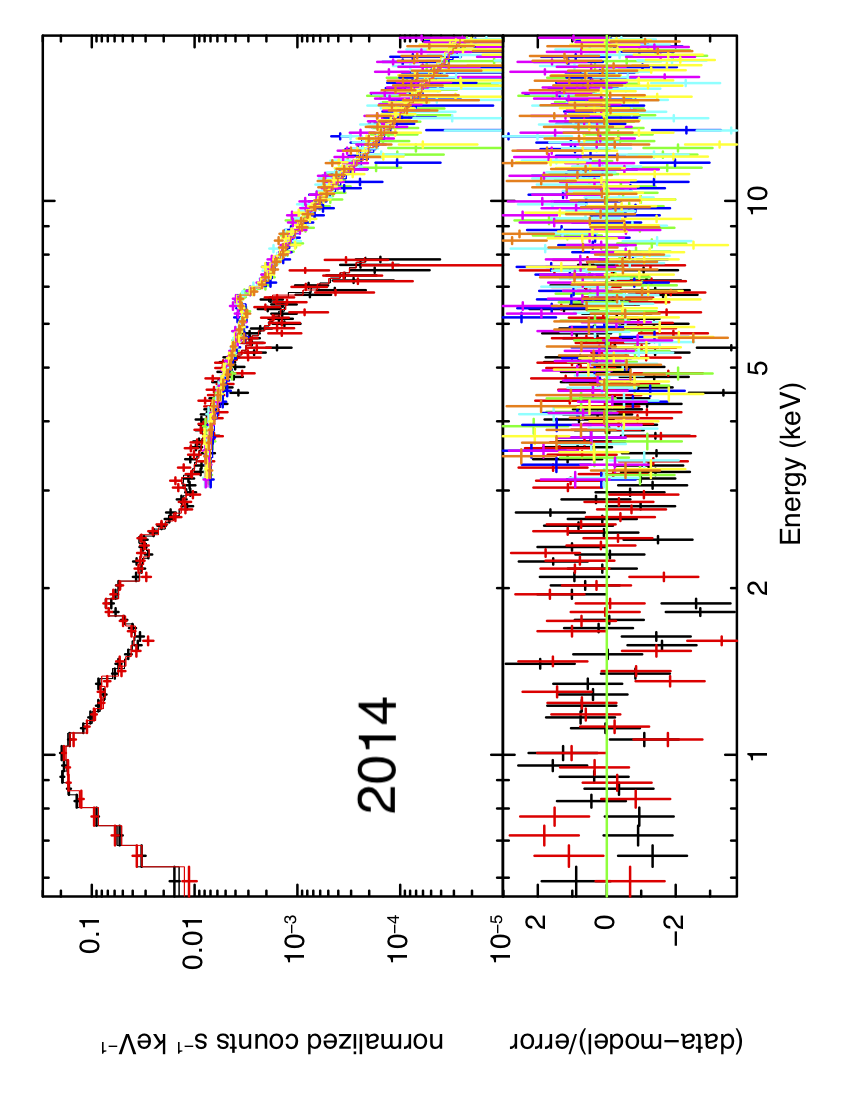}
 \end{minipage}
 \hfill
 \begin{minipage}{0.49\textwidth}
  \includegraphics[width=.72\textwidth,angle=270]{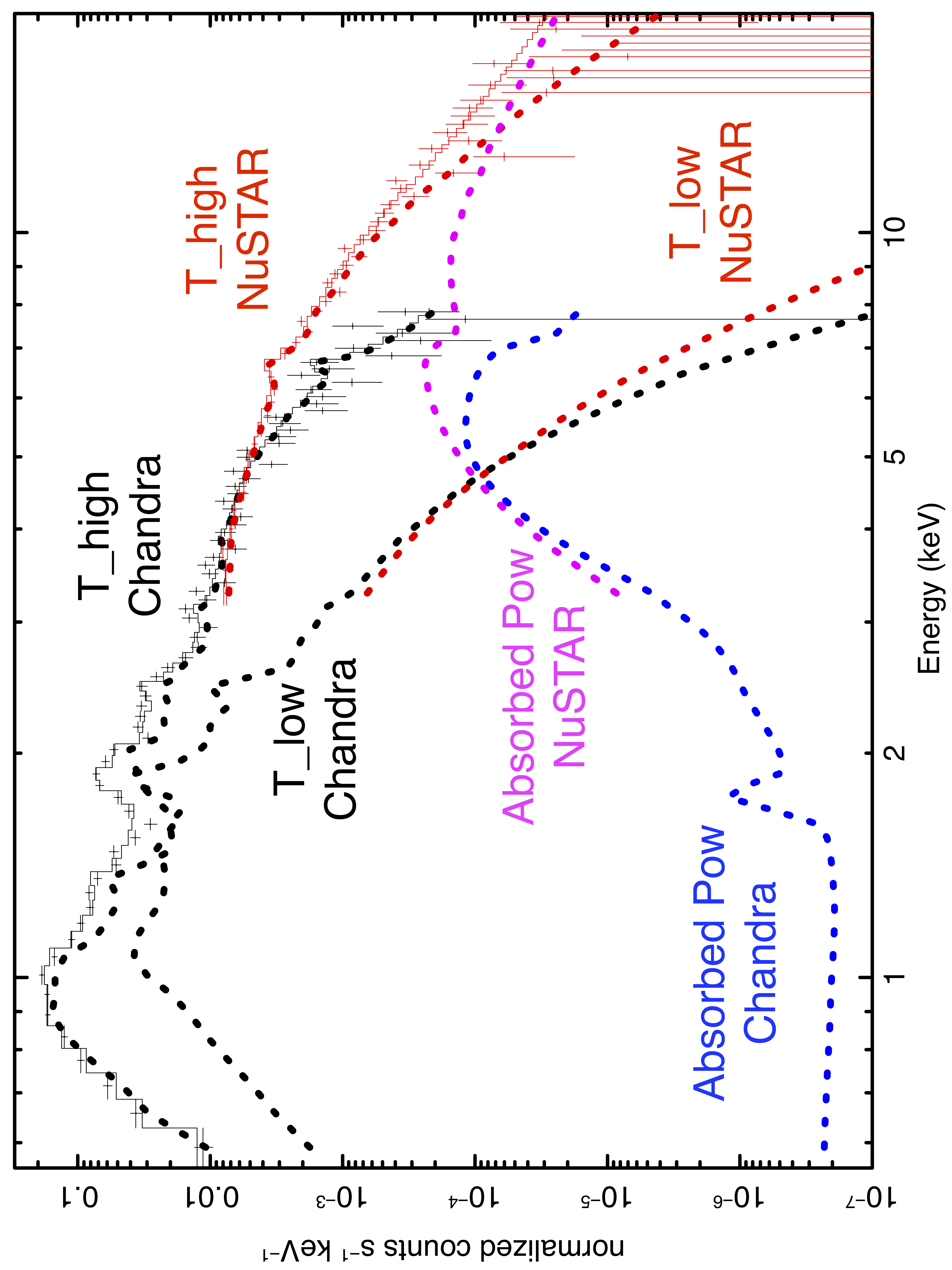}
 \end{minipage}
  \caption{Same as Fig. \ref{fig:chandra+nu_spectra_res} by adding an absorbed power-law, describing the PWN emission absorbed by the surrounding cold ejecta, to the best-fit model. Lower right panel shows the spectra extracted from obsIDs 15809 ({\it Chandra}/ACIS-S) and 40001014023 ({\it NuSTAR}/FPMA, see Table \ref{tab:obs}) and the contribution of the various components of the spectral model: {\it Chandra}/ACIS-S thermal components in black, {\it NuSTAR}/FPMA thermal components in red, {\it Chandra}/ACIS-S absorbed power-law in blue and {\it NuSTAR}/FPMA absorbed power-law in purple.}
  \label{fig:chandra+nu_spectra_pwn_single}
  \end{figure}
\begin{figure}[!h]
\centering
  \includegraphics[width=.55\textwidth,angle=270]{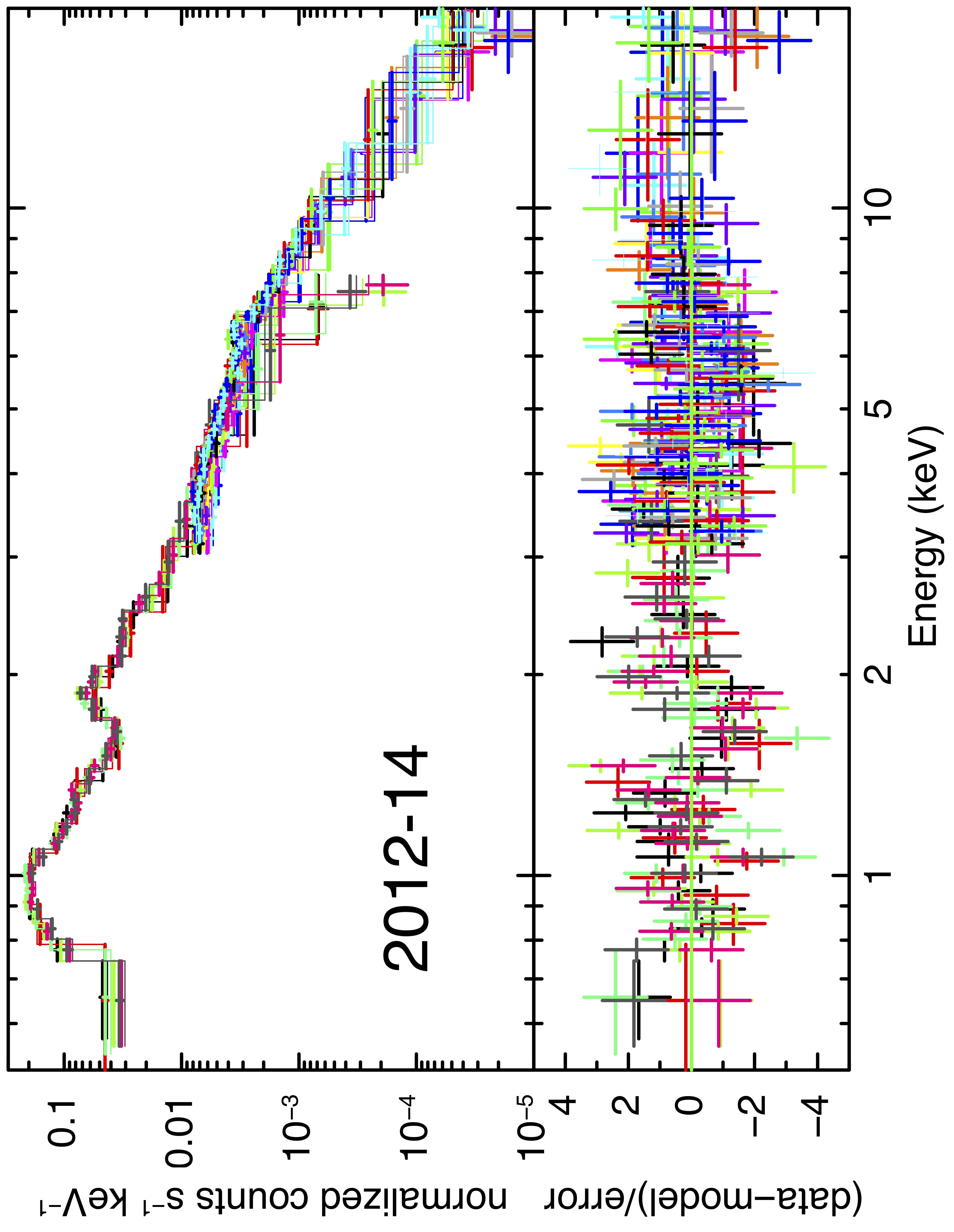}
\caption{Spectra extracted in the $0.5-20$ keV energy band from {\it Chandra}/ACIS-S and {\it NuSTAR}/FPMA,B in various years with the corresponding best-fit model and residuals. The best-fit model also takes into account the emission coming from a heavily absorbed PWN. A different color is associated with each of the twenty data sets. The spectra have been rebinned for presentation purposes} 
    \label{fig:chandra+nu_spectra_pwn}
\end{figure}

With this additional component, I obtained a very good description of the observed spectra over the whole $0.5-20$ keV energy band (Fig. \ref{fig:chandra+nu_spectra_pwn_single}). The resulting best-fit values of the photon index and the normalization of the absorbed power-law \emph{do not change significantly along the three years considered}. The bottom right panel of Fig. \ref{fig:chandra+nu_spectra_pwn_single} shows the contribution of the various components of the spectral model adopted, for the high-exposure observations 15809 ({\it Chandra}/ACIS) and 40001014023 ({\it NuSTAR}/FPMA). Therefore, I fitted simultaneously the 2012, 2013 and 2014 spectra in order to decrease the uncertainties of the best-fit parameters of the power-law component. Temperature, ionization parameter and normalization values of thermal emission were free to change in time (i.e. for different observation years), while normalization and photon index of nonthermal emission (as well as chemical abundances) were left free to vary in the fitting procedure, but forced to be the same over the three years. 
The resulting PWN best-fit photon index is $\Gamma = 2.5^{+0.3}_{-0.4} $ and the X-ray luminosity in the $[1-10]$ keV band is $L_{1-10} = (2.6 \pm 1.4) \times 10^{35}$ erg/s (Table \ref{tab:fit_whole}), being compatible with typical values found for PWNe (Sect. \ref{sect:pwnvsdsa}). Best-fit model and residuals are shown in Fig. \ref{fig:chandra+nu_spectra_pwn}.
\begin{table}[!ht]
    \caption{Best-fit parameters of the model adopted to describe {\it Chandra} and {\it NuSTAR} observations performed in 2012, 2013 and 2014.}
    \centering
    \begin{tabular}{c|c|c|c|c}
    \hline\hline
    Component& Parameter & 2012 & 2013 & 2014 \\
    \hline
    TBabs & n$_{\rm{H}}$ (10$^{22}$ cm$^{-2}$) & \multicolumn{3}{c}{0.235 (fixed)} \\
    \hline
    & kT (keV) & 2.85$_{-0.07}^{+0.08}$ & 2.87$\pm 0.07$ & 2.85$\pm 0.05$\\
    & O  & \multicolumn{3}{c}{0.32$_{-0.03}^{+0.04}$} \\
    & Ne & \multicolumn{3}{c}{0.57$ \pm 0.02$}\\
    VNEI & Mg & \multicolumn{3}{c}{0.36 $\pm 0.03$}\\
    & Si & \multicolumn{3}{c}{0.39$_{-0.02}^{+0.03}$}\\
    & S  & \multicolumn{3}{c}{0.65$ \pm 0.06$}\\
    & $\tau$ (10$^{11}$ s/cm$^3$)& 2.1$_{-0.5}^{+0.9}$& 1.7$\pm 0.3$& 1.8$_{-0.2}^{+0.4}$\\
    & EM ($10^{58} \mathrm{cm}^{-3}$) & 7.0$ \pm 0.3$& 7.3$\pm 0.3$ & 8.2$_{-0.2}^{+0.3}$ \\
    \hline
    & kT (keV)& 0.60$\pm 0.04$& 0.65$_{-0.03}^{+0.04}$& 0.63$_{-0.3}^{+0.4}$\\
    VNEI & $\tau$ (10$^{11}$ s/cm$^3$) & 1.8$_{-0.3}^{+0.5}$& 1.5$\pm{0.3}$& 1.7$\pm 0.3$ \\
    & EM ($10^{58} \mathrm{cm}^{-3}$) & 31$_{-3}^{+2}$ & 26$\pm 2$& 28$\pm 0.2$ \\
    \hline
    pow& Photon index $\Gamma$ & \multicolumn{3}{c}{2.5$_{-0.4}^{+0.3}$} \\
    \hline
    \multicolumn{2}{c|}{L$^{\rm{pwn}}_{0.5-8}$ (10$^{35}$ erg/s)} & \multicolumn{3}{c}{$4.1_{-2.8}^{+4}$} \\
    \multicolumn{2}{c|}{L$^{\rm{pwn}}_{1-10}$ (10$^{35}$ erg/s)} & \multicolumn{3}{c}{ $2.6 \pm 1.4 $} \\
    \multicolumn{2}{c|}{L$^{\rm{pwn}}_{10-20}$ (10$^{35}$ erg/s)}& \multicolumn{3}{c}{0.32$_{-0.01}^{+0.02}$}\\
    \hline
    \multicolumn{2}{c|}{Flux$_{0.5-8}$ (10$^{-13}$ erg/s/cm$^2$)}& 93$\pm 2$ & 92$_{-2}^{+4}$& 95$_{-1}^{+2}$\\
    \multicolumn{2}{c|}{Flux$_{10-20}$ (10$^{-13}$ erg/s/cm$^2$)}&1.5$\pm 0.2$ &1.6$_{-0.1}^{+0.6}$ &1.6$_{-0.1}^{+0.3}$\\
    \hline
    \multicolumn{2}{c|}{$\chi^2$ (d.o.f.)} & \multicolumn{3}{c}{1442 (1223)} \\
    \end{tabular}
    
         Chemical abundances and power law parameters are kept constant along the various years. Uncertainties are at 90\% confidence level.
    \label{tab:fit_whole}
\end{table}

I point out that, while the emission from the PWN is consistent with being constant, the flux  in the $8-20$ keV band increases by 10\% along 2012 and 2014. This is due to the increasing thermal emission arising from the interaction of the remnant with the HII region (e.g. \citealt[Or20]{fzp16}).

\begin{figure}[!htb]
\centering
\includegraphics[width=.55\textwidth,angle=270]{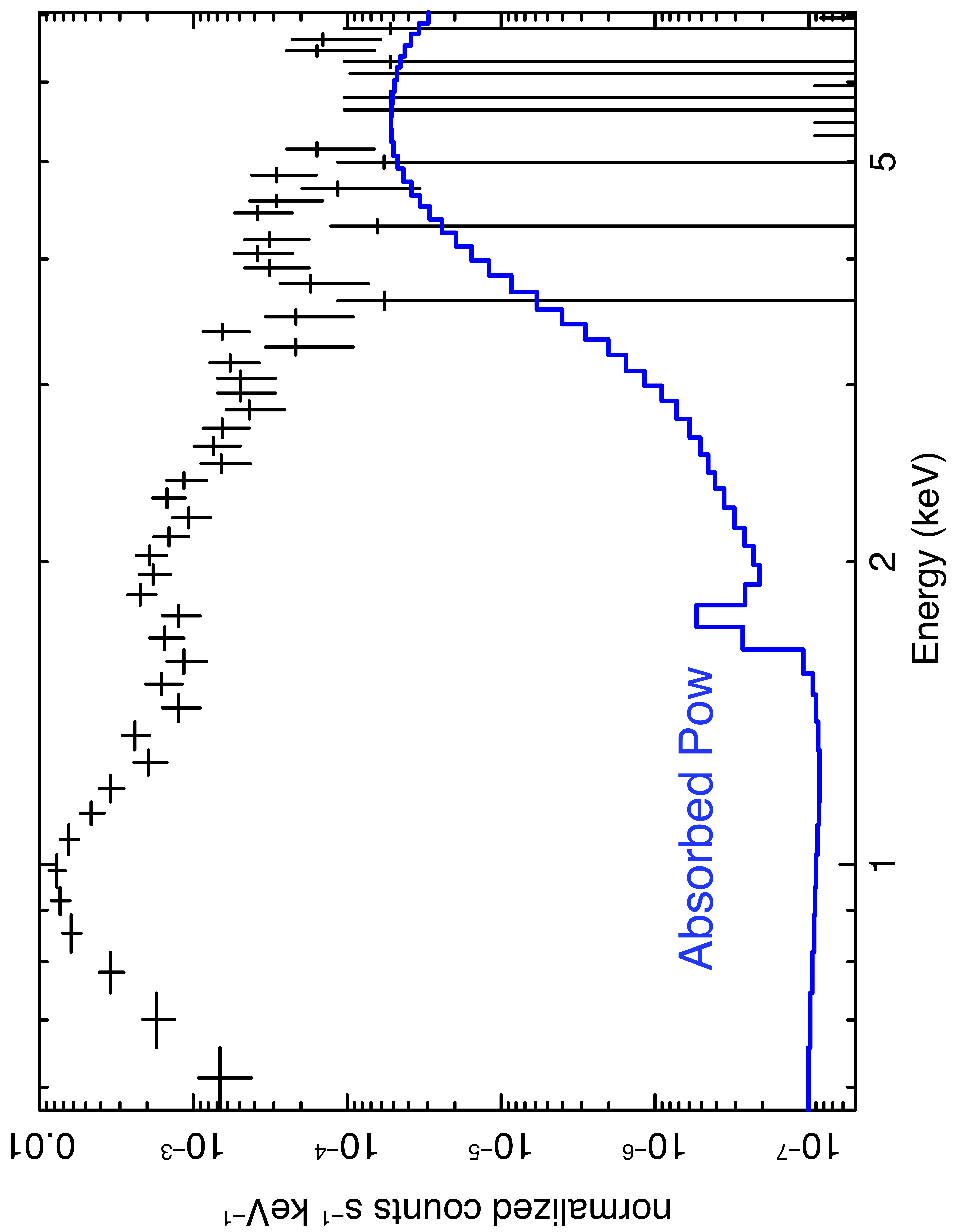}.
\caption{Spectrum of the central faint area of SN 1987A observed by {\it Chandra}/ACIS-S in 2014. In blue: contribution of the absorbed power law.}
\label{fig:faint_87A}
\end{figure}

Because of the high absorption by the ejecta, the PWN flux is strongly suppressed below 8 keV. I selected a region, identified by the black ring in the top left panel of Fig. \ref{fig:image_obs}, well within the ring of SN 1987A, where the X-ray flux is less contaminated by the interaction with the CSM and where the PWN should be. I show the corresponding {\it Chandra}/ACIS-S\footnote{{\it NuSTAR} is not able to resolve this area because of its wide PSF} spectrum in Fig. \ref{fig:faint_87A}. I performed a fit of this spectrum, finding that the emission in the central area of SN 1987A is mainly thermal. The flux of the putative PWN in this region, in 2014, is expected to be only $4\%$ of the total flux. Therefore, I conclude that, because of the heavy absorption by cold ejecta, the PWN is currently not detectable in the soft X-ray band, but only at energy $\gtrsim 10$ keV.

As an alternative scenario, I also considered the case in which the hard X-ray emission is associated with synchrotron radiation due to DSA  (see Sect. \ref{sect:synchro}) occurring in the outer layers of SN 1987A. I then removed the cold ejecta absorption and obtained a best-fit photon index of $\Gamma_{\rm{DSA}} = 2.0 \pm 0.4$, i.e. slightly lower than that obtained in the PWN scenario, though being still consistent with it. Analogously to what I obtained for the PWN scenario, I found that the photon index and the normalization of the nonthermal emission do not change significantly from 2012 to 2014. I then repeated the simultaneous fit of the data of all three years, obtaining a good description of the observed spectra ($\chi_{\rm{DSA}}^2=1457$ with 1221 d.o.f. to be compared with $\chi^2_{\rm{PWN}}=1442$ with 1221 d.o.f. of the PWN scenario). Though the PWN case provides a slightly lower $\chi^2$ (by 1\%, with the same number of d.o.f.), both scenarios provide a very good description of the data points and it is not possible to exclude any of them on a merely statistical basis. I also looked for pulsations and did not find any significant pulsed emission in the data. Therefore, I discuss physical implications of both scenarios in Sect. \ref{sect:pwnvsdsa}.

\section{Discussion} \label{sect:pwnvsdsa}
In Sect. \ref{sect:x-ray_data} I showed that a nonthermal component is needed to properly fit the {\it NuSTAR}/FPMA,B data in the $10-20$ keV band. As already mentioned, the possible origin of the physical mechanism responsible for such emission is twofold: either DSA or emission from a heavily absorbed PWN. Under a spectroscopic perspective, the only difference between the two scenarios is the presence of an additional absorption component in the PWN case, the \emph{VPHABS} model, because of the presence of cold ejecta surrounding the putative compact object. The heavy absorption leads to a negligible contribution of the power-law component at energies below 6 keV and thus to a steeper $\Gamma$. However, the photon index values in the two scenarios are compatible with each other taking into account the 90\% confidence error bars. Moreover, the very similar values of $\chi^2$ do not allow us to exclude one of the two possible emission mechanisms under a merely statistical point of view.

\subsection{DSA scenario} \label{sect:dsa}
In the DSA scenario, the flux is expected to vary with time in the same way both in the X-ray and radio bands. The synchrotron radio flux of SN 1987A increased by $\sim15\%$ between 2012 and 2014 (\citealt{cgn18}). Under the DSA hypothesis, this woud be at odd with my findings, since I observe a steady X-ray synchrotron emission, and a $15\%$ increase of the nonthermal X-ray flux is unlikely at the 90\% confidence level. 

To further investigate the emission nature in the DSA scenario, I replaced the power-law component with the XSPEC {\it SRCUT} model, which describes synchrotron emission from a power-law distribution of electrons energies with an exponentially cut-off in the assumption of homogeneous magnetic field \citep{rey98}. For each of the three years considered, I constrained the normalization of the \emph{SRCUT} component (i.e., its flux at 1 GHz, $S_{\mathrm{1GHz}}$), taking advantage of the corresponding values observed at 9 GHz by \citet{cgn18}, and taking into account the radio spectral index of SN 1987A ($\alpha=0.74$,  \citealt{zsn13}). I fixed $S_{\mathrm{1GHz}}$ and $\alpha$ in the {\it SRCUT} model and left the cut-off frequency, $\nu_{\mathrm{b}}$, free to vary. The radio-to-X-rays spectral index $\alpha$ can be assumed to be constant since the synchrotron characteristic energy loss time is longer than the age of the system, thus no significant decrease (Eq. \ref{eq:time:synch}) is expected as I explain below.

The synchrotron emission of electrons peaks at energy $h\nu_{\mathrm{b}}$, as defined in Eq. \ref{eq:nu_synch}. From my analisis, I obtained $h\nu_{\mathrm{b}}=2.4_{-0.4}^{+0.3}$ keV, $h\nu_{\mathrm{b}}=2.3_{-0.4}^{+0.3}$ keV and $h\nu_{\mathrm{b}}= 1.7_{-0.2}^{+0.4}$ keV, in 2012, 2013 and 2014, respectively. The cut-off energy is compatible with 2 keV in the three years considered, though the general trend seems to point towards a decrease with time. Considering the cut-off energy $h\nu_{\mathrm{b}}\sim2$ keV, I obtain values of $E_{100}$ spanning from $\sim 0.14$ to $\sim 0.33$, for $B_{100}$ ranging from 6 to 1, where $E_{100}$ and $B_{100}$ are the electron energy and the magnetic field expressed in units of 100 TeV and 100 $\mu$G, respectively. We can now estimate the time scale for synchrotron losses for these high energy electron from Eq. \ref{eq:time:synch}: $\tau_{\rm{sync}}\sim 40$ yr for $E_{100} = 0.33$ and $B_{100} = 1$. It should be noticed that the maximum electron energy seems to be quite high, especially considering the relatively low shock speed; in fact, the synchrotron radio emission originates in the HII region \citep{zsn13,cgn18,omp19}, where the shock velocity is of only 2000 km/s \citep[Or20]{cgn18}. 

In the DSA scenario, the acceleration time scale can be estimated as \citep{pmb06}:
\begin{equation}
\tau_{\rm{acc}}=124 \eta B^{-1}_{100} V_{\mathrm{s}}^{-2} E_{100} \frac{4}{3} \,\;\mathrm{yr}
\label{eq:tau_acc}
\end{equation}
where $V_{\mathrm{s}}$ is the shock velocity and $\eta$ is the Bohm factor (the acceleration efficiency) . In the hypothesis of maximum 
 efficiency ($\eta = 1$) and a standard magnetic field $B_{100} =1$, with $V_{\mathrm{s}}$ = 2000 km/s, it would be needed $\tau_{\mathrm{acc}}\sim390$ yr to accelerate the X-ray emitting electrons up to the observed maximum energy, i.e. much more than the age of SN 1987A. The observed maximum energy can be obtained in 25 yr only by assuming that the downstream magnetic field is amplified by the SN 1987A slow shock up to $\sim600$ $\mu$G (in this case the maximum electron energy would be of $\sim14$ TeV), and only assuming that the acceleration proceeds at the Bohm limit. The aforementioned issues (steady synchrotron flux and extremely large electron energy in a relatively slow shock) concur in making the DSA scenario unconvincing.
 
\subsection{PWN scenario} \label{sect:pwn}
On the other hand, the PWN scenario has a strong physical motivation, as I show below. In the absence of a direct identification of the compact object eventually powering PWN87A, the only possible way to constrain its properties is to use the X-ray luminosity obtained in the analysis to compare the putative PWN with the PWNe population.

The properties of a generic PWN can be associated with those of its progenitor SNR and surrounding ISM introducing the characteristic time and luminosity scales \citep{tm99}:
\begin{eqnarray}
 t_{\rm{ch}} &=& E_{\rm{sn}}^{-1/2} M_{\rm{ej}}^{5/6} \rho_{\rm{ism}}^{-1/3}\,, \label{eq:chscales1} \\ 
 L_{\rm{ch}} &=& E_{\rm{sn}}/ t_{\rm{ch}}\,, \label{eq:chscales2}\
\end{eqnarray}
where $E_{\rm{sn}}$ is the supernova explosion energy, whose characteristic value is $10^{51}$ erg, $M_{\rm{ej}}$ is the mass of the SNR ejecta and $\rho_{\rm{ism}}$ the mass density of the ISM.
These last parameters have been considered to vary uniformly in: $M_{\rm{ej}} \in [5-20] \, \rm{M_\odot}$ \citep{sec09} and $n_{\rm{ism}}\in[0.01-10]$ cm$^{-3}$ \citep{ber87,mpp97,lbw10,bp10,asv14}, where $\rho_{\rm{ism}}=m_{\mathrm{p}}p n_{\rm{ism}}$  and $m_{\mathrm{p}}$ is the proton mass. For the pulsar population, the best choice is to consider young $\gamma-$ray emitting pulsars \citep{wr11,jsk20}, better suited for describing pulsars powering PWNe than the old radio-emitting ones \citep{kr06}. 

Considering the pulsar parameters from that population (namely the initial spin-down time $\tau_0$ and luminosity $L_0$), with the choice of the canonical dipole braking index $n=3$, the PWNe population can be then constructed\footnote{The PWNe population is liable of changes in the parameters plane, especially if considering a different braking index value than the  standard  dipole  one  (\citealt{Parthasarathy:2020}.)} with the corresponding characteristic values defined in Eq.~\ref{eq:chscales1}-\ref{eq:chscales2}. In the $(\tau_0/t_{\rm{ch}},\,L_0/L_{\rm{ch}})$ plane the PWNe population appears as an ellipsoidal surface, where each point corresponds to various physical sources with different combinations of $\tau_0, L_0, M_{\rm{ej}}$ and $\rho_{\rm{ism}}$. 

\begin{figure}[!h]
\centering
\includegraphics[width=.7\textwidth]{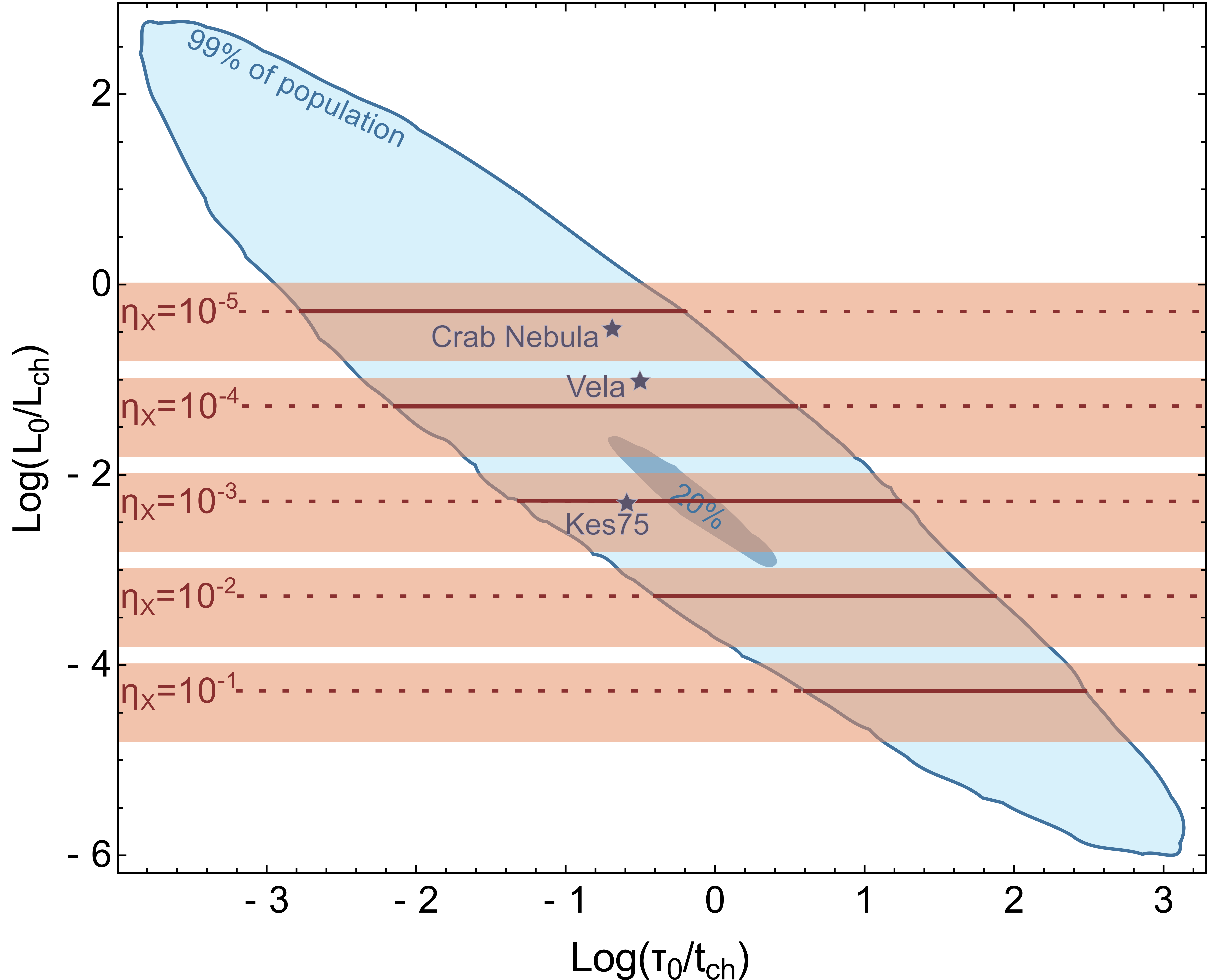}
\caption{Positioning of the putative PWN87A (brown lines) within the PWNe population (in light-blue) considering a density 100 particles/cm$^3$ of the HII region. The PWNe population is represented with contours enclosing the 99\% and the 20\% of the entire population. Different possible positions of the PWN have been obtained using the detected X-ray flux in the 0.5-8 keV band (with the best fit value shown as a solid-dashed line and the entire range of variation as a brown shaded area) and considering the widest possible variation for the X-ray efficiency $\eta_{\mathrm{X}}$ \citep{kp08}.}
\label{fig:pwnpop}
\end{figure}

Once the general population is placed in that parameters plane, I can use it to discuss the possible location of PWN87A and compare it with known sources (Fig. \ref{fig:pwnpop}). To do that, the measured X-ray luminosity must be scaled using the characteristic quantities specific for SN 1987A. The mass contained in the ejecta is well constrained to be $18\rm{M_\odot}$ \citep[and references therein]{oon20}, while the explosion energy is $E_{\rm{sn}} = 2\times10^{51}$ erg \citep[and references therein]{oon20}.
 Given the complex structure of the remnant, the density of the material in which the ejecta expand shows large variations across the system, with values ranging from the $\sim 0.1$ particles/cm$^3$ in the pre-shock blue supergiant wind, the $\sim 100$ cm$^{-3}$ in the HII region, up to $10^3$-$10^4$ cm$^{-3}$ in the dense cirmstellar ring. 
I consider a value of $\sim 100$ cm$^{-3}$, representative of the equatorial zone of the HII region.
Thus the two scalings are:
\begin{eqnarray}
t^{87\rm{A}}_{\rm{ch}}  =  807.6\,\rm{yr} \,\left( \frac{E_{\rm{sn}}}{2\times10^{51} \,\rm{erg}} \right)^{-1/2} \left( \frac{M_{\rm{ej}}}{ 18 \rm{M_\odot}} \right)^{5/6} \left( \frac{\rho_{\rm{ism}}}{ 100 m_{\mathrm{p}} \rm{cm}^{-3}} \right)^{-1/3}\,\\
L^{87\rm{A}}_{\rm{ch}} = 7.86\times 10^{40}\,\rm{erg/s} \left( \frac{E_{\rm{sn}}}{2\times10^{51} \,\rm{erg}} \right)
\end{eqnarray}\label{eq:chscales_1987A}
Taking into account the ejecta absorption, as calculated from the MHD simulation, I can derive from the spectral analysis the unabsorbed X-ray luminosity of the central source, thus finding L$_{\mathrm{X}}^{\rm{pwn}} = 4.1^{+4}_{-2.8} \times 10^{35}$ erg/s in the $0.5-8$ keV band, considering a distance to the source of $51.4$ kpc.
The X-ray luminosity must be converted in an estimate of the spin-down luminosity $\dot{E}$, only a part of it being converted into synchrotron radiation. 
The conversion efficiency between $\dot{E}$ and $L_{\mathrm{X}}^{\rm{pwn}}$ is defined by the parameter $\eta_{\mathrm{X}}=L_{\mathrm{X}}/\dot{E}$, that had been shown to have very large variations across the PWN population, from $\sim10^{-5}$ to $\sim10^{-1}$ \citep{kp08}.
A very high efficiency $\eta_{\mathrm{X}}=0.2$ was measured for the PWN in the Kes75 SNR \citep{hcg03}, characterized by the smallest known spin-down age ($\tau_c=726$ yr) and an extremely high magnetic field ($4.9\times 10^{13}$G). Another young PWN, the well known Crab nebula ($\tau_c=1230$ yr), shows an efficiency $\eta_{\mathrm{X}}\simeq 0.042$. The Vela nebula, with a spin-down age $\tau_c\sim11$ kyr has one of the lowest efficiency, $\eta_{\mathrm{X}}=2\times 10^{-5}$.
To maintain the analysis as general as possible, I considered the entire range of measured values for $\eta_{\mathrm{X}}$.

\begin{figure}[!ht]
\centering
  \includegraphics[width=.7\textwidth]{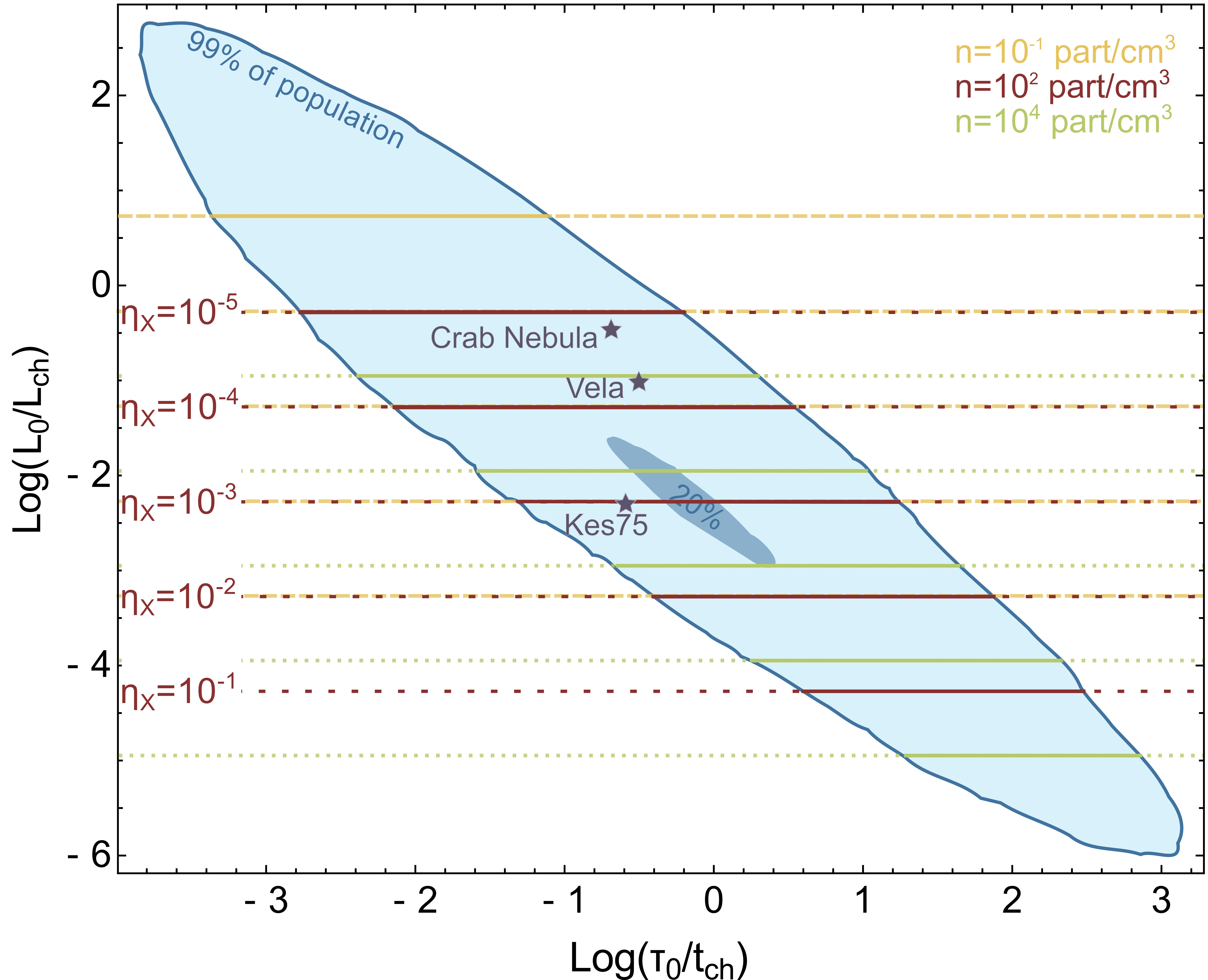}
  \caption{Variation of the predicted location of the PWN87A within the PWNe population, for different values of $\eta_{\mathrm{X}}$, considering an extreme variation of the ambient density, from $0.1$ cm$^{-3}$ (representative of the blue giant wind, in yellow) to $10^4$ cm$^{-3}$ (the maximum density in the dense ring in the HII region, in green).}
\label{fig:pwnpop_nvar}
\end{figure}

Fig.~\ref{fig:pwnpop} shows the population of PWNe as determined associating the $\gamma-$ray emitting pulsars with the discussed ranges of parameters for SNRs and ISM (in light blue).  Obviously, the initial spin-down time, in the absence of direct observations of the pulsar powering that PWN, cannot be constrained from available information. The positioning of the putative PWN in SN 1987A for varius $\eta_{\mathrm{X}}$ is shown as brown lines that intersect horizontally the distribution.
The three PWNe cited in the previous discussion are also shown for comparison.
As it can be easily seen, all the different possibilities lead to a location of PWN87A fully compatible with the population.
Moreover, this remains true even if I relax the assumption of an ambient density of 100 cm$^{-3}$. This is shown in Fig.~\ref{fig:pwnpop_nvar}, where the putative positions for varying $\eta_{\mathrm{X}}$ are derived considering the two extreme values for the ambient density ($0.1$--$10^4$ particles/cm$^3$).

An estimate (purely based on a statistical argument) of the most probable spin-down time for each $\eta_{\mathrm{X}}$ can be determined by combining the PWNe probability distribution with the estimates of the pulsar spin-down luminosity. The most probable values are reported in Table~\ref{tab:pwnVal}, with $n_{\rm{ism}}=100$ cm$^{-3}$. 
\begin{table}[!t]
\centering
\caption{Most probable values for the pulsar spin-down time and luminosity for different values of $\eta_{\mathrm{X}}$.}
    \centering
    \begin{tabular}{c|c|c}
    \hline\hline
    $\eta_{\mathrm{X}}$& L$_{\rm{0}}$ (erg/s)& $\tau_0$ (yr) \\
    \hline
    $10^{-5}$&   $4.1 \times 10^{40}$  &  18 \\
    $10^{-4}$&   $4.1 \times 10^{39}$  &  110\\
    $10^{-3}$&   $4.1 \times 10^{38}$  &  480 \\
    $10^{-2}$&   $4.1 \times 10^{37}$  &  3160\\
    $10^{-1}$&   $4.1 \times 10^{36}$  &  25000\\
    \hline
    \end{tabular}
      
     All values are given considering $n_{\rm{ism}}=100$ cm$^{-3}$.
  \label{tab:pwnVal}
\end{table}

\subsection{Comparison with results by Alp et al. 2021}

Recently, \citet{alf21} (hereafter A21) reported on the analysis of {\it XMM-Newton}/RGS and {\it NuSTAR} data of SN 1987A. The authors state that the X-ray broadband spectrum from 0.5 to 24 keV can be fitted with three thermal components and that a non-thermal component is not statistically significant. I here discuss and compare their best-fit model with the {\it Chandra}/ACIS-S and {\it NuSTAR}/FPMA,B observations.

The best-fit model reported by A21 is made of three \emph{VPSHOCK} components with temperatures kT$_{\rm{1}} \sim 0.5$ keV, kT$_{\rm{2}} \sim 1$ keV and kT$_{\rm{3}} \sim 4$ keV. The corresponding ionization parameters are $\tau_{\rm{1}} \sim 9 \times 10^{11}$ s/cm$^3$, $\tau_{\rm{2}} \gtrsim 5 \times 10^{12}$ s/cm$^3$ and $\tau_{\rm{3}} \sim 3 \times 10^{11}$ s/cm$^3$. The values of $\tau_{\rm{1}}$ and $\tau_{\rm{2}}$ indicate that the plasma is in collisional ionization equilibrium, at odds with previous findings (e.g. \citealt{zmd09, svc21}). The intermediate component is associated with plasma in CIE and implies a density $> 6000$ cm$^{-3}$ (the lower limit being obtained in the unrealistic assumption that the plasma has been shocked immediately after the explosion). This high density may be reached within the equatorial ring, where, however, the analysis of {\it Chandra}/HETG,LETG, {\it XMM-Newton}/RGS and EPIC data indicates that the plasma is not in CIE and has lower temperatures (e.g., \citealt{zmd09, svc21}). Moreover, the ionization parameter of the 0.9 keV component is higher by a factor of $\sim$5 than that of the 0.5 keV component, thus suggesting that the plasma at 1 keV is denser than that at 0.5 keV\footnote{If both components originate from the equatorial ring, the time elapsed from the shock impact should be almost the same for the two components.}. This is at odds with expectations, since the post shock temperature should decrease by increasing the particle density. This makes the A21 model difficult to understand from a physical point of view.

It is also worth mentioning that A21 do not compare their best fit model with the model adopted in this thesis. A21 only show that a non-thermal component is not necessary when three thermal components are adopted, but they do not verify whether the model with three thermal components fits the data better than that with two thermal components and a power-law. In the following, I compare the model by A21 with the results presented in this Chapter, by jointly fitting the {\it Chandra} and {\it NuSTAR} multi-epoch data.

I reproduced the model by A21 (three \emph{VPSHOCK} components within XSPEC) by fixing the chemical abundances to their best-fit values and letting the $\tau$ and kT of the three components free to vary within the error bars reported in Table E1 by A21. Moreover, I let the normalization of the three components completely free to vary, to account for possible issues of cross calibration between {\it XMM-Newton} and {\it Chandra}. For all years considered in this analysis (2012, 2013, 2014) the model by A21 is systematically worse than the model I adopted in this thesis, despite the higher number of free parameters used by A21. In particular, the model by A21 provides higher values of $\chi^2$ with respect to my model in all the epochs considered, as shown in detail in Table \ref{Alp}.
\begin{table}[!h]
    \centering
    \caption{Comparison between the PWN model and the best-fit model by A21.}
    \begin{tabular}{c|c|c|c}
    \hline\hline
    & 2012 & 2013 & 2014 \\
    \hline
    $\chi^2$(d.o.f.) A21 & 580 (460) & 363 (268) & 591 (497) \\
    $\chi^2$(d.o.f.) PWN (this thesis) & 548 (461) & 337 (269) & 568 (498)\\
    \end{tabular}
    \label{Alp}
\end{table}

\begin{figure}[!ht]
\centering
\includegraphics[width=.47\textwidth,angle=270]{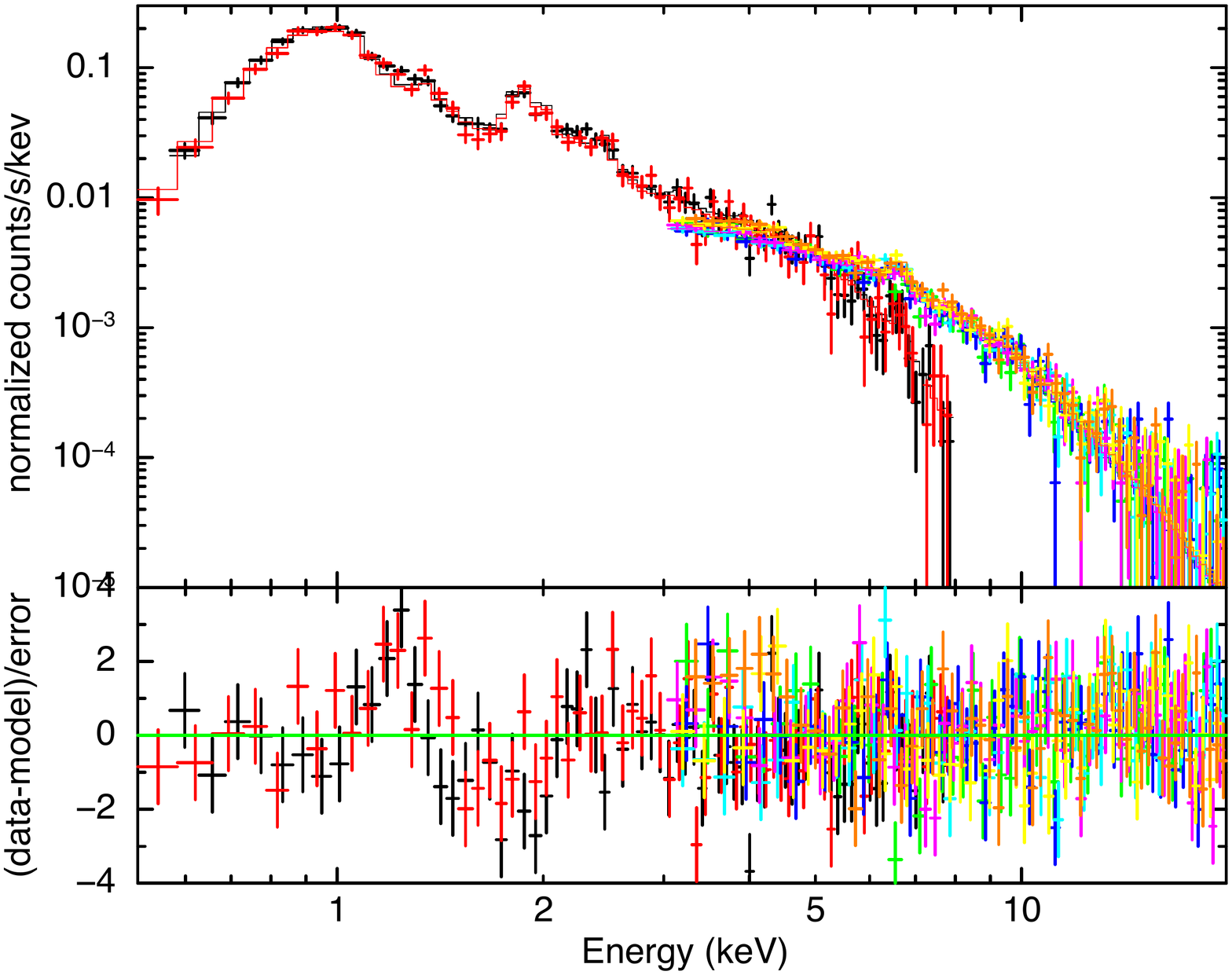}
\includegraphics[width=.47\textwidth,angle=270]{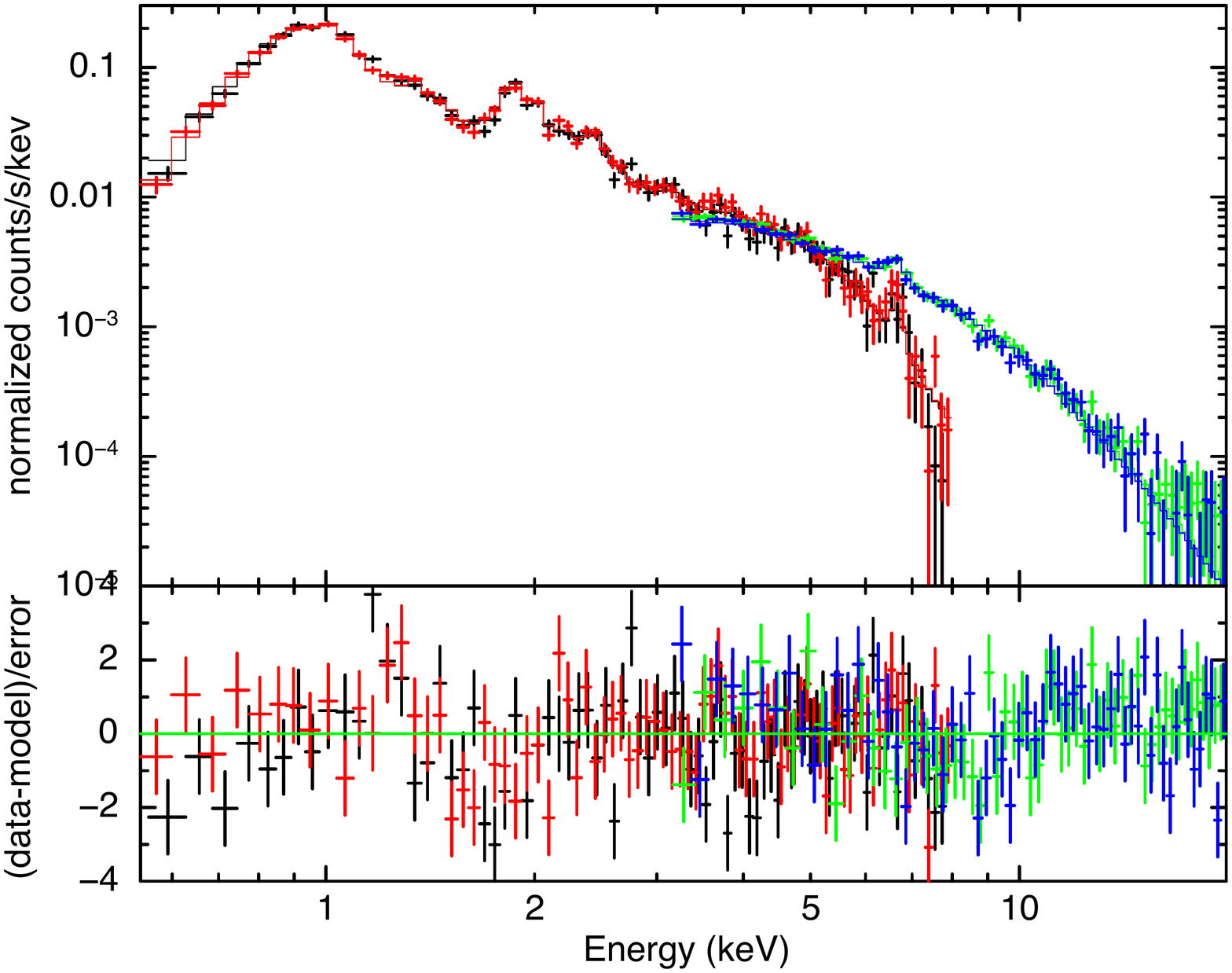}
\includegraphics[width=.47\textwidth,angle=270]{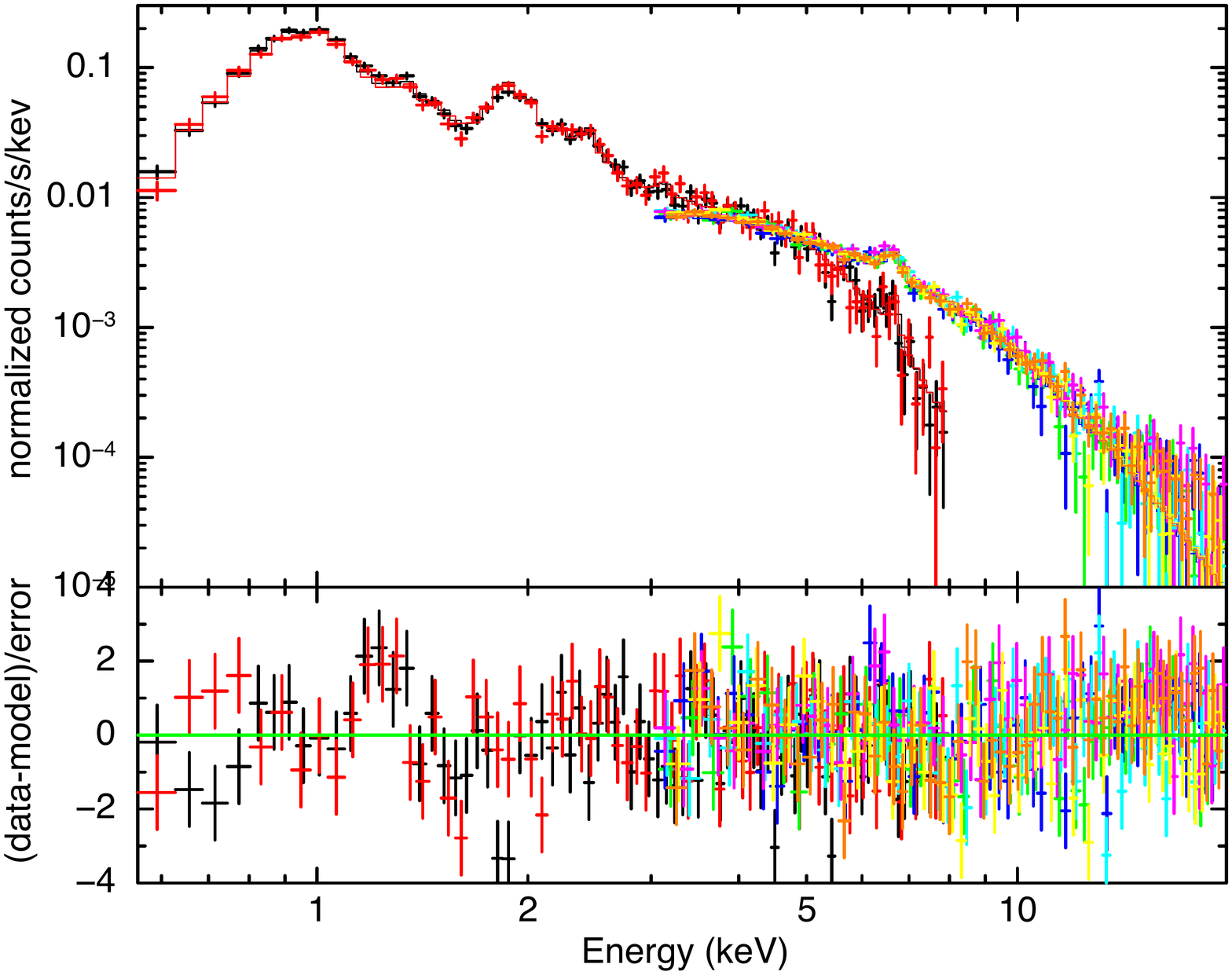}
\caption{{\it Chandra} and {\it NuSTAR} spectra fitted with the three-thermal component model by A21 in the three years: 2012 (top), 2013 (center), 2014 (bottom).}
\label{fig:alp}
\end{figure}

Fig. \ref{fig:alp} shows that the model by A21 leads to large residuals in the 1-1.3 keV energy band for each epoch. I note that one of the two {\it XMM-Newton}/RGS detector does not work in the 0.9-1.2 keV energy bands and A21 did not include {\it XMM-Newton}/EPIC data in the simultaneous fit with the {\it NuSTAR} spectra. Therefore, the low sensitivity of the {\it XMM-Newton}/RGS in the crucial band between 0.9 and 1.2 keV most likely hid these residuals. Conversely, the PWN model correctly fit these data points (see Fig. \ref{fig:chandra+nu_spectra_pwn_single}).

In summary, the model by A21 requires the presence of a plasma component at 10$^7$ K which is supposed to be in equilibrium of ionization and denser than a cold component at $5\times 10^6$ K. Furthermore, it provides a poorer fit to {\it Chandra} and {\it NuSTAR} spectra although it has more free parameters. My model provides a better description of the {\it Chandra} and {\it NuSTAR} spectra in 2012, 2013, and in 2014, with best fit parameters that are physically sound and well suited to describe the physical consitions of the plasma in the shocked ring (low temperature component), HII region (high temperature component) and in the PWN (non-thermal component; Fig. \ref{fig:pwnpop}).
Finally, by analyzing a proprietary {\it NuSTAR} observation performed in 2020, A21 find that the X-ray flux in the 10-24 keV energy band is almost constant between 2012 and 2020 (grey points in Fig. 1 in A21). This is in agreement with what expected in the PWN scenario proposed in this thesis.

\subsection{Synthetic {\it Lynx} spectrum} \label{app:lynx}

I produced a synthetic observation of SN 1987A, as predicted with the MHD model for year 2037. In particular, I synthesized a \emph{Lynx}/X-ray Microcalorimeter (LXM) spectrum. \emph{Lynx} is a NASA large mission concept study for the next generation of X-ray telescopes, proposed to be launced in 2036 (see Appendix \ref{app:telescopes} for details). 

\begin{figure}[!h]
    \centering
    \includegraphics[width=0.7\textwidth,angle=270]{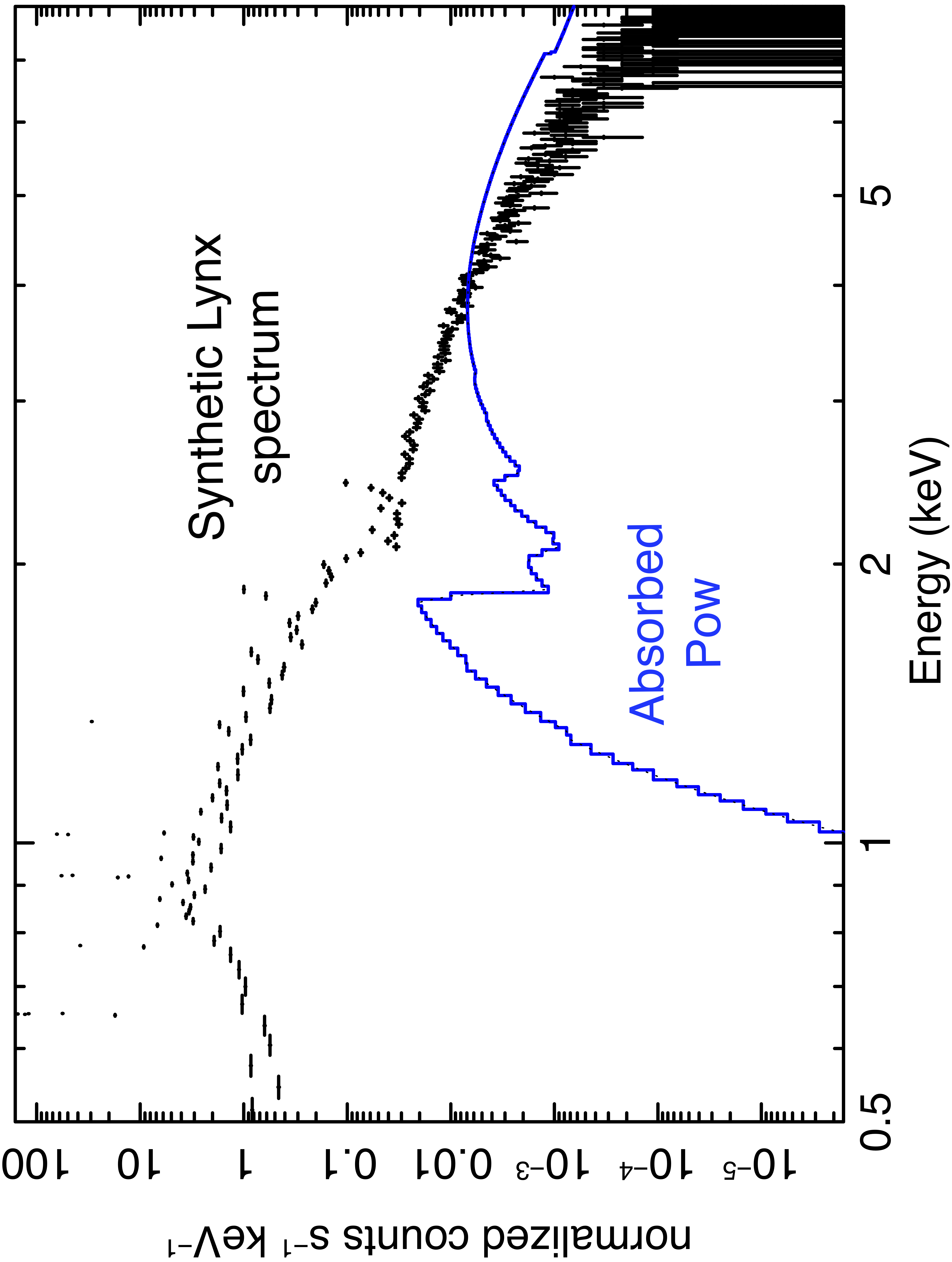}
    \caption{Black points: synthetic {\it Lynx} spectrum of the emission observed in 2018 in the central, faint region of SN 1987A. Blue line: power-law component absorbed by the ejecta as predicted taking into account the MHD model in 2037.}
    \label{fig:lynx_synth}
\end{figure}

The MHD simulations performed by Or20 predict that the thermal emission stays almost constant in the next 15-20 years. Therefore, I synthesized the 2037 spectrum under this assumption, though ejecta contribution and/or interaction of the remnant with other inhomogeneities beyond the ring may affect these predictions. Fig. \ref{fig:lynx_synth} shows the synthetic spectrum extracted from a circular region with radius $R=0.3''$, well within the bright ring of SN 1987A (i.e. the black circular region in the upper right panel of Fig. \ref{fig:image_obs}), assuming an exposure time of 300 ks. 

Under the assumption of a steady X-ray flux for the PWN, I calculated its emission in 2037, by taking into account the new absorption pattern of the cold ejecta\footnote{The ejecta expansion systematically reduces their capability to absorb X-ray photons}, as predicted by the MHD model in 2037. I found that the PWN emission would emerge over thermal emission above 4 keV thus becoming detectable in the soft x-ray band (see Fig. \ref{fig:lynx_synth}.). Further details about the future detectability of the PWN and on the thermal emission of the putative NS will be described in a forthcoming paper (Greco et al., in prep.).

\section{Conclusions}
\label{sect:conc_1987A}

In conclusion, a PWN seems to be the most likely source of the synchrotron radiation in hard X-rays of SN 1987A, even though the DSA scenario cannot be firmly excluded. A more conclusive way to discern between the two scenarios will be provided by future observations. An increase in the hard ($>10$ keV) X-ray flux, similar to that observed in radio, would be easily detectable with {\it NuSTAR}, thus supporting the DSA scenario. On the other hand, the rapid ejecta expansion and rarefaction will reduce the soft X-ray absorption of an inner source. In particular, I estimated from the MHD simulation that the PWN will become detectable by \emph{Chandra}/ACIS-S and/or \emph{Lynx}/LXM in the 2030s, definitely confirming the PWN scenario. With the present data and based on my findings, the PWN scenario seems the most likely (and appealing) to account for the nonthermal X-ray emission that I have detected.

%% file: Conclusions.tex
\chapter{Summary and conclusions}
\label{ch:final}

X-ray spectroscopy allows us to scrutiny high energy processes in astrophysical environments and is crucial to shed light on the physical and chemical conditions of SNRs.

In this thesis I have exploited the diagnostic potential provided by X-ray spectroscopy to tackle various open issues in SNR physics. 
\begin{itemize}
\item A detailed knowledge of abundances and masses of chemical elements in SNRs is needed to correctly compare the measured values with the yields predicted by explosive nucleosynthesis models. However, analyses of CCD X-ray spectra are affected by a degeneracy between the abundances and the plasma emission measure, because of the moderate spectral resolution. The only, partial, solution to this issue is to estimate relative abundances and compare them with the corresponding predictions. This solution, however, hampers the measure of masses and can be misleading in the comparison with nucleosynthesis models.

Metal-rich ejecta, with moderate or negligible mixing with the ISM, are expected to be present in SNRs and to emit in X-rays when shocked by the reverse shock. The set of spectral simulations presented in this thesis shows that the high-abundance regime, called pure-metal ejecta regime, is characterized by a prominent recombination edge in the X-ray spectra at the ionization energy of the given species. I show that such feature is not detectable with current CCD detectors. However, my analysis clearly shows that the spectral signature of pure-metal plasma will be clearly detectable with a higher spectral resolution, like that provided by microcalorimeter spectrometers. In particular, I find that the enhanced RRC of pure ejecta will be revealed in the near future by {\it XRISM} and, later, by the {\it ATHENA} X-ray telescopes. 

I have developed a tool able that can self-consistently synthesize thermal x-ray spectra from a multi-D HD/MHD simulation. The tool coherently reconstructs the chemical abundances of plasma in each computational cell. The tool is flexible and can be applied to different physical scenarios involving SNRs. In particular, I also adopted it to reconstruct the X-ray emission distribution from an HD model of an SNR with a large-scale anisotropy in the ejecta distribution \citep{tom20}.

In this way, I verified the observability of pure-ejecta emission in high resolution spectra of SNRs by considering the specific case of Cas A. Cas A is characterized by a high number of dense clumps of ejecta, which are the ideal regions for the search of pure-metal ejecta plasma. I synthesized spectra from a 3D HD simulation performed by \citet{omp15}, in particular, focusing on the Fe-rich southeastern part of the shell. The simulations predicts the presence of pure-Fe ejecta knots in this region. I synthesized a {\it XRISM} X-ray spectrum of the southeastern part of Cas A. The resulting spectrum shows prominent Fe recombination edges, as expected by the set of spectral simulations discussed above. Thanks to the high spectral resolutions of microcalorimeters, spectral fittings on the synthetic XRISM spectra recover the correct mass and abundance estimates of the considered elements, thus removing the degeneracy between emission measure and abundances. I then conclude that future {\it XRISM} and {\it ATHENA} observations of SNRs will be able to discover pure-metal ejecta emission and to derive the correct masses of the yields of SN explosions.
\item The MM-SNR IC 443 is placed in one of the most complex environments known for a SNR. It interacts with a molecular cloud in north-west and south-east and with an atomic cloud in the north-east. Its internal shells have been intensively studied \citet{tbr06,tbm08,mtu17}, and overionized plasma has been detected. Another peculiarity of IC 443 is the off-center position of the PWN CXOU J061705.3+222127, presumed to belong to the remnant but also possibly being a rambling one seen in projection. In fact, the proper motion of such object does not point towards the geometric center of the SNR. The proposed age of IC 443 spans a wide range, between $\sim 3000$ yr \citet{tbm08} and $\sim 20000$ yr \citet{che99}.

 In this thesis, I presented the analysis of {\it XMM-Newton}/EPIC archived observations of IC 443. I revealed a jet-like structure, distorted by the interaction with a dense molecular cloud, whose projection towards the remnant matches the original position of the progenitor star, estimated by taking into account the proper motion of the PWN. This result is the first evidence of a possible link between the PWN and the remnant, also confirmed by the rough alignment between the PWN's jet and the ejecta collimated structure. The jet's plasma is made of shocked overionized ejecta. The most likely scenario that explains this configuration is the adiabatic expansion subsequent the interaction with the reflected shock, which in turn is generated by the early collision of the primary shock wave with the dense molecular cloud in south-east, close to the explosion site. 

A set of HD simulations performed by \citet{uog20} supports the link between the PWN and IC443. In particular, the model well reproduces the observed distribution of ISM and ejecta by assuming that the original position of the progenitor star corresponds to the position of the PWN 8000 yr ago. To verify the agreement between the HD model and the data under a spectroscopic perspective, I synthesized {\it XMM-Newton}/EPIC-MOS spectra from the HD model and compared them with the the actual data. I performed this analysis on two different regions: one dominated by the ISM emission, in the northern limb; the other dominated by ejecta, in the central area of IC 443. This comparison shows that the spectral main features observed in X-rays are well described by the model.

\item SN 1987A was a core-collapse SN discovered on 1987 February 23, occurred at 51.4 kpc from Earth in the Large Magellanic Cloud. Its dynamical evolution is strictly related to the inhomogeneous medium, made by a dense ring-like structure with a diffuse HII region. Despite the unique consideration granted with deep and continuous observations, and the neutrinos detection, strongly indicating the formation of a NS, the elusive compact object of SN 1987A has not been detected yet. The most likely explanation for this non-detection is ascribable to the absorption due to the dense and cold ejecta, that have not been heated by the reverse shock yet. In this scenario, the photo-electric absorption can easily hide the X-ray emission of the putative compact object. So far, the only constraints available are the upper limits on the luminosity of such an object, either considering thermal or non-thermal radiation.

Here, I reported on the detection of synchrotron radiation between 10 keV and 20 keV in {\it NuSTAR}/FPMA,B spectra of SN 1987A. This radiation may be powered by two physical processes: either DSA or emission from a PWN. Though both processes lead to the same nonthermal emission, the presence of cold ejecta in the internal shell of SN 1987A would significantly absorb part of the radiation emitted by the PWN. Therefore, by using a 3D MHD simulation I reconstructed the absorption pattern encountered by the radiation in the case of the PWN and included it in the analysis of the {\it Chandra}/ACIS-S and {\it NuSTAR}/FPMA,B spectra .

 I found that the PWN scenario is perfectly compatible with the data, though also the DSA case can not be firmly excluded under a merely statistical point of view. In any case, the physical implications in the DSA imply quite unconvincing assumptions on the intensity of the magnetic field and on the energy of accelerated electrons, especially considering the relatively low shock velocity in SN 1987A ($\sim 2000$ km/s). On the other hand, the PWN scenario is physically very sound and points towards a likely detection of the elusive compact remnant of SN 1987A. Future observations with {\it NuSTAR} and, especially, with the next generation of X-ray telescopes in the 2030s, such as {\it Lynx}, will provide stronger constraints on this issue.

\end{itemize}




%% file: Appendix1.tex
\chapter{The X-ray telescopes}
\label{app:telescopes}
In this appendix I review the main characteristics of the X-ray telescopes and detectors used during my PhD. In Sect. \ref{sect:actual_telescopes}, I consider the currently operating cameras:

\begin{itemize}
\item {\it Chandra} Advanced CCD Imaging Spectrometer (ACIS) detector, used to investigate Cas A in Ch. \ref{ch:rrc}  and SN 1987A in Ch. \ref{ch:pwn_87A}.
\item {\it XMM-Newton} European Photon Imaging Camera (EPIC), used to analyze data of IC 443 in Ch. \ref{ch:ic443}.\item {\it NuSTAR} Focal Plane Module A and B (FPMA and FPMB), used to study SN 1987A in Ch. \ref{ch:pwn_87A}. 
\end{itemize}
In Sect. \ref{sect:future_telescopes} the features of future X-ray telescopes, such as {\it XRISM} and {\it Lynx}, are presented.

\section{Currently active observatories} \label{sect:actual_telescopes}

\subsection{{\it Chandra}} \label{sect:chandra}
Details of the {\it Chandra} mission can be found at \url{https://cxc.harvard.edu/proposer/POG/pdf/MPOG.pdf}. Here I recall the main features. 

The {\it Chandra} X-ray Observatory was launched on July 23, 1999. The telescope system is made of four pairs of nested Wolter-I mirrors (forming the  High Resolution Mirror Assembly, HRMA) and has a focal lenght of 10 m. Beside the ACIS camera, described below, {\it Chandra} also benefits of:
\begin{itemize}
\item The High Resolution Camera (HRC) which has the highest spatial resolution, matching the point spread function of the HRMA most closely. 
\item The High/Low Energy Transmission Gratings (HETG/LETG) used for high resolution spectroscopy, achieving resolving power (E/$\Delta E$) up to 1000 in the band between 0.4 and 10 keV (for HETG) and between 0.08 and 0.2 keV (for LETG). 
 \end{itemize}

\begin{figure}[!htb]
\centering
\includegraphics[width=0.7\columnwidth]{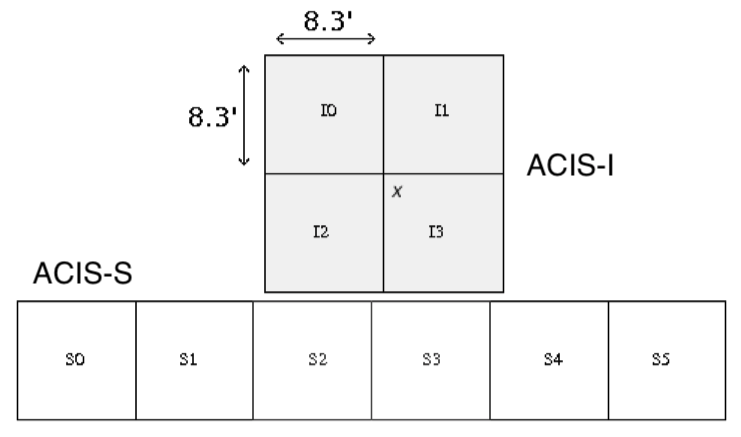}
\caption{Schematic view of the ACIS focal plane, not to scale (from \url{https://cxc.harvard.edu/proposer/POG/pdf/MPOG.pdf}).}
\label{fig:acis_scheme}
\end{figure}

\begin{figure}[!ht]
\centering
\includegraphics[width=0.6\columnwidth]{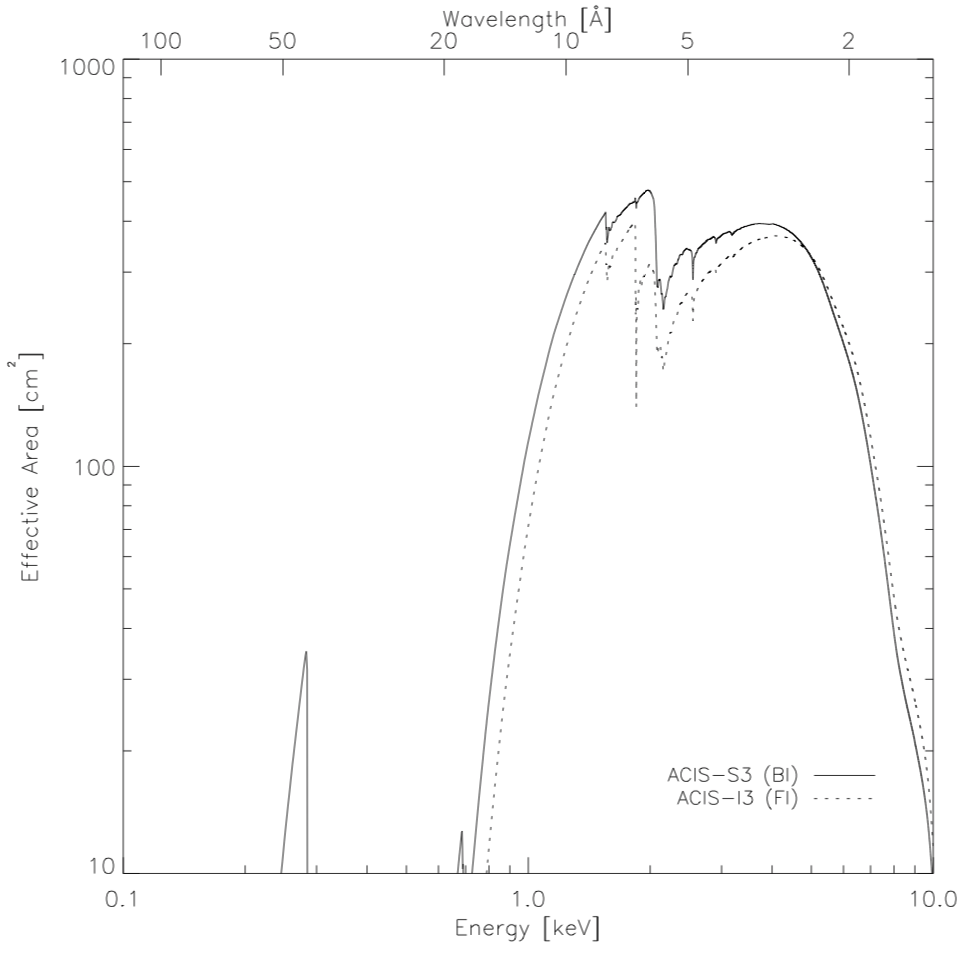}
\caption{HRMA/ACIS effective area versus energy for {\it Chandra} Cycle 23. Dashed line marks the Front-Illuminated CCD I3 and the solid line the Back-Illuminated CCD S3. These curves include the effects of molecular contamination  (from \url{https://cxc.harvard.edu/proposer/POG/pdf/MPOG.pdf}).}
\label{fig:effarea_acis}
\end{figure}

\begin{figure}[!ht]
\centering
\includegraphics[width=0.6\columnwidth]{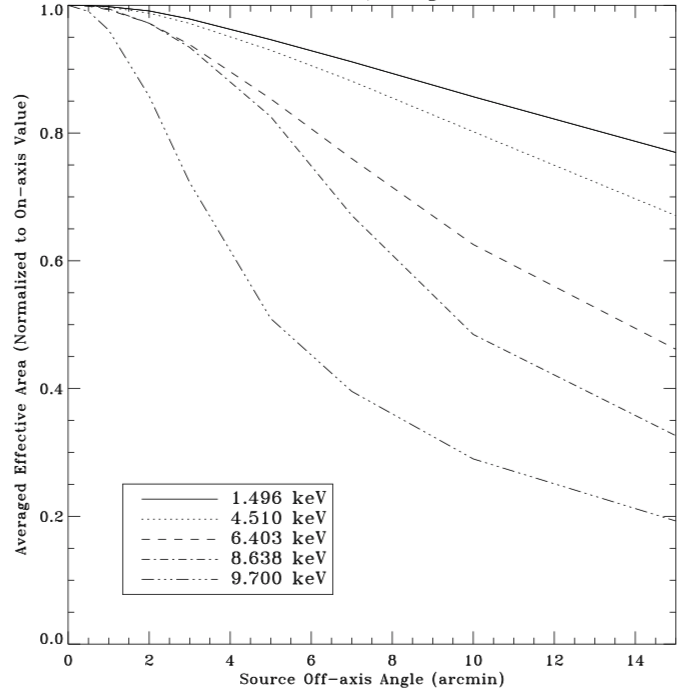}
\caption{Averaged Effective Area, i.e. the vignetting, as a function of the off axis angle at various energies for {\it Chandra}/HRMA  (from \url{https://cxc.harvard.edu/proposer/POG/pdf/MPOG.pdf}).}
\label{fig:vig_chandra}
\end{figure}

ACIS is composed of two CCD arrays, ACIS-I and ACIS-S, a 4-chip and 6-chip array, respectively (see Fig. \ref{fig:acis_scheme}). The former is better suited for observations requiring a wider Field of View (FOV) ($16' \times 16'$), while the latter is preferred for smaller FOVs ($8' \times 8 arcmin$) and for dispersed spectra. ACIS cameras offer high spatial resolution ($0.492''$) and moderate spectral resolution (E/$\Delta E = \sim 20/50$ at 1-6 keV) in a large energy band (0.1-10 keV). Unfortunately, the efficiency of ACIS cameras at energies $< 1$ keV is degraded as a result of molecular contamination on the optical blocking filter. In  Fig \ref{fig:effarea_acis} I show the predicted effective area for {\it Chandra} cycle 23. The variation of the effective area with the source off-axis angle is shown in Fig. \ref{fig:vig_chandra}. Table \ref{tab:chandra_characteristics} summaryzes  the main characteristics of the HRMA and of the ACIS-S camera.

\begin{table}[!hb]
\centering
\caption{Main characteristics of {\it Chandra} HRMA and ACIS}
\begin{tabular}{c|c}
\hline\hline
\multicolumn{2}{c}{HRMA} \\
\hline
Focal lenght & 10 m\\
PSF FWHM& $< 0.5''$ \\
Effective area (0.25 keV)& 800 cm$^{2}$\\
Effective area (5.0 keV)& 400 cm$^{2}$\\
Effective area (8.0 keV)& 100 cm$^{2}$\\
\hline
\multicolumn{2}{c}{ACIS-S} \\
\hline
Pixel size& 24 $\mu$m ($0.492''$)\\
Effective area (0.5 keV) & 110 cm$^{2}$\\
Effective area (1.5 keV) & 600 cm$^{2}$\\
Effective area (8.0 keV) & 40 cm$^{2}$\\
Spectral resolution (1.5 keV)& 140 eV \\
Spectra resolution (5.9 keV)& 270 eV \\
\hline
\end{tabular}
\label{tab:chandra_characteristics}
\end{table}

\begin{figure}[!htb]
\centering
\includegraphics[width=0.7\columnwidth]{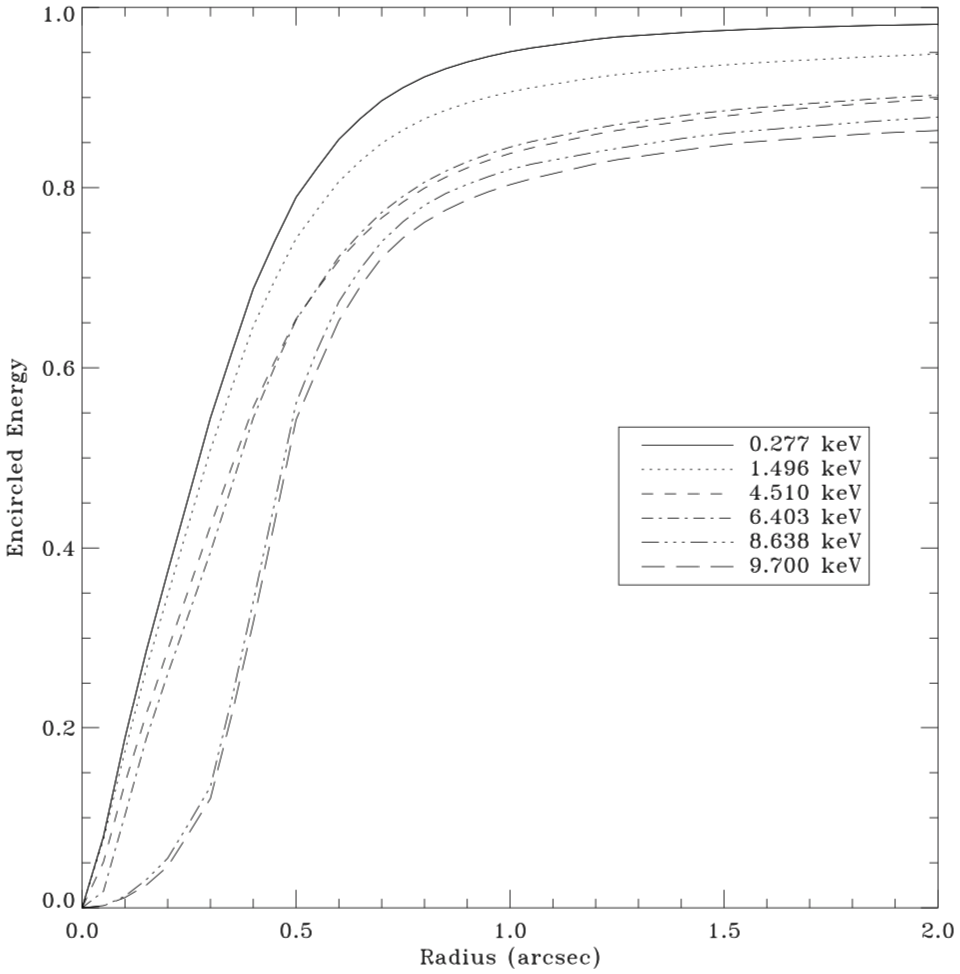}
\caption{Fractional encircled energy vs the radius for an on-axis point source, at different X-ray energies for {\it Chandra}/ACIS  (from \url{https://cxc.harvard.edu/proposer/POG/pdf/MPOG.pdf}).}
\label{fig:psf_chandra}
\end{figure}

It is possible to improve the spatial resolution of the {\it Chandra}/ACIS cameras thanks to the dithering of the spacecraft, thus obtaining subpixel spatial resolution (the pixel size is $0.492''$). Therefore, it is possible to analyze regions which are resolved by {\it Chandra} even if their characteristic dimension is lower than the {\it Chandra} PSF. When analyzing X-ray emission from small regions, it is necessary to correct the observed flux by taking into account the PSF effects, which depend on the dimension of the selected region and on the energy (see Fig. \ref{fig:psf_chandra}).

\subsection{{\it XMM-Newton}} 
All the information shown here and further details about {\it XMM-Newton} can be found on the web site \url{https://www.cosmos.esa.int/web/xmm-newton}, in \citet{taa01} and \citet{sab01}.

{\it XMM-Newton} is a satellite developed by the European Space Agency (ESA) and launched on 1999 December 10. Three different telescopes are present in the observatory, each of them being composed by a system of 58 co-axials and co-focal mirrors with grazing incidence, disposed in configuration Wolter I. Here, I will focus on EPIC, the CCD camera I used to analyse IC 443, but other instruments are present: two \emph{Reflection Grating Spectrometers} (\emph{RGS}), which perform X-ray spectroscopy with a relatively high energy resolution ($E/ \Delta E \approx 100- 500$) in the 0.3 - 2.1 keV energy band; and an \emph{Optical Monitor (OM)}, namely a telescope with a diameter of 30 cm which operates in the $180-600$ nm band allowing image and spectral analysis.

\begin{figure}[!h]
\centering
\includegraphics[width=.7\columnwidth]{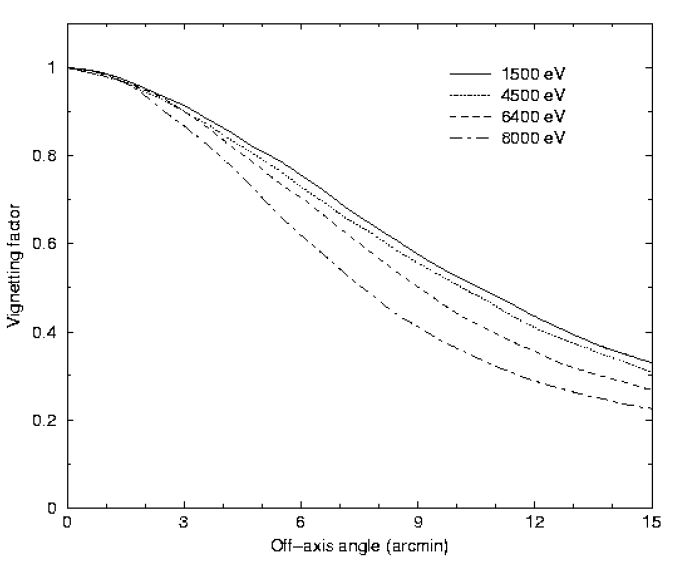}
\caption{Dependence of the vignetting factor for {\it XMM-Newton} mirrors on the angular distance from the optical axis at different energies. (from \url{https://www.cosmos.esa.int/web/xmm-newton})}
\label{Vignetting}
\end{figure}

The focal length of {\it XMM-Newton} is 7.5 m and the effective area, reported in Table \ref{parametristrumenti}, is the highest ever reached for a telescope operating in the 0.1-10 keV energy band (see Fig. \ref{fig:eff_area_comparison}). Fig. \ref{Vignetting}) shows the telescope vignetting function at different energies. 


\begin{figure}[!h]
\begin{minipage}{0.5\columnwidth}
\centering
\includegraphics[width=0.9\columnwidth]{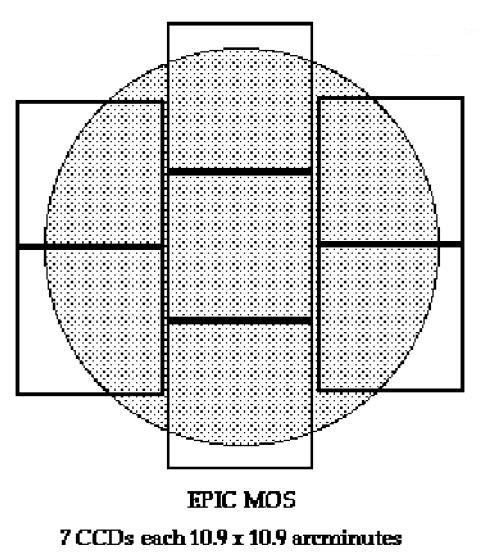}
\end{minipage}
\hfill
\begin{minipage}{0.5\columnwidth}
\includegraphics[width=0.9\columnwidth]{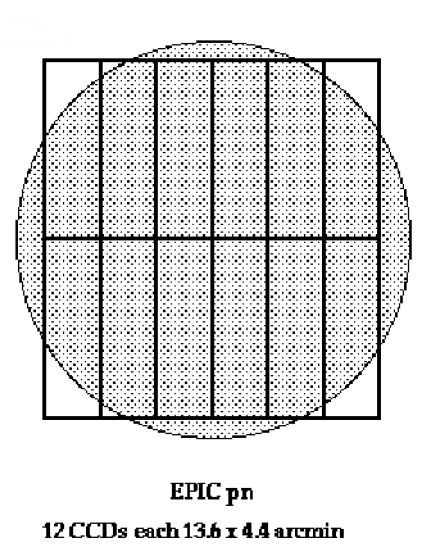}
\end{minipage}
\caption{Disposition of the MOS (left) and pn (right) CCDs onboard {\it XMM-Newton}. The shaded circle indicates the telescope field of view. (from \url{https://www.cosmos.esa.int/web/xmm-newton})}
\label{ccdstructure}
\end{figure}

EPIC consists of three cameras, two MOS and one pn, and of the \emph{Radiation Monitor System (RMS)} which controls the spurious radiation and turns off the instruments when this flux is so high that it might damage the cameras. The two MOS cameras are made of seven MOS-CCDs, while the pn camera is made of twelve pn-CCDs (Fig. \ref{ccdstructure}). In front of each detector a calibration source and a system of four filters for visible and ultraviolet radiation (\emph{thin1, thin2, medium, thick}) are placed. Table \ref{parametristrumenti} shows the main features of the detectors.

\begin{table}[!h]
\caption{Main characteristics of {\it XMM-Newton}'s cameras.}
\centering
\begin{tabular}{c|c|c|c|c}
\hline
& pn& MOS& RGS& OM\\
\hline
Passing band& 0.15-15 keV& 0.15-12 keV& 0.3-2 keV$^1$& 160-600 nm\\
\hline
Effective area$^2$ (cm$^2$)& $\sim$ 1400 &   $\sim$ 450 &  $\sim$ 55& -\\
\hline
Field of View$^{3}$ & $30'$& $30'$& $5'$& $17'$\\
\hline
PSF$^4$ (arcsec)& 6.6/15& 5/14& - & $\sim$ 60\\
\hline
Pixels size& 150 $\mu$m (4.1'')& 40 $\mu$m (1.1'')& 81 $\mu$m$^5$& 1''\\
\hline
Spectral resolution$^5$ & 80 (eV)& 70 (eV)& 0.04 A& 0.5/1.0 nm\\ 
\hline
\end{tabular}

 $^{1}$ First order of dispersion. $^{2}$ The value takes into account the effective area (at 1 keV and without filters) of the optics and detector. $^{3}$Diameter. $^{4}$ FWHM/HEW. $^{5}$ At 1 keV.
\label{parametristrumenti}
\end{table}

\begin{figure}[!h]
\centering
\includegraphics[scale=0.6]{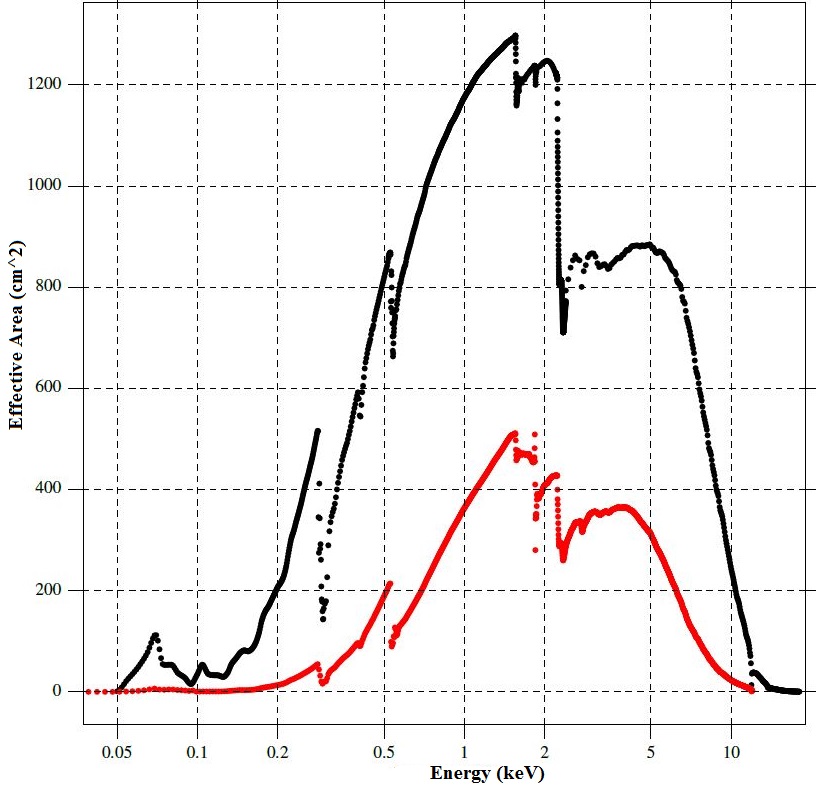}
\caption{Black line: pn on axis effective area. Red line: MOS1 on axis effective area. For both curves the filter used is \emph{medium}. (from \url{https://www.cosmos.esa.int/web/xmm-newton})}
\label{areaefficace}
\end{figure}

The {\it XMM-Newton}/EPIC field of view is circular with a diameter of $\sim 30'$. The two MOS detectors are rotated by 90 deg in order to cover gaps between the CCDs (with the exception of the gaps surrounding the central one). A significant portion of CCDs is out of the field of view (indicated by the shaded circle in Fig. \ref{ccdstructure}). Subsequently, in this area the only contribution detectable is the non photonic one: it is possible to estimate the non photonic contribution simply looking at data revealed by pixels out of the FOV. Fig \ref{areaefficace} shows on axis effective area as a function of energy for EPIC MOS and EPIC pn. MOS Effective Area is lower than that of pn because of two factors: i) the presence of an RGS, in each optical path from the mirrors to the focal plane, that deviates  $\sim 40 \%$ of the radiation towards its own detector and not reaching MOS-CCDs; ii) the lower quantum efficiency of MOS CCDs with respect to pn ones. 

\subsection{{\it NuSTAR}}
All the information shown here and additional details can be found at \url{https://heasarc.gsfc.nasa.gov/docs/nustar/nustar\_obsguide.pdf}.

\begin{figure}[!ht]
\centering
\includegraphics[width=0.9\columnwidth]{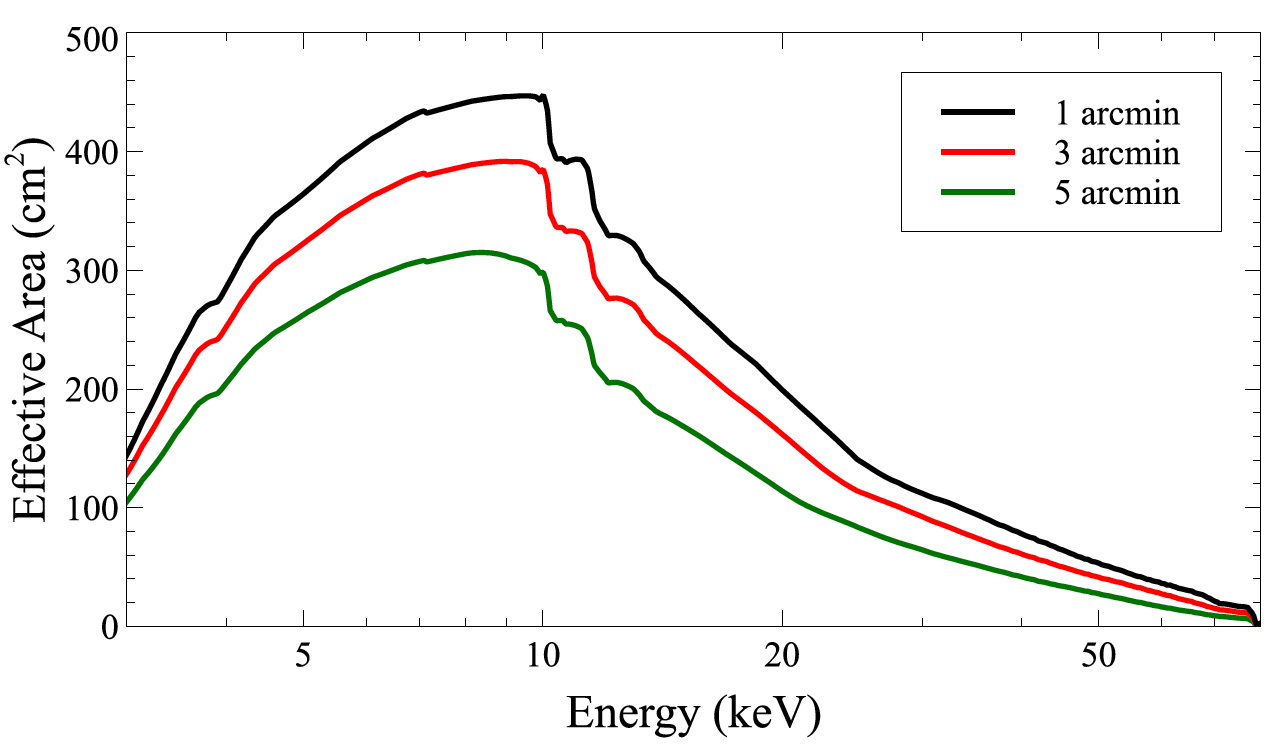}
\caption{Dependence of effective area (of one {\it NuSTAR} telescope) on energy and off-axis angle (from \url{https://heasarc.gsfc.nasa.gov/docs/nustar/NuSTAR\_observatory\_guide-v1.0.pdf}).}
\label{fig:eff_area_nustar}
\end{figure}

\begin{figure}[!ht]
\centering
\includegraphics[width=0.8\columnwidth]{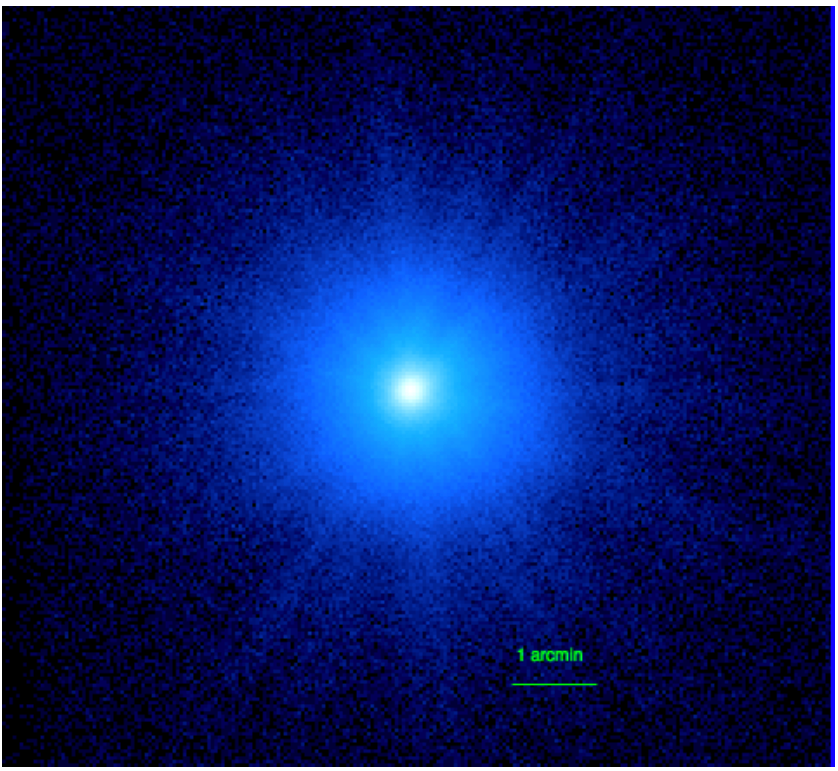}
\caption{Image of the \emph{NuSTAR} PSF for FPMA (from \url{https://heasarc.gsfc.nasa.gov/docs/nustar/NuSTAR\_observatory\_guide-v1.0.pdf}) in logarithmic scale.}
\label{fig:psf_nustar}
\end{figure}

The \emph{NuSTAR} observatory was launched on 2012 June 13, and is the first mission focused on high energy X-rays (3-79 keV, see Fig. \ref{fig:eff_area_nustar}). It is made of two co-aligned hard X-ray telescopes in Wolter-I configuration, each with 133 shells. In the fully deployed, configuration the X-ray optics and detector benches are separated by a stiff mast that permits to achieve a 10-meter focal length. The focal plane bench consists of two independent solid state photon counting detector modules (FPMA \& B), each composed by a 2x2 array of CdZnTe (CZT) crystal detectors, that operates at a temperature of 15$^{\circ}$ C. Main features of the optics and cameras are shown in Table \ref{tab:nustar} and an image of the PSF for the detector FPMA is shown in Fig. \ref{fig:psf_nustar}. 

\begin{table}[!h]
\caption{{\it NuSTAR} telescope and detectors characteristics}
\centering
\begin{tabular}{c|c}
\hline\hline
Field of view& 12.2$\times$12.2 arcmin\\
Angular resolution$^{*}$& 58'' HPD (18'' FWHM)\\
\hline
Detectors size& 2$\times$2 cm \\
Pixel number& 32$\times$32 \\
Pixel pitch& 605 $\mu$m \\
Spectral resolution& 400/900 ev at 10/68 keV \\
Sensitivity in 6-10 keV& 2$\times$10$^{-15}$ erg \, cm$^{-2}$ \, $s^{-1}$  \\
Sensitivity in 10-30 keV& 1$\times$10$^{-14}$ erg \, cm$^{-2}$ \, $s^{-1}$  \\
\hline
\end{tabular}
\label{tab:nustar}
\end{table}

\begin{figure}[!ht]
\centering
\includegraphics[width=0.9\columnwidth]{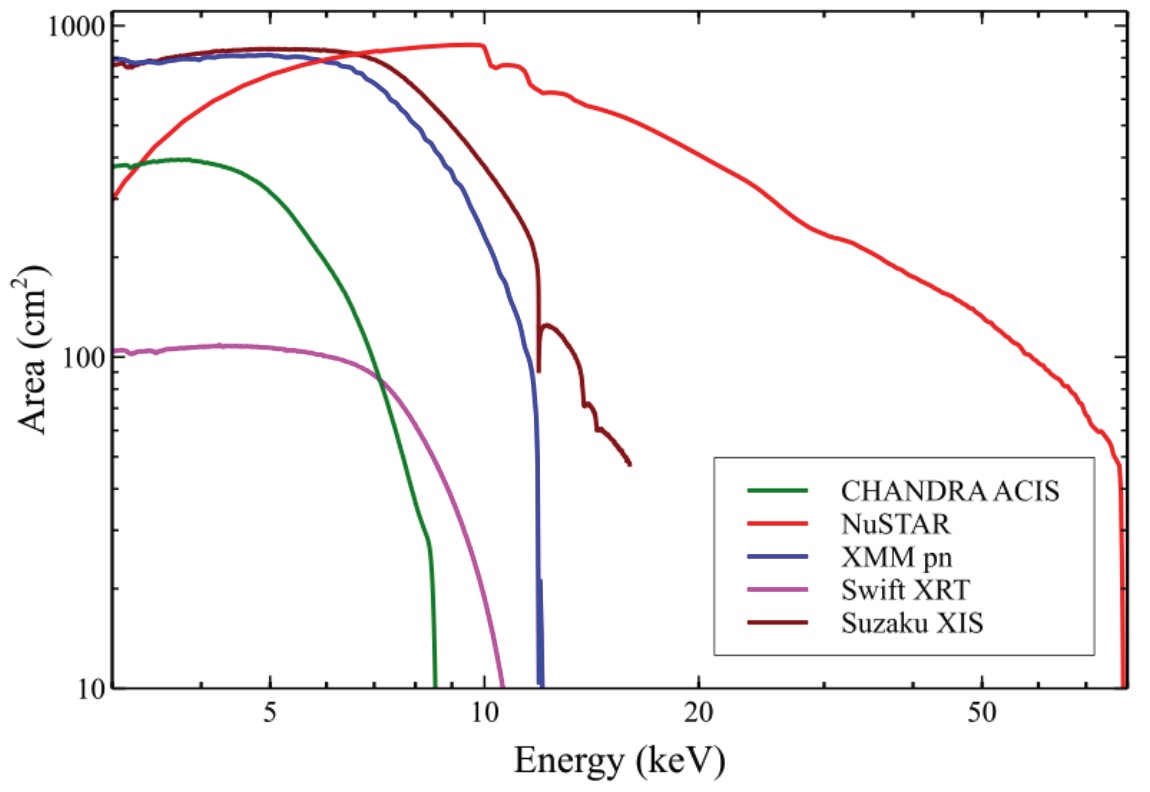}
\caption{On-axis effective collecting area of various X-ray telescopes in the 3-80 keV band  (from \url{https://heasarc.gsfc.nasa.gov/docs/nustar/NuSTAR\_observatory\_guide-v1.0.pdf)}}
\label{fig:eff_area_comparison}
\end{figure}

Fig. \ref{fig:eff_area_comparison} shows the effective area of different X-ray telescopes between 3 and 80 keV. It can be seen that pn camera, onboard the {\it XMM-Newton} telescope, has particularly high effective area below 10 keV, {\it NuSTAR} has sensitivity in a very wide range and {\it Chandra} has a relatively low effective area but, as shown in Sect. \ref{sect:chandra}, provides the best spatial resolution. 

\section{Future missions} \label{sect:future_telescopes}
In this section I shortly review the main features of two future X-ray telescopes, {\it XRISM} and {\it Lynx}, expected to be launched in 2022 and 2030s, respectively. I synthesized spectra of these future telescopes in Ch. \ref{ch:rrc} and \ref{ch:pwn_87A} by using their response and ancillary files available online\footnote{For {\it XRISM}: \url{https://heasarc.gsfc.nasa.gov/docs/xrism/proposals/}}$^{,}$\footnote{For {\it Lynx}: \url{https://hea-www.cfa.harvard.edu/soxs/responses.html}}.

\subsection{{\it XRISM}}
The X-ray Imaging and Spectroscopy Mission ({\it XRISM}) is a JAXA/NASA collaborative mission with the aim to investigate celestial X-ray celestial objects with high-resolution spectroscopy. {\it XRISM} is expected to launch in early 2022. The role of {\it XRISM} is to resume most of the science capability lost with the Hitomi mishap.

The {\it XRISM} payload will consist of two instruments:
\begin{itemize}
\item Resolve, a soft X-ray spectrometer, which combines an X-Ray Mirror Assembly and an X-ray calorimeter spectrometer, providing a 5-7 eV energy resolution in the 0.3-12 keV bandpass with a FOV of about 3 arcmin. Resolve is a 6$\times$6 pixel microcalorimeter array. Each pixel is $30''$ in size, and the entire detector must be cooled to 50 mK in order to detect the heat imparted by each individual X-ray photon that is focused on the detector.
\item Xtend, a soft X-ray imager, which is a CCD detector that extends the FOV of the observatory to 38 arcmin$^2$ over the energy range 0.4-13 keV, using the identical X-Ray Mirror Assembly.
\end{itemize}

Table \ref{tab:xrism_characteristics} shows the requirements and the goals expected to be achieved thanks to the {\it XRISM} X-ray Mirror Assembly and the microcalorimeter Resolve. The white paper describing the capabilities of {\it XRISM} and a sample of the science topics that the mission will look into can be found at \url{https://heasarc.gsfc.nasa.gov/docs/xrism/about/XRISM\_White\_Paper.pdf}. 

\begin{table}[!htb]
\caption{{\it XRISM} telescope and detectors requirement and goals}
\centering
\begin{tabular}{c|c|c}
\hline\hline
Parameter & Requirement & Goal \\
\hline
Field of view& \multicolumn{2}{c}{2.9$\times$2.9 arcmin}\\
Angular resolution$^{*}$& 1.7 (HPD)& 1.2 (HPD)\\
Spectral resolution& 7 eV (FWHM) & 5 eV (FWHM) \\
Effective area (1 keV) & $> 160$ cm$^{2}$ &  250 cm$^2$ \\
Effective area (6 keV) & $> 210$ cm$^{2}$ & $> 312$ cm$^2$ \\
Cryogen-mode Lifetime & 3 years & 4+ years \\
\hline
\end{tabular}

Information taken from \url{https://heasarc.gsfc.nasa.gov/docs/xrism/about/XRISM\_White\_Paper.pdf} and \url{https://heasarc.gsfc.nasa.gov/docs/xrism/about/XRISM\_Factssheet\_04.pdf}.
\label{tab:xrism_characteristics}
\end{table}

\subsection{{\it Lynx}}
The {\it Lynx} X-ray Observatory is currently a NASA Large Mission concept study. If top prioritized, the launch of {\it Lynx} is forecasted to occur in mid 2030s. Its aims are to investigate the dawn of black holes, the drivers of galaxy evolution and the energetic properties of stellar evolution and stellar ecosystems, as reported by the {\it Lynx} Final Report\footnote{Available at \url{https://www.lynxobservatory.com}}.

The {\it Lynx} payload will consist of four main elements:
\begin{itemize}
\item The {\it Lynx} X-ray Mirror Assembly (LMA), the central element of the observatory which will be based on a new technology called Silicon Metashell Optics, allowing significant improvements in sensitivity, spectroscopic potential and imaging with respect to {\it Chandra}. It will have a $0.5''$ PSF and a 2m$^2$ effective area at 1 keV.
\item The High Definition X-ray Imager (HDXI), an active pixel array of pixels covering a wide (22$\times$ x 22 arcmin) FOV with a pixel size of $0.3''$, providing sub-arcsecond imaging resolution and moderate spectral resolution. 
\item The {\it Lynx} X-ray Microcalorimeter (LXM), an imaging spectrometer with a resolving power $R \sim 2000$  and a non-dispersive spatial resolution ~ $0.5''$. It will be composed by an array of $1''$ pixels covering a $5\times5$' FOV and providing spectral resolution of 3 eV. There will be two additional arrays optimized for even finer imaging ($0.5''$ pixels) and better spectral resolution (0.3 eV in the soft band).
\item The X-ray Grating Spectrometer (XGS), which will provide an even higher resolving power R ~ 5000 to 70000 in the soft X-ray band for point sources, significantly improving the current level reached by gratings on {\it Chandra}.
\end{itemize}

%% file: Acronym.tex
\chapter{Acronym Index}
\noindent BH = Black Hole

\noindent BP = Particle Background

\noindent BS = Blanck Sky

\noindent CCD = Charged-Coupled Device

\noindent CIE = Collisional Ionization Equilibrium

\noindent CMB = Cosmic Microwave Background

\noindent CSM = CircumStellar Medium

\noindent d.o.f. = degrees of freedom

\noindent DSA = Diffusive Shock Acceleration

\noindent EM = Emission Measure

\noindent EPIC = European Photon Imaging Camera

\noindent EW = Equivalent Width

\noindent ISM = InterStellar Medium

\noindent FB = Free-Bound

\noindent FF = Free-Free

\noindent FOV = Field Of View

\noindent FWHM = Full Width Half Maximum

\noindent HPD = Half Power Diameter

\noindent HD = HydroDynamical

\noindent LXM = Lynx X-ray Microcalorimeter

\noindent MHD = Magneto-HydroDynamical

\noindent MM = Mixed Morphology

\noindent NEI = Non Equilibrium Ionization

\noindent NS = Neutron Star

\noindent PWN = Pulsar Wind Nebula

\noindent RRC = Radiative Recombination Continua

\noindent SN = SuperNova

\noindent SNR = SuperNova Remnant

\noindent ZAMS = Zero Age Main Sequence


%

%